\documentclass{aa}
\usepackage{txfonts}
\usepackage{graphicx}
\usepackage{longtable, lscape, subfigure}

\begin{document}

\title{A systematic search for novae in M\,31 on a large set of digitized 
archival Schmidt plates \thanks{Tables \ref{tab:lightcurves1} - \ref{tab:lightcurves10} and Figures \ref{fig:chart1} - \ref{fig:chart84} are only available in electronic form via http://www.edpsciences.org}}

\author{M. Henze\inst{1}
	\and H. Meusinger\inst{1}
	\and W. Pietsch\inst{2}}

\institute{Th\"uringer Landessternwarte Tautenburg, D-07778 Tautenburg, 
         Germany \\
	email: henze@tls-tautenburg.de
	\and Max-Planck-Institut f\"ur extraterrestrische Physik, 
	D-85748 Garching, Germany}

\date{Received 17 July 2007 / Accepted 14 September 2007}

\abstract
{}
{This paper reports on the detection of optical novae in our neighbour 
galaxy M\,31 based on digitized historical Tautenburg Schmidt plates. The accurate
positions of the detected novae lead to a much larger database when searching 
for recurrent novae in M\,31.}
{We conducted a systematic search for novae on 306 digitized 
Tautenburg Schmidt plates covering a time span of 36 years from 1960 to 1996. 
From the database of both $\sim 3 \times 10^5$ light curves and 
$\gtrsim 10^6$ detections on only one plate per colour band, 
nova candidates were efficiently selected by automated algorithms and 
subsequently individually inspected by eye.}
{We report the detection of 84 nova candidates. We found 55 
nova candidates from the automated analysis of the light curves. Among these, 22 were previously unknown, 
12 were known but not identified on Tautenburg Schmidt plates 
before, and 21 novae had been discovered previously on Tautenburg plates. An additional 
29 known novae could be confirmed by the detailed investigation of single detections. 
One of our newly discovered nova candidates shows a high 
position coincidence with a nova detected about 30 years earlier. Therefore, this object is 
likely to be a recurrent nova. Furthermore, we re-investigated all 41 nova 
candidates previously found on Tautenburg plates and confirm all but two. Positions 
are given for all nova candidates with a typical accuracy of $\sim 0\,\farcs4$. We present 
light curves and finding charts as online material. 
}
{The analysis of the plates has shown the wealth of information
still buried in old plate archives. Extrapolating from this survey,
digitization of other historical M\,31 plate 
archives (e.g. from the Mount Wilson or Asiago observatories) for a systematic nova search looks very
promising.}


\titlerunning{Search for novae in M31 on digitized Schmidt plates}

\maketitle

%
%
\section{Introduction}
%
Classical novae (CN) are thermonuclear explosions on the surface of white dwarfs (WD) 
in cataclysmic binary systems. These explosions are caused by the transfer of matter from the
companion to the WD. The transferred hydrogen-rich matter is accumulated on 
the surface of the WD until hydrogen ignition starts as a thermonuclear 
runaway process in the degenerated matter of the envelope. The resulting
expansion of the hot envelope causes the luminosity of the star to rise 
by more than 9 magnitudes within a few days 
(see Hernanz \cite{Hernanz05}; Warner \cite{Warner95}; and 
references therein).

A comprehensive compilation of novae in nearby galaxies is desirable for
various issues such as nova physics (e.g., Pietsch et al. \cite{Pietsch05})
or the distributions of stellar populations (e.g., Ciardullo et al. 
\cite{Ciardullo87}; Hatano et al. \cite{Hatano97}; Yungelson et al.
\cite{Yungelson97}). In our galaxy, 
the investigation of the nova population is hampered by 
the large area (namely the whole sky) to be scrutinised and by 
our unfavourable position close to the Galactic Plane. For nearby 
extragalactic systems, the situation is comparatively more promising. 
In particular, our huge neighbour galaxy M\,31 is very suitable for 
different kinds of statistical studies of novae because of its proximity and the high number of optical nova outbursts per year.

Pietsch et al. (\cite{Pietsch05}) find that optical novae represent the 
major class of supersoft X-ray sources (SSS) in M\,31. More than 30\% of the
optical novae in the M\,31 centre area show an SSS phase within a year
(Pietsch et al. \cite{Pietsch07}; hereafter PHS07). The X-ray monitoring of optical novae allows the mass of the ejecta, the burned mass, and the mass of the WD in the system to be constrained.

There have been various systematical surveys of novae in M\,31 for more than 
80 years starting with the pioneering work of Hubble (\cite{Hubble29}) and Arp (\cite{Arp56}). 
Among the many other contributors, we refer here only to the work of
Rosino et al. (\cite{Rosino64}, \cite{Rosino73}, 
\cite{Rosino89}), Ciardullo and co-workers (e.g., \cite{Ciardullo87}),
Sharov \& Alksnis (\cite{Sharov91}, \cite{Sharov92}, \cite{Sharov97}),
and Shafter and co-workers (e.g., Shafter \& Irby \cite{Shafter01}). 
More recent efforts include, in particular, the AGAPE project 
(Ansari et al. \cite{Ansari04}),
the POINT-AGAPE project (Darnley et al. \cite{Darnley04}), and the 
WeCAPP project (Fliri et al. \cite{Fliri06}).
The result of this huge effort is a relatively large number of known novae 
in M\,31. Recently, PHS07 collected 719 optical 
novae and nova candidates in M\,31 with outbursts before the end of 2005
based on their own research and a search in the literature. This catalogue 
is regularly updated\footnote{available via the
Internet at \hfill \\ http://www.mpe.mpg.de/~m31novae/opt/m31}. PHS07 point
out that position information (for some also time of outburst) is poor for many of the early nova detections. The positions could be significantly improved by re-analysing of the original plates. Accurate positions - also for historical novae
- are important for identifying novae in different wavelength regimes, and especially for secure identification of recurrent novae.

A successful search for optical novae over the whole body of M\,31 requires 
a large set of observations that have to be deep enough and have to 
cover a sufficiently large 
($\sim\,10$ square degrees) field. For such an aim, the plate archives of 
large Schmidt telescopes provide very useful observational material. 
The present study is concerned with the M\,31~plates in the archive of 
the Tautenburg Schmidt telescope (Sect.\,2). Some plates, in particular the oldest ones, have 
been used already for nova searches by eye 
(Moffat \cite{Moffat67}; B\"orngen \cite{Boerngen68}; Meinunger \cite{Meinunger75}). 
Here we describe the first automised, systematic nova search in the M\,31
field on a large and complete selection of digitized 
Tautenburg Schmidt plates of the M\,31 field.

Preliminary results of the present study have already been briefly summarised (Henze et al. \cite{Henze06}). The present paper includes a more detailed description 
of the observational data (Sect.\,2), the data reduction (Sect.\,3), the calibration 
(Sect.\,4), and the procedures for selecting nova candidates (Sect.\,5).
The results are presented in Sect.\,6 and discussed in Sect.\,7.
Finally, a summary is given in Sect.\,8.  \\

%
%
\section{Observations}
\begin{figure*}
	\resizebox{\hsize}{!}{\includegraphics[angle=270]{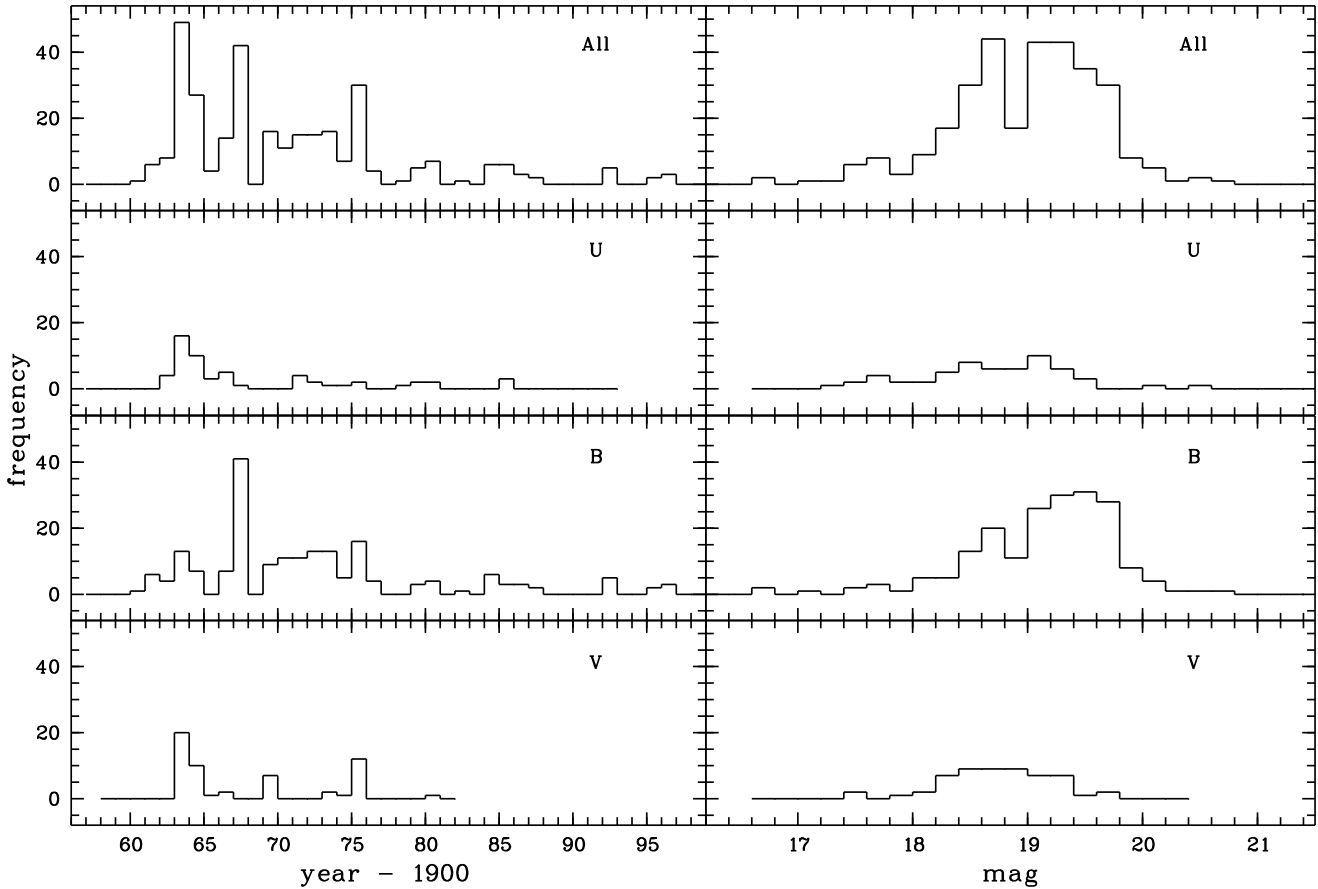}}
	\caption{{\bf Left}: Distribution of epochs of the sample plates for all plates and 
	$U$, $B$, $V$ plates, respectively. {\bf Right}: Distribution of limiting magnitude for all sample plates, 
	as well as for $U$, $B$, and $V$ plates separately. For the definition of 
	limiting magnitude in this work see text and Fig.\,\ref{fig:maghist}.}
	\label{fig:platedistrib}
\end{figure*}

%
This work is based upon photographic plates taken in the years 1960 to 1996
with the Tautenburg Schmidt telescope (free aperture 1.34 m, focal length 4 m). A single Tautenburg plate 
covers an unvignetted field of $3\,\fdg3 \times 3\,\fdg3$ with a plate scale of 
$51\,\farcs4$ per mm. With 554 plates in total, 
the M\,31 field is the most frequently observed field in the Tautenburg Schmidt 
archive. We selected a sample of 289 plates in the $UBV$ bands, plus
17 blue-sensitized plates without filter (NFBS). The main selection criterion 
was the exposure time, which was chosen to be $\geq 10$ min as a basic requirement 
for a sufficiently deep plate limit. 
Figure\,\ref{fig:platedistrib} shows the distribution of epochs of the 
sample plates and Table\,\ref{table:plates} gives the number of plates per colour band and summarises the emulsions and filters used. Parameters for the individual plates are summarised in Tables \ref{tab:lightcurves1} - \ref{tab:lightcurves10}. The Tautenburg colour bands $UBV$ are almost 
identical to the Johnson system (e.g., van den Bergh \cite{vandenBergh64}; 
Andruk et al. \cite{Andruk94}; Brunzendorf \cite{Brunzendorf01}). 
The NFBS system is close to the B system and, in the context of the present paper, 
the two systems are considered to be nearly identical.

%

\begin{figure}
	\resizebox{\hsize}{!}{\includegraphics[angle=270]{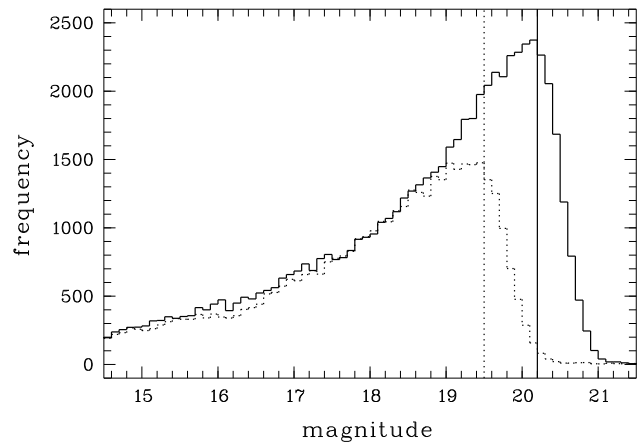}}
	\caption{Distribution of calibrated magnitudes (number per 
	0.1\,mag interval) for all objects detected on 
	photometric reference plates ({\bf solid:} B plate 8171, 
	{\bf dotted:} U plate 1101). 
	Straight vertical lines indicate the limiting magnitudes defined 
	here by the peaks in the distributions 
	($B_{\rm lim} = 20.2,\, U_{\rm lim} = 19.5$).}
	\label{fig:maghist}
\end{figure}

The distributions of the limiting magnitudes of the selected plates are shown in 
Fig.\,\ref{fig:platedistrib}, where the limiting magnitude is defined here as the 
{\it peak} of the magnitude distribution as shown 
in Fig.\,\ref{fig:maghist} (i.e., the faintest objects detected on a 
plate are typically $\sim 1$ mag fainter). 
Note that the limiting magnitude varies significantly across the field due 
to the bright and spatially variable background from M\,31. Especially for objects in 
the bulge, the limiting magnitude is significantly brighter, and inside a 
certain limiting isophote, the probability detecting stellar objects 
approaches zero (see Sect.\,\ref{sec:reduction} for the object detection). 
From the statistics of the detected stars from the external 
calibration catalogue (Sect.\,\ref{sec:calibration}), 
we estimate the semi-major axes of these limiting isophotes to  
$\sim3\arcmin$ in $B$ and $\sim6\arcmin$ in $V$, 
respectively. Based on the surface photometry of M\,31 given by 
Walterbos \& Kennicutt (\cite{Walterbos87}), these limits correspond to
a limiting surface brightness of 19.80 mag arcsec$^{-2}$ ($B$) and 
19.73 mag arcsec$^{-2}$ ($V$). In the $U$ band, the semi-major axis of 
the limiting isophote is as small as $\sim1\arcmin$ where no surface brightness 
is available from Walterbos \& Kennicutt (\cite{Walterbos87}).

Obviously, these plates constitute valuable observational material suited to searching for bright variables 
in our neighbour galaxy. We recall that the spectral properties of the 
plate emulsions, the filters, and the optics have not changed significantly 
over the years; hence, the plate material is remarkably homogeneous over this 
long time base. Note, however, that the majority of these plates were not 
taken as a part of a systematic survey and that the plates thus do not cover 
exactly the same field. \\
%
\begin{table}[ht]
\caption{Numbers of plates per colour band.}
\label{table:plates}
\centering
\begin{tabular}{lllr}\hline\hline
	Colour band &  Emulsions                   & Filter    & Number \\ \hline
	U           & 103a-O, ZU2, IIa-0	   & UG 2      &  57  \\
	B           & 103a-O, ZU2, ZU21, IIa-O, AS & GG 13     &  176 \\
	NFBS        & ZU2, ZU21, AS  	           & no filter &  17  \\
	V           & 103a-D, ZO1, IIa-D	   & GG 11     &  56  \\ \hline
\end{tabular}
\end{table}
%
%
%
\section{Data reduction}
\label{sec:reduction}
%
All plates were digitized with the Tautenburg Plate Scanner (see 
Brunzendorf \& Meusinger \cite{Brunz99} for a brief description). The 
digital images have a pixel size of 10 $\mu$m $\times$ 
10 $\mu$m ($0\,\farcs5 \times 0\,\farcs5$) and a resolution depth 
of 12 bit per pixel. The images were reduced with the source detection software package 
{\it Source Extractor} (SE; Bertin \& Arnouts \cite{Bertin96}).
The SE creates output tables containing internal x-, y-positions, internal 
magnitudes, and additional object parameters. The subsequent data reduction 
and analysis were done using the ESO MIDAS package.

We used the SE in the Photo-Mode (DETECT\_TYPE: PHOTO), which takes the 
nonlinear response of the photographic emulsions into consideration. 
Fluxes were measured with a flexible elliptical aperture (SE mode AUTO).
Special care was taken to consider the strongly fluctuating background 
surface brightness.  
To do so, we configured the SE to model a detailed background image (SE parameters 
BACK\_SIZE = 7, BACK\_FILTERSIZE = 7) using median filters. 
In an area of BACK\_SIZE, the mean and the standard deviation $\sigma$ were 
computed, and all pixel values exceeding $\pm 3\, \sigma$ rejected. 
This procedure was repeated iteratively until all remaining pixel values 
lay within $\pm 3\, \sigma$ of the current mean value. A background map 
was built from all areas of BACK\_SIZE using the current mean of the area 
as a value for the background. Additionally, a smoothing median filter was 
applied to groups of BACK\_FILTERSIZE areas to model the 
final background image. This image was subtracted from the 
original image and object detection performed on the background-subtracted
image. We decided to use a very low detection threshold 
(SE parameters DETECT\_THRESH = 1.3, DETECT\_MINAREA = 5) in order to 
reach a high completeness still at the plate limit. This typically yields $\sim 50\,000$ detected objects per plate. The big drawback 
to the high completeness is, of course, a strong contamination by spurious 
detections that have to be considered carefully in the 
subsequent data analysis. 

To judge the reliability of the detections, we used the SE output parameters 
FWHM\_IMAGE (computed in units of pixel assuming a Gaussian PSF; stars have FWHM $\sim7\ldots15$), ELLIPTICITY (stars: $0.05\ldots0.1$), FLAGS (image reduction errors), and CLASS\_STAR. The last 
parameter is useful for distinguishing between unresolved objects (``stars''; CLASS\_STAR close to 1) 
and extended objects (mainly background galaxies; CLASS\_STAR close to 0). This classification is based on a Neural Network. Since the reliability of the CLASS\_STAR parameter
breaks down for fainter objects (CLASS\_STAR distributed randomly between $0\ldots1$), we computed the additional 
index NONSTELLAR, which is based solely on the deviation of the measured 
object profile from the mean profile:
\begin{displaymath}
\mbox{NONSTELLAR} = \frac{r - \tilde r}{\sigma_{r}},
\label{eqn:nonstellar}
\end{displaymath}
where $r$ is the radius of the object, $\tilde r$ the median radius 
of all objects with similar (internal) magnitudes, and $\sigma_{r}$ the 
standard deviation of the radii of all these objects with $r\leq \tilde r$. 
Objects with NONSTELLAR $>$ 3 are not considered as stellar.

%
\section{Calibration}
\label{sec:calibration}

The astrometric and photometric calibrations follow a similar 
procedure to previous studies based on digitized 
Tautenburg Schmidt plates 
(e.g., Scholz et al. \cite{Scholz97}; Brunzendorf \& Meusinger \cite{Brunz01}). 
First, deep reference plates are selected for each filter band and are then calibrated by means of an external catalogue. All other plates are calibrated using the large number of objects from the respective reference plate. 

\subsection{Astrometric calibration}
\label{sec:cal_astro}
The astrometric calibration of the reference plate (2588) is done in the following steps:  
(1) selecting an appropriate reference catalogue, (2) matching 
the SE output table from the plate to the reference catalogue, and 
(3) transforming the internal plate coordinates into the reference system 
defined by the catalogue. For the first step, the USNO-B1.0 
catalogue (Monet et al. \cite{Monet03}) was used. The catalogue query 
routine VizieR (Ochsenbein et al. \cite{Ochsenbein00}) at the CDS, Strasbourg, 
yields $\sim55\,000$ objects in the field with a position accuracy of 
$\leq 0\,\farcs1$. For these objects rectangular coordinates were 
computed by VizieR and rescaled to the plate coordinate system.

\begin{figure}
	\resizebox{\hsize}{!}{\includegraphics[angle=270]{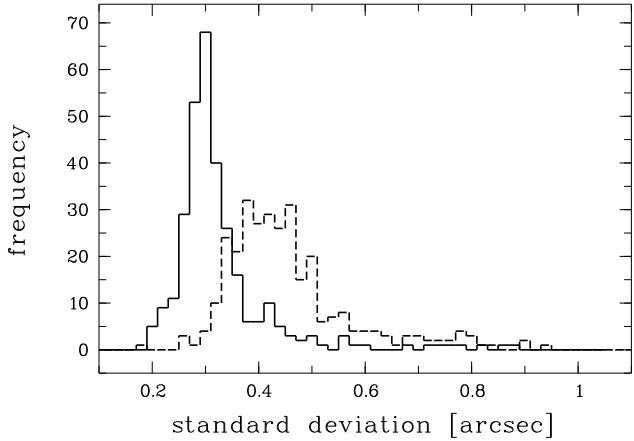}}
	\caption{Distribution of the standard deviations of coordinate differences
	between all astrometric calibrated plates and the USNO-B1.0 catalogue. 
	Solid curve: objects with $m_{\rm cal} = 16\ldots18$ mag; dotted curve: objects with $m_{\rm cal} >18$ mag.}
	\label{fig:astrometry}
\end{figure}

To match the external catalogue to the catalogue of objects detected on the 
plate, we used a routine developed at Tautenburg (Brunzendorf \cite{Brunzendorf01}),
which is based on the MIDAS routine \texttt{find/pair}. 
First the 100 brightest objects in both tables are matched. In case of 
success, the coordinate systems are adjusted. This 
procedure is iteratively repeated with an increasing number of objects 
(up to $\sim$ 20\,000 matches) and with increased precision.
For calibration purposes, we did not use all the matched objects but selected 
a sample of $\sim$ 7\,000 objects with (a) zero proper motion according to the USNO 
catalogue and (b) star-like image profile. With those objects of high 
reliability, we computed transformation relations from pixel coordinates 
to equatorial coordinates using two-dimensional polynomials of third order:\\
\begin{displaymath}
\alpha = \sum_{i,j=1}^3 a_{ij} x^i y^j,     \qquad 
\qquad   \delta = \sum_{i,j=1}^3 b_{ij} x^i y^j  .
\label{eqn:regastro}
\end{displaymath}
The calibration of all other plates using the reference plate essentially 
follows the same method. Figure \ref{fig:astrometry} shows the distribution of astrometric standard errors for all plates. The resulting 
mean position accuracy is estimated to $0\,\farcs3$ for objects with calibrated magnitudes (see Sect. \ref{sec:photcal}) between 16 and 18 and to $0\,\farcs45$ for objects with calibrated magnitudes greater than 18. The overall astrometric accuracy is estimated to $0\,\farcs4$\\
%
\subsection{Photometric calibration}
\label{sec:photcal}
For the photometric calibration, separate reference plates were used for 
each of the three filter bands. The external catalogue used here is the M\,31 part of 
the Local Group Survey (LGS; Massey et al. 
\cite{Massey06}). The LGS maps 2.2 square degrees along the major 
axis of M\,31 using a mosaic of CCD images. With $UBV$ measurements for 
20\,000 to 30\,000 stars (uncertainties $\leq 0.01$ mag), the LGS provides an 
exceptionally large calibration sample. Because the plates have already been 
astrometrically calibrated in the previous step, the equatorial 
coordinates could be used for matching the catalogue of objects on the 
reference plate with the LGS.

We find that the relation between the internal magnitudes and the catalogue 
magnitudes is not satisfactorily modelled by a single polynomial over the 
whole magnitude range (calibrated magnitudes $\sim 12 \dots 21$). Therefore, 
a relation is used consisting of separate regressions for three different 
magnitude intervals with smooth transitions between these intervals. In this 
way a good and robust fit is achieved by using combinations of linear 
and quadratic regressions. Increasing the order of the polynomials, on 
the other hand, does not produce better results.
For each magnitude interval, the calibrated magnitudes $m_{\rm cal}$ 
are modelled by
\begin{displaymath}
m_{\rm cal} = \sum_{\rm i,j}^2 c_{\rm i,j} \, m_{\rm int}^{\rm i} \, b^{\rm j},
\label{eqn:regphoto}
\end{displaymath}
where $m_{\rm int}$ is the internal magnitude of the object and $b$ its background intensity as measured by the SE. 
The inclusion of the local background as a parameter takes into 
account that the correct background subtraction is always difficult for a non-linear
detector, especially when the background is bright and has a high spatial
frequency. We find that the photometric accuracy is indeed significantly
improved in this way ($0.1 \ldots 0.2$ mag better).  
Furthermore, we confirm earlier results (e.g., Brunzendorf \cite{Brunzendorf01}) 
that including colour terms does not yield a substantial improvement of the fit over including the background.

\begin{figure}
	\resizebox{\hsize}{!}{\includegraphics[angle=270]{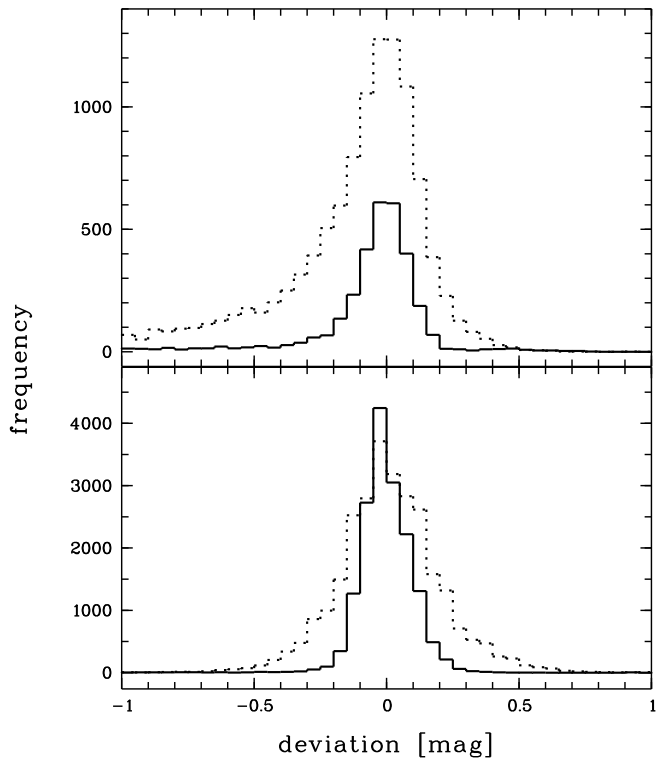}}
	\caption{{\bf Top}: Distribution of the magnitude differences for all 
	objects on the B reference plate 8171 matched with LGS objects. $U$ and $V$ 
	reference plates show similar statistics. Solid line: $m_{\rm cal} < 18$ mag; dotted line: $m_{\rm cal} >18$ mag. {\bf Bottom}. The same for typical magnitude differences for photometric calibrated plates shown with matched objects of plates 8171 and 8089.}
	\label{fig:photometry}
\end{figure}

For the calibration of all the other plates, the field was subdivided into two
different areas that are complementary and separated by an ellipse 
approximating the contours of the outer spiral arms of M\,31. The region 
inside this ellipse contains the area with the strong and spatially varying
background of the disk. The outer region is characterised by a much 
fainter and relatively homogeneous background light distribution. This separate calibration approach yields a further improvement in
the photometric accuracy. The calibration itself is done 
in the same way as described above for the reference plates. Both procedures work semi-interactively, where every single plate 
needs an individual setting of the magnitude intervals and of 
the order of polynomials used for the regression. 
Figure\,\ref{fig:photometry} illustrates the results of the photometric 
calibration. 
%
\section{Data analysis}
\subsection{Creation of light curves}

To create light curves for all detected objects, it is first necessary to 
match the object catalogues from the different plates. Previous searches for variable 
sources on digitized Schmidt plates (Scholz et al. \cite{Scholz97}; 
Brunzendorf \& Meusinger \cite{Brunz01}) started from a basic object sample 
defined as the sample of all objects detected on a minimum number of plates. 
However, such an approach cannot be applied here because the time span
of the observability of extragalactical novae can be as short as a few days. 
Also, the irregular sampling of the observations has to be taken into account. Thus it is possible 
that novae are detected on only two plates of similar epochs. To get light curves of all objects detected on at least two plates, every plate has to be correlated with every other plate.
This was done here for every colour band separately. Of course a nova can also be detected on only one plate of a colour band. We define the detection in this case not as a light curve and discuss these objects in Sect. \ref{sec:analysis_single}.

The correlation procedure works as follows. The field actually 
considered is the field of the astrometric reference plate. For every other plate, only the 
overlap region with the reference plate field was used for correlation. 
The reference field was subdivided into segments with a size of 
$0\,\fdg1 \times 0\,\fdg1$. For each of these sub-areas, the coordinates of all 
objects in the first plate table were matched with the coordinates of all 
other objects in the respective part of the field of all other plates. Since the typical position accuracy for a single plate in comparison to the reference catalogue is $\sim 0\,\farcs4$ (see Sect.\,\ref{sec:cal_astro} and  Fig.\,\ref{fig:astrometry}), the typical position error for the comparison of two plates is $\sigma \sim \sqrt{2}\times 0\,\farcs4$. Hence we used an identification radius of $2\arcsec$ corresponding to $\sim 3.5\,\sigma$. This correlation procedure was repeated 
iteratively for all objects of the remaining plates that could not be 
identified in previous steps. By this means, the highest completeness could be achieved 
since every object on every plate was, in principle, compared with all objects 
on all other plates. The segmenting of the field was done with the large 
number of plates and the huge number of $\sim 45\,000$ detections for every 
plate in mind, in order to speed up the correlation procedure. Matching 
problems at borders of segments could be solved using overlap regions and 
selection criteria avoiding double matching.

For each colour band, the correlation procedure produced tables (a) for 
all objects detected on at least two plates (``multi-detection objects") 
and (b) for all objects detected on only one plate (``single-detection 
objects"). Both tables contain object coordinates and 
magnitudes for every plate the object was detected on. For multi-detection objects, the coordinates on the reference plate were used, 
if existing, or the coordinates on the first plate the object was 
detected on. Knowing the associated Julian Date (JD) for every plate, we were able to create light curves for every object. The analysis 
of the multi-detection light curves is described in 
Sect.\,\ref{sec:analysis_light}. Single-detection objects 
need to be treated differently and are discussed in 
Sect.\,\ref{sec:analysis_single}.
\subsection{Selection of nova candidates from light curves}
\label{sec:analysis_light}
In a first step, the non-stellar objects were removed from the tables by a routine 
based on the geometric parameters (see Sect.\,\ref{sec:reduction}). 
As criteria for a star-like appearance we used: 
CLASS\_STAR $\geq 0.33$, NONSTELLAR $\leq 6$, FWHM\_IMAGE $\leq 18$ (to remove extended objects and very bright foreground stars), ELLIPTICITY $\leq 0.15$ (to remove background galaxies and the most obvious plate defects), and FLAGS $\leq 3$ (to remove serious reduction errors). Every object 
not satisfying at least one of these constraints was removed.
These criteria are quite weak and each of them defines a lower limit.
The reason for using weak constraints is to avoid a strong reduction of the 
sample at faint magnitudes where all these parameters are uncertain. 
Although this preselection only removes the obviously non-star like 
objects, the total number of objects is reduced by up 
to 30\% (see Table \ref{table:multi_selection}). Note that in this step overlapping stellar images are also rejected, which could not be deblended correctly by the Source Extractor.

\begin{table}
\caption{Total number of objects in multi-detection tables.}
\label{table:multi_selection}
\centering
\begin{tabular}{crr}\hline\hline
	Colour band & \# all objects &  \# star like objects \\ \hline
	B & 116\,267 & 80\,424\\
	U & 61\,899  & 45\,573\\
	V & 100\,931 & 71\,970\\ \hline
\end{tabular}
\end{table}

For the remaining objects, we computed light curve parameters 
specifically designed to select nova candidates. These 
parameters are (a) the time lag between the first and the last 
detections of the object (\textit{jddiff}), (b) the number of 
plates (\textit{nop}) the object was detected on, (c) the magnitude 
difference of the limiting magnitude for the particular plate that the light curve peak was detected on to this peak magnitude (\textit{limitdiff}), and (d) the probability that this peak is not due to a random photometric fluctuation 
(\textit{peakdiff}). The last parameter 
is computed as follows: the peak measurement is removed from the light 
curve and the mean magnitude is computed from this modified light curve. 
\textit{Peakdiff} now gives the difference between this modified mean 
and the actual peak value in terms of the standard deviation of the respective 
plate in the respective magnitude interval. The standard deviation 
is computed in the following way: the sample of multi-detection 
objects is binned into 0.5 mag intervals according to 
their magnitude on the reference plate. Then the differences between the object magnitudes on  
this plate and on the reference plate are computed for each calibrated 
plate. The distribution of these 
differences is represented by a Gaussian (see Fig.\,\ref{fig:photometry} for an example)
and the standard deviation can be used to make individual estimates of 
the uncertainties on the photometric calibration. The mean standard 
deviation over all plates and all magnitude intervals is 
$\sim 0.2\ldots 0.3$ mag.
We decided to use low values for \textit{peakdiff} in order to be able 
to discover nova light curves that have not been detected close to the 
maximum. This, of course, results in a large number of objects having to be checked carefully.

\begin{figure}
	\resizebox{\hsize}{!}{\includegraphics[angle=270]{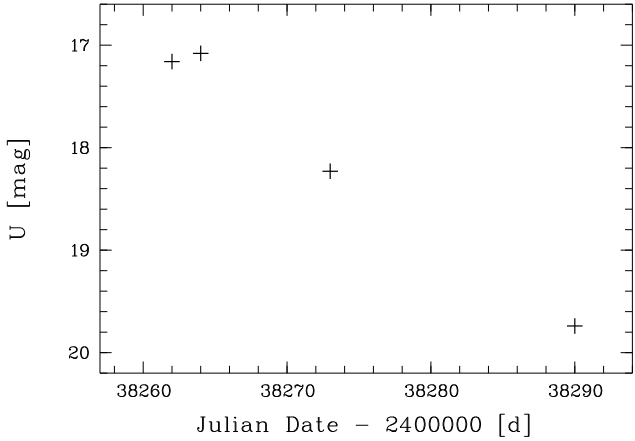}}
	\caption{Light curve of nova candidate 4 = Nova Moffat 2 = M31N~1963-08a.
	The light curve parameters are \textit{jddiff} = 28 d, \textit{limitdiff} = 2.0 mag, 
	and \textit{peakdiff} = 8.1. Detection images of this object are shown in Fig.\,\ref{fig:novalight_ima} below.}
	\label{fig:novalight}
\end{figure}

To interpret these parameters we analysed the data for the novae 
detected by Moffat (\cite{Moffat67}) 
and B\"orngen (\cite{Boerngen68}) on Tautenburg plates. A sample 
of 20 typical nova light curves was constructed (Fig.\,\ref{fig:novalight} for 
illustration), and their parameters were used 
to define the region of the parameter space where novae should be searched. 
Not all of the known novae were included, and the remaining objects were used to 
check the approach. We consider four different cases in detail:
\begin{itemize}
\item[1)] \textbf{nop $>2$}: the analysis of these 
light curves does not require the parameter \textit{limitdiff}, which would 
remove too many faint objects 
\item[2)] \textbf{nop $=2$}: the parameter \textit{peakdiff} is not useful 
here so we use instead \textit{limitdiff} to remove spurious objects.
In fact the actual values of \textit{limitdiff} are quite low and lead to
$\sim 100$ selected objects where nova candidates constitute only a small
percentage. The whole sample (without exclusion of the faintest objects)
yields several hundred objects in which noise detections and plate
faults are the vast majority.
\item[3)] \textbf{recurr}: 
here we search for recurrent novae defined by light curves consisting 
of two parts separated by more than 360 days where at least one part 
satisfies the selection criteria for case 1 (i.e., \textit{nop}$\geq$2).  
\item[4)] \textbf{recurr2}: as case 3 but for \textit{nop}=2.
\end{itemize}

\begin{figure}
	\resizebox{\hsize}{!}{\includegraphics[angle=0]{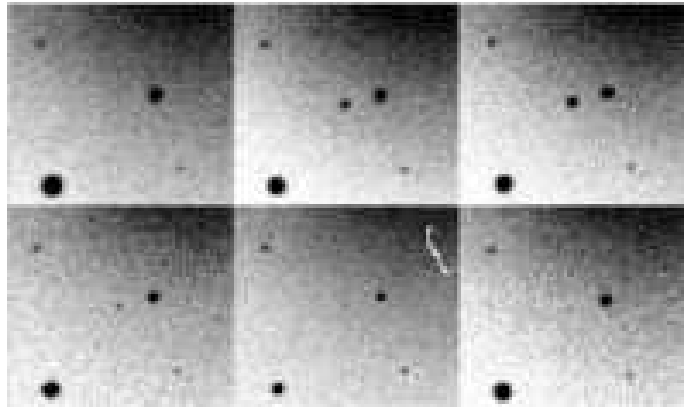}}
	\caption{Images of nova candidate 4 on Tautenburg $U$ plates. Frames are sorted chronologically from left to right and from top to bottom. Frame size is $\sim 2\arcmin \times 2\arcmin$, and north is up and east is left. The object appears in the centre of the four middle frames. The first and the last frames show plates 732 and 1260 that surround the nova detections (see also Tables \ref{tab:lightcurves1} - \ref{tab:lightcurves10}). The resulting light curve is shown in Fig.\,\ref{fig:novalight} above.}
	\label{fig:novalight_ima}
\end{figure}

The selection criteria are similar for the three colour bands, but
the values of the parameters were individually configured for each 
band. The final selection parameters are summarised in 
Table\,\ref{table:novamodel}. The last two columns give the numbers of selected objects and of 
final nova candidates, respectively. Note that the criteria for recurrent novae are applied to light curves consisting of two parts separated by more than 360 days.

Two types of objects were automatically rejected by the selection 
procedure: (1) objects from those parts of the field that are covered 
by only a small number of plates with small epoch differences and 
(2) nova candidates obviously mimicked by high-proper motion
foreground stars (where a star has moved out of the identification radius 
over the long time base of our plate material). All other objects 
satisfying the light curve selection criteria are regarded as ``selected 
objects'', and their numbers are given in Table\,\ref{table:novamodel}.

In a next step, the light curves of all selected objects are checked 
individually. In particular, 
each selected object is correlated with the object lists from 
the other two colour bands. Several objects selected in one band
were removed because the light curve in another band clearly 
indicates that the object is not a nova.  On the other hand, the independent 
selection of the same object in two or even three colour bands is a 
very strong argument for a nova candidate.
Each object from the final sample of selections was inspected by eye 
on the digitized plates, mainly to directly check the stellar appearance 
by comparing its image structure with the  
nearby stars of similar magnitude. Furthermore, we checked via the 
Aladin Sky Atlas at the CDS, Strasbourg, if the objects are visible on more
than one DSS POSS II plate (of course not expected for a nova) and if the objects already have entries in the databases of
SIMBAD and NED. All objects passing all tests successfully constitute our
sample of nova candidates. The corresponding numbers from the 
three bands are listed in Table\,\ref{table:novamodel}.

\begin{table}
\caption{Nova selection criteria for the different colour bands.}
\label{table:novamodel}
\centering
\begin{tabular}{cllrr}\hline\hline
	Colour band & Case & Parameters & \#sel & \#cand\\ \hline
	U & nop $> 2$ & jddiff $\leq$ 50, peakdiff $\geq$ 1 & 28 & 11\\
	  & nop = 2 & jddiff $\leq$ 25, limitdiff $\geq$ 0 & 139 & 10\\
	  & recurr & jddiff $\leq$ 50, peakdiff $\geq$ 1 & 43 & 0\\
          & recurr2 & jddiff $\leq$ 25, limitdiff $\geq$ 0 & 139 & 1\\
	B & nop $> 2$ & jddiff $\leq$ 70, peakdiff $\geq$ 0.5 & 119 & 20\\
	  & nop = 2 & jddiff $\leq$ 5, limitdiff $\geq$ 0 & 107 & 15\\ 
	  & recurr & jddiff $\leq$ 70, peakdiff $\geq$ 0.5 & 227 & 0\\
          & recurr2 & jddiff $\leq$ 5, limitdiff $\geq$ 0 & 77 & 0\\
	V & nop $> 2$ & jddiff $\leq$ 50, peakdiff $\geq$ 0.5 & 70 & 2\\ 
	  & nop = 2 & jddiff $\leq$ 20, limitdiff $\geq$ 0.5 & 69 & 5\\
	  & recurr & jddiff $\leq$ 50, peakdiff $\geq$ 0.5 & 445 & 0\\
          & recurr2 & jddiff $\leq$ 20, limitdiff $\geq$ 0.5 & 122 & 0\\ \hline
\end{tabular}
\end{table}

\subsection{Selection of nova candidates among the single detections}
\label{sec:analysis_single}
The number of single-detection objects is much greater than the number of 
multi-detection objects (see Tables\,\ref{table:multi_selection} and 
\ref{table:singletab}). Reasons for this are (a) a large number of noise 
detections due to the low detection thresholds, 
(b) various ``plate defects'',
and (c) the fact that several of the older plates had been broken in 
the past. These plates have been repaired, which results, however, in a 
significant loss of astrometric accuracy and consequently in matching problems 
and wrong identifications in the context of creating the light curves.

To select objects of high reliability, the measurements on heavily
damaged plates were completely removed from the analysis of the single 
detections. For the remaining plates we considered only objects brighter 
than 1 mag below the plate limit and applied the following selection criteria 
based on the image structure: CLASS\_STAR $\geq$ 0.8, 
ELLIPTICITY $\leq$ 0.1, FLAGS $\leq$ 3, FWHM\_IMAGE $\leq$ 15, and 
NONSTELLAR $\leq$ 3. The objects satisfying all these criteria constitute 
the sample of high priority in contrast to the whole sample of single 
detections that have only normal priority. The constraints are much stronger
here than for the multi-detection objects. The high-priority sample therefore 
contains only about 0.1\% of all single detections (Table\,\ref{table:singletab}). 
However, even this high-priority sample is still heavily contaminated by plate defects that 
can perfectly mimic star-like objects. Although there are elaborate 
methods known to distinguish between plate defects and stars
(e.g., Greiner et al. \cite{Greiner90}), such techniques are too expensive 
to be applied to the several hundred objects in our study.
A single detection therefore is not enough for classifying an object as a nova 
candidate. In the context of the present work, the only way to restrict 
the large number of single detections to a sizable candidate sample is 
to correlate the table of single-detection objects with other tables in 
order to get additional information on these objects. Note that the terms multi-detection object and single-detection object were defined in the context of the light curve analysis. Since light curves were constructed for every colour band independently, it is possible that e.g. two single detections from different colour bands show a position coincidence. Therefore, we 
defined identification criteria for nova candidates by referring to the position 
difference ({\it dist}) and the time lag ({\it diff}) of two records in different
tables to be identified as the same object. We matched the high-priority
samples of each colour band with the tables of multi detections or high 
priority single detections in the other colour bands. The identification criterion
in these cases is dist $\leq 2\arcsec$ \& $\mid$diff$\mid$ $\leq$ 50d. Additionally,
we matched the normal priority single detections with the table of the already
known novae to find out which novae we can confirm and to search for
recurrent novae. Here we used (a) dist $\leq 10\arcsec$ \& $\mid$diff$\mid$
$\leq$ 100d for identifying novae to confirm and (b) dist $\leq 10\arcsec$ \& $\mid$diff$\mid$ $>$ 100d
for identifying recurrent nova candidates. Table\,\ref{table:singlecorr}
shows the results of these correlations. The first two columns give the
matched tables and the third column gives the number of nova 
candidates discovered or confirmed using correlations.

Furthermore, we matched the normal
priority single detections in a given colour band with the normal priority single
detections in the other colour band. Because of the high number of objects, we
used stricter identification criteria (dist $\leq 1\arcsec$ \&
$\mid$diff$\mid$ $\leq$ 50d). We did not find any additional nova candidates with these selection criteria.

%
\begin{table}
\caption{Number of all single-detection objects (\#normal priority)
and of objects in the high-priority sample, respectively (last column).}
\label{table:singletab}
\centering
\begin{tabular}{crrr}\hline\hline
	Colour band & Plates & \#normal priority & \#high priority\\ \hline
	U & 57 & 253\,240 & 294\\
	B & 193 &  582\,211 & 573\\
	V & 56 & 255\,058 & 182\\ \hline
\end{tabular}
\end{table}

\begin{table}
\caption{Correlations of single detections.}
\label{table:singlecorr}
\centering
\begin{tabular}{llr}\hline\hline
	First table & Second table & \#cand\\ \hline
	single high priority & multi & 16\\
	single high priority & single high priority & 1\\
	single normal priority & PHS07 & 24\\ \hline
\end{tabular}
\end{table}

%
\section{Results}
\label{sec:results}
%
Using the criteria detailed in the previous section we found in total 55 
nova candidates based solely on data from the Tautenburg plates; details are given 
in Tables \ref{tab:tauno} - \ref{tab:newno}. Table \ref{tab:tauno} contains 
all objects that were previously discovered on Tautenburg Schmidt plates by 
Moffat (\cite{Moffat67}), B\"orngen (\cite{Boerngen68}), and Meinunger 
(\cite{Meinunger75}) and that were confirmed here. Table \ref{tab:oldno} 
contains all selected nova candidates that match already known novae 
from the literature that have not been measured before on Tautenburg plates. 
Especially for the candidates in Table\,\ref{tab:oldno}, the independent detection 
and classification as nova candidates makes them very likely to actually 
be novae. Table \ref{tab:newno} presents all nova candidates that were 
newly detected in this study. For nova candidates in Table\,\ref{tab:newno},
no counterparts were found in the nova catalogue (PHS07) having a position 
difference of $\leq 60\arcsec$ and a time lag of less than 100 days. 

Furthermore, we confirm 29 known novae mostly found by analysing matches of 
the single-detection objects with PHS07 (Table\,\ref{tab:confno}). Some of 
these objects have more than one single detection, which could not be correlated, however, because one or more of these detections were made on broken plates having large astrometric errors. Hence these objects were 
not found independently using the analysis of the light curves, but they do
confirm already known nova candidates. For all of these objects, we present significantly improved positions.
A second confirmation exists for each of these objects, either
a second single detection, a faint but not detected object found 
by eye on one of our sample plates, an object found by eye on a plate 
not included in the original plate sample (e.g. an $R$ plate), or an entry in PHS07. 
To judge the possibility that bright nova candidates appear only as one single 
detection, we analysed the high-priority sample of the $U$ band single detections 
(see Table \ref{table:singletab}) and checked the digitized images of the objects. 
All but two objects were characterised as plate defects due to their geometry in 
comparison to nearby stars or because the objects cannot be found 
on subsequent neighbour plates with small epoch differences to the detection plate. 
We found the remaining two objects to be nova 
candidates using correlations described in Sect.\,\ref{sec:analysis_single}. 
Therefore, we assume that no bright nova candidate was missed that only appears 
as a single detection. Some nova candidates in Tables \ref{tab:tauno} - \ref{tab:confno} 
(indicated by (*)) show specific features in their identification. We 
discuss them individually in Sect.\,\ref{sec:dis_rem}.

\begin{figure*}
\resizebox{\hsize}{!}{\includegraphics[angle=0]{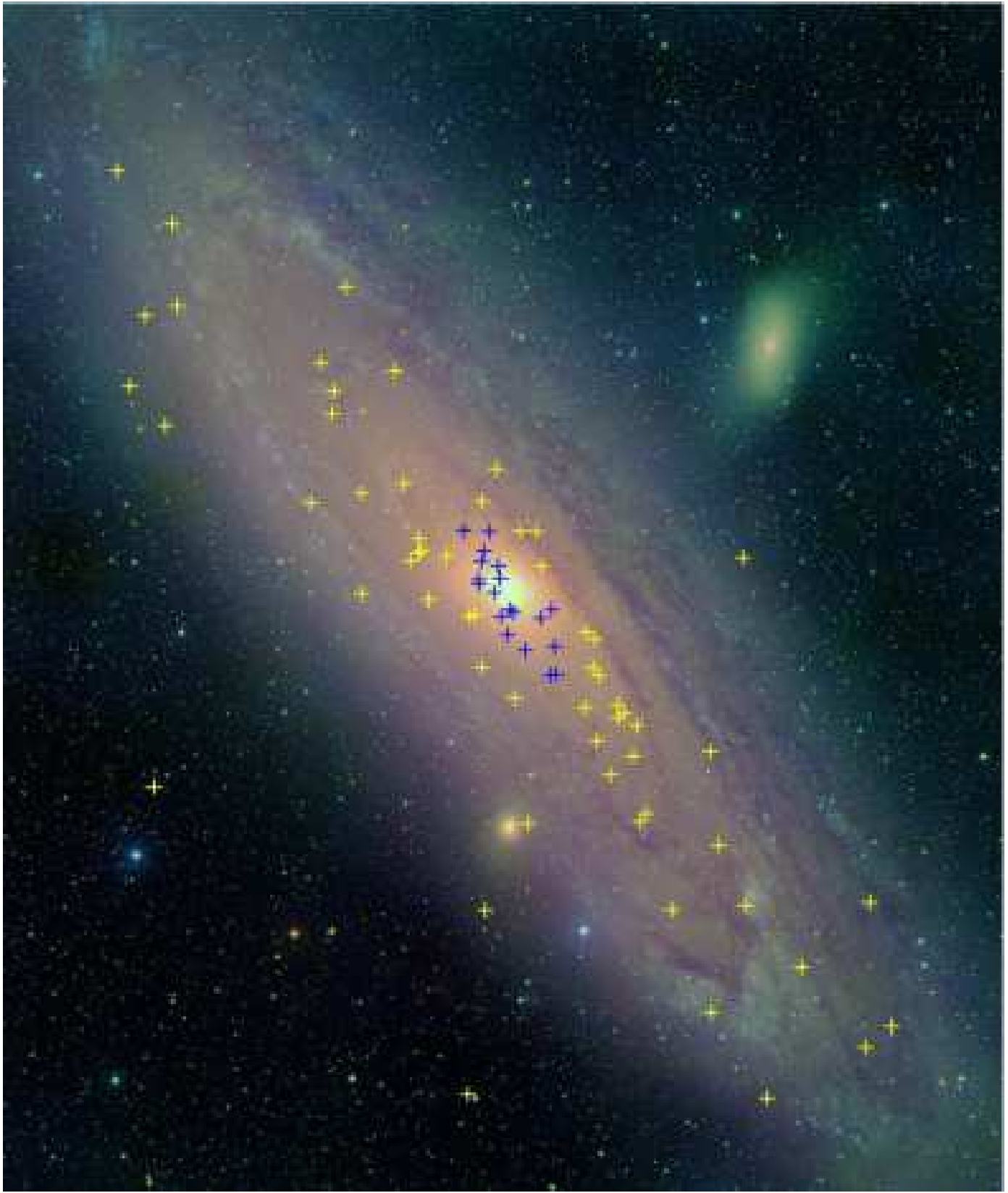}}
\caption{Distribution of nova candidates found in the present study. Five objects lie outside the displayed field shown by a colour composite based on Tautenburg $UBV$ plates. The field size is $1\,\fdg7 \times 2\,\fdg0$. North is up and east is left. Blue crosses mark nova candidates inside the bulge region with a semi-major axis of $\sim700\arcsec$ (e.g. Beaton et al. \protect\cite{Beaton07}). }
\label{fig:foundnovae}
\end{figure*}

We present finding charts for all 84 novae in Figs.\,\ref{fig:chart1} - \ref{fig:chart84}. 
All these charts have a size of $4\arcmin \times 4\arcmin$ and are oriented north up and east left. 
Light curve data points of 82 nova candidates detected on the plates (see Sect.\,\ref{sec:dis_rem} 
for the remaining two) are given in Tables \ref{tab:lightcurves1} - \ref{tab:lightcurves10}. 
Figure \ref{fig:foundnovae} shows the spatial distribution of the novae over the disk of M\,31.

We found no light curve unambiguously showing a recurrent nova in our 
data base. One candidate was selected using the model for recurrent 
novae shown in Table\,\ref{table:novamodel}, but the subsequent visual 
inspection revealed an incorrect identification with a nearby star on a broken plate. However,
one of our newly found novae correlates with a known nova and is therefore
likely to be a recurrent nova. We also discuss this object in Sect.\,\ref{sec:dis_rem}.

The observed maximum brightness of all nova candidates is distributed as follows: 3 novae 
are brighter than 16 mag, 6 novae range between 16 and 17 mag, 29 novae range between 17 
and 18 mag, and 44 novae are fainter than 18 mag. For 7 novae, we estimated $t_{2}$, i.e. 
the time (in days) taken for the luminosity to fall 2 mag below the (observed) maximum. 
We give the speed class for these objects according to Payne-Gaposchkin (\cite{Payne-Gaposchkin57}) 
and discuss these objects in Sect.\,\ref{sec:dis_rem}. Most of the previously known novae were 
detected on Tautenburg plates after the already known outburst. However, note that 9 novae were 
detected $\leq 5$ days before the previously known outburst and 3 novae even earlier 
(up to 84 days before). These 3 objects are also discussed in Sect.\,\ref{sec:dis_rem}.

\section{Discussion of individual nova candidates}
\label{sec:discuss}
%
Before we discuss individual nova candidates, we want to emphasise that we searched 
explicitly for objects that were visible only on a short time span. Consequently 
this means that the seldom novae in globular clusters (like in Quimby et al. \cite{Quimby07}) 
are not recognised by this procedure if those clusters are bright enough to be detected. 
A search for such novae could be made in the context of an analysis of long-term variable 
objects, which we plan to conduct in the future.

In addition, we want to point out that the results presented in the present paper 
contain a few modifications compared to the preliminary results published by Henze et al. 
(\cite{Henze06}) where 19 new historical nova candidates in M\,31 have been announced. 
Meanwhile, we identified one of these objects (nova 32, discussed below) with an 
already known nova and found four additional new nova candidates (novae 35, 36, 42, and 43).

\subsection{Rejected novae}
%
We have re-investigated all nova candidates previously found on 
Tautenburg plates in considerable detail. Most of them were confirmed as nova candidates and are 
listed in Tables\,\ref{tab:tauno} or \ref{tab:confno}. However, for two objects 
we found that they have very likely been
misclassified as novae, whatever they are instead. These objects will be discussed
briefly below.

\subsubsection*{Nova M31N 1966-10a = Nova B\"orngen 21}
This nova candidate is described by B\"orngen (1968) as ``nova like, 
weakly confirmed" and was detected on only three plates. We find 
that there is a very faint object visible on all of these three plates;
however, we noticed position differences of up to $4\,\farcs3$ between 
the three objects and therefore do not consider this object as a nova. 
(There is also no pattern recognisable in this position's differences 
that can be interpreted as caused by a high-proper motion star.)

\subsubsection*{Nova M31N 1967-08c = Nova Meinunger S10753}
This object was classified as a nova by Meinunger (\cite{Meinunger75})
because of a significant ($\geq 1$ mag) outburst that is
confirmed in our data. However, there is a very faint object visible at that position 
on many other plates of different epochs.
Moreover, there is clearly an object visible on POSS II plates. We checked the 
possibility of this object being a nova in a globular cluster (similar to the one 
reported by Quimby et al. \cite{Quimby07}), but there is no entry in the catalogue 
of M\,31 globular clusters (Galetti et al. \cite{Galleti06}) at this position. 
Consequently
we consider this object not to be a nova in M\,31.

\subsection{Remarks on individual nova candidates}
\label{sec:dis_rem}
%
Here we briefly present remarks on individual objects. We discuss a candidate 
recurrent nova and a rejected candidate recurrent nova, give details on two novae 
with large position corrections, and discuss three objects that we detected much 
earlier than the previously known outbursts. Furthermore, we note a nova that 
might belong to M\,32, discuss three novae with lower position accuracy, and 
give remarks on six very fast or very bright novae and on one very slow nova.

\subsubsection*{Nova 4 = M31N 1963-08a = Nova Moffat 2}
Fast nova, $t_2 \sim 20$ days.

\subsubsection*{Nova 5 = M31N 1963-09b = Nova Moffat 1}
Moffat (\cite{Moffat67}) gave wrong coordinates for this object.
In fact, the nova Moffat 1 coincides with nova Rosino 47 
(Rosino \cite{Rosino73}). This is clearly seen on the plates where the nova candidate originally was 
found by Moffat. The date of outburst is 2 days earlier than given by Rosino and is very well constrained (1 day). This nova is very bright (15.4 mag) and very fast ($t_2 \sim 3$ days).

\subsubsection*{Nova 7 = M31N 1963-10b = Nova B\"orngen 8}
Fast nova, $t_2 \sim 13$ days.

\subsubsection*{Nova 17 = M31N 1967-10b = Nova Meinunger S10731}
Very fast nova, $t_2 \sim 8$ days.

\subsubsection*{Nova 18 = M31N 1967-10c = Nova B\"orngen 26}
Brightest nova found in this work (15.2 mag).

\subsubsection*{Nova 24 = M31N 1973-10b = Nova ShAl 15}
Very fast nova, $t_2 \sim 8$ days.

\subsubsection*{Nova 28 = M31N 1985-09g = Nova Rosino 139}
We detected this nova 21 days before the outburst date reported in Rosino et al. (\cite{Rosino89}). But Rosino also noted a \textit{possible} detection of this nova 2 days before our detection date. Because of the good agreement with our data, let us suppose that this possible detection indeed was real and that Rosino detected the outburst of the nova at nearly the same date as we did.

\subsubsection*{Nova 31 = M31N 1992-12b = Nova Shafter \& Irby 1992-01}
This nova was found on Tautenburg plates even 84 days before the outburst reported by Shafter \& Irby (\cite{Shafter01}). First we note that Shafter \& Irby found this nova at the beginning of an observation period with previous observations one year ago. Furthermore, Shafter \& Irby conducted their survey in the H$\alpha$ band, in which the brightness of novae declines more slowly (Shafter \& Irby \cite{Shafter01}). Therefore, we assume that this nova is a slow one and still visible in H$\alpha$ 100 days after outburst. Unfortunately there were no M\,31 plates taken in Tautenburg at this time (see Table \ref{tab:lightcurves9}).

\subsubsection*{Nova 32 = M31N 1996-08d = Nova ShAl 48}
Despite a position difference of $46\,\farcs1$, this nova candidate is 
identified with nova ShAl 48, since the dates of outburst coincide very 
well and also the finding chart given by Sharov \& Alksnis (1997) 
evidently shows the same object. In agreement with Alksnis (2006), we 
assume that the large position difference is due to a typing error 
in Sharov \& Alksnis (\cite{Sharov97}). 

\subsubsection*{Nova 40 = M31N 1972-01b}
Very fast nova, $t_2 \sim 7$ days.

\subsubsection*{Nova 43 = M31N 1975-09a}
This nova candidate was found in our search for recurrent novae using
correlations of single-detection objects with the PHS07 catalogue. Originally we identified this
object with nova M31N 1999-01a found by Rector et al. (\cite{Rector99}),
because the distance of the two objects is only $3\,\farcs4$. However,
a finding chart provided by Rector (\cite{Rector07}) clearly shows that both nova candidates have different positions. However, we confirmed this object as a new nova candidate using additional plates.

\subsubsection*{Nova 44 = M31N 1975-11a}
The position of this newly found candidate correlates 
(distance $\sim$ $1\,\farcs0$) with nova M31N 1945-09c detected by Baade \& Arp 
(\cite{Baade64}). Hence we classify this nova as recurrent, although the position 
accuracy of $0\,\farcs3$ given by Baade \& Arp may be too optimistic, and 
there are no finding charts available for this object to confirm this 
assumption. Unfortunately Baade \& Arp gave no date of outburst for their nova 32/Table 4. For naming purposes, PHS07 set the date of outburst to 1 September 1945 but the actual date can be between 1945 and 1949. Therefore, the recurrence time is estimated to 26 - 30 years. As for nova 41, we classify this object as a new nova because the particular
outburst in 1975 was not known before.

\subsubsection*{Nova 57 = M31N 1962-08a = Nova Meinunger S10730}
This candidate is near to the centre of M\,32 (distance $\sim$ $1\,\farcm8$). Therefore, we note that this nova might belong to the galaxy M\,32.

\subsubsection*{Nova 58 = M31N 1962-07a = Nova B\"orngen 2}
After the detected outburst in September 1962, this object became visible for a year. In agreement with Rosino (\cite{Rosino64}), we note that this is a very slow nova ($t_2 \sim 350$ days).

\subsubsection*{Nova 62 = M31N 1963-10a = Nova B\"orngen 7}
This object was not detected in the automatic search routine due to the proximity to a bright star. For the same reason, no photometry is given here. We confirmed this nova candidate by eye and computed the coordinates separately. Therefore, the position accuracy is only $\sim 1\arcsec$.

\subsubsection*{Nova 64 = M31N 1964-08b = Nova B\"orngen 12}
This object was not detected in the automatic search routine because the plates it is detected on are of lower photometric quality and have therefore not been included in the 306 selected plates of the present study. We confirmed this nova candidate by eye and computed the coordinates separately. Therefore, the position accuracy is only $\sim 1\,\farcs5$.

\subsubsection*{Nova 68 = M31N 1966-09c = Nova B\"orngen 18}
This object was detected only on broken plates. Therefore, the position accuracy is only $\sim 0\,\farcs8$

\subsubsection*{Nova 82 = M31N 1985-09b = Nova Ciardullo 20}
We detected this nova 19 days before the outburst reported by Ciardullo et al. (\cite{Ciardullo87}). Similar to Nova 31, the date of outburst reported by Ciardullo et al. is at the beginning of an observation period of their H$\alpha$ survey. Therefore, we interpret this object as a moderately fast nova that is visible much longer in the H$\alpha$ band. 
%

\section{Summary}
%
Performing a systematic search on a large number of 306 archival Schmidt plates, we found or confirmed in total 84 novae in M\,31. We found 22 new nova candidates by performing light curve analysis, one of them a candidate for a recurrent nova, 12 nova candidates that are known but not found on Tautenburg plates before, and 21 nova candidates that are known to have been previously discovered on Tautenburg plates. Furthermore, using single detections, we can confirm 29 nova candidates reported in the literature. For all objects we give significantly improved positions. Even though most plates were not taken in the context of a systematic nova survey and the sampling is very irregular, this database has proved to be very valuable by the high number of novae detected on the plates. Moreover, the number of 22 newly-found nova candidates and the fact that one of them is a candidate for a recurrent nova impressively shows the potential of ``old" plate archives. Our study was made possible by the ongoing digitization of the Tautenburg plate archive. The re-analysis of archival plates makes it possible to tap the full potential of long-term databases.
\begin{acknowledgements}
We thank Andrejs Alksnis, Halton C. Arp, and Travis Rector for their kind support in finding out 
details about novae and candidate recurrent novae. We are grateful to Philipp Schalldach 
for his extensive help in checking and confirming the nova candidates. 
The anonymous referee is acknowledged for her/his constructive comments and suggestions. 
This research has made use of the SIMBAD database, the VizieR catalogue access tool, and the Aladin sky atlas, which 
are operated at the CDS, Strasbourg, France. We acknowledge use of the NASA/IPAC Extragalactic Database (NED), which 
is operated by the Jet Propulsion Laboratory, California Institute of Technology, under 
contract with the National Aeronautics and Space Administration. This research has made use of the General catalogue 
of Variable Stars Volume V Extragalactic Variable Stars (GCVS Vol.~V) which is operated at 
Sternberg Astronomical Institute, Moscow, Russia. We acknowledge use of the Digitized Sky 
Survey that was produced at the Space Telescope Science Institute under U.S. Government 
grant NAG W-2166.
\end{acknowledgements}
%

%

%
\begin{table*}
\begin{center}
\caption{Basic data for the previously known Tautenburg nova candidates}
\label{tab:tauno}
\begin{tabular}{lccclrrcc}\hline\hline\noalign{\smallskip}
Nr$^a$. & $\alpha$ [$\degr$]$^b$ & $\delta$ [$\degr$]$^b$ & Mag$^c$ & Name$^d$ & $d$ [$\arcsec$]$^e$ & $t$ [d]$^e$ & JD [d]$^f$& Year$^g$\\ \noalign{\smallskip}\hline\noalign{\smallskip}
$\;\;$1 & 9.46035 & 41.35378 & 19.3 (B) & M31N1962-08b ([M75],S10754)  & 2.0 & 30 & 2437913 & 1962\\
$\;\;$2 & 10.41910 & 41.05943 & 17.2 (U) & M31N1962-11a ([B68],N05)  & 5.4 & 0 & 2437988 & 1962\\
$\;\;$3 & 10.14533 & 40.73047 & 17.8 (U) & M31N1962-09b ([B68],N03)  & 9.5 & 75 & 2437989 & 1962\\
$\;\;$4(*) & 10.69604 & 41.22367 & 17.1 (U) & M31N1963-08a ([M67],N2)  & 0.3 & 0 & 2438264 & 1963\\
$\;\;$5(*) & 10.88359 & 41.35630 & 15.4 (U) & M31N1963-09b ([M67],N1,R047)  & 4.6 & -2 & 2438284 & 1963\\
$\;\;$6 & 11.11041 & 41.65485 & 19.1 (U) & M31N1963-09a ([M67],N3)  & 19.5 & 1 & 2438284 & 1963\\
$\;\;$7(*) & 10.22257 & 40.99426 & 17.4 (U) & M31N1963-10b ([B68],N08)  & 2.4 & 12 & 2438322 & 1963\\
$\;\;$8 & 10.78286 & 41.37024 & 17.6 (U) & M31N1963-10e ([B68],N09)  & 1.2 & 3 & 2438328 & 1963\\
$\;\;$9 & 10.39708 & 40.98540 & 18.0 (V) & M31N1963-12a ([B68],N11)  & 1.7 & 6 & 2438373 & 1963\\
10 & 10.57277 & 41.12444 & 17.1 (U) & M31N1963-10d ([B68],N10)  & 1.0 & 51 & 2438373 & 1963\\
11 & 11.01297 & 41.26195 & 17.6 (U) & M31N1964-09a ([B68],N13)  & 0.9 & 23 & 2438675 & 1964\\
12 & 10.69894 & 41.28864 & 17.7 (U) & M31N1964-10a ([B68],N14)  & 2.3 & 2 & 2438672 & 1964\\
13 & 10.72437 & 41.36870 & 17.6 (U) & M31N1965-06a ([B68],N15)  & 2.2 & 68 & 2439007 & 1965\\
14 & 10.64249 & 41.16618 & 17.2 (U) & M31N1966-09b ([B68],N20)  & 0.4 & 11 & 2439389 & 1966\\
15 & 10.48746 & 41.13616 & 18.7 (B) & M31N1967-08a ([B68],N22)  & 1.3 & 29 & 2439735 & 1967\\
16 & 10.50658 & 41.06845 & 19.0 (B) & M31N1967-09a ([B68],N24)  & 3.2 & 28 & 2439767 & 1967\\
17(*) & 10.90417 & 41.31848 & 16.9 (B) & M31N1967-10b ([M75],S10731)  & 13.4 & 0 & 2439789 & 1967\\
18(*) & 11.08395 & 41.56903 & 15.2 (B) & M31N1967-10c ([B68],N26)  & 2.4 & 0 & 2439793 & 1967\\
19 & 10.82215 & 41.32761 & 17.3 (B) & M31N1967-10a ([B68],N25)  & 1.6 & 18 & 2439793 & 1967\\
20 & 8.68975 & 40.84419 & 18.6 (B) & M31N1967-10d ([M75],S10752)  & 30.5 & -2 & 2439798 & 1967\\
21 & 9.86471 & 40.73456 & 19.5 (B) & M31N1967-08b ([B68],N23)  & 3.5 & 67 & 2439798 & 1967\\ \hline
\end{tabular}
\end{center}
Notes:\hspace{0.3cm} $^a $: identification number; (*) indicates objects discussed in Sect.\,\ref{sec:dis_rem}\\
\hspace*{1.1cm} $^b $: right ascension and declination are given in J2000.0\\
\hspace*{1.1cm} $^c $: magnitude and colour band of the detected maximum\\
\hspace*{1.1cm} $^d $: names of the objects following the naming scheme of PHS07; in parentheses we give the original names where M67 = Moffat (\cite{Moffat67}),\\
\hspace*{1.5cm} B68 = B\"orngen (\cite{Boerngen68}), M75 = Meinunger (\cite{Meinunger75}), R = Rosino (\cite{Rosino73})\\
\hspace*{1.1cm} $^e $: position difference $d$ and time lag $t$ between the present study and the previously published data; the time lag is computed as\\
\hspace*{1.5cm} $t = jd_{\mbox{here}} - jd_{\mbox{previous}}$, i.e. negative values mean an earlier detection on our plates than in previous works\\
\hspace*{1.1cm} $^f $: Julian Date of maximum luminosity on Tautenburg plates; for the first detection on Tautenburg plates see Tables \ref{tab:lightcurves1} - \ref{tab:lightcurves10}\\
\hspace*{1.1cm} $^g $: year of the outburst\\
\end{table*}

%
\begin{table*}
\begin{center}
\caption{As for Table\,\ref{tab:tauno} but for the previously known nova candidates 
which have not been reported before on Tautenburg plates.}
\label{tab:oldno}
\centering
\begin{tabular}{lccclrrcc}\hline\hline\noalign{\smallskip}
Nr. & $\alpha$ [$\degr$] & $\delta$ [$\degr$] & Mag & Name$^a$ & $d$ [$\arcsec$] & $t$ [d] & JD [d] & Year\\ 
\noalign{\smallskip}\hline\noalign{\smallskip}
22 & 10.67795 & 41.23536 & 18.2 (B) & M31N1969-10a (R086)  & 0.6 & 2 & 2440508 & 1969\\
23 & 10.43157 & 41.05191 & 18.5 (B) & M31N1972-11a (R097)  & 6.1 & -2 & 2441621 & 1972\\
24(*) & 10.20390 & 40.83336 & 16.1 (B) & M31N1973-10b (ShAl,N15)  & 13.5 & -3 & 2441980 & 1973\\
25 & 10.66535 & 41.08436 & 18.9 (B) & M31N1973-10a (R102)  & 0.2 & 6 & 2441984 & 1973\\
26 & 10.50494 & 41.19690 & 18.3 (B) & M31N1974-08a (R105)  & 1.1 & 4 & 2442276 & 1974\\
27 & 10.76844 & 41.22392 & 19.1 (B) & M31N1975-10a (R110)  & 1.4 & 19 & 2442718 & 1975\\
28(*) & 10.88516 & 41.33082 & 18.2 (U) & M31N1985-09g (R139)  & 0.2 & -21 & 2446302 & 1985\\
29 & 10.38892 & 41.03891 & 17.4 (U) & M31N1985-09h (ShAl,N18)  & 23.5 & -5 & 2446322 & 1985\\
30 & 10.60396 & 41.30883 & 17.9 (B) & M31N1987-08a (ShAl,N32)  & 1.9 & 4 & 2447039 & 1987\\
31(*) & 10.47412 & 41.12277 & 18.0 (B) & M31N1992-12b ([SI2001],N1992-01)  & 0.6 & -84 & 2448893 & 1992\\
32(*) & 10.76761 & 40.40784 & 17.3 (B) & M31N1996-08d (ShAl,N48)  & 46.1 & 2 & 2450316 & 1996\\
33 & 10.58225 & 41.23554 & 16.4 (B) & M31N1996-08g (ShAl,N49)  & 1.2 & 0 & 2450317 & 1996\\ \hline
\end{tabular}
\end{center}
Note:\hspace{0.32cm} $^a $: Names in parentheses refer to R = Rosino (\cite{Rosino64}, \cite{Rosino73}, \cite{Rosino89}),
ShAl = Sharov \& Alksnis (\cite{Sharov91}, \cite{Sharov92}, \cite{Sharov97}),\\
\hspace*{1.5cm}and SI2001 = Shafter \& Irby (\cite{Shafter01}).
\end{table*}

\begin{table*}
\caption{
As for Tables\,\ref{tab:tauno} and \ref{tab:oldno} but for the new nova candidates from the present study.
}
\label{tab:newno}
\centering
\begin{tabular}{lccclcc}\hline\hline\noalign{\smallskip}
Nr. & $\alpha$ [$\degr$] & $\delta$ [$\degr$] & Mag & Name & JD [d] & Year\\ 
\noalign{\smallskip}\hline\noalign{\smallskip}
34 & 11.90708 & 41.75834 & 19.5 (U) & M31N1962-09a & 2437911 & 1962\\
35 & 10.70393 & 41.30974 & 17.7 (U) & M31N1963-08b & 2438258 & 1963\\
36 & 11.45452 & 41.89021 & 19.1 (U) & M31N1963-08c & 2438262 & 1963\\
37 & 10.01956 & 40.62433 & 18.4 (V) & M31N1963-12b & 2438373 & 1963\\
38 & 11.43947 & 41.75058 & 17.5 (U) & M31N1966-10b & 2439417 & 1966\\
39 & 9.33152 & 40.52856 & 18.8 (B) & M31N1970-11a & 2440917 & 1970\\
40(*) & 10.38148 & 40.87432 & 16.7 (B) & M31N1972-01b & 2441328 & 1972\\
41 & 9.81823 & 40.52356 & 18.8 (B) & M31N1972-12b & 2441680 & 1972\\
42 & 10.48303 & 41.18581 & 15.6 (B) & M31N1975-02a & 2442455 & 1975\\
43(*) & 10.68148 & 41.19253 & 18.5 (B) & M31N1975-09a & 2442665 & 1975\\
44(*) & 10.36907 & 40.88704 & 18.6 (V) & M31N1975-11a & 2442741 & 1975\\
45 & 10.44820 & 40.95410 & 18.6 (B) & M31N1975-12b & 2442775 & 1975\\
46 & 11.47816 & 40.92837 & 19.0 (B) & M31N1979-11a & 2444194 & 1979\\
47 & 10.47874 & 41.01253 & 18.0 (U) & M31N1980-09b & 2444490 & 1980\\
48 & 10.14823 & 41.32396 & 17.6 (U) & M31N1980-09a & 2444490 & 1980\\
49 & 10.74837 & 41.28688 & 18.0 (U) & M31N1980-09c & 2444490 & 1980\\
50 & 11.08057 & 41.60674 & 19.3 (B) & M31N1984-08a & 2445940 & 1984\\
51 & 11.51205 & 41.73243 & 18.6 (U) & M31N1985-08a & 2446299 & 1985\\
52 & 11.58009 & 41.97954 & 18.5 (U) & M31N1985-08b & 2446299 & 1985\\
53 & 12.51510 & 41.42756 & 18.8 (U) & M31N1985-08c & 2446299 & 1985\\
54 & 10.43303 & 41.07269 & 16.5 (B) & M31N1992-09a & 2448893 & 1992\\
55 & 11.54533 & 41.61147 & 19.5 (B) & M31N1996-08e & 2450317 & 1996\\ \hline	
\end{tabular}
\end{table*}

\begin{table*}
\begin{center}
\caption{
As for Tables\,\ref{tab:tauno} and \ref{tab:oldno} but for the confirmed nova 
candidates from the present study.
}
\label{tab:confno}
\centering
\begin{tabular}{lccclrrcc}\hline\hline\noalign{\smallskip}
Nr. & $\alpha$ [$\degr$] & $\delta$ [$\degr$] & Mag & Name$^a$ & $d$ [$\arcsec$] & $t$ [d] & JD [d] & Year\\ 
\noalign{\smallskip}\hline\noalign{\smallskip}
56 & 10.70850 & 41.47538 & 17.1 (B) & M31N1960-11a ([B68],N01)  & 0.9 & 8 & 2437261 & 1960\\
57(*) & 10.63423 & 40.87034 & 18.9 (U) & M31N1962-08a ([M75],S10730)  & 2.3 & 27 & 2437911 & 1962\\
58(*) & 10.86169 & 41.25254 & 17.8 (U) & M31N1962-07a ([B68],N02)  & 0.5 & 41 & 2437911 & 1962\\
59 & 10.61778 & 41.36867 & 17.5 (B) & M31N1962-09d ([B68],N04)  & 1.0 & 30 & 2437961 & 1962\\
60 & 10.66868 & 41.22996 & 18.7 (U) & M31N1962-10a ([B68],N06)  & 0.9 & 28 & 2437989 & 1962\\
61 & 10.74055 & 41.13677 & 18.0 (U) & M31N1963-09c (R048(079))  & 0.5 & 2 & 2438290 & 1963\\
62(*) & 10.94180 & 41.64252 & - & M31N1963-10a ([B68],N07)  & 1.0 & 0 & 2438311 & 1963\\
63 & 11.01505 & 41.43409 & 20.0 (B) & M31N1963-10c ([M75],S10733)  & 3.3 & 4 & 2438319 & 1963\\
64(*) & 10.65329 & 41.36931 & - & M31N1964-08b ([B68],N12)  & 2.3 & 7 & 2438621 & 1964\\
65 & 12.43549 & 41.57251 & 18.9 (V) & M31N1964-09b ([M75],S10755)  & 18.5 & 2 & 2438642 & 1964\\
66 & 10.30955 & 40.72338 & 18.9 (U) & M31N1965-09c ([B68],N16)  & 2.7 & -1 & 2439025 & 1965\\
67 & 11.05491 & 41.78507 & 16.5 (V) & M31N1965-11c ([B68],N17)  & 4.6 & 0 & 2439088 & 1965\\
68(*) & 10.60744 & 41.22068 & 19.5 (U) & M31N1966-09c ([B68],N18)  & 2.0 & 9 & 2439389 & 1966\\
69 & 10.09839 & 40.40145 & 17.5 (U) & M31N1966-09f ([B68],N19)  & 3.8 & 0 & 2439390 & 1966\\
70 & 11.13049 & 41.41552 & 19.4 (U) & M31N1966-09d ([M75],S10734)  & 2.7 & 2 & 2439390 & 1966\\
71 & 9.87626 & 40.48731 & 18.8 (U) & M31N1966-09e ([M75],S10729)  & 5.7 & 1 & 2439390 & 1966\\
72 & 10.91867 & 41.44963 & 19.3 (B) & M31N1967-09b ([M75],S10732)  & 5.3 & 1 & 2439763 & 1967\\
73 & 10.71251 & 41.26450 & 18.4 (U) & M31N1967-11a ([B68],N27)  & 0.8 & 1 & 2439802 & 1967\\
74 & 11.46352 & 41.54582 & 18.7 (B) & M31N1970-08a ([M75],S10735)  & 2.3 & 14 & 2440840 & 1970\\
75 & 10.75543 & 41.22672 & 19.3 (B) & M31N1971-08a (R091)  & 0.9 & 2 & 2441185 & 1971\\
76 & 10.73660 & 41.33528 & 17.5 (B) & M31N1972-10a (R095)  & 1.1 & -2 & 2441605 & 1972\\
77 & 10.72987 & 40.72172 & 18.3 (U) & M31N1973-09b (ShAl,N14)  & 1.7 & 20 & 2441976 & 1973\\
78 & 10.74249 & 41.31710 & 18.5 (U) & M31N1974-07a (R103)  & 0.5 & 84 & 2442335 & 1974\\
79 & 10.58411 & 41.12401 & 18.4 (U) & M31N1974-08b (R106)  & 0.8 & 57 & 2442335 & 1974\\
80 & 10.74133 & 41.42048 & 17.6 (U) & M31N1978-10a (R120)  & 1.1 & 11 & 2443816 & 1978\\
81 & 10.87900 & 41.33578 & 18.5 (V) & M31N1980-10a (R123)  & 0.1 & 3 & 2444523 & 1980\\
82(*) & 10.57499 & 41.17245 & 18.6 (U) & M31N1985-09b ([CFN87],N20)  & 1.6 & -19 & 2446299 & 1985\\
83 & 10.74728 & 41.27864 & 20.4 (B) & M31N1995-11c ([SI2001],N1995-05)  & 1.2 & -1 & 2450048 & 1995\\
84 & 10.22216 & 40.54756 & 17.8 (B) & M31N1996-12b (ShAl,N54)  & 0.9 & -2 & 2450432 & 1996\\ \hline
\end{tabular}
\end{center}
Note:\hspace{0.32cm} $^a $: Additionally, CFN87 refers to Ciardullo et al. (\cite{Ciardullo87})
\end{table*}

%
%

%
\onltab{10}{
\begin{table*}
\caption{Nova light curves continued. See notes below for a more detailed description.}
\label{tab:lightcurves1}
\begin{center}
\begin{tabular}{rrrrcrrrrrrrrrrr}\hline\hline\noalign{\smallskip}
PlNr & Date & RJD & Limit & Band & 60-11a & 62-08a& 62-08b & 62-09a & 62-07a & 62-09d & 62-09b & 62-11a & 62-10a\\ \hline
10 & 1960-11-22 & 37261.34 & 18.5 & NFBS & 17.1 &  &   &   &   &   &   &   &  \\
35 & 1961-01-12 & 37312.28 & 19.2 & NFBS &   &  &   &   &   &   &   &   &  \\
61 & 1961-01-19 & 37319.29 & 18.6 & NFBS &   &  &   &   &   &   &   &   &  \\
62 & 1961-01-19 & 37319.30 & 18.5 & NFBS &   &  &   &   &   &   &   &   &  \\
192 & 1961-08-06 & 37517.49 & 19.7 & NFBS &   &  &   &   &   &   &   &   &  \\
197 & 1961-08-07 & 37519.45 & 19.0 & NFBS &   &  &   &   &   &   &   &   &  \\
205 & 1961-09-01 & 37544.40 & 19.3 & NFBS &   &  &   &   &   &   &   &   &  \\
471 & 1962-09-03 & 37910.55 & 19.5 & U &   & 18.9& 18.8 & 19.5 & 17.8 &   &   &   &  \\
472 & 1962-09-03 & 37910.58 & 19.7 & B &   & 19.8& 19.6 & 20.1 & 19.1 &   &   &   &  \\
490 & 1962-09-05 & 37913.47 & 19.8 & NFBS &   &  & 19.3 & 19.6 & 18.2 &   &   &   &  \\
669 & 1962-10-23 & 37961.27 & 19.7 & B &   &  &   &   &   & 17.5 &   &   &  \\
717 & 1962-11-19 & 37988.31 & 19.6 & B &   &  &   &   &   &   & 18.4 & 18.8 &  \\
718 & 1962-11-19 & 37988.38 & 17.6 & U &   &  &   &   &   &   &   & 17.2 &  \\
721 & 1962-11-20 & 37989.34 & 19.2 & U &   &  &   & 19.8 & 19.4 &   & 17.8 & 18.1 & 18.7\\
732 & 1962-11-24 & 37993.32 & 18.9 & U &   &  &   &   &   &   & 17.9 & 18.4 &  \\
792 & 1963-01-13 & 38043.29 & 17.7 & NFBS &   &  &   &   &   &   &   &   &  \\
1079 & 1963-08-13 & 38255.44 & 19.8 & B &   &  &   &   &   &   &   &   &  \\
1080 & 1963-08-13 & 38255.46 & 18.7 & V &   &  &   &   &   &   &   &   &  \\
1085 & 1963-08-14 & 38256.41 & 18.3 & V &   &  &   &   &   &   &   &   &  \\
1086 & 1963-08-14 & 38256.44 & 19.2 & B &   &  &   &   &   &   &   &   &  \\
1090 & 1963-08-16 & 38257.50 & 17.9 & U &   &  &   &   &   &   &   &   &  \\
1092 & 1963-08-17 & 38258.48 & 19.5 & V &   &  &   &   &   &   &   &   &  \\
1093 & 1963-08-17 & 38258.55 & 19.0 & U &   &  &   &   &   &   &   &   &  \\
1096 & 1963-08-20 & 38261.48 & 19.1 & B &   &  &   &   &   &   &   &   &  \\
1101 & 1963-08-20 & 38262.46 & 19.5 & U &   &  &   &   & 19.6 &   &   &   &  \\
1102 & 1963-08-21 & 38262.52 & 19.3 & V &   &  &   &   &   &   &   &   &  \\
1107 & 1963-08-22 & 38263.52 & 19.5 & B &   &  &   &   &   &   &   &   &  \\
1108 & 1963-08-22 & 38263.56 & 19.1 & U &   &  &   &   &   &   &   &   &  \\
1110 & 1963-08-23 & 38265.41 & 19.3 & B &   &  &   &   &   &   &   &   &  \\
1124 & 1963-08-26 & 38268.45 & 20.1 & B &   &  &   &   &   &   &   &   &  \\
1125 & 1963-08-27 & 38268.47 & 19.3 & V &   &  &   &   &   &   &   &   &  \\
1133 & 1963-08-31 & 38272.53 & 18.4 & U &   &  &   &   &   &   &   &   &  \\
1134 & 1963-08-31 & 38272.58 & 19.0 & B &   &  &   &   &   &   &   &   &  \\
1135 & 1963-08-31 & 38272.60 & 17.9 & V &   &  &   &   &   &   &   &   &  \\
1175 & 1963-09-09 & 38282.37 & 18.6 & V &   &  &   &   &   &   &   &   &  \\
1181 & 1963-09-10 & 38283.40 & 19.3 & B &   &  &   &   &   &   &   &   &  \\
1187 & 1963-09-11 & 38284.40 & 19.2 & U &   &  &   &   &   &   &   &   &  \\
1198 & 1963-09-12 & 38285.41 & 19.2 & U &   &  &   &   &   &   &   &   &  \\
1203 & 1963-09-13 & 38286.45 & 19.1 & U &   &  &   &   &   &   &   &   &  \\
1207 & 1963-09-14 & 38286.62 & 17.4 & V &   &  &   &   &   &   &   &   &  \\
1212 & 1963-09-15 & 38287.47 & 18.7 & V &   &  &   &   &   &   &   &   &  \\
1224 & 1963-09-15 & 38288.42 & 19.9 & B &   &  &   &   &   &   &   &   &  \\
1238 & 1963-09-17 & 38290.40 & 20.1 & U &   &  &   &   & 19.1 &   &   &   &  \\ \hline
\end{tabular}
\end{center}
Notes:\hspace{0.3cm} $^a $: Tautenburg plate number; note that Tables \ref{tab:lightcurves1} - \ref{tab:lightcurves10} list the properties of all 306 plates used in this work\\
\hspace*{1.1cm} $^b $: Date and Julian Date of plate exposure; RJD = JD - $2\,400\,000$\\
\hspace*{1.1cm} $^c $: limiting magnitude of plate; for a definition of the limiting magnitude in this work see Fig.\,\ref{fig:maghist}\\
\hspace*{1.1cm} $^d $: colour band according to Table\,\ref{table:plates}\\
\hspace*{1.1cm} $^e $: and subsequent columns give the light curve data; columns are indicated by their M31N 19$\sim$ name following the naming scheme of\\
\hspace*{1.5cm} PHS07 \\
\end{table*}
}

%
\onltab{11}{
\begin{table*}
\caption{Nova light curves continued. See notes below Table\,\ref{tab:lightcurves1} for a more detailed description.}
\label{tab:lightcurves2}
\centering
\begin{tabular}{rrrrcrrrrrrrrrrr}\hline\hline
PlNr & Date & RJD & Limit & Band & 63-08b& 63-08c & 63-08a & 63-09a & 63-09b & 63-09c & 63-10b & 63-10c & 63-10d\\ \hline
1090 & 1963-08-16 & 38257.50 & 17.9 & U & 17.7&   &   &   &   &   &   &   &  \\
1092 & 1963-08-17 & 38258.48 & 19.5 & V &  &   &   &   &   &   &   &   &  \\
1093 & 1963-08-17 & 38258.55 & 19.0 & U & 18.3& 19.4 &   &   &   &   &   &   &  \\
1096 & 1963-08-20 & 38261.48 & 19.1 & B &  &   &   &   &   &   &   &   &  \\
1101 & 1963-08-20 & 38262.46 & 19.5 & U & 18.5& 19.1 & 17.2 &   &   &   &   &   &  \\
1102 & 1963-08-21 & 38262.52 & 19.3 & V &  &   &   &   &   &   &   &   &  \\
1107 & 1963-08-22 & 38263.52 & 19.5 & B &  &   & 17.6 &   &   &   &   &   &  \\
1108 & 1963-08-22 & 38263.56 & 19.1 & U & 18.8& 19.7 & 17.1 &   &   &   &   &   &  \\
1110 & 1963-08-23 & 38265.41 & 19.3 & B &  &   & 17.8 &   &   &   &   &   &  \\
1124 & 1963-08-26 & 38268.45 & 20.1 & B &  &   &   &   &   &   &   &   &  \\
1125 & 1963-08-27 & 38268.47 & 19.3 & V &  &   &   &   &   &   &   &   &  \\
1133 & 1963-08-31 & 38272.53 & 18.4 & U & 18.6&   & 18.2 &   &   &   &   &   &  \\
1134 & 1963-08-31 & 38272.58 & 19.0 & B &  &   &   &   &   &   &   &   &  \\
1135 & 1963-08-31 & 38272.60 & 17.9 & V &  &   &   &   &   &   &   &   &  \\
1175 & 1963-09-09 & 38282.37 & 18.6 & V &  &   &   & 18.5 &   &   &   &   &  \\
1181 & 1963-09-10 & 38283.40 & 19.3 & B &  &   &   &   &   &   &   &   &  \\
1187 & 1963-09-11 & 38284.40 & 19.2 & U &  &   &   & 19.1 & 15.4 &   &   &   &  \\
1198 & 1963-09-12 & 38285.41 & 19.2 & U &  &   &   & 19.5 & 16.4 &   &   &   &  \\
1203 & 1963-09-13 & 38286.45 & 19.1 & U &  &   &   & 19.5 & 17.0 &   &   &   &  \\
1207 & 1963-09-14 & 38286.62 & 17.4 & V &  &   &   &   & 17.7 &   &   &   &  \\
1212 & 1963-09-15 & 38287.47 & 18.7 & V &  &   &   &   &   &   &   &   &  \\
1224 & 1963-09-15 & 38288.42 & 19.9 & B &  &   &   &   & 18.9 & 18.9 &   &   &  \\
1238 & 1963-09-17 & 38290.40 & 20.1 & U &  &   & 19.7 &   & 18.5 & 18.0 &   &   &  \\
1239 & 1963-09-17 & 38290.45 & 18.7 & V &  &   &   &   &   &   &   &   &  \\
1240 & 1963-09-18 & 38290.46 & 18.5 & V &  &   &   &   &   &   &   &   &  \\
1254 & 1963-10-08 & 38311.32 & 18.4 & V &  &   &   &   &   &   & 17.2 &   &  \\
1260 & 1963-10-12 & 38315.36 & 18.7 & U &  &   &   &   &   &   & 17.8 &   &  \\
1261 & 1963-10-12 & 38315.43 & 18.3 & V &  &   &   &   &   &   & 17.1 &   &  \\
1267 & 1963-10-15 & 38318.40 & 20.4 & U &  &   &   &   &   &   & 17.7 &   &  \\
1268 & 1963-10-16 & 38319.37 & 19.9 & B &  &   &   &   &   &   & 18.0 & 20.0 &  \\
1269 & 1963-10-16 & 38319.40 & 19.0 & V &  &   &   &   &   &   & 17.5 &   &  \\
1272 & 1963-10-19 & 38322.31 & 19.1 & U &  &   &   &   &   &   & 17.4 &   & 19.4\\
1275 & 1963-10-20 & 38323.32 & 19.0 & U &  &   &   &   &   &   & 17.6 &   & 19.0\\
1277 & 1963-10-21 & 38324.47 & 18.9 & V &  &   &   &   &   &   & 18.3 &   &  \\
1278 & 1963-10-22 & 38324.50 & 18.6 & V &  &   &   &   &   &   & 18.5 &   &  \\
1288 & 1963-10-24 & 38327.43 & 19.5 & B &  &   &   &   &   &   & 18.5 &   &  \\
1289 & 1963-10-24 & 38327.46 & 19.5 & B &  &   &   &   &   &   & 18.5 &   &  \\
1290 & 1963-10-25 & 38327.54 & 18.6 & U &  &   &   &   &   &   & 17.7 &   &  \\
1295 & 1963-10-25 & 38328.42 & 18.9 & V &  &   &   &   &   &   & 19.0 &   &  \\
1296 & 1963-10-25 & 38328.45 & 19.2 & V &  &   &   &   &   &   & 18.9 &   &  \\
1297 & 1963-10-26 & 38328.52 & 18.8 & U &  &   &   &   &   &   & 17.6 &   &  \\
1398 & 1963-12-09 & 38373.22 & 19.0 & V &  &   &   &   &   &   &   &   & 16.7\\
1399 & 1963-12-09 & 38373.26 & 19.0 & V &  &   &   &   &   &   &   &   & 16.8\\
1400 & 1963-12-09 & 38373.33 & 19.0 & U &  &   &   &   &   &   &   &   & 17.1\\
1729 & 1964-09-02 & 38641.43 & 19.9 & B &  &   &   &   &   &   &   &   &  \\ \hline
\end{tabular}
\end{table*}
}

%
\onltab{12}{
\begin{table*}
\caption{Nova light curves continued. See notes below Table\,\ref{tab:lightcurves1} for a more detailed description.}
\label{tab:lightcurves3}
\centering
\begin{tabular}{rrrrcrrrrrrrrrrr}\hline\hline
PlNr & Date & RJD & Limit & Band & 63-10e & 63-12a & 63-12b& 64-09b & 64-09a & 64-10a & 65-06a & 65-09c & 65-11c\\ \hline
1278 & 1963-10-22 & 38324.50 & 18.6 & V &   &   &  &   &   &   &   &   &  \\
1288 & 1963-10-24 & 38327.43 & 19.5 & B & 17.7 &   &  &   &   &   &   &   &  \\
1289 & 1963-10-24 & 38327.46 & 19.5 & B & 17.8 &   &  &   &   &   &   &   &  \\
1290 & 1963-10-25 & 38327.54 & 18.6 & U & 17.6 &   &  &   &   &   &   &   &  \\
1295 & 1963-10-25 & 38328.42 & 18.9 & V & 18.4 &   &  &   &   &   &   &   &  \\
1296 & 1963-10-25 & 38328.45 & 19.2 & V & 18.6 &   &  &   &   &   &   &   &  \\
1297 & 1963-10-26 & 38328.52 & 18.8 & U & 17.6 &   &  &   &   &   &   &   &  \\
1398 & 1963-12-09 & 38373.22 & 19.0 & V &   & 18.0 & 18.6&   &   &   &   &   &  \\
1399 & 1963-12-09 & 38373.26 & 19.0 & V &   & 18.2 & 18.4&   &   &   &   &   &  \\
1400 & 1963-12-09 & 38373.33 & 19.0 & U &   & 18.2 & 18.3&   &   &   &   &   &  \\
1729 & 1964-09-02 & 38641.43 & 19.9 & B &   &   &  & 19.5 &   &   &   &   &  \\
1735 & 1964-09-03 & 38642.44 & 18.5 & V &   &   &  & 18.9 &   &   &   &   &  \\
1736 & 1964-09-04 & 38642.47 & 19.9 & B &   &   &  & 19.2 &   &   &   &   &  \\
1750 & 1964-09-13 & 38651.56 & 17.5 & V &   &   &  & 19.9 & 17.3 &   &   &   &  \\
1770 & 1964-09-28 & 38667.33 & 18.3 & V &   &   &  & 19.6 &   &   &   &   &  \\
1771 & 1964-09-28 & 38667.36 & 20.5 & B &   &   &  &   & 18.5 &   &   &   &  \\
1772 & 1964-09-28 & 38667.39 & 18.8 & V &   &   &  &   & 18.8 &   &   &   &  \\
1778 & 1964-09-30 & 38669.38 & 19.7 & B &   &   &  &   & 18.1 &   &   &   &  \\
1779 & 1964-09-30 & 38669.42 & 19.0 & V &   &   &  &   & 18.4 &   &   &   &  \\
1780 & 1964-10-01 & 38669.49 & 19.5 & B &   &   &  &   & 18.2 &   &   &   &  \\
1781 & 1964-10-01 & 38669.53 & 18.7 & V &   &   &  &   & 18.2 &   &   &   &  \\
1784 & 1964-10-01 & 38670.32 & 18.8 & U &   &   &  &   & 17.7 & 18.4 &   &   &  \\
1786 & 1964-10-01 & 38670.42 & 18.8 & V &   &   &  &   & 18.5 &   &   &   &  \\
1788 & 1964-10-02 & 38670.51 & 19.7 & B &   &   &  &   & 18.1 &   &   &   &  \\
1789 & 1964-10-02 & 38670.55 & 18.7 & V &   &   &  &   & 18.5 &   &   &   &  \\
1790 & 1964-10-02 & 38670.61 & 18.5 & U &   &   &  &   & 17.7 & 18.6 &   &   &  \\
1794 & 1964-10-03 & 38671.51 & 18.1 & V &   &   &  &   &   &   &   &   &  \\
1795 & 1964-10-03 & 38671.57 & 17.7 & U &   &   &  &   & 17.7 & 17.7 &   &   &  \\
1796 & 1964-10-03 & 38672.33 & 18.3 & U &   &   &  &   & 18.2 & 18.4 &   &   &  \\
1798 & 1964-10-04 & 38672.55 & 17.5 & U &   &   &  &   & 17.9 &   &   &   &  \\
1801 & 1964-10-05 & 38673.56 & 18.4 & U &   &   &  &   & 17.9 & 18.8 &   &   &  \\
1802 & 1964-10-05 & 38674.31 & 18.7 & U &   &   &  &   & 17.8 & 19.0 &   &   &  \\
1803 & 1964-10-05 & 38674.38 & 18.5 & V &   &   &  &   &   &   &   &   &  \\
1804 & 1964-10-05 & 38674.41 & 20.0 & B &   &   &  &   & 18.4 &   &   &   &  \\
1811 & 1964-10-06 & 38674.61 & 18.4 & U &   &   &  &   & 17.6 & 18.5 &   &   &  \\
1812 & 1964-10-06 & 38675.29 & 19.0 & U &   &   &  &   & 17.9 & 18.3 &   &   &  \\
1813 & 1964-10-06 & 38675.37 & 18.3 & U &   &   &  &   & 17.9 & 18.3 &   &   &  \\
2047 & 1965-09-04 & 39007.48 & 18.9 & U &   &   &  &   &   &   & 17.6 &   &  \\
2068 & 1965-09-21 & 39025.46 & 18.8 & U &   &   &  &   &   &   & 19.2 & 18.9 &  \\
2140 & 1965-11-23 & 39088.31 & 18.3 & V &   &   &  &   &   &   &   &   & 16.5\\
2144 & 1965-12-15 & 39110.32 & 18.5 & U &   &   &  &   &   &   &   &   &  \\ 
2256 & 1966-08-19 & 39356.54 & 18.8 & V &   &   &  &   &   &   &   &   &  \\
2257 & 1966-08-19 & 39356.56 & 18.5 & V &   &   &  &   &   &   &   &   &  \\
2315 & 1966-09-20 & 39389.47 & 19.3 & U &   &   &  &   &   &   &   &   &  \\ \hline
\end{tabular}
\end{table*}
}

%
\onltab{13}{
\begin{table*}
\caption{Nova light curves continued. See notes below Table\,\ref{tab:lightcurves1} for a more detailed description.}
\label{tab:lightcurves4}
\centering
\begin{tabular}{rrrrcrrrrrrrrrrr}\hline\hline
PlNr & Date & RJD & Limit & Band & 66-09c & 66-09d & 66-09e& 66-09f & 66-09b & 66-10b & 67-08a & 67-08b & 67-09b\\ \hline
2257 & 1966-08-19 & 39356.56 & 18.5 & V &   &   &  &   &   &   &   &   &  \\
2315 & 1966-09-20 & 39389.47 & 19.3 & U & 19.5 & 19.8 & 18.8& 17.5 & 17.2 &   &   &   &  \\
2317 & 1966-09-21 & 39389.59 & 19.3 & U &   & 19.4 & 19.1& 17.7 & 17.5 &   &   &   &  \\
2328 & 1966-09-23 & 39391.58 & 19.5 & B &   &   &  &   & 18.2 &   &   &   &  \\
2329 & 1966-09-23 & 39391.62 & 19.2 & B &   &   &  &   & 18.1 &   &   &   &  \\
2338 & 1966-10-08 & 39407.38 & 19.0 & B &   &   &  &   &   &   &   &   &  \\
2340 & 1966-10-11 & 39410.38 & 19.6 & B &   &   &  &   & 18.8 &   &   &   &  \\
2341 & 1966-10-11 & 39410.42 & 18.4 & B &   &   &  &   &   &   &   &   &  \\
2342 & 1966-10-15 & 39414.42 & 18.3 & B &   &   &  &   &   &   &   &   &  \\
2344 & 1966-10-17 & 39416.33 & 18.7 & B &   &   &  &   &   & 18.8 &   &   &  \\
2345 & 1966-10-17 & 39416.38 & 17.7 & U &   &   &  &   &   & 18.0 &   &   &  \\
2346 & 1966-10-18 & 39416.64 & 17.5 & U &   &   &  &   &   & 17.5 &   &   &  \\
2348 & 1966-10-20 & 39419.41 & 18.1 & U &   &   &  &   &   & 17.8 &   &   &  \\
2526 & 1967-08-28 & 39731.39 & 18.1 & B &   &   &  &   &   &   &   &   &  \\
2527 & 1967-08-29 & 39732.40 & 18.4 & B &   &   &  &   &   &   &   &   &  \\
2529 & 1967-09-01 & 39734.54 & 19.4 & B &   &   &  &   &   &   & 18.9 & 19.9 &  \\
2530 & 1967-09-01 & 39734.56 & 18.9 & B &   &   &  &   &   &   & 18.7 &   &  \\
2531 & 1967-09-01 & 39734.59 & 18.6 & B &   &   &  &   &   &   &   &   &  \\
2535 & 1967-09-05 & 39739.41 & 19.3 & B &   &   &  &   &   &   &   & 19.6 &  \\
2536 & 1967-09-05 & 39739.45 & 19.5 & B &   &   &  &   &   &   & 19.6 & 19.8 &  \\
2537 & 1967-09-06 & 39739.47 & 19.3 & B &   &   &  &   &   &   & 19.6 & 19.5 &  \\
2538 & 1967-09-06 & 39739.50 & 18.9 & B &   &   &  &   &   &   &   &   &  \\
2558 & 1967-09-29 & 39763.35 & 19.8 & B &   &   &  &   &   &   &   & 19.6 & 19.3\\
2559 & 1967-09-29 & 39763.38 & 19.1 & B &   &   &  &   &   &   &   &   &  \\
2560 & 1967-10-01 & 39764.66 & 16.8 & B &   &   &  &   &   &   &   &   &  \\
2565 & 1967-10-02 & 39765.50 & 19.0 & B &   &   &  &   &   &   &   & 20.0 &  \\
2566 & 1967-10-02 & 39765.53 & 19.2 & B &   &   &  &   &   &   &   & 20.1 &  \\
2574 & 1967-10-03 & 39767.37 & 19.0 & B &   &   &  &   &   &   &   &   &  \\
2576 & 1967-10-04 & 39767.48 & 19.3 & B &   &   &  &   &   &   &   & 19.5 &  \\
2588 & 1967-10-11 & 39775.45 & 19.1 & B &   &   &  &   &   &   &   & 19.6 &  \\
2590 & 1967-10-12 & 39775.50 & 19.7 & B &   &   &  &   &   &   &   & 19.6 &  \\
2604 & 1967-10-13 & 39776.52 & 19.0 & B &   &   &  &   &   &   &   &   &  \\
2605 & 1967-10-13 & 39776.58 & 18.5 & B &   &   &  &   &   &   &   &   &  \\
2609 & 1967-10-25 & 39789.28 & 19.1 & B &   &   &  &   &   &   &   & 20.1 &  \\
2611 & 1967-10-25 & 39789.34 & 19.3 & B &   &   &  &   &   &   &   & 19.6 &  \\
2615 & 1967-10-26 & 39790.40 & 19.3 & B &   &   &  &   &   &   &   & 19.6 &  \\
2620 & 1967-10-29 & 39792.51 & 19.4 & B &   &   &  &   &   &   &   & 19.8 &  \\
2623 & 1967-10-29 & 39793.42 & 19.2 & B &   &   &  &   &   &   &   & 19.5 &  \\
2624 & 1967-10-29 & 39793.45 & 18.9 & B &   &   &  &   &   &   &   & 19.5 &  \\
2631 & 1967-10-30 & 39794.47 & 19.2 & B &   &   &  &   &   &   &   & 19.7 &  \\
2632 & 1967-10-31 & 39794.49 & 18.9 & B &   &   &  &   &   &   &   &   &  \\
2636 & 1967-11-01 & 39796.31 & 17.5 & B &   &   &  &   &   &   &   &   &  \\
2638 & 1967-11-03 & 39797.55 & 19.4 & B &   &   &  &   &   &   &   & 19.5 &  \\
2639 & 1967-11-03 & 39797.57 & 18.6 & B &   &   &  &   &   &   &   &   &  \\
2644 & 1967-11-03 & 39798.42 & 19.5 & B &   &   &  &   &   &   &   & 19.6 &  \\
2645 & 1967-11-03 & 39798.45 & 19.4 & B &   &   &  &   &   &   &   & 19.7 &  \\
2652 & 1967-11-06 & 39800.57 & 19.8 & B &   &   &  &   &   &   &   & 19.8 &  \\
2653 & 1967-11-06 & 39800.59 & 18.1 & B &   &   &  &   &   &   &   &   &  \\
2661 & 1967-11-06 & 39801.45 & 19.5 & B &   &   &  &   &   &   &   & 19.9 &  \\
2662 & 1967-11-07 & 39801.48 & 19.6 & B &   &   &  &   &   &   &   & 19.7 &  \\
2670 & 1967-11-07 & 39802.37 & 19.5 & B &   &   &  &   &   &   &   & 19.5 &  \\
2671 & 1967-11-07 & 39802.39 & 19.4 & B &   &   &  &   &   &   &   & 19.5 &  \\
2672 & 1967-11-07 & 39802.43 & 19.2 & U &   &   &  &   &   &   &   &   &  \\ \hline
\end{tabular}
\end{table*}
}

%
\onltab{14}{
\begin{table*}
\caption{Nova light curves continued. See notes below Table\,\ref{tab:lightcurves1} for a more detailed description.}
\label{tab:lightcurves5}
\centering
\begin{tabular}{rrrrcrrrrrrrrrrr}\hline\hline
PlNr & Date & RJD & Limit & Band & 67-09a & 67-10a & 67-10b& 67-10c & 67-10d & 67-11a & 69-10a & 70-08a & 70-11a\\ \hline
2538 & 1967-09-06 & 39739.50 & 18.9 & B &   &   &  &   &   &   &   &   &  \\
2558 & 1967-09-29 & 39763.35 & 19.8 & B & 19.7 &   &  &   &   &   &   &   &  \\
2559 & 1967-09-29 & 39763.38 & 19.1 & B &   &   &  &   &   &   &   &   &  \\
2560 & 1967-10-01 & 39764.66 & 16.8 & B &   &   &  &   &   &   &   &   &  \\
2565 & 1967-10-02 & 39765.50 & 19.0 & B & 19.2 &   &  &   &   &   &   &   &  \\
2566 & 1967-10-02 & 39765.53 & 19.2 & B &   &   &  &   &   &   &   &   &  \\
2574 & 1967-10-03 & 39767.37 & 19.0 & B & 19.0 &   &  &   &   &   &   &   &  \\
2576 & 1967-10-04 & 39767.48 & 19.3 & B & 19.3 &   &  &   &   &   &   &   &  \\
2588 & 1967-10-11 & 39775.45 & 19.1 & B &   & 19.0 &  &   &   &   &   &   &  \\
2590 & 1967-10-12 & 39775.50 & 19.7 & B &   & 18.0 &  &   &   &   &   &   &  \\
2604 & 1967-10-13 & 39776.52 & 19.0 & B &   & 18.0 &  &   &   &   &   &   &  \\
2605 & 1967-10-13 & 39776.58 & 18.5 & B &   & 17.5 &  &   &   &   &   &   &  \\
2609 & 1967-10-25 & 39789.28 & 19.1 & B &   & 17.7 & 17.0&   &   &   &   &   &  \\
2611 & 1967-10-25 & 39789.34 & 19.3 & B &   & 17.9 & 16.9&   &   &   &   &   &  \\
2615 & 1967-10-26 & 39790.40 & 19.3 & B &   & 17.9 & 17.5&   &   &   &   &   &  \\
2620 & 1967-10-29 & 39792.51 & 19.4 & B &   & 18.4 & 18.0&   &   &   &   &   &  \\
2623 & 1967-10-29 & 39793.42 & 19.2 & B &   & 17.3 & 18.1& 15.2 &   &   &   &   &  \\
2624 & 1967-10-29 & 39793.45 & 18.9 & B &   & 17.3 & 18.2& 15.2 &   &   &   &   &  \\
2631 & 1967-10-30 & 39794.47 & 19.2 & B &   & 18.1 & 18.6& 15.2 &   &   &   &   &  \\
2632 & 1967-10-31 & 39794.49 & 18.9 & B &   & 18.0 & 18.5& 15.4 &   &   &   &   &  \\
2636 & 1967-11-01 & 39796.31 & 17.5 & B &   &   &  & 15.4 &   &   &   &   &  \\
2638 & 1967-11-03 & 39797.55 & 19.4 & B &   & 18.1 & 19.2& 15.6 & 18.7 &   &   &   &  \\
2639 & 1967-11-03 & 39797.57 & 18.6 & B &   & 18.5 &  & 15.6 &   &   &   &   &  \\
2644 & 1967-11-03 & 39798.42 & 19.5 & B & 19.8 & 18.5 & 19.2& 15.7 & 18.6 &   &   &   &  \\
2645 & 1967-11-03 & 39798.45 & 19.4 & B & 19.8 & 18.7 & 18.8& 15.7 &   &   &   &   &  \\
2652 & 1967-11-06 & 39800.57 & 19.8 & B & 19.4 &   & 19.3& 15.8 &   & 18.1 &   &   &  \\
2653 & 1967-11-06 & 39800.59 & 18.1 & B &   & 18.2 &  & 15.7 &   &   &   &   &  \\
2661 & 1967-11-06 & 39801.45 & 19.5 & B &   & 18.5 &  & 15.6 &   &   &   &   &  \\
2662 & 1967-11-07 & 39801.48 & 19.6 & B & 19.7 & 18.1 & 19.4& 15.7 &   &   &   &   &  \\
2670 & 1967-11-07 & 39802.37 & 19.5 & B &   & 18.2 & 19.9& 15.7 & 19.2 &   &   &   &  \\
2671 & 1967-11-07 & 39802.39 & 19.4 & B &   & 18.7 & 19.6& 15.6 & 19.1 &   &   &   &  \\
2672 & 1967-11-07 & 39802.43 & 19.2 & U & 19.3 & 17.5 & 19.0& 15.1 & 18.4 & 18.4 &   &   &  \\
2677 & 1967-11-09 & 39804.45 & 17.6 & B &   &   &  & 15.8 &   &   &   &   &  \\
2678 & 1967-11-10 & 39804.49 & 18.0 & B &   &   &  & 15.9 &   &   &   &   &  \\
2915 & 1969-07-25 & 40427.53 & 18.3 & B &   &   &  &   &   &   &   &   &  \\
2923 & 1969-10-04 & 40499.35 & 19.6 & V &   &   &  &   &   &   &   &   &  \\
2928 & 1969-10-04 & 40499.42 & 19.6 & V &   &   &  &   &   &   &   &   &  \\
2934 & 1969-10-05 & 40500.36 & 18.8 & V &   &   &  &   &   &   &   &   &  \\
2936 & 1969-10-05 & 40500.42 & 19.0 & V &   &   &  &   &   &   &   &   &  \\
2946 & 1969-10-07 & 40501.50 & 19.0 & V &   &   &  &   &   &   &   &   &  \\
2947 & 1969-10-07 & 40501.52 & 19.2 & V &   &   &  &   &   &   &   &   &  \\
2957 & 1969-10-08 & 40503.39 & 18.4 & B &   &   &  &   &   &   &   &   &  \\
2958 & 1969-10-08 & 40503.41 & 19.3 & B &   &   &  &   &   &   &   &   &  \\
2968 & 1969-10-09 & 40504.37 & 19.5 & B &   &   &  &   &   &   &   &   &  \\
2969 & 1969-10-09 & 40504.39 & 19.8 & B &   &   &  &   &   &   &   &   &  \\
2985 & 1969-10-11 & 40506.40 & 19.0 & V &   &   &  &   &   &   &   &   &  \\
2990 & 1969-10-13 & 40508.35 & 19.1 & NFBS &   &   &  &   &   &   &   &   &  \\
2994 & 1969-10-13 & 40508.42 & 19.5 & B &   &   &  &   &   &   & 18.4 &   &  \\
2995 & 1969-10-13 & 40508.44 & 18.7 & B &   &   &  &   &   &   &   &   &  \\
2997 & 1969-10-14 & 40508.48 & 19.4 & NFBS &   &   &  &   &   &   & 18.2 &   &  \\
3013 & 1970-02-06 & 40624.26 & 18.7 & B &   &   &  &   &   &   &   &   &  \\
3014 & 1970-02-06 & 40624.29 & 18.8 & B &   &   &  &   &   &   &   &   &  \\
3024 & 1970-02-09 & 40627.27 & 18.6 & B &   &   &  &   &   &   &   &   &  \\
3035 & 1970-03-03 & 40649.30 & 18.9 & B &   &   &  &   &   &   &   &   &  \\
3059 & 1970-03-09 & 40655.29 & 17.8 & B &   &   &  &   &   &   &   &   &  \\
3138 & 1970-09-10 & 40839.58 & 19.0 & B &   &   &  &   &   &   &   & 18.7 &  \\
3139 & 1970-09-10 & 40839.61 & 18.3 & B &   &   &  &   &   &   &   &   &  \\
3171 & 1970-11-25 & 40916.22 & 18.9 & B &   &   &  &   &   &   &   &   & 18.9\\
3172 & 1970-11-25 & 40916.25 & 19.5 & B &   &   &  &   &   &   &   &   & 19.1\\
3190 & 1970-11-26 & 40917.29 & 19.0 & B &   &   &  &   &   &   &   &   & 18.8\\
3191 & 1970-11-26 & 40917.32 & 19.2 & B &   &   &  &   &   &   &   &   & 19.1\\
3224 & 1971-08-21 & 41184.60 & 18.8 & B &   &   &  &   &   &   &   &   &  \\ \hline
\end{tabular}
\end{table*}
}

%
\onltab{15}{
\begin{table*}
\caption{Nova light curves continued. See notes below Table\,\ref{tab:lightcurves1} for a more detailed description.}
\label{tab:lightcurves6}
\centering
\begin{tabular}{rrrrcrrrrrrrrrrr}\hline\hline
PlNr & Date & RJD & Limit & Band & 71-08a & 72-01b & 72-10a& 72-11a & 72-12b & 73-10a & 73-10b & 73-09b & 74-08a\\ \hline
3191 & 1970-11-26 & 40917.32 & 19.2 & B &   &   &  &   &   &   &   &   &  \\
3224 & 1971-08-21 & 41184.60 & 18.8 & B & 19.3 &   &  &   &   &   &   &   &  \\
3243 & 1971-08-29 & 41193.45 & 19.9 & B &   &   &  &   &   &   &   &   &  \\
3244 & 1971-08-30 & 41193.51 & 18.6 & U &   &   &  &   &   &   &   &   &  \\
3245 & 1971-08-30 & 41193.56 & 20.0 & B &   &   &  &   &   &   &   &   &  \\
3253 & 1971-09-22 & 41217.41 & 19.0 & U &   &   &  &   &   &   &   &   &  \\
3254 & 1971-09-23 & 41217.51 & 19.7 & B &   &   &  &   &   &   &   &   &  \\
3255 & 1971-09-23 & 41217.56 & 19.1 & U &   &   &  &   &   &   &   &   &  \\
3256 & 1971-09-23 & 41217.63 & 19.7 & B &   &   &  &   &   &   &   &   &  \\
3263 & 1971-11-17 & 41273.43 & 18.0 & B &   &   &  &   &   &   &   &   &  \\
3267 & 1971-11-20 & 41275.56 & 18.0 & B &   &   &  &   &   &   &   &   &  \\
3272 & 1971-11-20 & 41276.25 & 19.4 & B &   &   &  &   &   &   &   &   &  \\
3273 & 1971-11-20 & 41276.27 & 19.1 & B &   &   &  &   &   &   &   &   &  \\
3274 & 1971-11-20 & 41276.30 & 17.7 & U &   &   &  &   &   &   &   &   &  \\
3275 & 1971-11-20 & 41276.38 & 18.6 & B &   &   &  &   &   &   &   &   &  \\
3277 & 1971-11-23 & 41278.51 & 19.0 & B &   &   &  &   &   &   &   &   &  \\
3282 & 1972-01-11 & 41328.40 & 17.1 & B &   & 16.7 &  &   &   &   &   &   &  \\
3300 & 1972-01-15 & 41332.24 & 18.2 & U &   & 16.9 &  &   &   &   &   &   &  \\
3309 & 1972-01-16 & 41333.22 & 18.4 & B &   & 17.8 &  &   &   &   &   &   &  \\
3315 & 1972-01-17 & 41334.23 & 17.5 & B &   &   &  &   &   &   &   &   &  \\
3319 & 1972-01-18 & 41335.23 & 18.7 & B &   & 18.7 &  &   &   &   &   &   &  \\
3333 & 1972-01-19 & 41336.26 & 18.3 & B &   &   &  &   &   &   &   &   &  \\
3339 & 1972-01-20 & 41337.32 & 17.8 & U &   &   &  &   &   &   &   &   &  \\
3673 & 1972-10-14 & 41605.43 & 19.2 & B &   &   & 17.5& 18.6 &   &   &   &   &  \\
3675 & 1972-10-30 & 41621.45 & 19.3 & B &   &   &  & 18.5 &   &   &   &   &  \\
3676 & 1972-10-31 & 41621.48 & 18.9 & B &   &   &  & 18.6 &   &   &   &   &  \\
3680 & 1972-11-01 & 41623.31 & 19.3 & B &   &   &  & 18.9 &   &   &   &   &  \\
3703 & 1972-11-14 & 41635.52 & 19.0 & B &   &   &  &   &   &   &   &   &  \\
3734 & 1972-12-14 & 41665.50 & 16.6 & B &   &   &  &   &   &   &   &   &  \\
3746 & 1972-12-28 & 41680.28 & 19.2 & B &   &   &  &   & 18.8 &   &   &   &  \\
3758 & 1972-12-29 & 41681.34 & 18.5 & B &   &   &  &   & 19.0 &   &   &   &  \\
3826 & 1973-10-20 & 41976.33 & 18.6 & U &   &   &  &   &   & 18.5 & 18.2 & 18.3 &  \\
3827 & 1973-10-20 & 41976.38 & 19.3 & B &   &   &  &   &   & 20.0 & 17.7 & 20.2 &  \\
3830 & 1973-10-24 & 41980.33 & 19.2 & B &   &   &  &   &   &   & 16.1 &   &  \\
3837 & 1973-10-27 & 41982.48 & 19.7 & B &   &   &  &   &   &   & 16.8 &   &  \\
3838 & 1973-10-27 & 41982.52 & 19.8 & B &   &   &  &   &   & 19.5 & 17.4 &   &  \\
3839 & 1973-10-27 & 41982.56 & 19.4 & B &   &   &  &   &   &   & 17.0 &   &  \\
3840 & 1973-10-27 & 41982.59 & 18.4 & B &   &   &  &   &   &   & 16.9 &   &  \\
3846 & 1973-10-27 & 41983.42 & 18.5 & V &   &   &  &   &   &   & 16.5 &   &  \\
3849 & 1973-10-28 & 41983.50 & 19.5 & B &   &   &  &   &   & 18.9 & 17.3 &   &  \\
3861 & 1973-10-28 & 41984.42 & 18.5 & V &   &   &  &   &   &   & 16.9 &   &  \\
3868 & 1973-10-31 & 41987.43 & 19.3 & B &   &   &  &   &   & 19.7 & 18.3 & 19.4 &  \\
3870 & 1973-11-01 & 41987.53 & 18.8 & B &   &   &  &   &   &   & 18.1 &   &  \\
3872 & 1973-11-01 & 41987.60 & 18.8 & B &   &   &  &   &   &   & 18.0 &   &  \\
3907 & 1973-11-21 & 42008.43 & 18.7 & B &   &   &  &   &   &   &   &   &  \\
3953 & 1973-12-19 & 42036.33 & 18.4 & B &   &   &  &   &   &   &   &   &  \\
3958 & 1973-12-20 & 42037.33 & 19.0 & B &   &   &  &   &   &   &   & 18.9 &  \\
4172 & 1974-08-16 & 42275.51 & 18.6 & B &   &   &  &   &   &   &   &   & 18.3\\
4175 & 1974-08-16 & 42275.58 & 18.9 & B &   &   &  &   &   &   &   &   & 18.7\\
4188 & 1974-08-23 & 42282.54 & 18.8 & B &   &   &  &   &   &   &   &   &  \\ \hline
\end{tabular}
\end{table*}
}

%
\onltab{16}{
\begin{table*}
\caption{Nova light curves continued. See notes below Table\,\ref{tab:lightcurves1} for a more detailed description.}
\label{tab:lightcurves7}
\centering
\begin{tabular}{rrrrcrrrrrrrrrrr}\hline\hline
PlNr & Date & RJD & Limit & Band & 74-08b & 74-07a & 75-02a& 75-09a & 75-10a & 75-11a & 75-12b & 78-10a & 79-11a\\ \hline
4175 & 1974-08-16 & 42275.58 & 18.9 & B &   &   &  &   &   &   &   &   &  \\
4188 & 1974-08-23 & 42282.54 & 18.8 & B & 18.3 &   &  &   &   &   &   &   &  \\
4196 & 1974-09-11 & 42302.42 & 19.8 & B &   &   &  &   &   &   &   &   &  \\
4197 & 1974-09-11 & 42302.45 & 20.6 & B &   &   &  &   &   &   &   &   &  \\
4198 & 1974-09-12 & 42302.48 & 19.2 & V &   &   &  &   &   &   &   &   &  \\
4209 & 1974-10-14 & 42334.51 & 18.3 & U & 18.4 & 18.5 &  &   &   &   &   &   &  \\
4289 & 1975-02-10 & 42454.25 & 19.0 & B &   &   & 15.7&   &   &   &   &   &  \\
4294 & 1975-02-11 & 42455.28 & 18.4 & B &   &   & 15.6&   &   &   &   &   &  \\
4429 & 1975-08-04 & 42628.56 & 19.3 & B &   &   &  &   &   &   &   &   &  \\
4506 & 1975-09-09 & 42665.40 & 19.3 & B &   &   &  & 18.5 &   &   &   &   &  \\
4511 & 1975-09-29 & 42685.42 & 19.6 & B &   &   &  &   &   &   &   &   &  \\
4512 & 1975-09-29 & 42685.45 & 19.7 & B &   &   &  &   &   &   &   &   &  \\
4518 & 1975-10-11 & 42696.48 & 18.3 & V &   &   &  &   &   &   &   &   &  \\
4528 & 1975-10-29 & 42715.41 & 19.1 & B &   &   &  &   &   &   &   &   &  \\
4532 & 1975-10-30 & 42716.31 & 18.5 & V &   &   &  &   &   &   &   &   &  \\
4533 & 1975-10-30 & 42716.33 & 19.6 & B &   &   &  &   & 19.4 &   &   &   &  \\
4534 & 1975-10-30 & 42716.35 & 18.9 & B &   &   &  &   &   &   &   &   &  \\
4537 & 1975-10-30 & 42716.46 & 18.9 & V &   &   &  &   &   &   &   &   &  \\
4545 & 1975-10-31 & 42717.26 & 19.1 & B &   &   &  &   &   &   &   &   &  \\
4547 & 1975-10-31 & 42717.30 & 18.7 & V &   &   &  &   &   &   &   &   &  \\
4550 & 1975-10-31 & 42717.36 & 18.8 & V &   &   &  &   &   &   &   &   &  \\
4560 & 1975-11-01 & 42718.26 & 19.4 & B &   &   &  &   & 19.1 &   &   &   &  \\
4563 & 1975-11-02 & 42719.27 & 19.3 & B &   &   &  &   &   &   &   &   &  \\
4572 & 1975-11-24 & 42741.25 & 19.2 & V &   &   &  &   &   & 18.6 &   &   &  \\
4575 & 1975-11-24 & 42741.33 & 18.6 & B &   &   &  &   &   & 18.7 &   &   &  \\
4578 & 1975-11-24 & 42741.39 & 18.0 & V &   &   &  &   &   &   &   &   &  \\
4579 & 1975-11-25 & 42742.22 & 18.9 & B &   &   &  &   &   &   &   &   &  \\
4580 & 1975-11-25 & 42742.24 & 18.3 & V &   &   &  &   &   &   &   &   &  \\
4581 & 1975-11-25 & 42742.26 & 18.2 & V &   &   &  &   &   &   &   &   &  \\
4584 & 1975-11-25 & 42742.32 & 18.7 & V &   &   &  &   &   & 18.9 &   &   &  \\
4610 & 1975-12-28 & 42775.22 & 18.5 & V &   &   &  &   &   &   &   &   &  \\
4611 & 1975-12-28 & 42775.24 & 18.8 & V &   &   &  &   &   &   &   &   &  \\
4612 & 1975-12-28 & 42775.26 & 19.4 & B &   &   &  &   &   &   & 18.6 &   &  \\
4613 & 1975-12-28 & 42775.28 & 19.4 & B &   &   &  &   &   &   & 18.7 &   &  \\
4614 & 1975-12-28 & 42775.31 & 18.4 & U &   &   &  &   &   &   & 17.7 &   &  \\
4615 & 1975-12-28 & 42775.35 & 18.7 & U &   &   &  &   &   &   & 17.7 &   &  \\
4651 & 1976-02-02 & 42811.28 & 18.2 & B &   &   &  &   &   &   &   &   &  \\
4824 & 1976-07-30 & 42989.53 & 19.4 & B &   &   &  &   &   &   &   &   &  \\
4889 & 1976-09-23 & 43044.53 & 19.4 & B &   &   &  &   &   &   &   &   &  \\
4898 & 1976-10-21 & 43073.43 & 19.3 & B &   &   &  &   &   &   &   &   &  \\
5200 & 1978-11-03 & 43816.27 & 18.1 & U &   &   &  &   &   &   &   & 17.6 &  \\
5391 & 1979-10-26 & 44172.50 & 18.5 & B &   &   &  &   &   &   &   &   &  \\
5392 & 1979-10-26 & 44172.53 & 17.2 & U &   &   &  &   &   &   &   &   &  \\
5410 & 1979-11-16 & 44194.39 & 19.0 & B &   &   &  &   &   &   &   &   & 19.0\\
5411 & 1979-11-16 & 44194.41 & 18.5 & U &   &   &  &   &   &   &   &   & 18.4\\
5414 & 1979-12-14 & 44222.35 & 19.5 & B &   &   &  &   &   &   &   &   & 19.6\\
5428 & 1980-01-14 & 44253.30 & 19.0 & B &   &   &  &   &   &   &   &   & 19.2\\
5547 & 1980-09-07 & 44489.52 & 19.1 & B &   &   &  &   &   &   &   &   &  \\ \hline
\end{tabular}
\end{table*}
}

%
\onltab{17}{
\begin{table*}
\caption{Nova light curves continued. See notes below Table\,\ref{tab:lightcurves1} for a more detailed description.}
\label{tab:lightcurves8}
\centering
\begin{tabular}{rrrrcrrrrrrrrrrr}\hline\hline
PlNr & Date & RJD & Limit & Band & 80-09b & 80-09a & 80-09c& 80-10a & 84-08a & 85-08b & 85-08a & 85-09g & 85-09b\\ \hline
5428 & 1980-01-14 & 44253.30 & 19.0 & B &   &   &  &   &   &   &   &   &  \\
5547 & 1980-09-07 & 44489.52 & 19.1 & B & 19.0 & 17.7 &  &   &   &   &   &   &  \\
5548 & 1980-09-07 & 44489.54 & 18.2 & U & 18.0 & 17.6 & 18.0&   &   &   &   &   &  \\
5560 & 1980-09-07 & 44490.45 & 18.4 & U & 18.2 & 18.0 & 18.1&   &   &   &   &   &  \\
5595 & 1980-10-10 & 44523.33 & 19.2 & V &   &   &  & 18.5 &   &   &   &   &  \\
5606 & 1980-10-31 & 44544.34 & 18.7 & B &   &   &  &   &   &   &   &   &  \\
5608 & 1980-11-01 & 44545.33 & 19.1 & B &   &   &  &   &   &   &   &   &  \\
5843 & 1982-08-15 & 45196.49 & 18.5 & B &   &   &  &   &   &   &   &   &  \\
6040 & 1984-07-30 & 45911.52 & 18.8 & NFBS &   &   &  &   &   &   &   &   &  \\
6045 & 1984-08-01 & 45913.50 & 19.2 & NFBS &   &   &  &   &   &   &   &   &  \\
6055 & 1984-08-27 & 45940.45 & 19.6 & B &   &   &  &   & 19.3 &   &   &   &  \\
6057 & 1984-08-28 & 45940.52 & 19.6 & B &   &   &  &   & 19.5 &   &   &   &  \\
6061 & 1984-09-28 & 45971.51 & 19.0 & NFBS &   &   &  &   &   &   &   &   &  \\
6063 & 1984-09-28 & 45971.55 & 19.0 & NFBS &   &   &  &   &   &   &   &   &  \\
6125 & 1985-08-21 & 46299.45 & 19.5 & B &   &   &  &   &   & 19.1 & 19.3 & 19.4 &  \\
6126 & 1985-08-22 & 46299.48 & 19.5 & U &   &   &  &   &   & 18.5 & 18.6 & 18.4 & 18.6\\
6132 & 1985-08-24 & 46301.50 & 19.6 & B &   &   &  &   &   & 19.5 & 19.7 & 19.2 &  \\
6133 & 1985-08-24 & 46301.54 & 19.0 & U &   &   &  &   &   & 18.6 & 19.1 & 18.2 & 19.2\\
6137 & 1985-09-13 & 46321.54 & 19.8 & B &   &   &  &   &   & 20.4 & 19.9 & 19.9 &  \\
6138 & 1985-09-13 & 46321.58 & 18.9 & U &   &   &  &   &   &   & 19.1 & 19.0 &  \\
6277 & 1986-09-10 & 46683.47 & 19.5 & NFBS &   &   &  &   &   &   &   &   &  \\ \hline
\end{tabular}
\end{table*}
}

%
\onltab{18}{
\begin{table*}
\caption{Nova light curves continued. See notes below Table\,\ref{tab:lightcurves1} for a more detailed description.}
\label{tab:lightcurves9}
\centering
\begin{tabular}{rrrrcrrrrrrrrrrr}\hline\hline
PlNr & Date & RJD & Limit & Band & 85-08c & 85-09h & 87-08a & 92-12b & 92-09a & 95-11c & 96-08g & 96-08e & 96-08d\\ \hline
6125 & 1985-08-21 & 46299.45 & 19.5 & B &   &   &   &   &   &   &   &   &  \\
6126 & 1985-08-22 & 46299.48 & 19.5 & U & 18.8 &   &   &   &   &   &   &   &  \\
6132 & 1985-08-24 & 46301.50 & 19.6 & B & 20.5 &   &   &   &   &   &   &   &  \\
6133 & 1985-08-24 & 46301.54 & 19.0 & U & 19.0 &   &   &   &   &   &   &   &  \\
6137 & 1985-09-13 & 46321.54 & 19.8 & B &   & 17.9 &   &   &   &   &   &   &  \\
6138 & 1985-09-13 & 46321.58 & 18.9 & U &   & 17.4 &   &   &   &   &   &   &  \\
6277 & 1986-09-10 & 46683.47 & 19.5 & NFBS &   &   &   &   &   &   &   &   &  \\
6278 & 1986-09-10 & 46683.49 & 19.7 & NFBS &   &   &   &   &   &   &   &   &  \\
6379 & 1986-11-29 & 46764.45 & 18.9 & B &   &   &   &   &   &   &   &   &  \\
6518 & 1987-08-31 & 47039.43 & 19.9 & B &   &   & 17.9 &   &   &   &   &   &  \\
6523 & 1987-09-04 & 47042.57 & 19.9 & B &   &   & 19.0 &   &   &   &   &   &  \\
8089 & 1992-08-06 & 48840.51 & 20.0 & B &   &   &   &   &   &   &   &   &  \\
8095 & 1992-08-27 & 48862.44 & 19.3 & B &   &   &   &   &   &   &   &   &  \\
8149 & 1992-09-25 & 48891.39 & 19.3 & B &   &   &   & 18.3 &   &   &   &   &  \\
8171 & 1992-09-28 & 48893.49 & 20.2 & B &   &   &   & 18.0 & 16.5 &   &   &   &  \\
8183 & 1992-09-29 & 48894.47 & 19.9 & B &   &   &   & 18.3 & 17.4 &   &   &   &  \\
8840 & 1995-01-27 & 49745.30 & 17.9 & B &   &   &   &   &   &   &   &   &  \\
9090 & 1995-11-26 & 50048.34 & 19.8 & B &   &   &   &   &   & 20.4 &   &   &  \\
9145 & 1996-08-20 & 50315.51 & 19.4 & B &   &   &   &   &   &   & 17.3 & 19.5 & 17.3\\
9150 & 1996-08-21 & 50316.53 & 19.5 & B &   &   &   &   &   &   & 16.4 & 19.5 & 17.8\\
9162 & 1996-12-14 & 50432.35 & 18.7 & B &   &   &   &   &   &   &   &   &  \\ \hline
\end{tabular}
\end{table*}
}

%
\onltab{19}{
\begin{table*}
\caption{Nova light curves continued. See notes below Table\,\ref{tab:lightcurves1} for a more detailed description.}
\label{tab:lightcurves10}
\begin{tabular}{rrrrcrrrrrrrrrrr}\hline\hline
PlNr & Date & RJD & Limit & Band & 96-12b\\ \hline
9150 & 1996-08-21 & 50316.53 & 19.5 & B &  \\
9162 & 1996-12-14 & 50432.35 & 18.7 & B & 17.8\\ \hline
\end{tabular}
\end{table*}
}

\onlfig{8}{
\begin{figure*}[t]
\centering
\subfigure[Nova 1 (B)\label{fig:chart1}]{\includegraphics[scale=.26, angle=0]{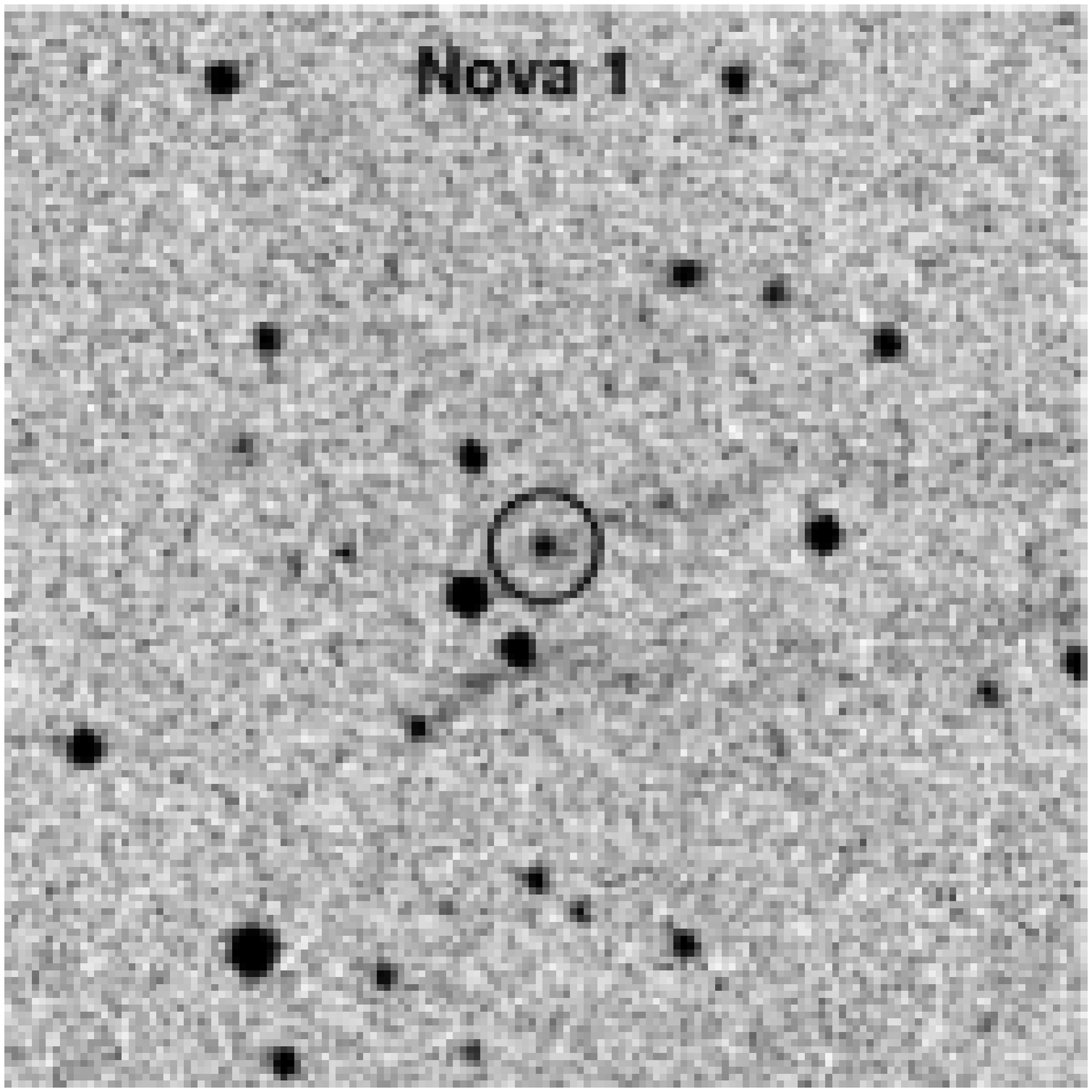}}\qquad
\subfigure[Nova 2 (B)]{\includegraphics[scale=.26, angle=0]{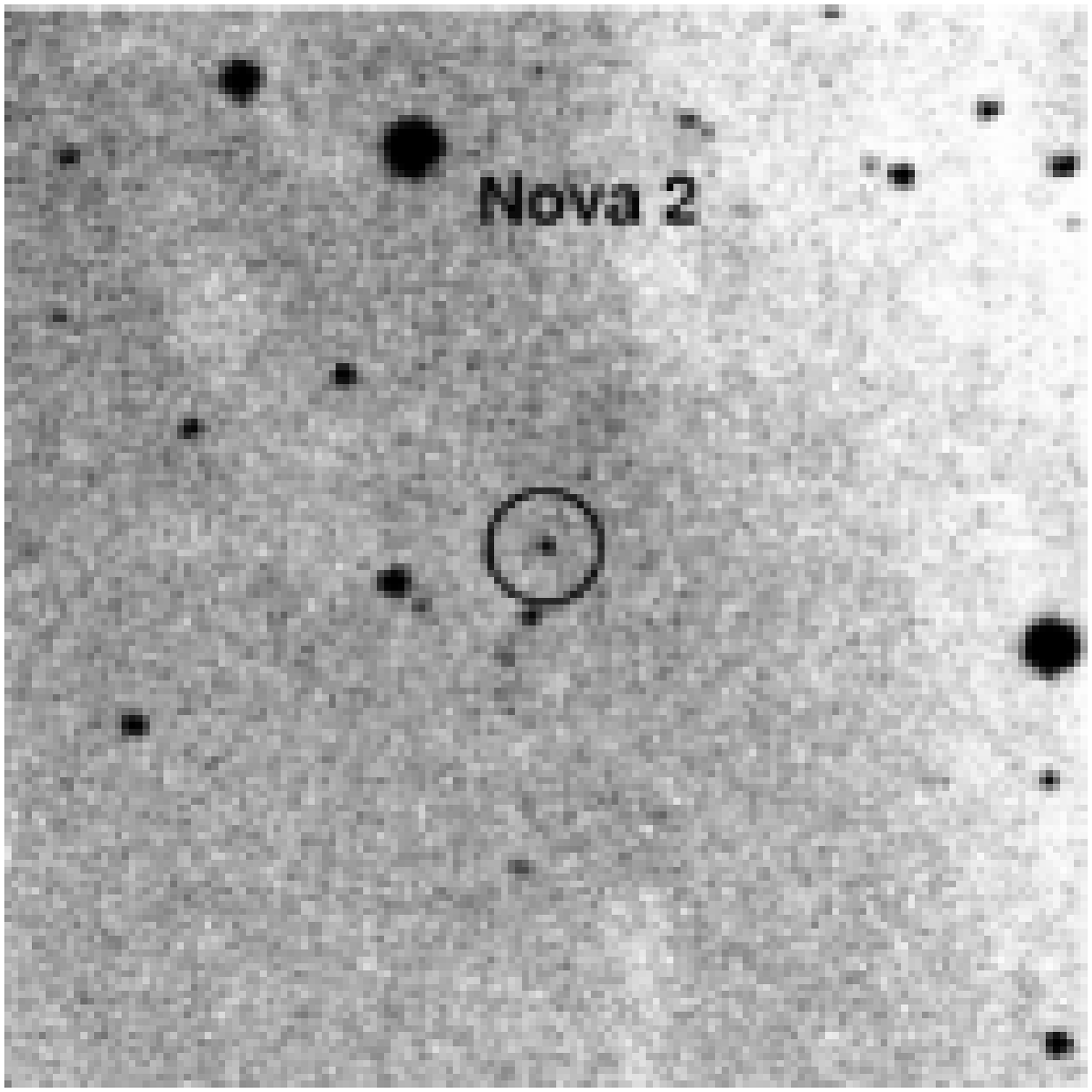}}\qquad
\subfigure[Nova 3 (U)]{\includegraphics[scale=.26, angle=0]{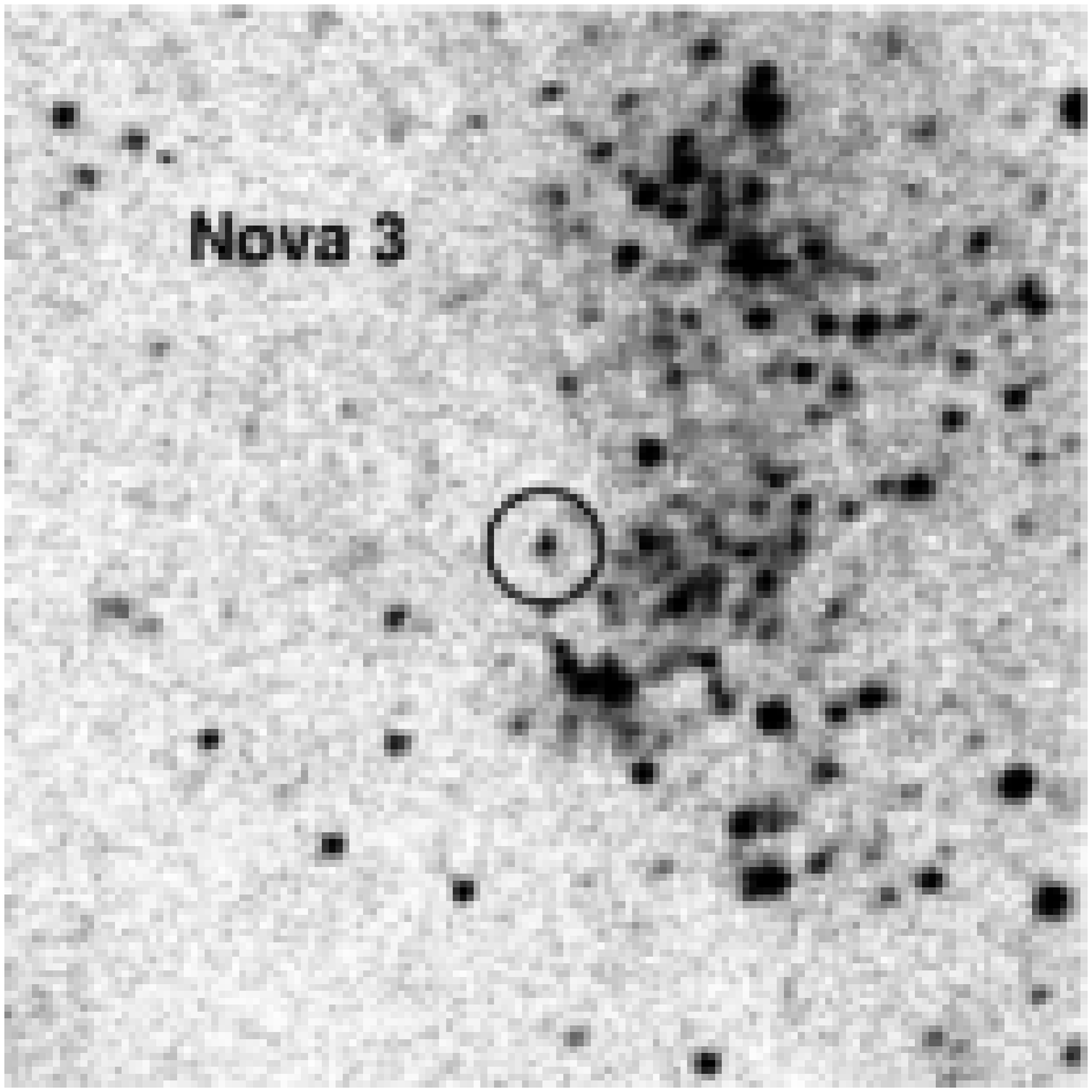}}\\
\subfigure[Nova 4 (U)]{\includegraphics[scale=.26, angle=0]{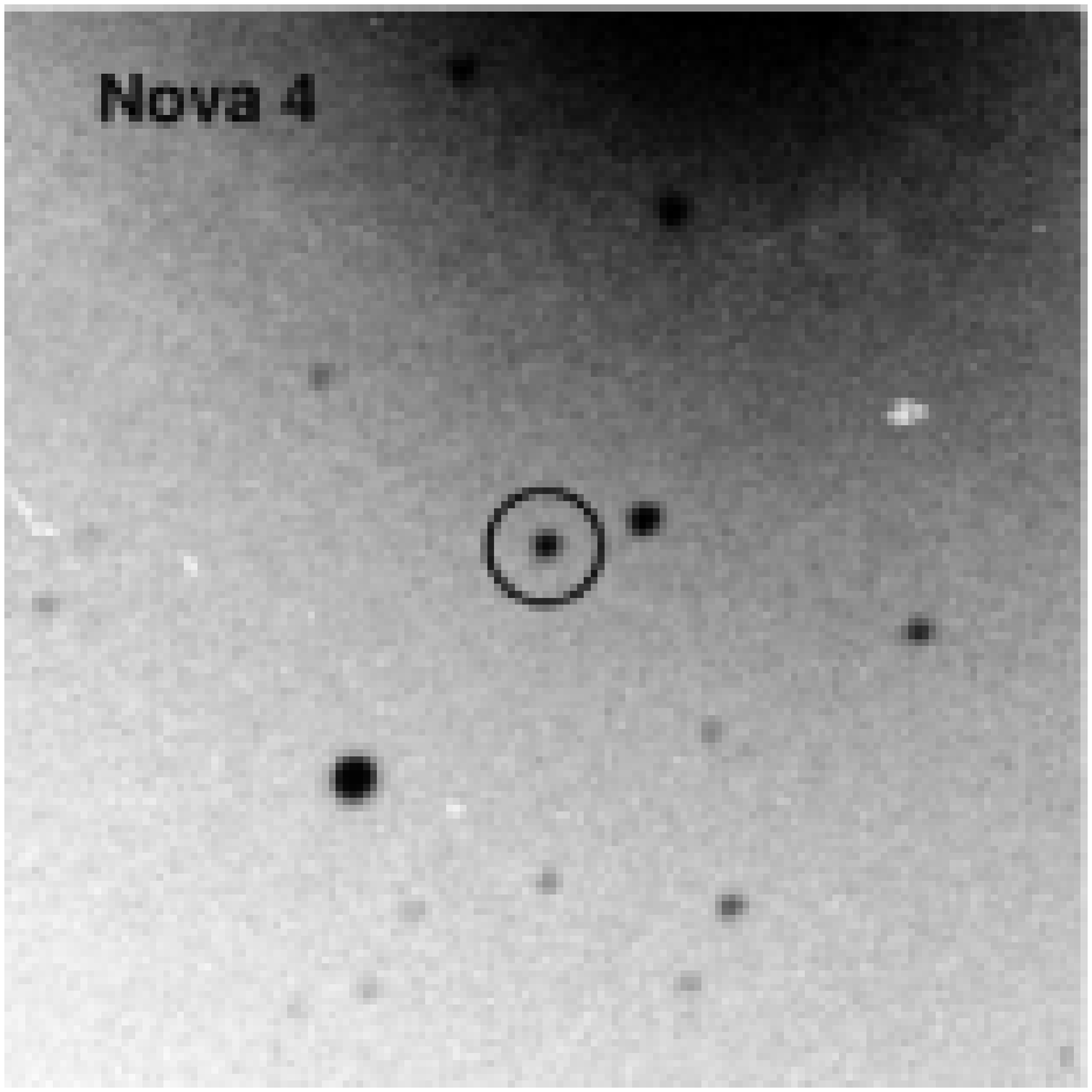}}\qquad
\subfigure[Nova 5 (U)]{\includegraphics[scale=.26, angle=0]{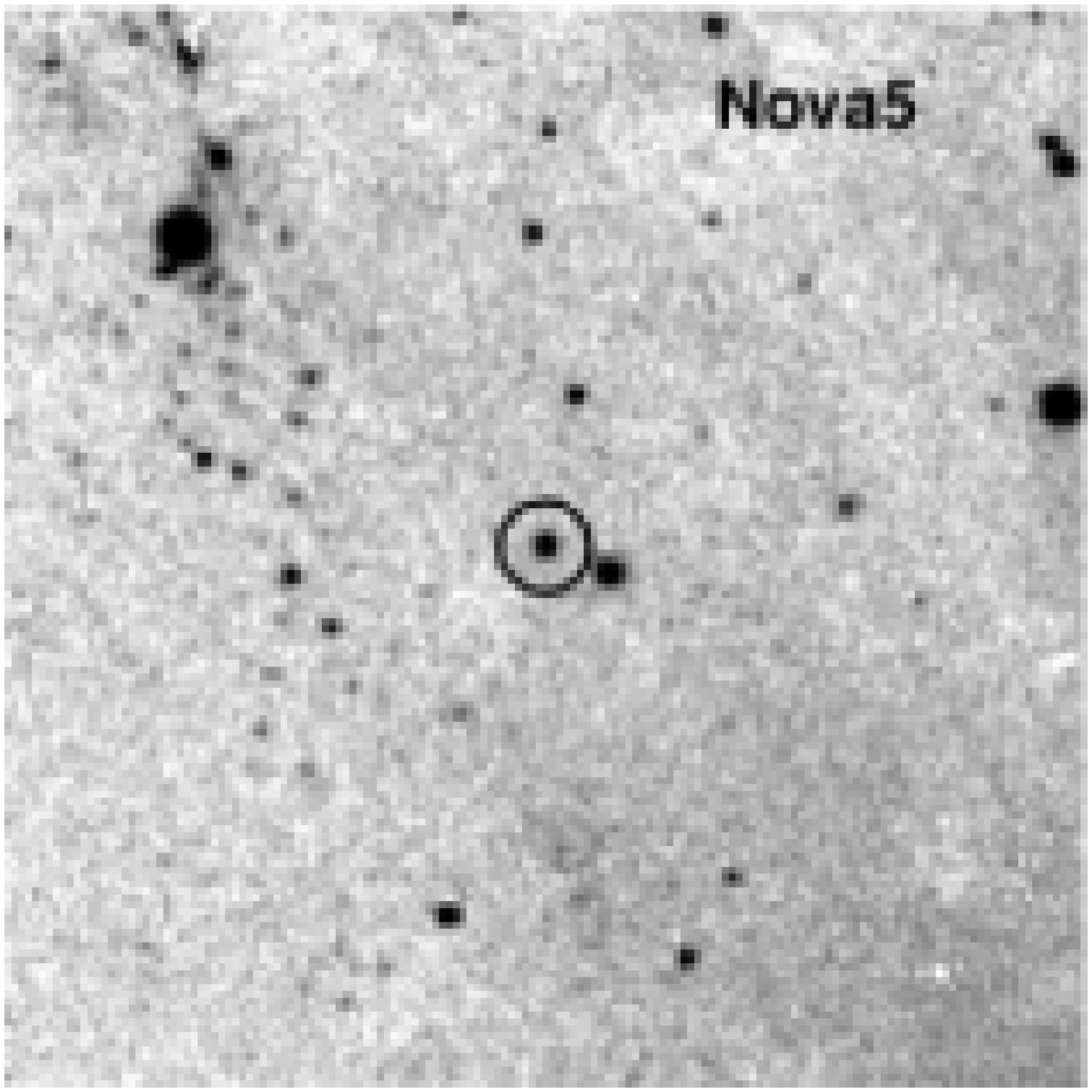}}\qquad
\subfigure[Nova 6 (V)]{\includegraphics[scale=.26, angle=0]{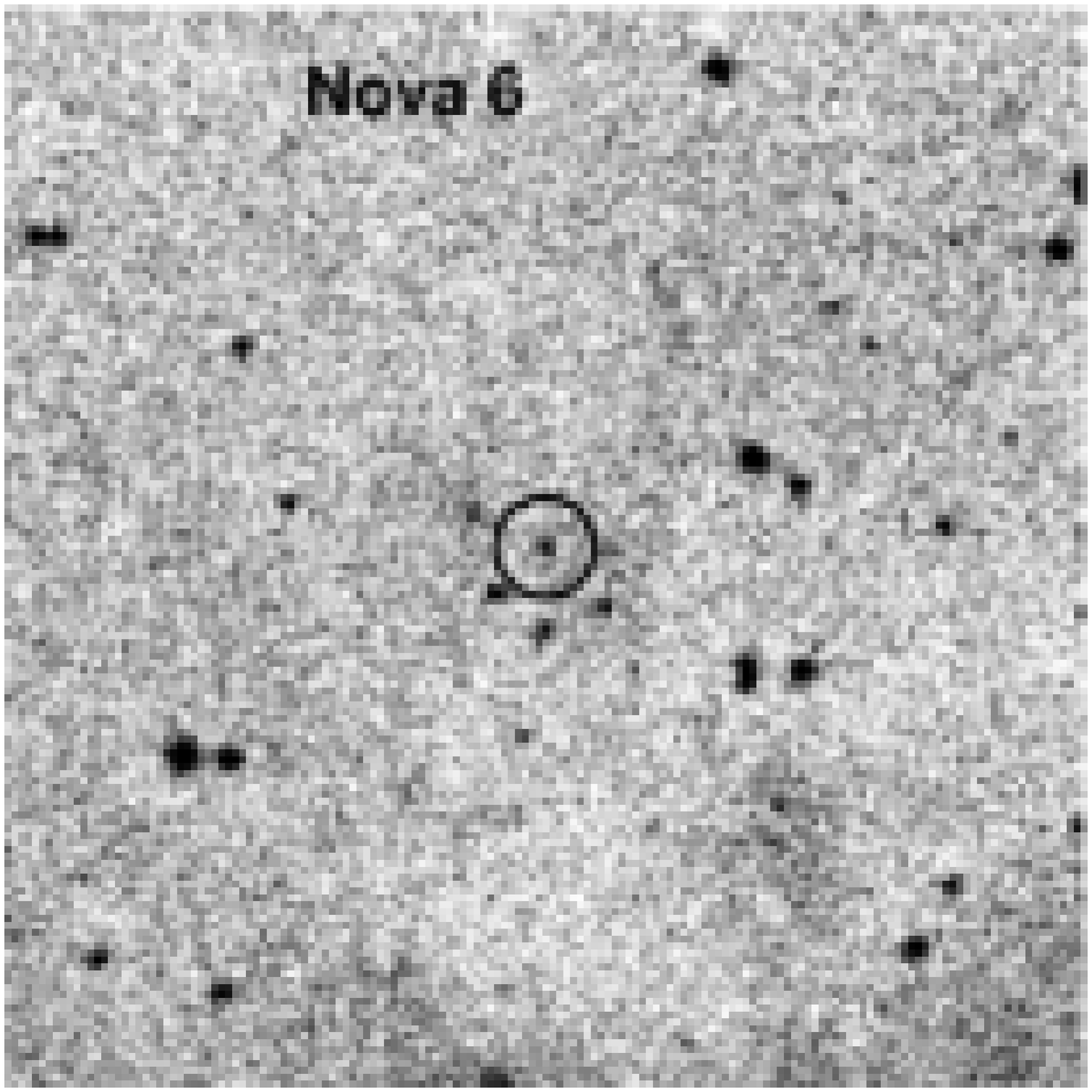}}\\
\subfigure[Nova 7 (B)]{\includegraphics[scale=.26, angle=0]{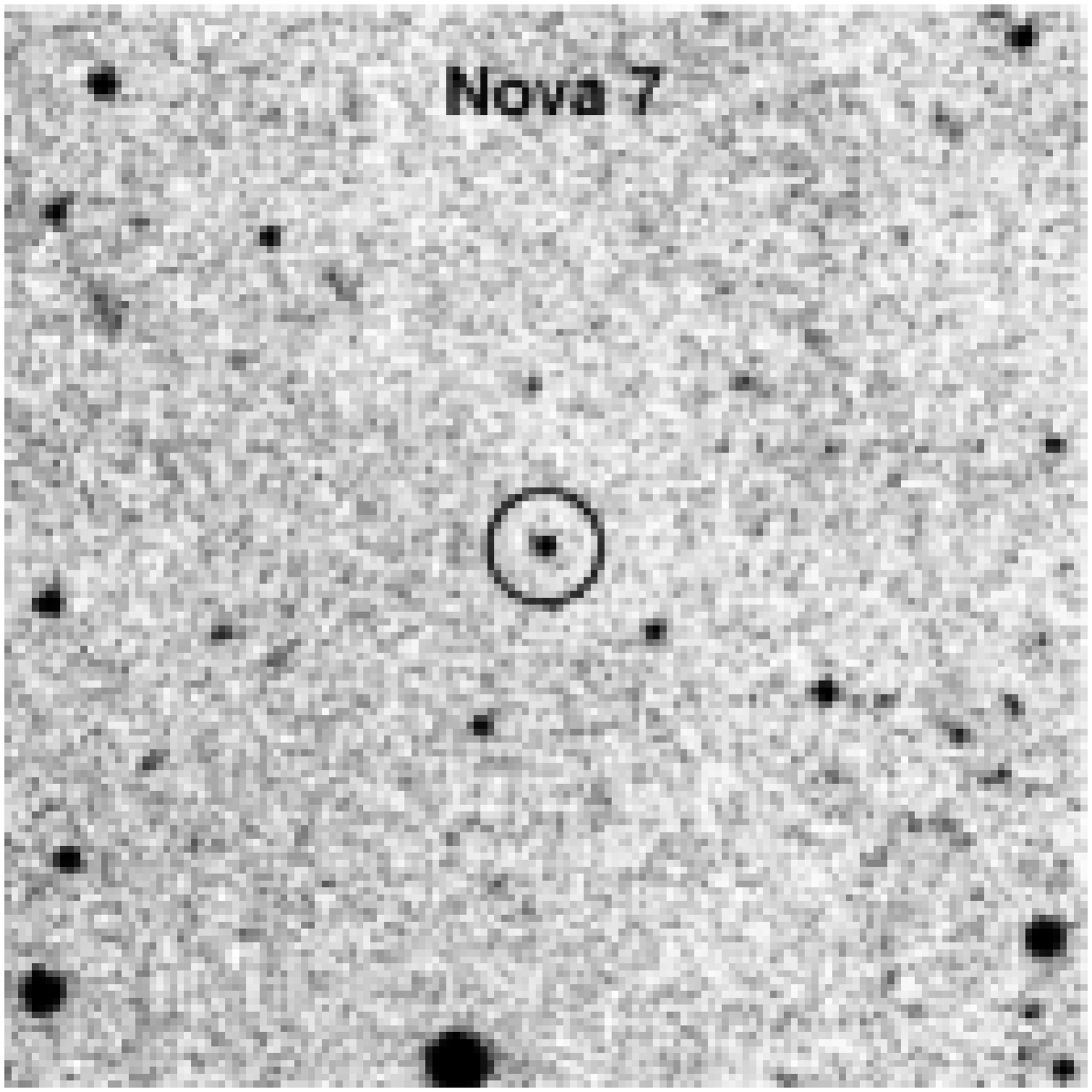}}\qquad
\subfigure[Nova 8 (B)]{\includegraphics[scale=.26, angle=0]{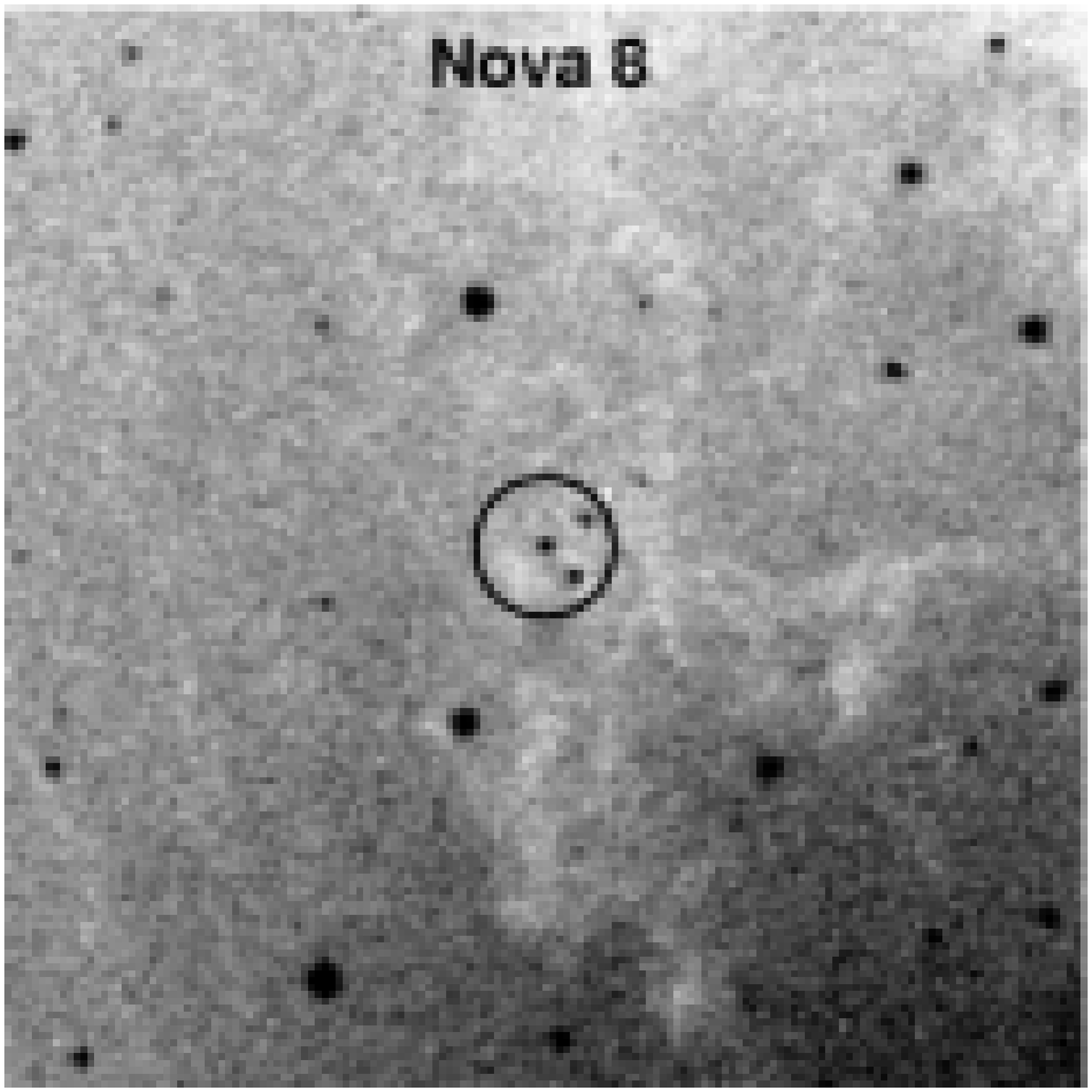}}\qquad
\subfigure[Nova 9 (V)]{\includegraphics[scale=.26, angle=0]{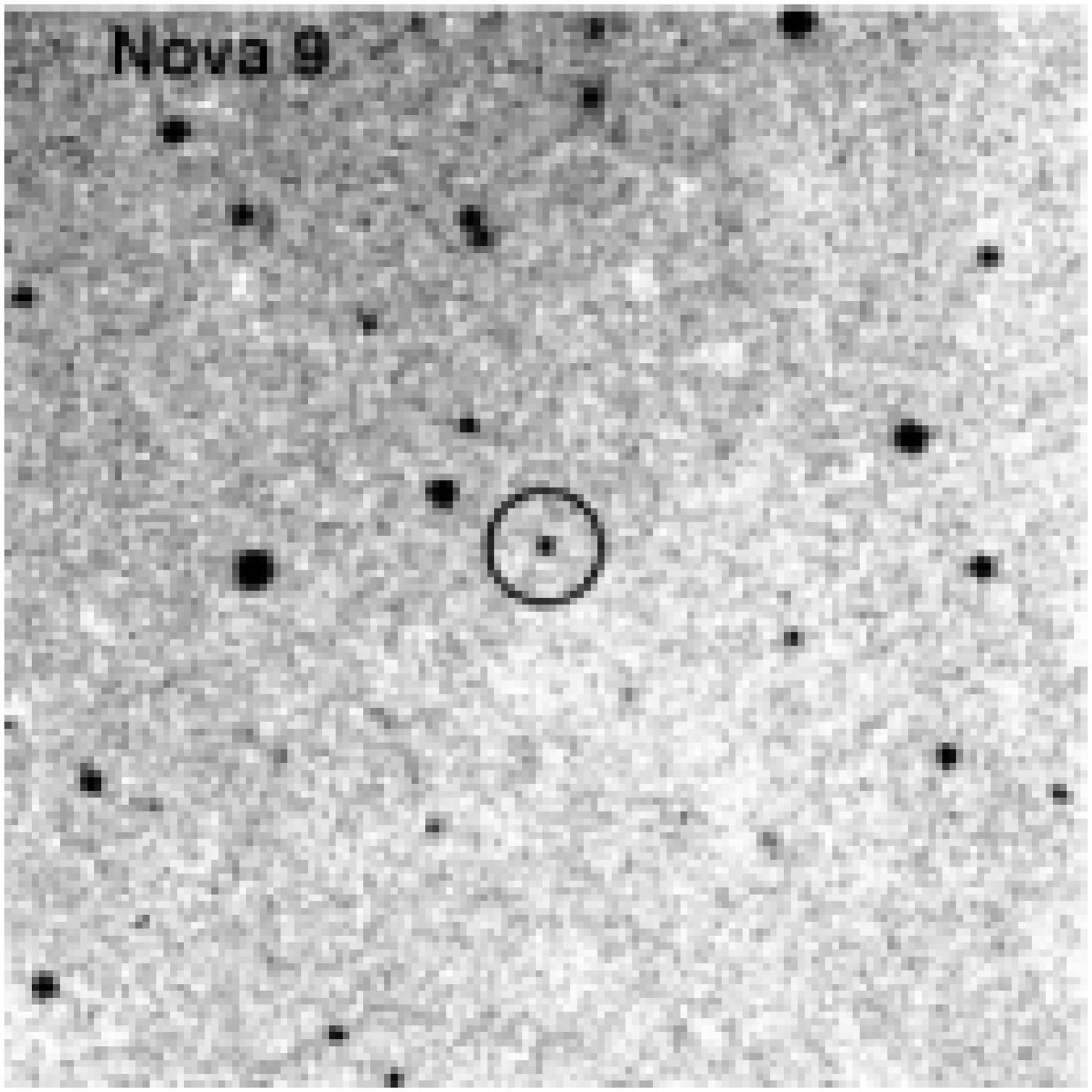}}\\
\subfigure[Nova 10 (U)]{\includegraphics[scale=.26, angle=0]{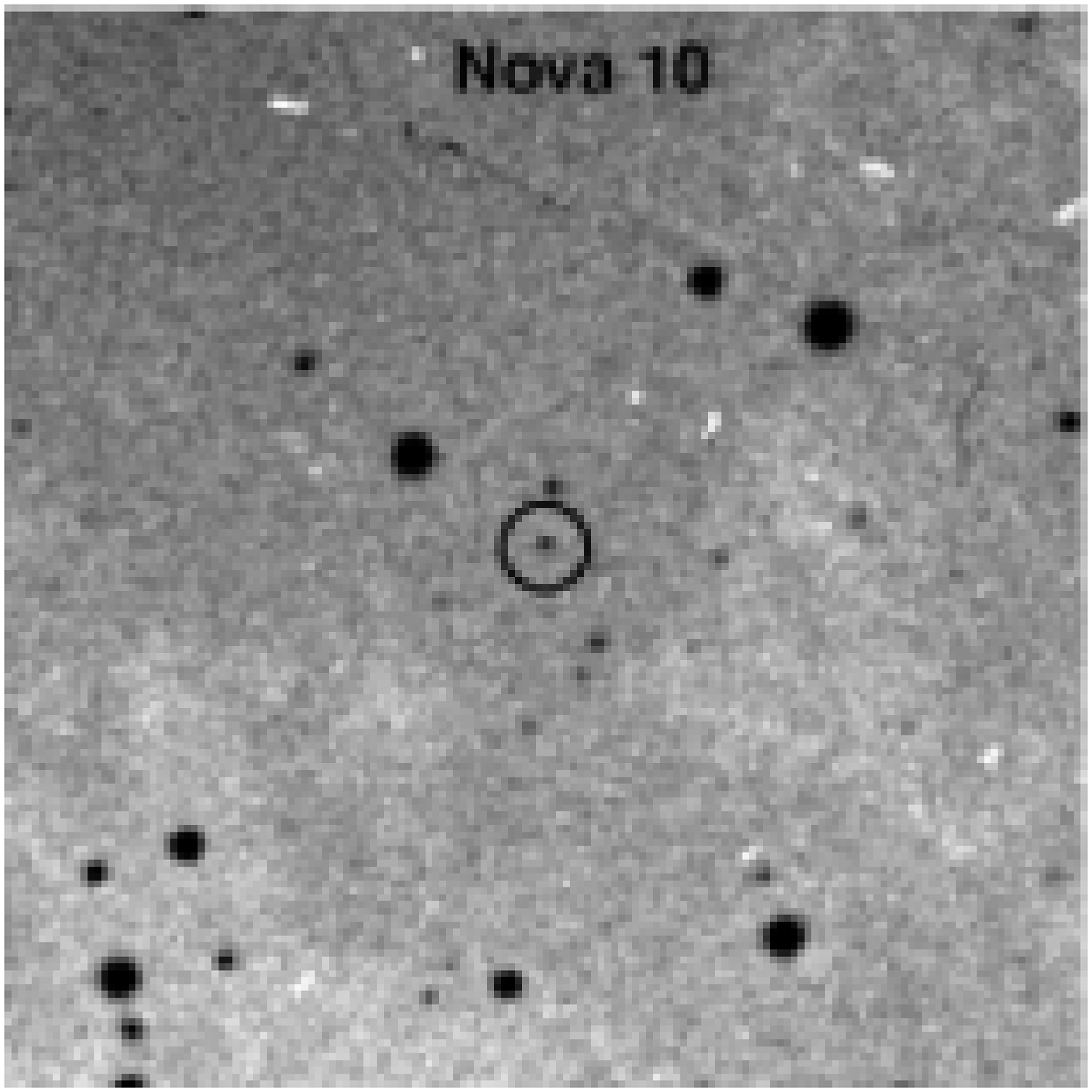}}\qquad
\subfigure[Nova 11 (B)]{\includegraphics[scale=.26, angle=0]{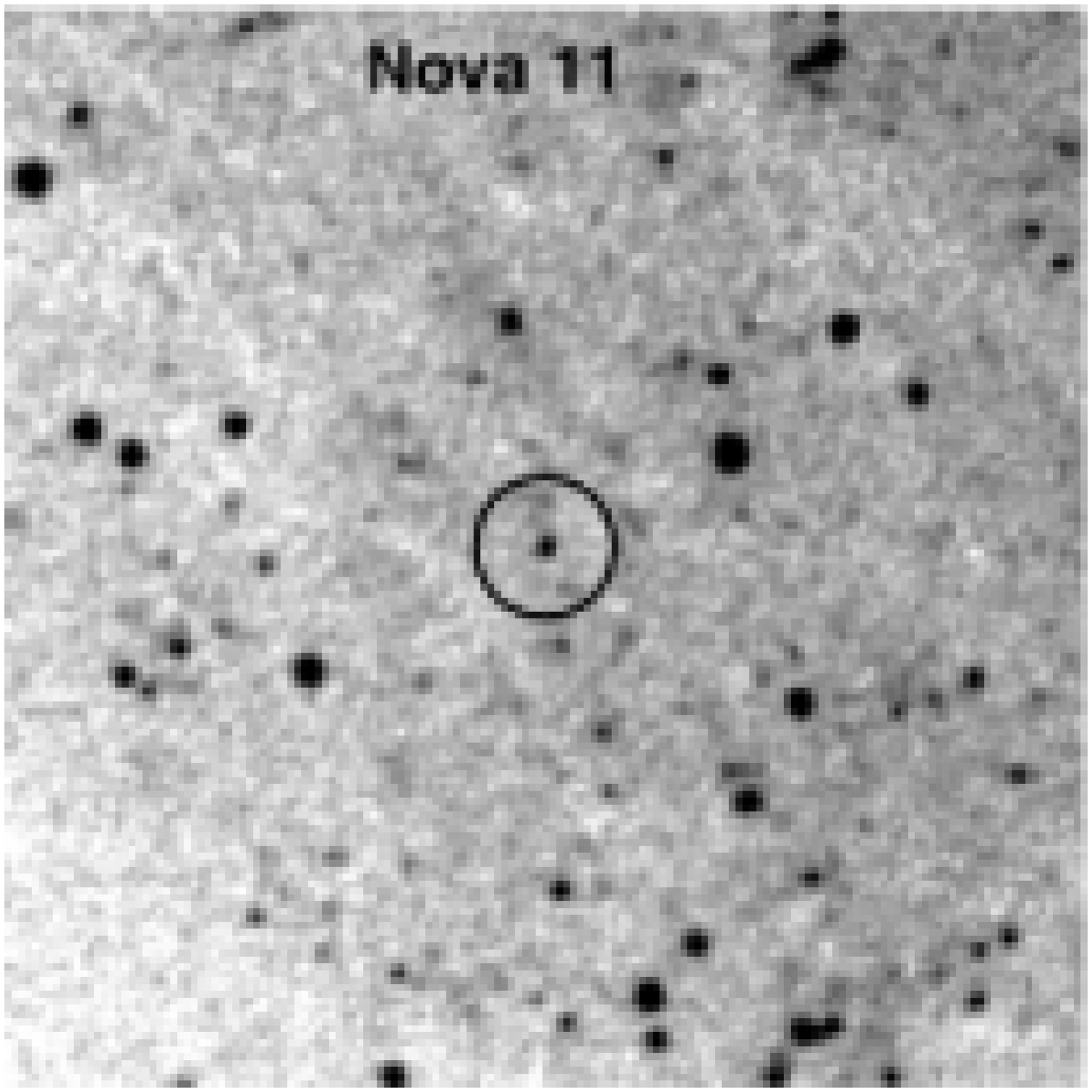}}\qquad
\subfigure[Nova 12 (U)]{\includegraphics[scale=.26, angle=0]{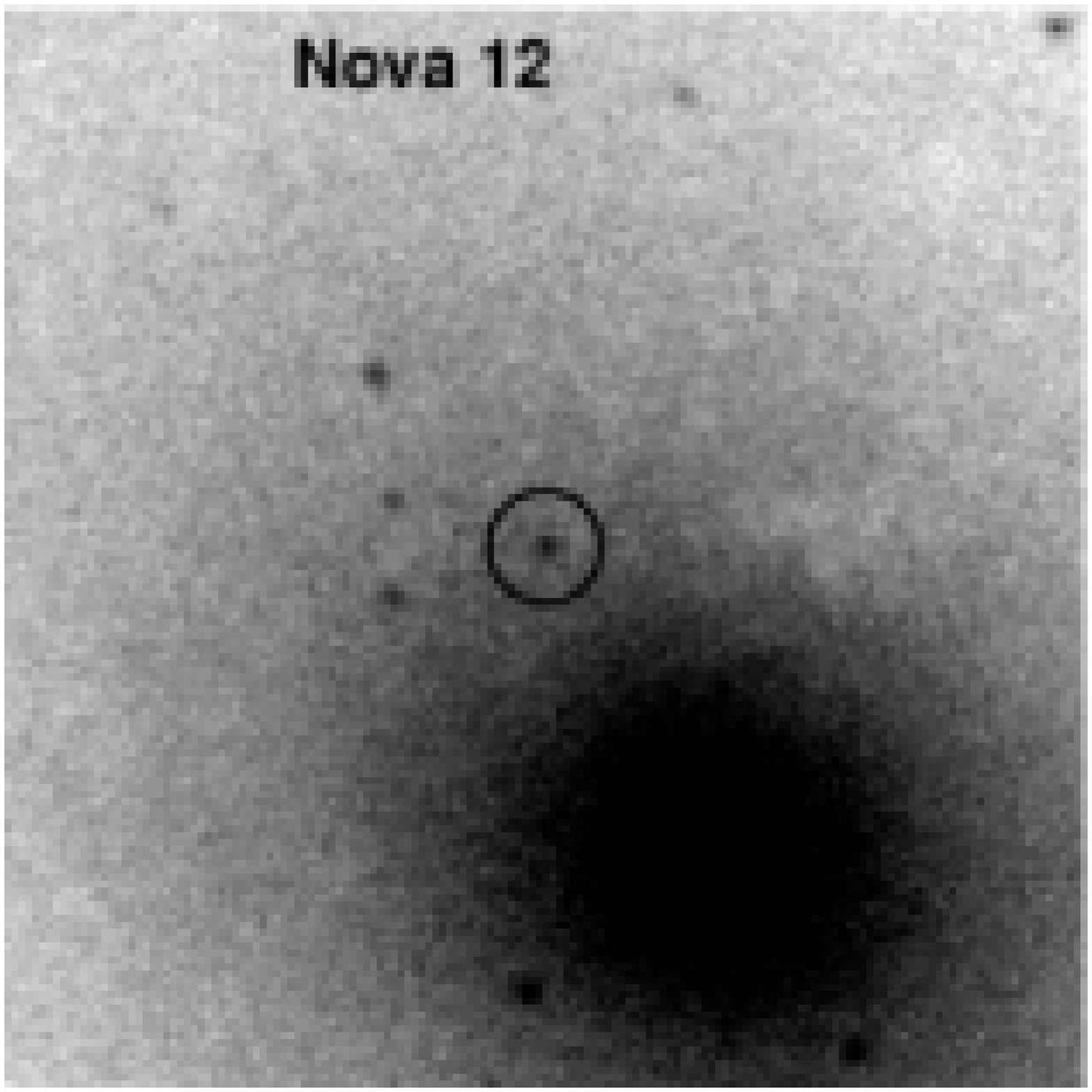}}
\caption{Finding charts for novae 1 - 12.}
\end{figure*}
}

\onlfig{9}{
\begin{figure*}[t]
\centering
\subfigure[Nova 13 (U)]{\includegraphics[scale=.26, angle=0]{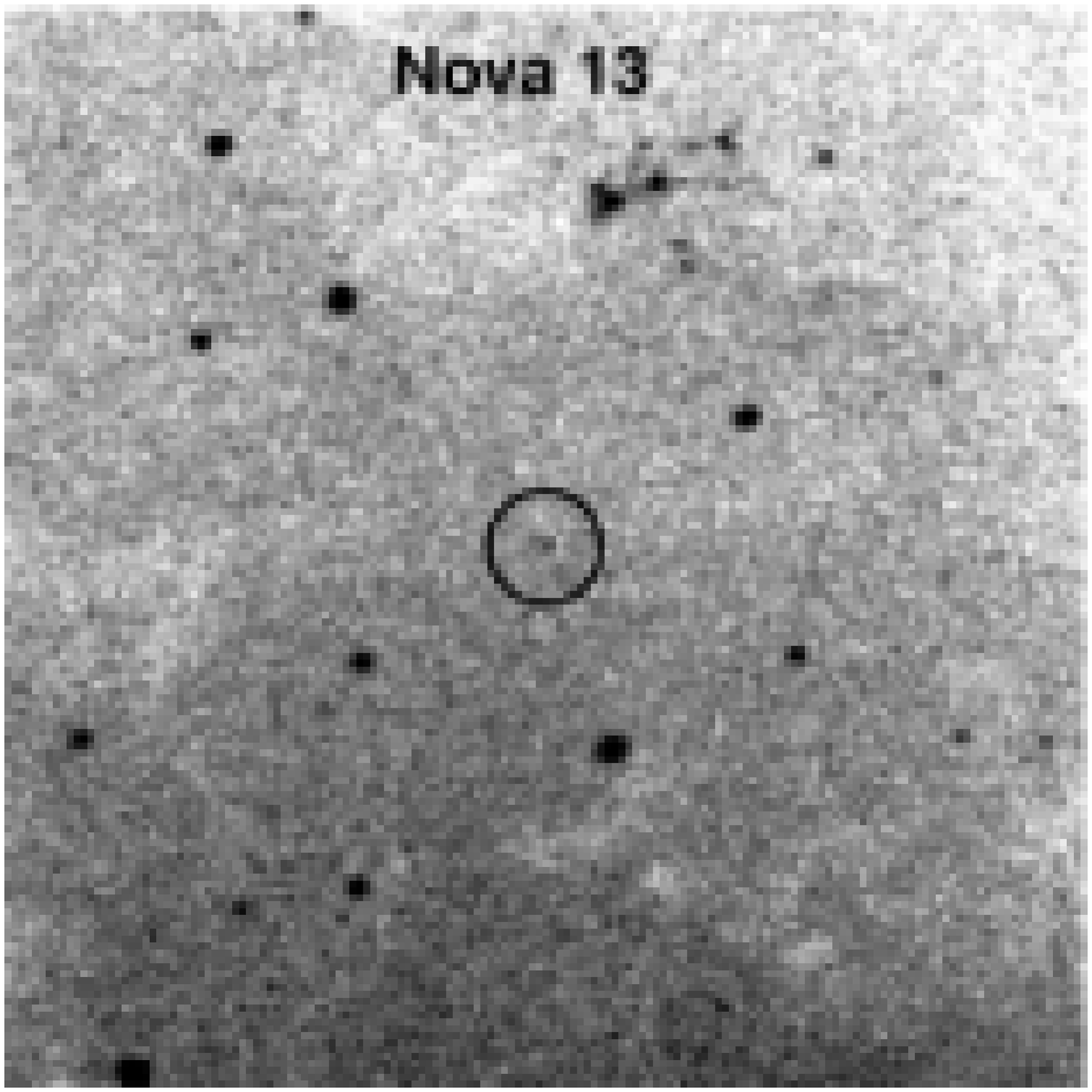}}\qquad
\subfigure[Nova 14 (U)]{\includegraphics[scale=.26, angle=0]{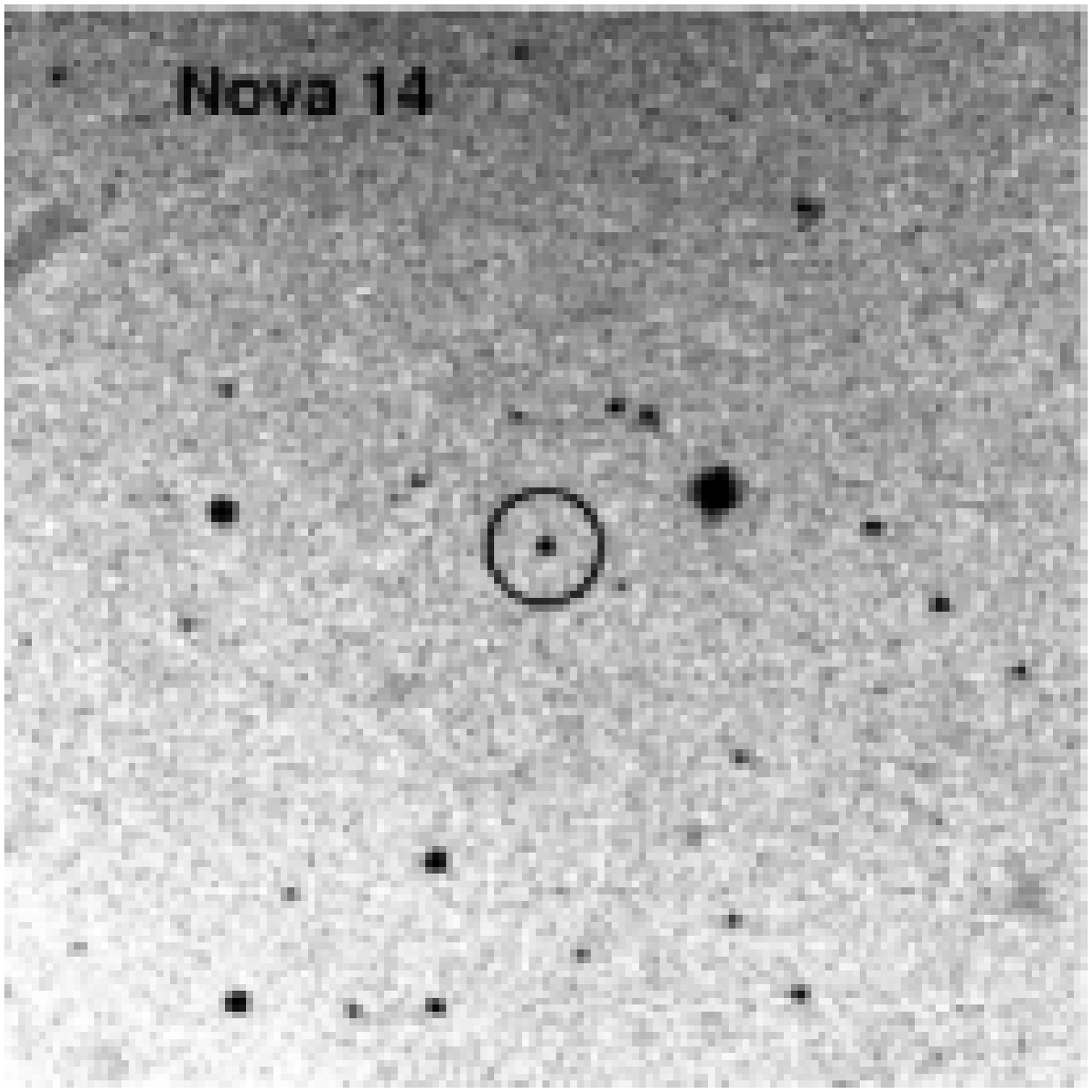}}\qquad
\subfigure[Nova 15 (B)]{\includegraphics[scale=.26, angle=0]{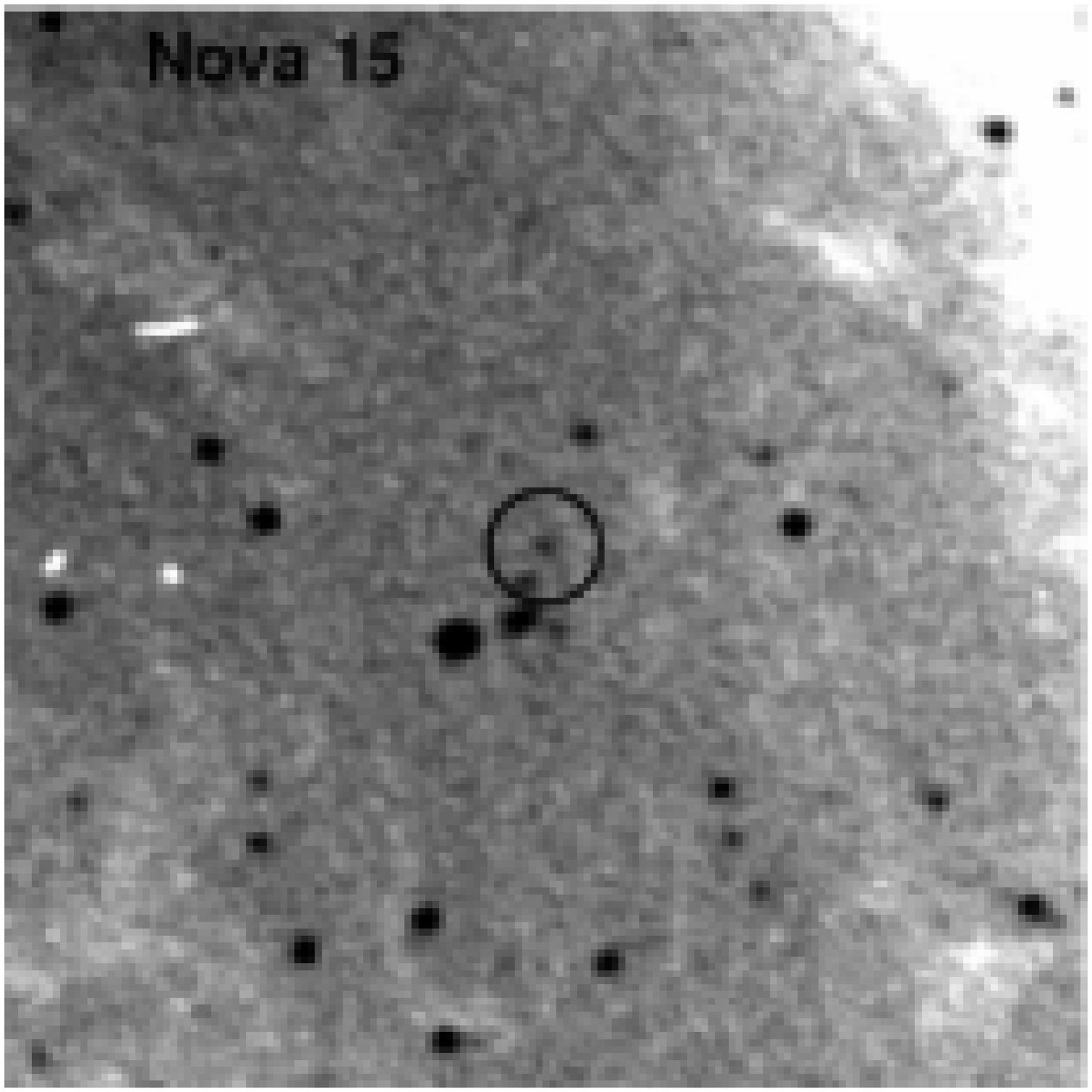}}\\
\subfigure[Nova 16 (U)]{\includegraphics[scale=.26, angle=0]{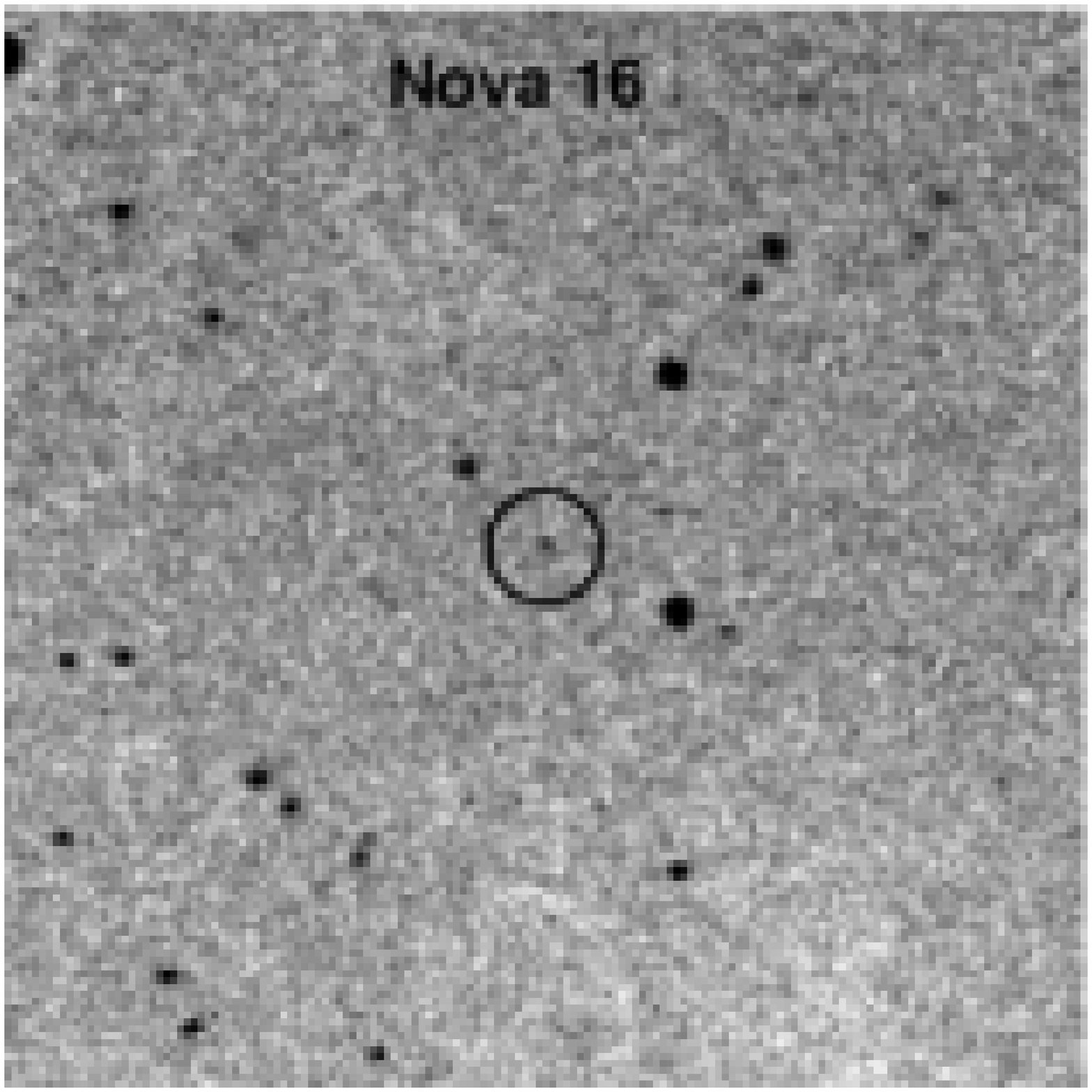}}\qquad
\subfigure[Nova 17 (B)]{\includegraphics[scale=.26, angle=0]{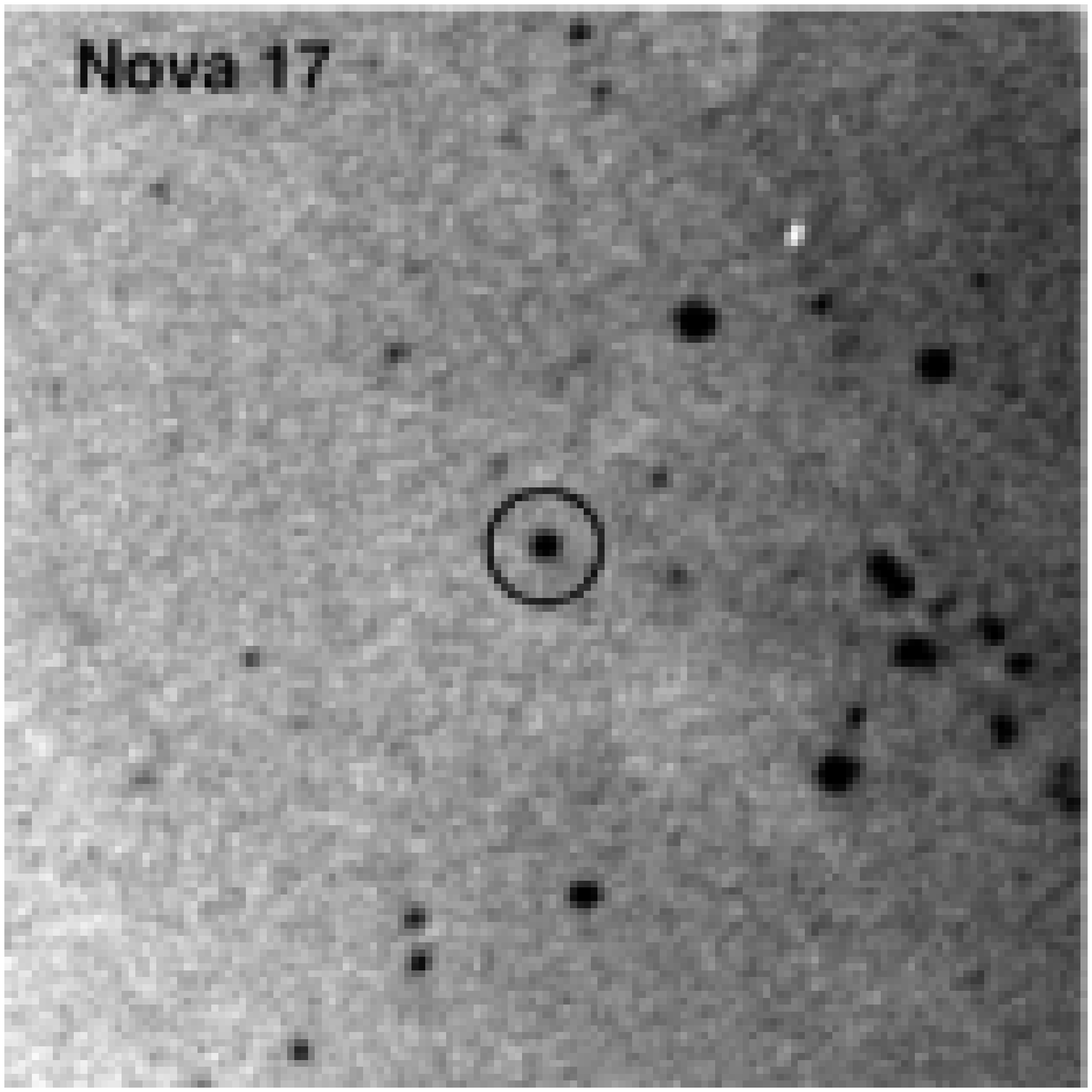}}\qquad
\subfigure[Nova 18 (U)]{\includegraphics[scale=.26, angle=0]{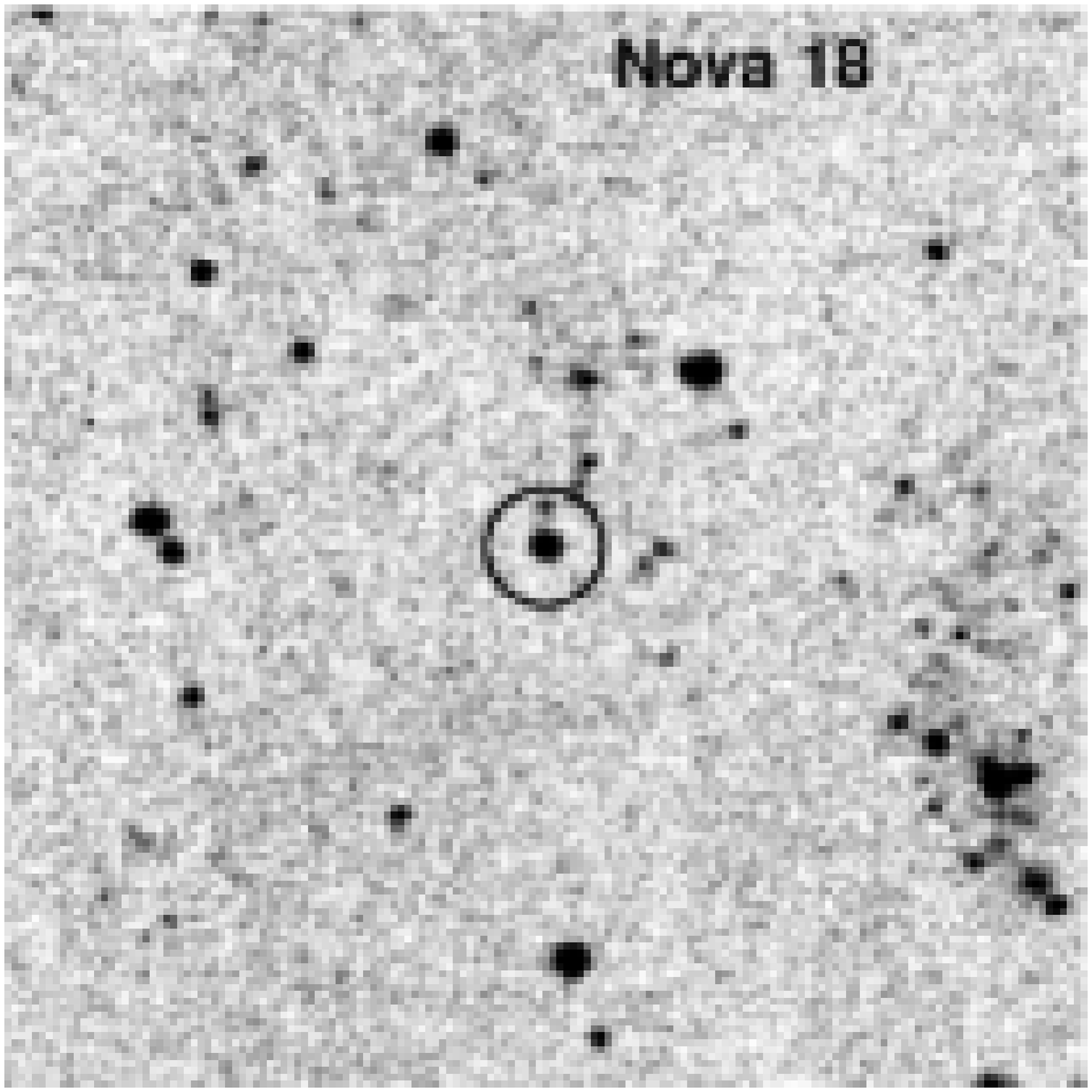}}\\
\subfigure[Nova 19 (B)]{\includegraphics[scale=.26, angle=0]{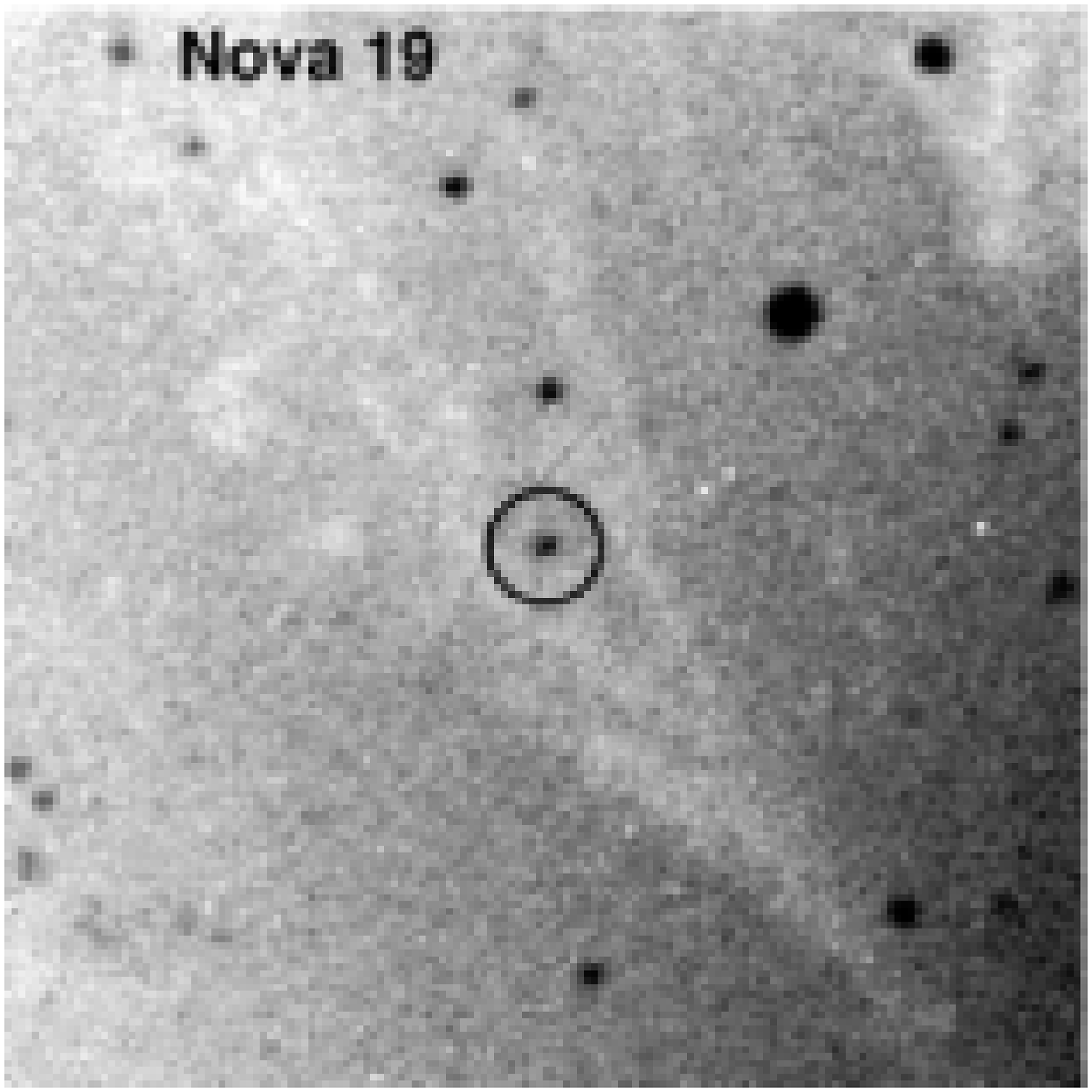}}\qquad
\subfigure[Nova 20 (B)]{\includegraphics[scale=.26, angle=0]{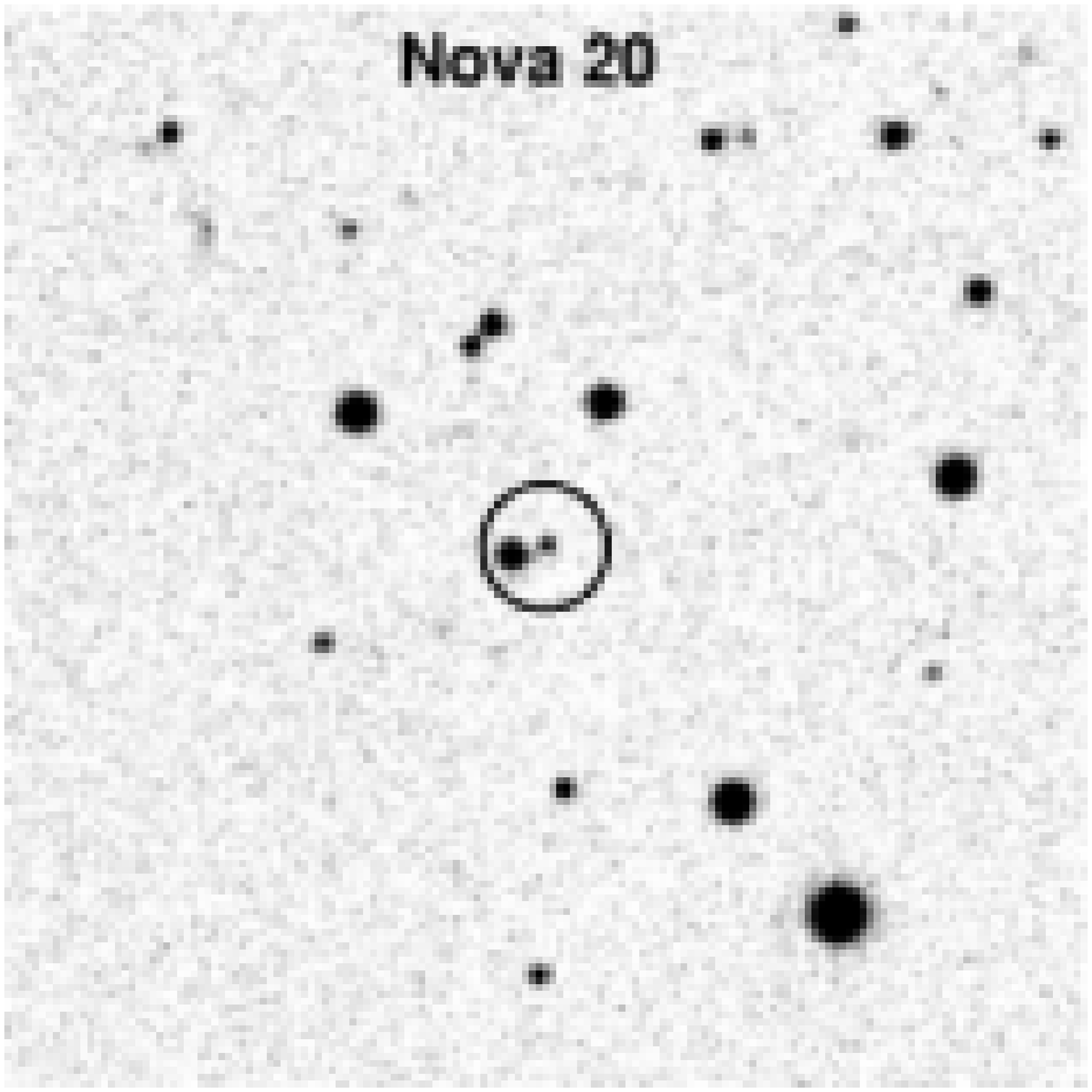}}\qquad
\subfigure[Nova 21 (B)]{\includegraphics[scale=.26, angle=0]{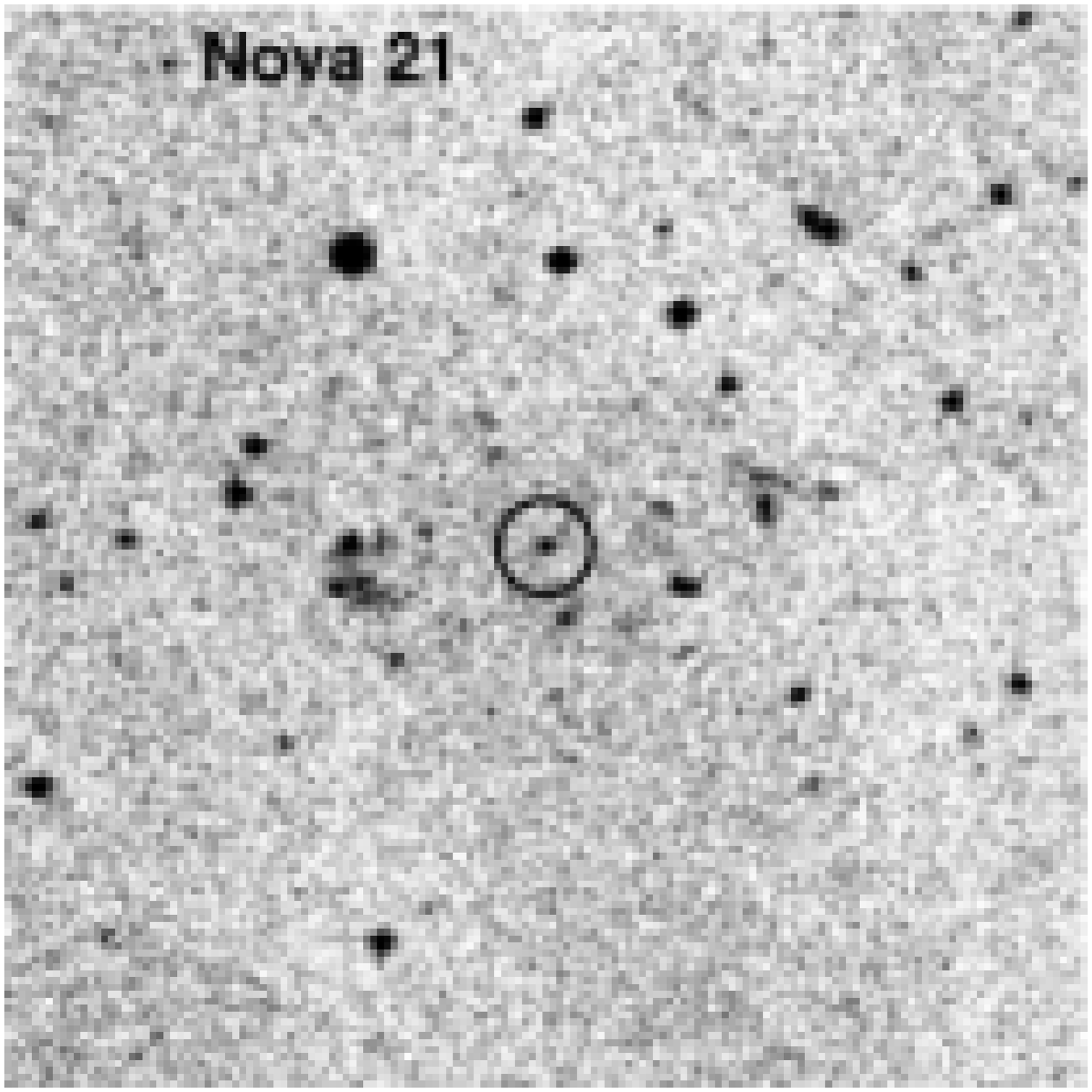}}\\
\subfigure[Nova 22 (B)]{\includegraphics[scale=.26, angle=0]{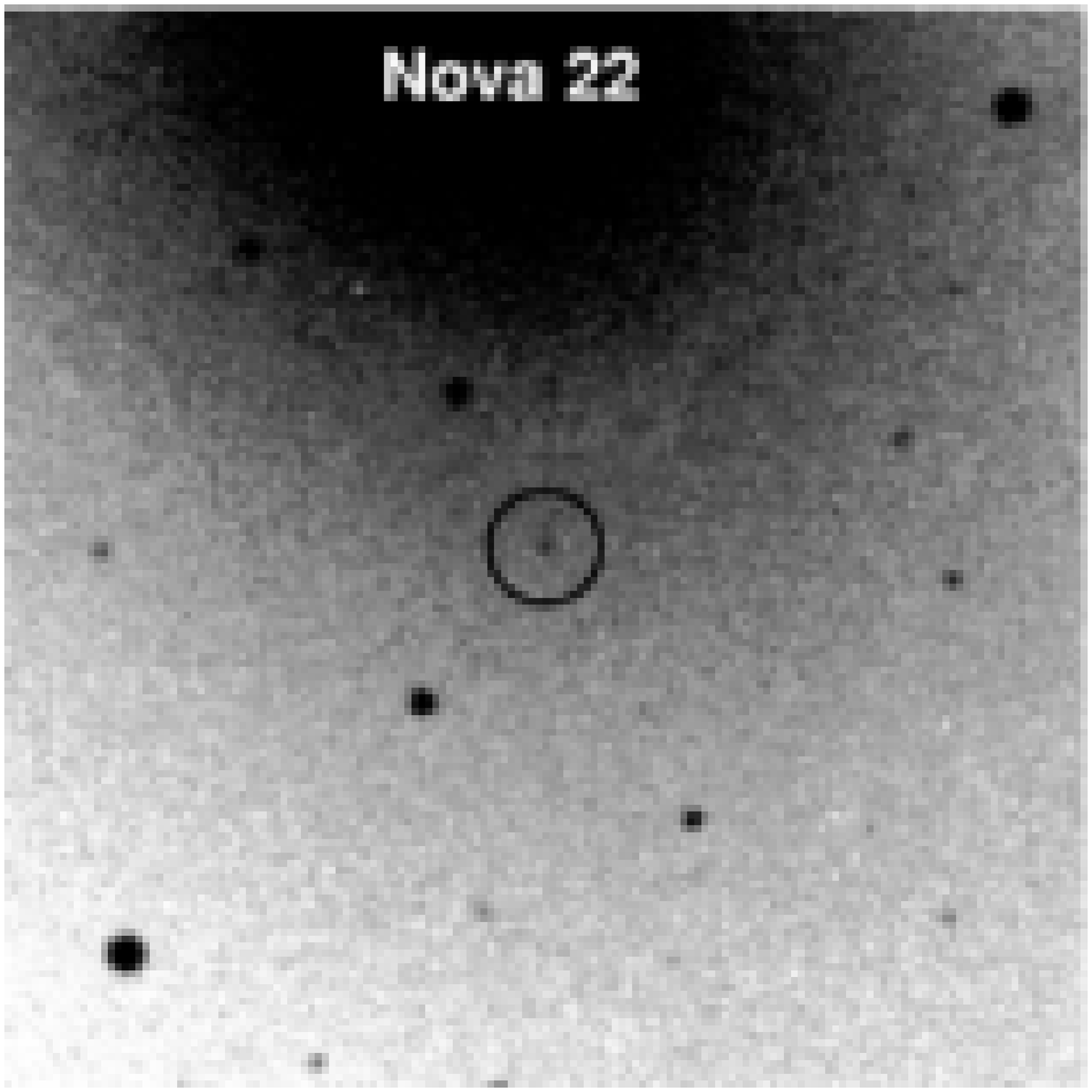}}\qquad
\subfigure[Nova 23 (B)]{\includegraphics[scale=.26, angle=0]{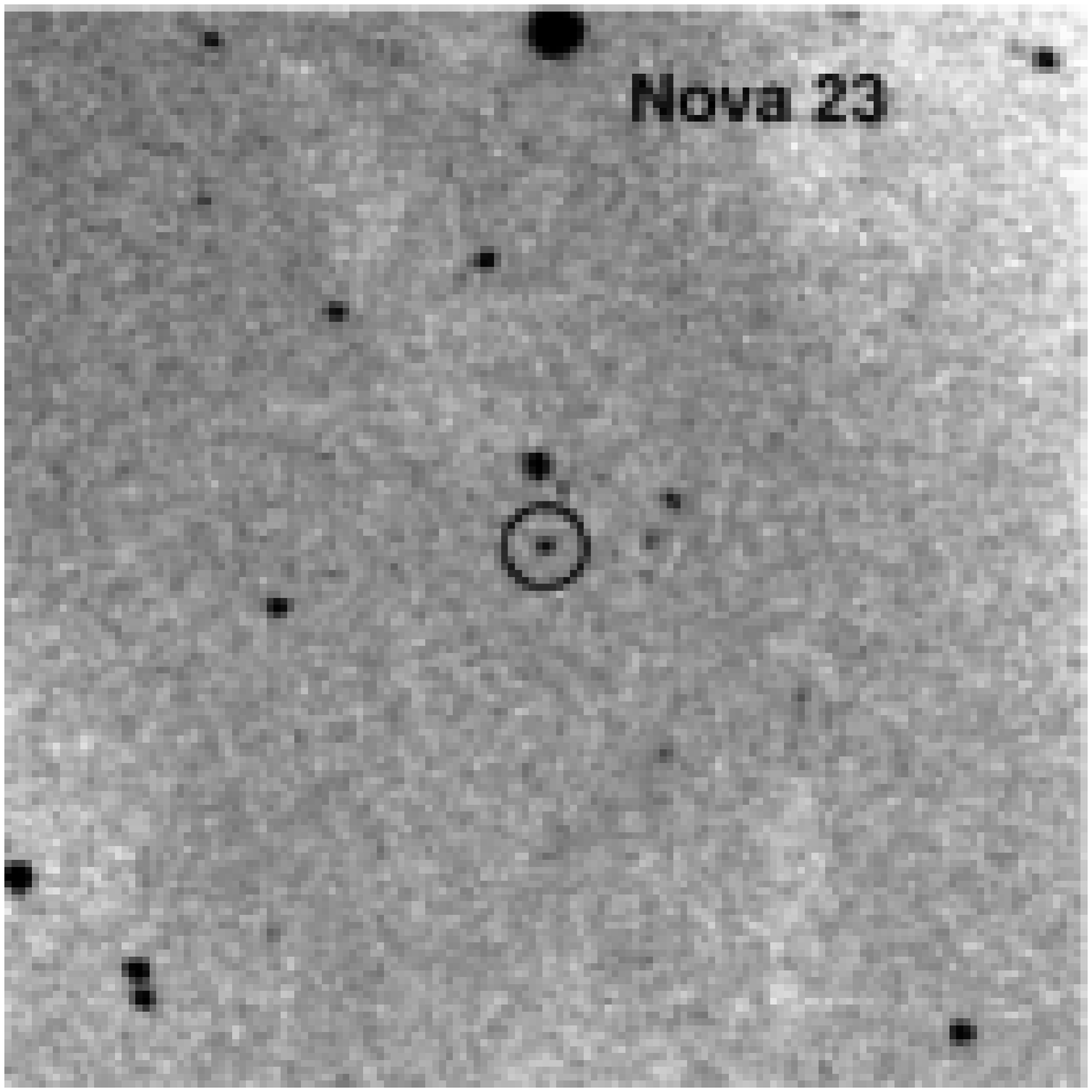}}\qquad
\subfigure[Nova 24 (B)]{\includegraphics[scale=.26, angle=0]{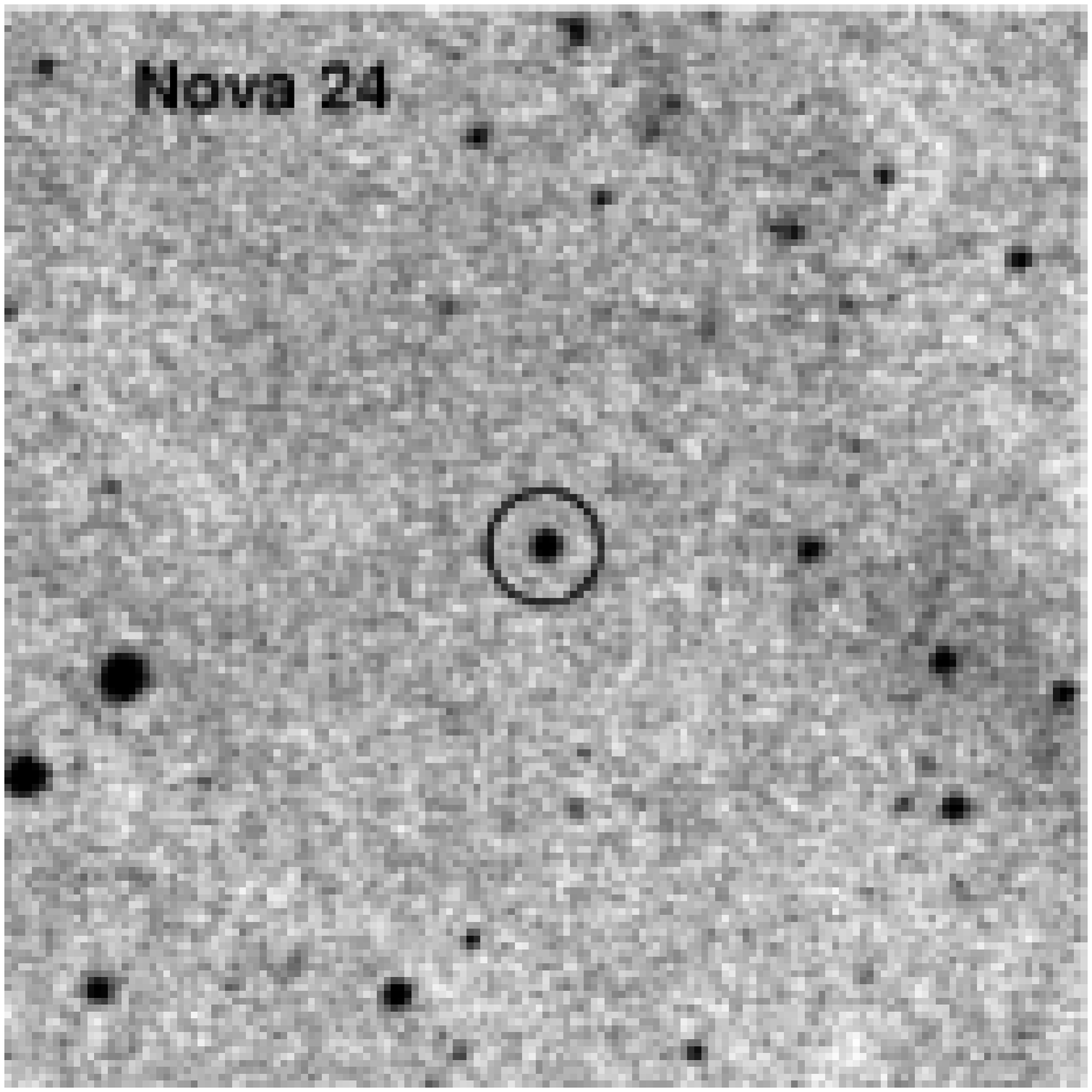}}
\caption{Finding charts for novae 13 - 24.}
\end{figure*}
}

\onlfig{10}{
\begin{figure*}[t]
\centering
\subfigure[Nova 25 (U)]{\includegraphics[scale=.26, angle=0]{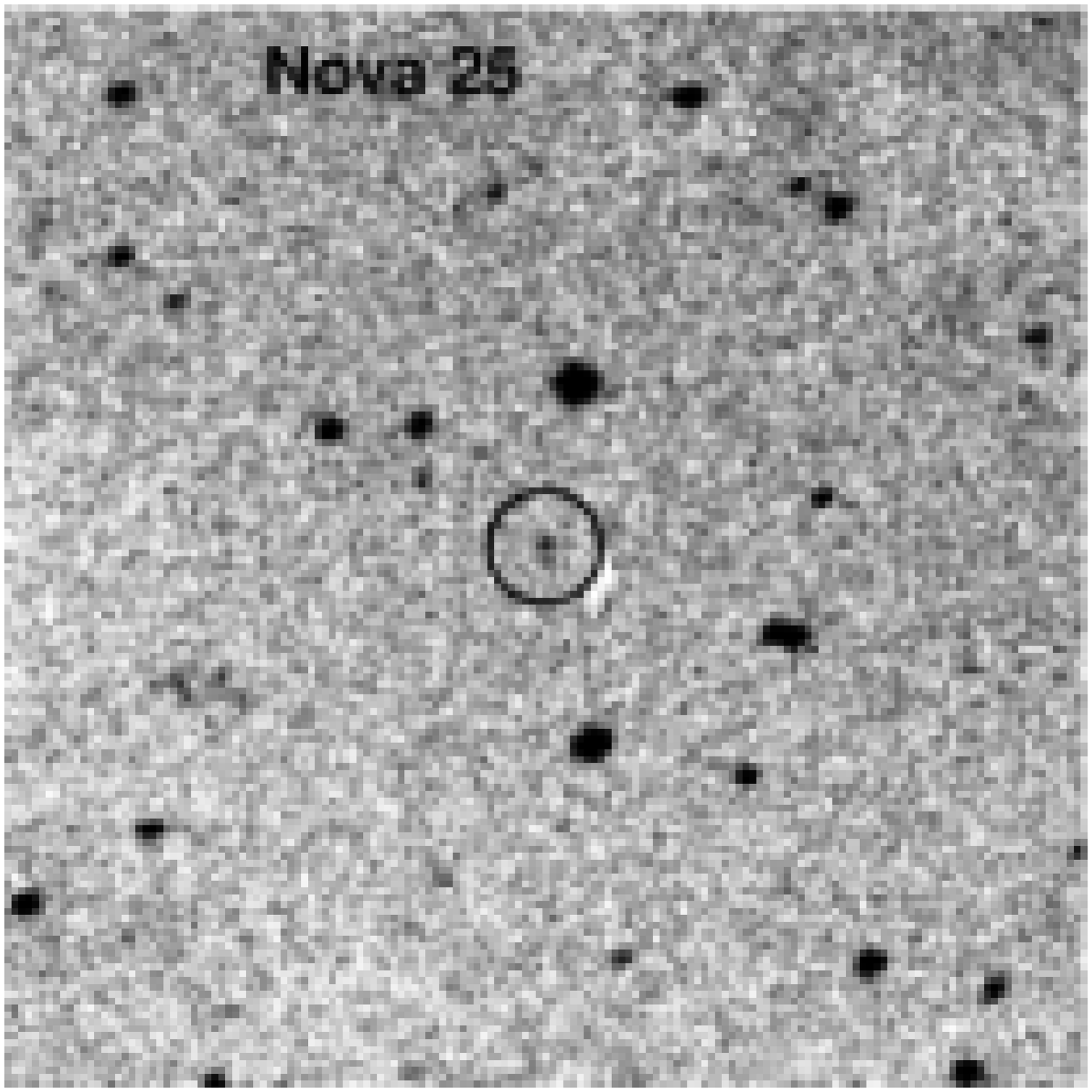}}\qquad
\subfigure[Nova 26 (B)]{\includegraphics[scale=.26, angle=0]{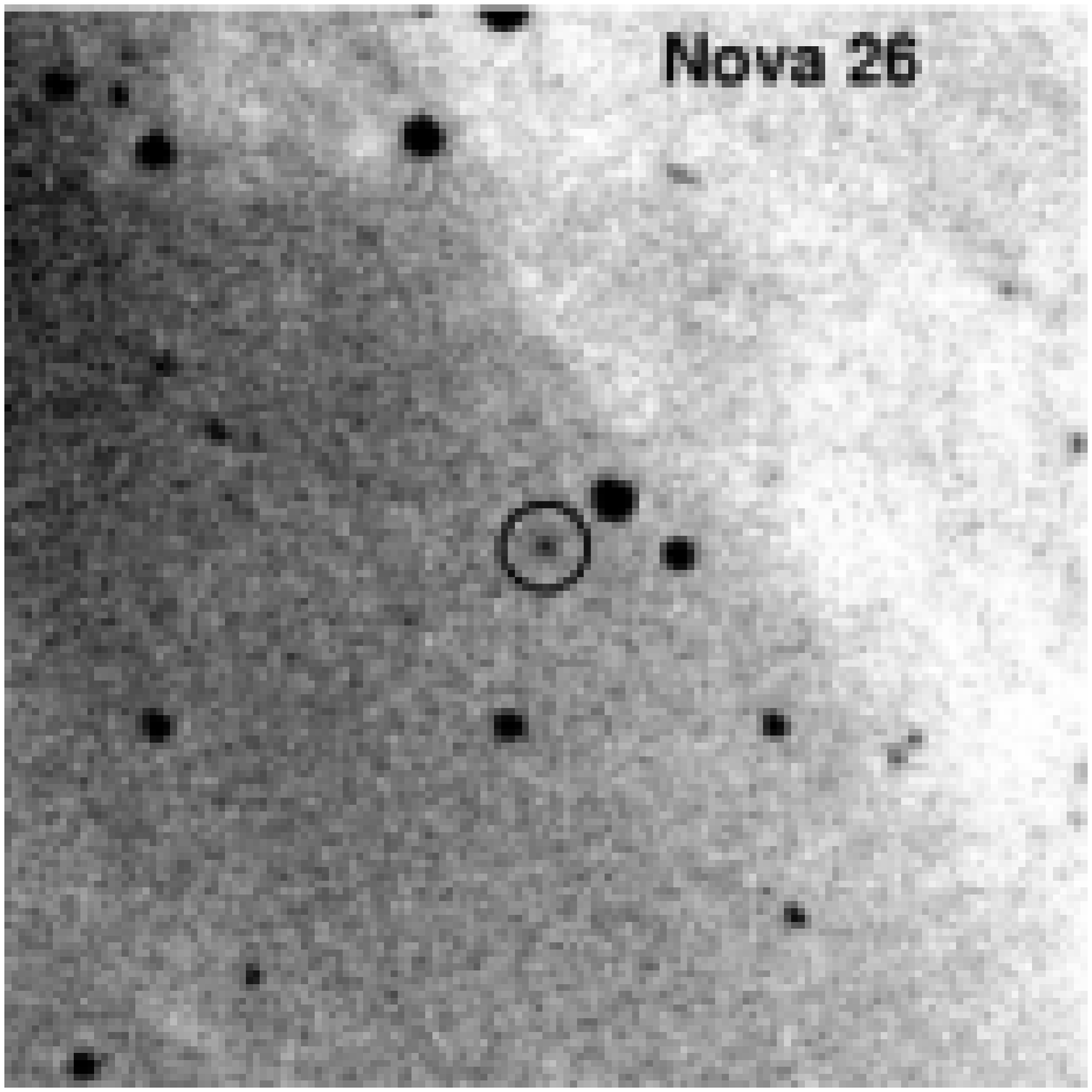}}\qquad
\subfigure[Nova 27 (B)]{\includegraphics[scale=.26, angle=0]{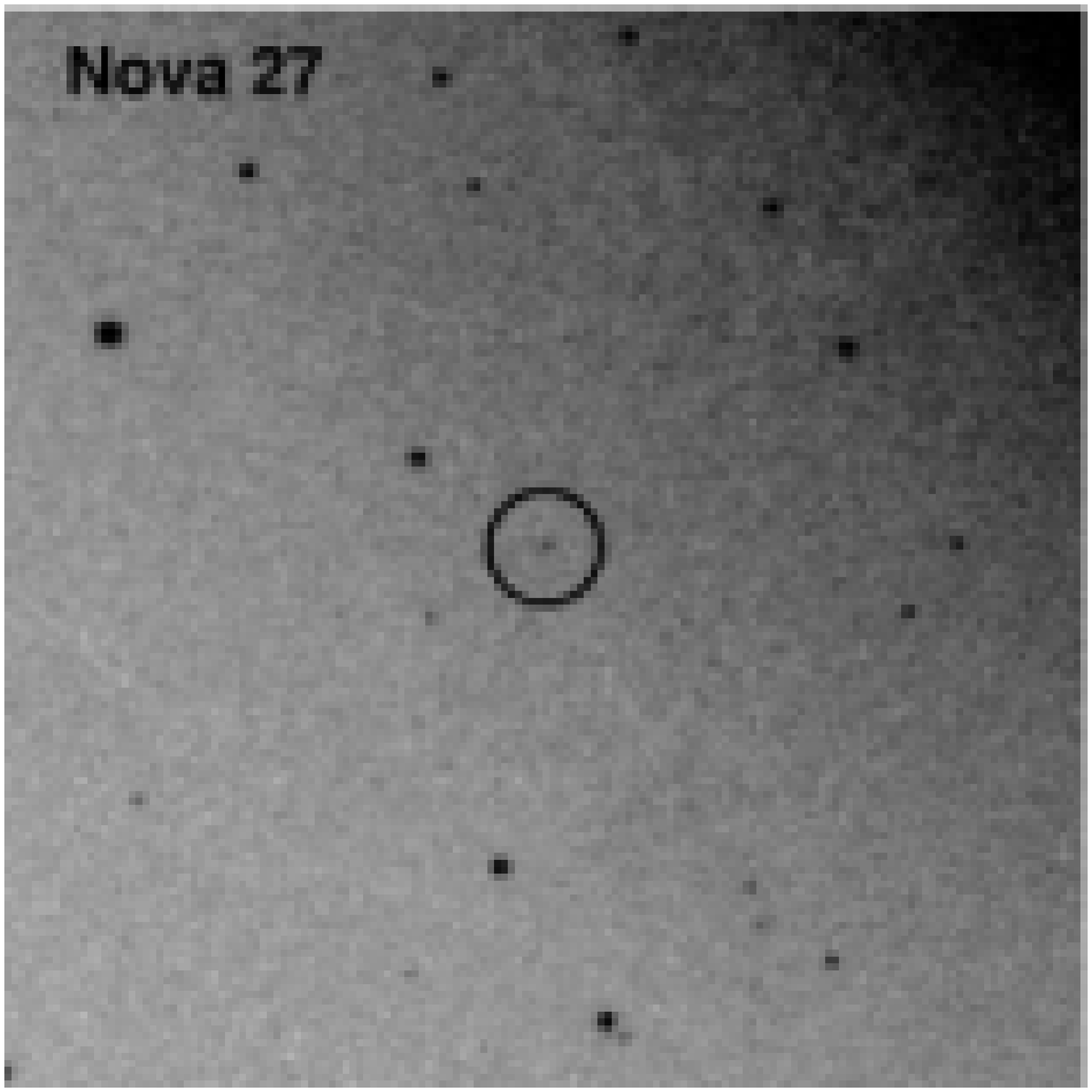}}\\
\subfigure[Nova 28 (B)]{\includegraphics[scale=.26, angle=0]{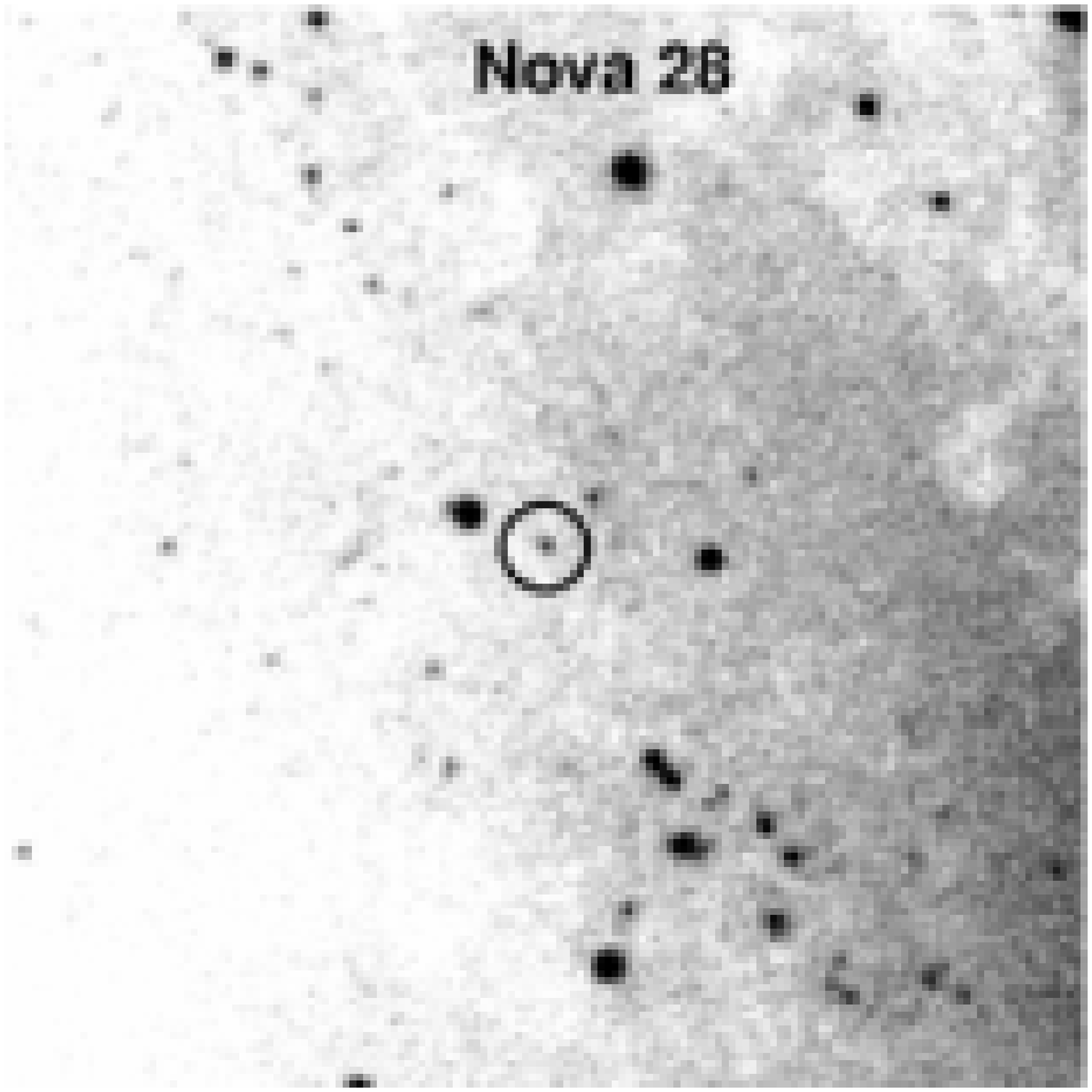}}\qquad
\subfigure[Nova 29 (B)]{\includegraphics[scale=.26, angle=0]{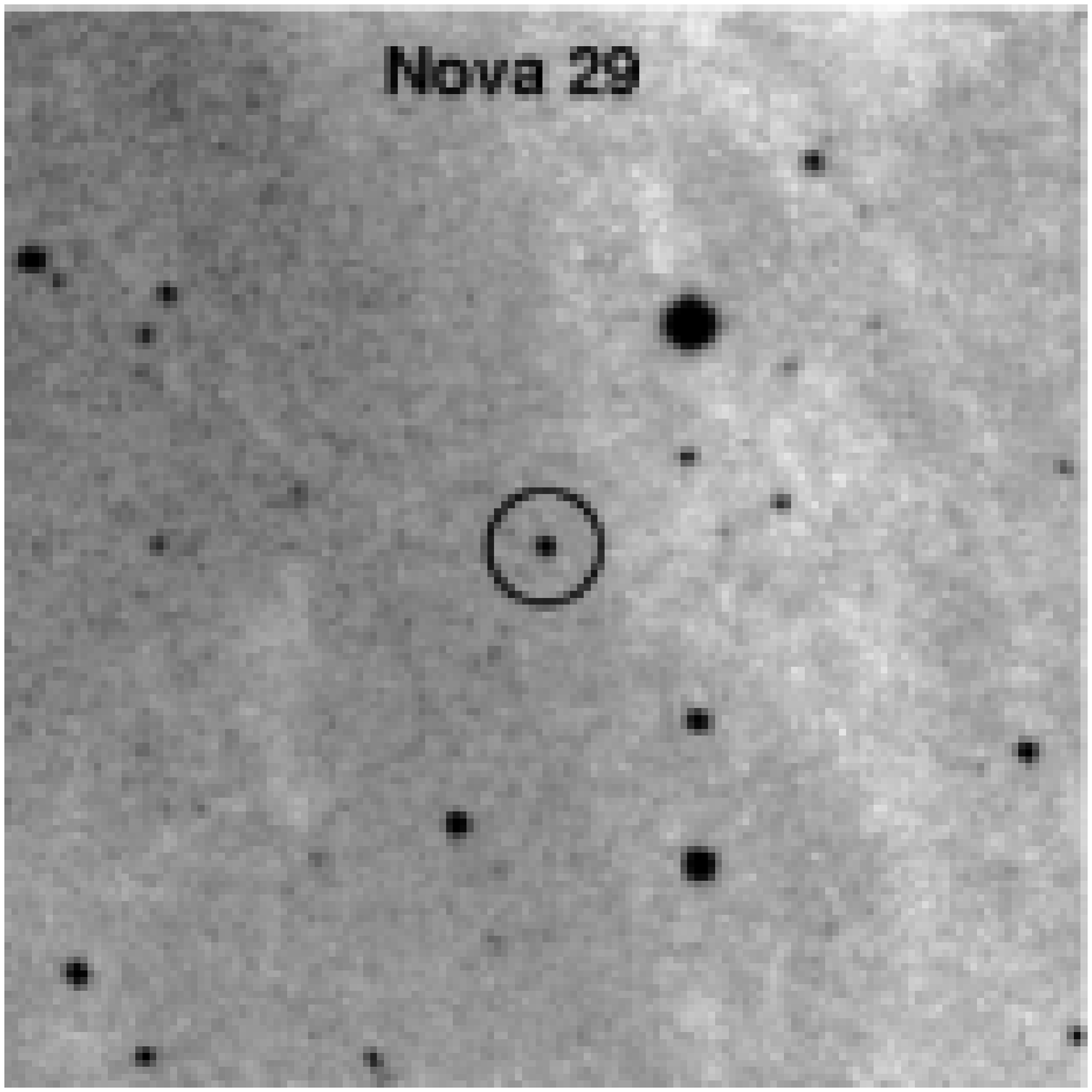}}\qquad
\subfigure[Nova 30 (B)]{\includegraphics[scale=.26, angle=0]{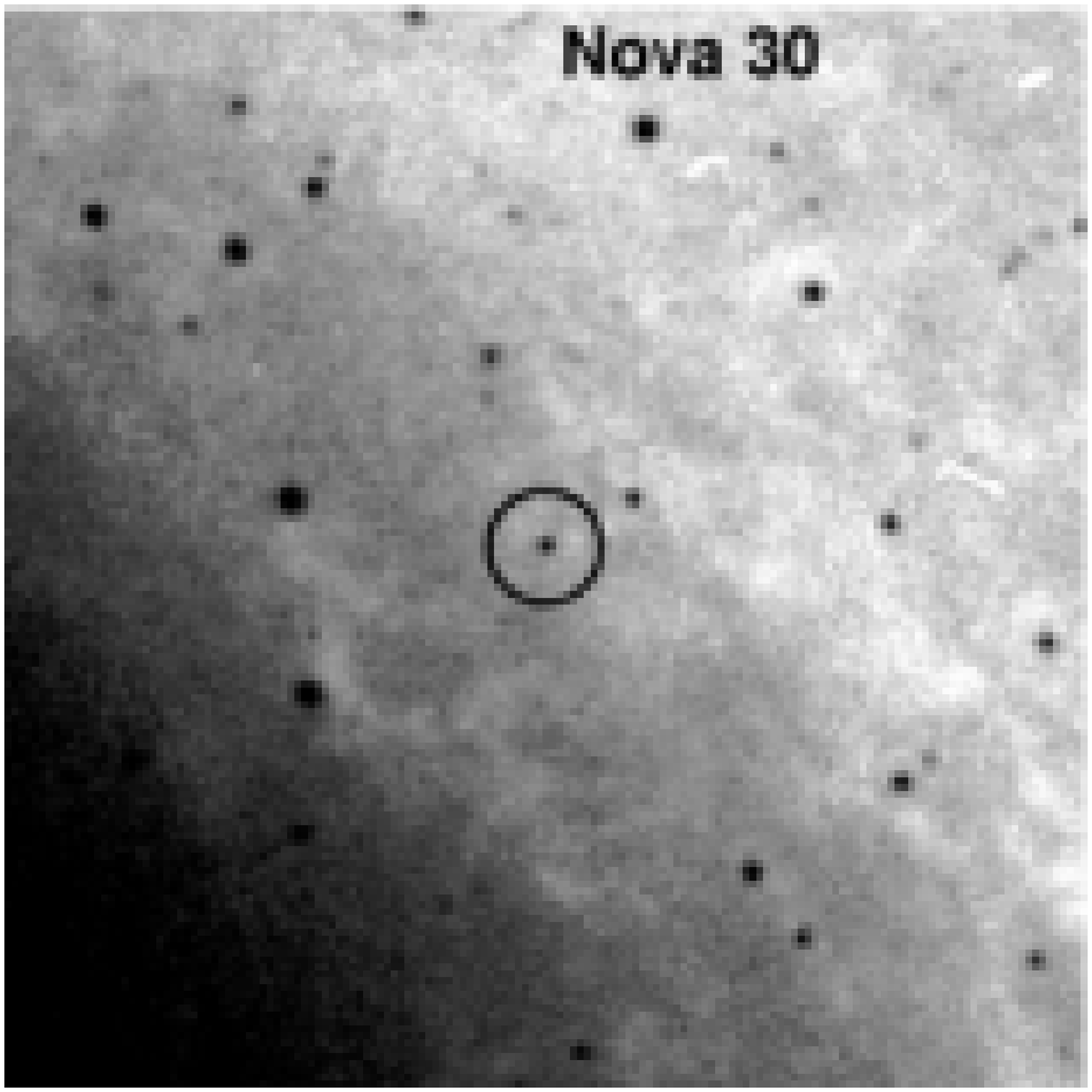}}\\
\subfigure[Nova 31 (B)]{\includegraphics[scale=.26, angle=0]{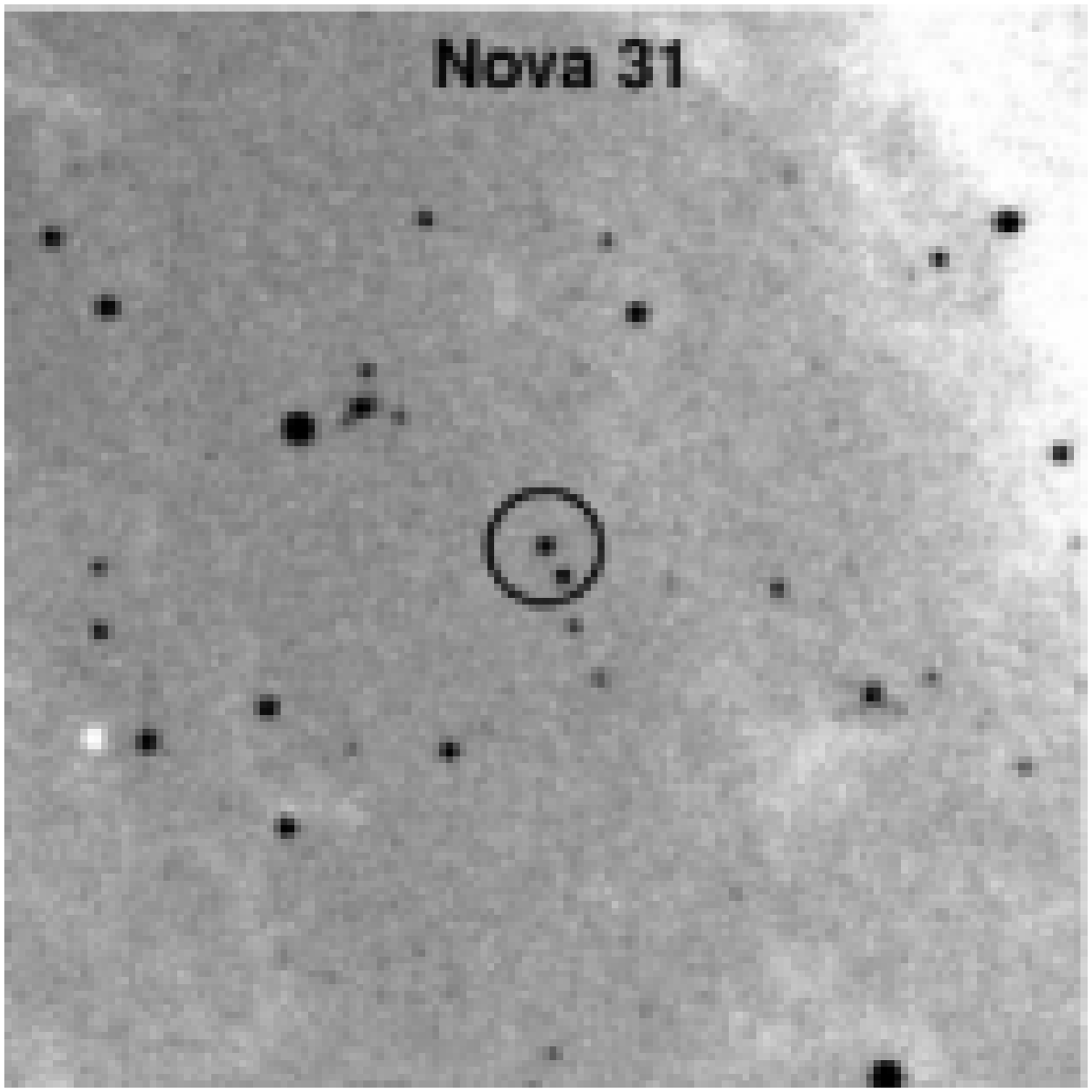}}\qquad
\subfigure[Nova 32 (B)]{\includegraphics[scale=.26, angle=0]{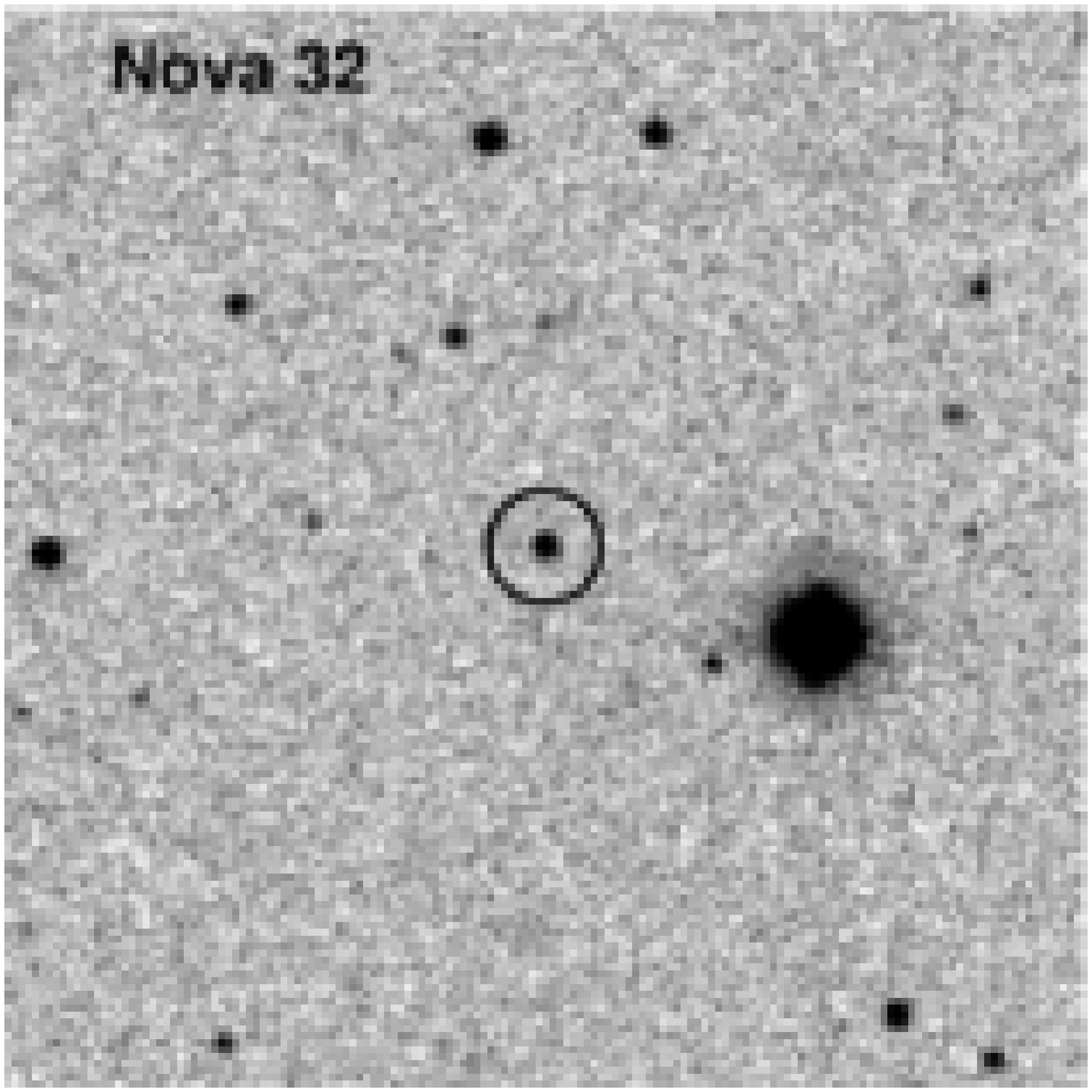}}\qquad
\subfigure[Nova 33 (B)]{\includegraphics[scale=.26, angle=0]{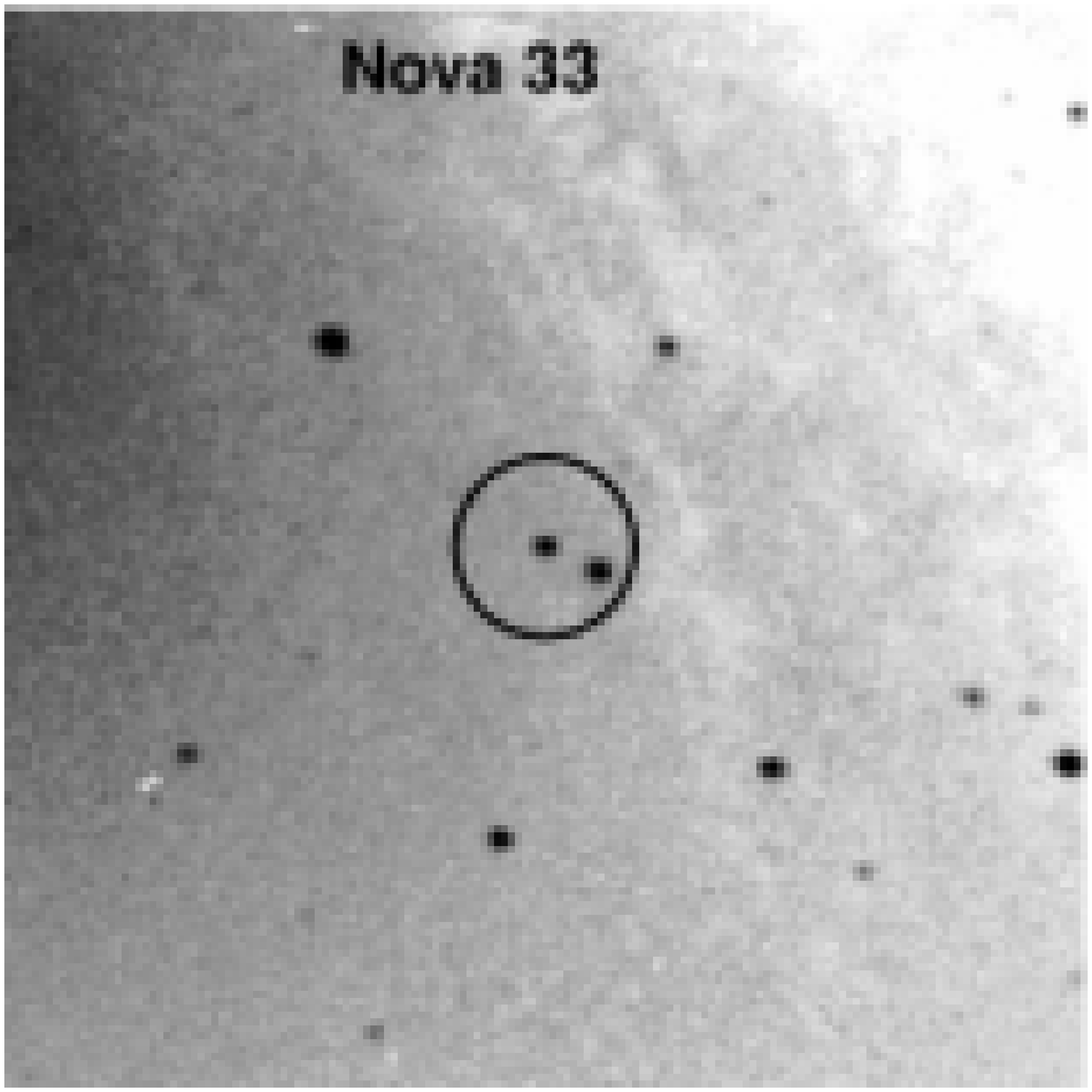}}\\
\subfigure[Nova 34 (B)]{\includegraphics[scale=.26, angle=0]{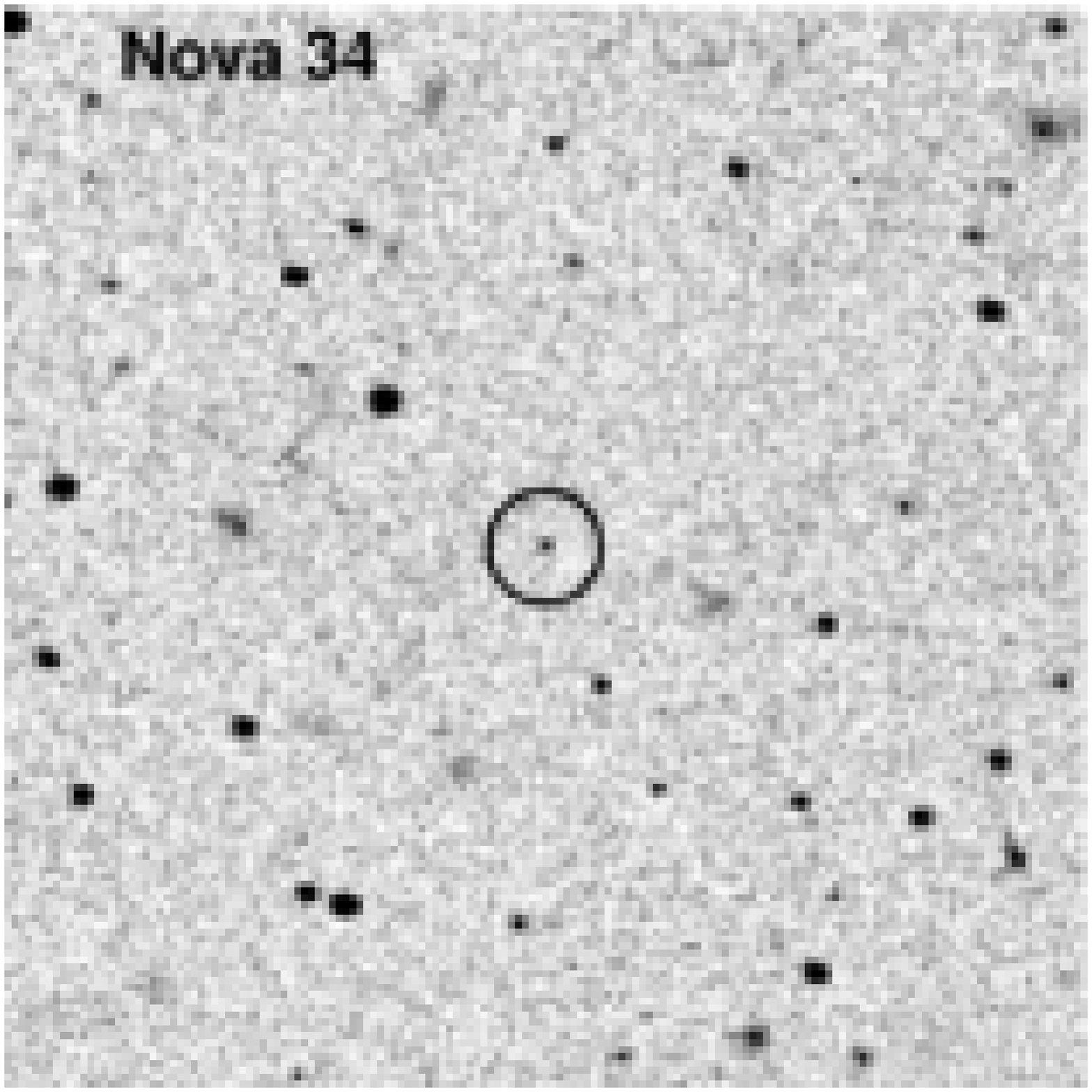}}\qquad
\subfigure[Nova 35 (U)]{\includegraphics[scale=.26, angle=0]{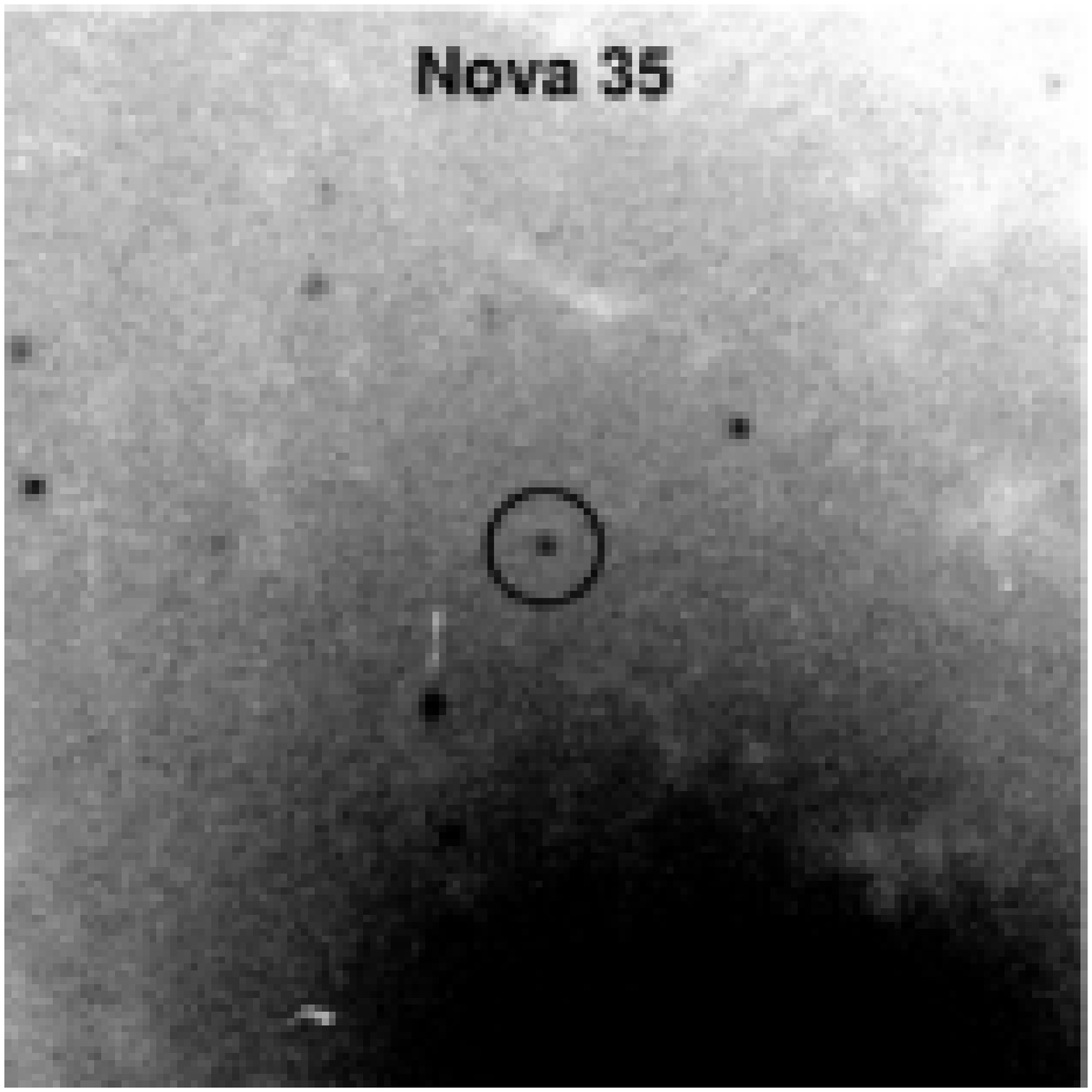}}\qquad
\subfigure[Nova 36 (U)]{\includegraphics[scale=.26, angle=0]{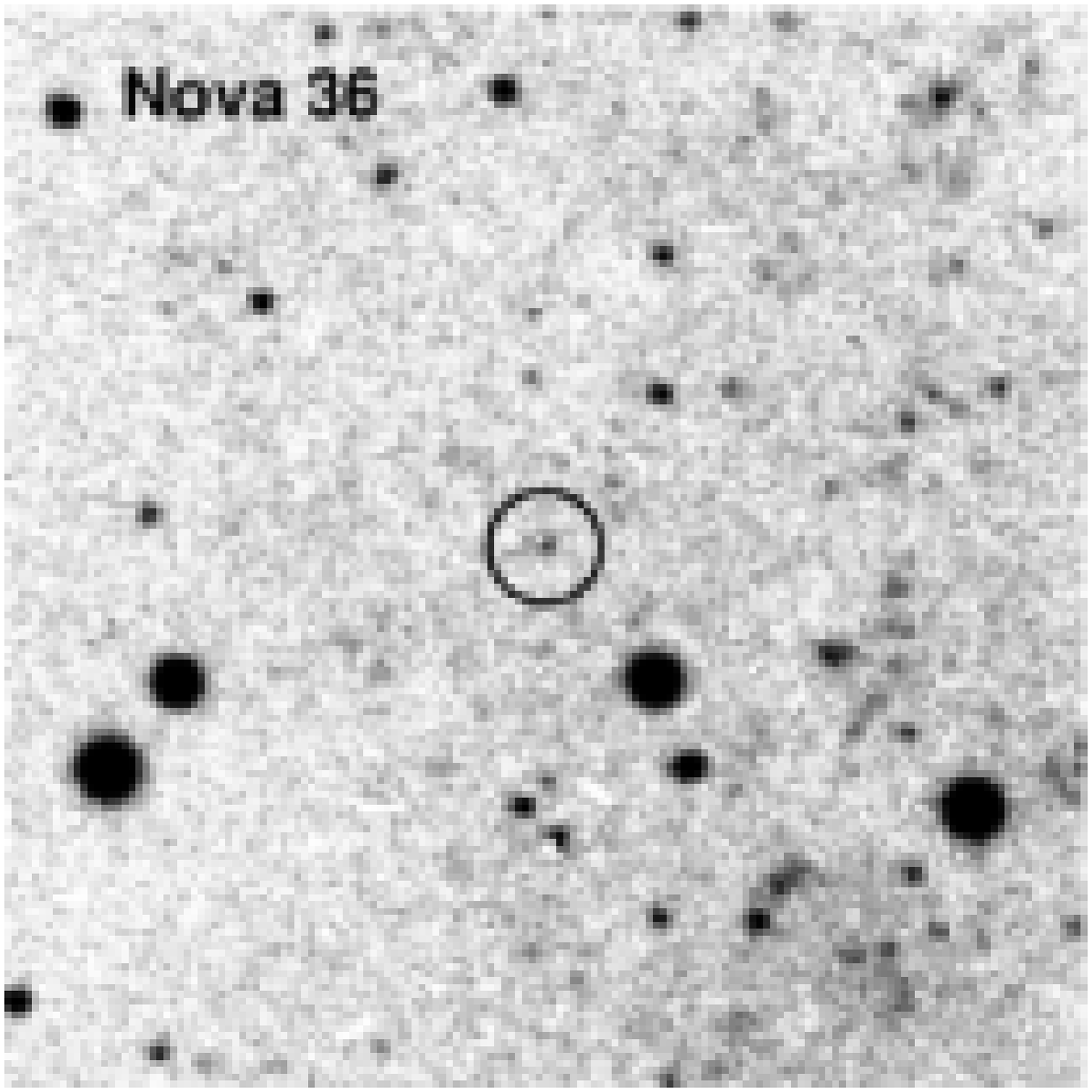}}
\caption{Finding charts for novae 25 - 36.}
\end{figure*}
}

\onlfig{11}{
\begin{figure*}[t]
\centering
\subfigure[Nova 37 (V)]{\includegraphics[scale=.26, angle=0]{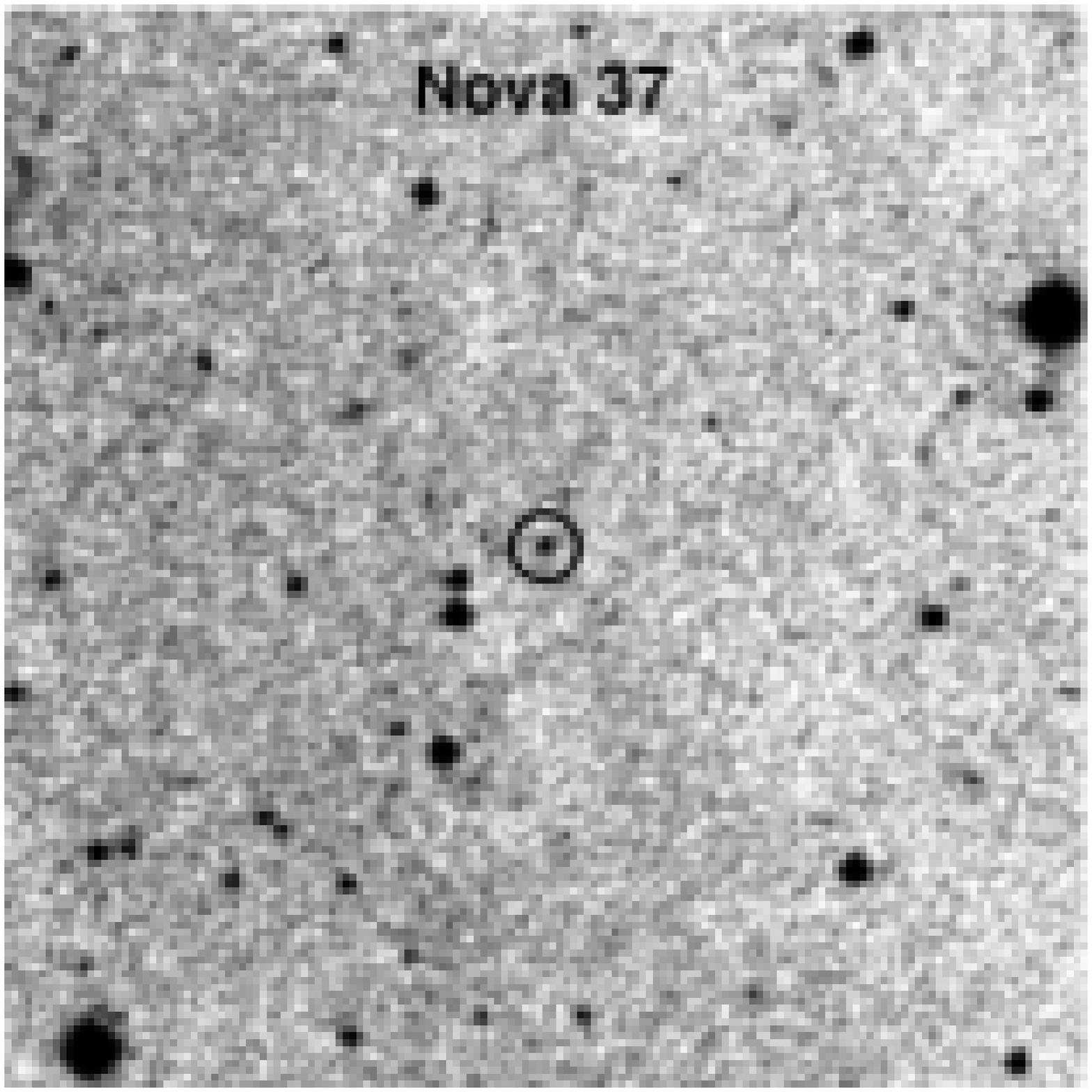}}\qquad
\subfigure[Nova 38 (B)]{\includegraphics[scale=.26, angle=0]{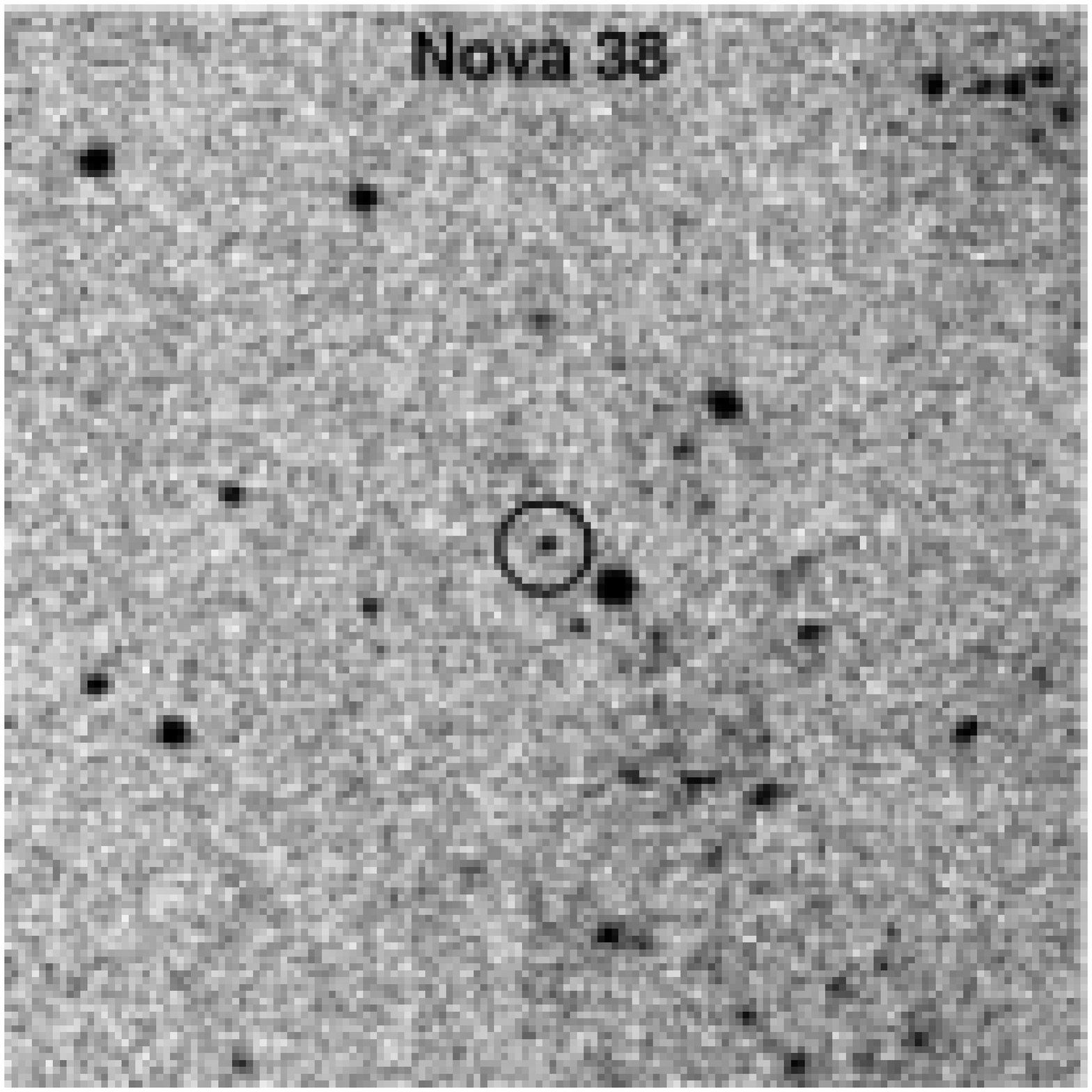}}\qquad
\subfigure[Nova 39 (B)]{\includegraphics[scale=.26, angle=0]{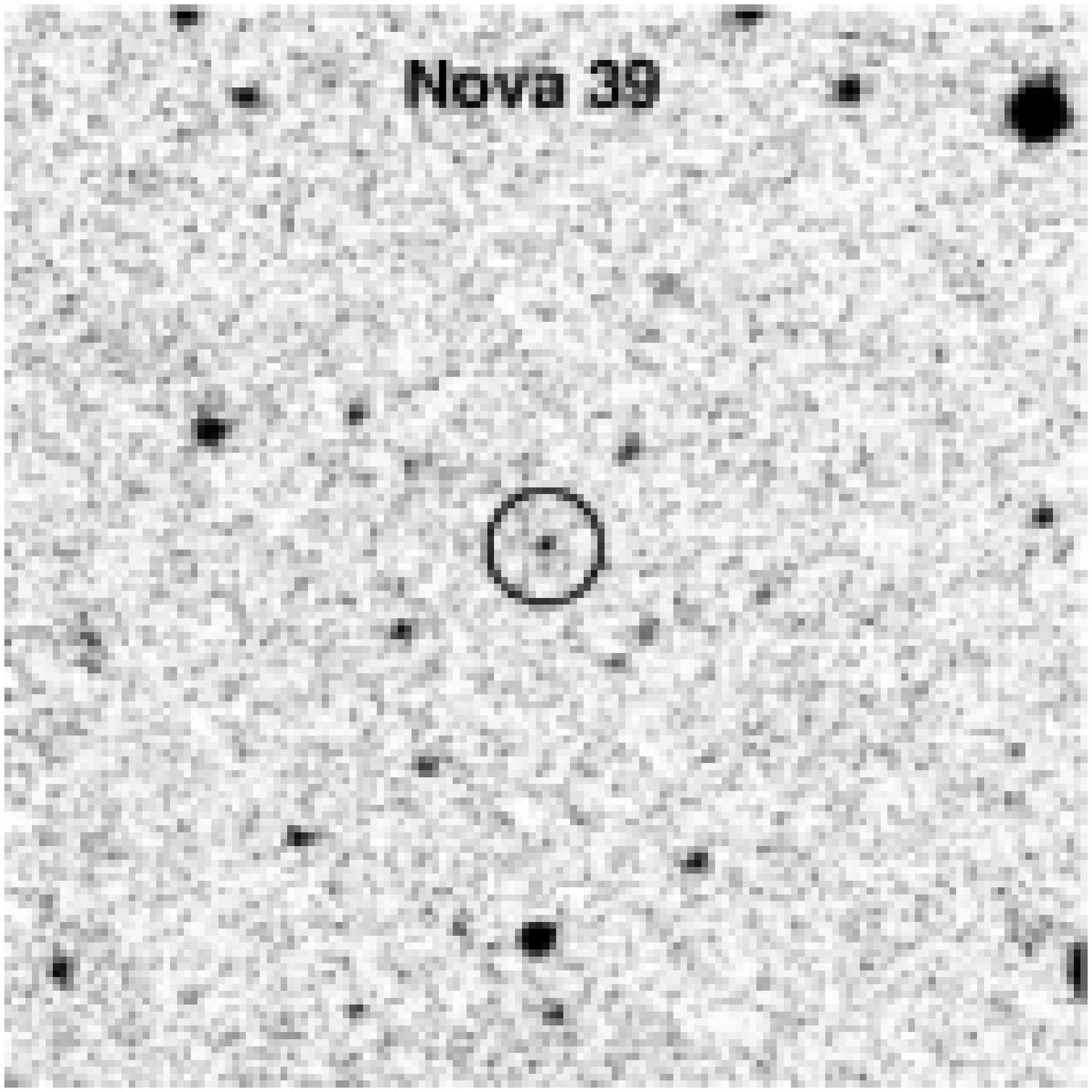}}\\
\subfigure[Nova 40 (B)]{\includegraphics[scale=.26, angle=0]{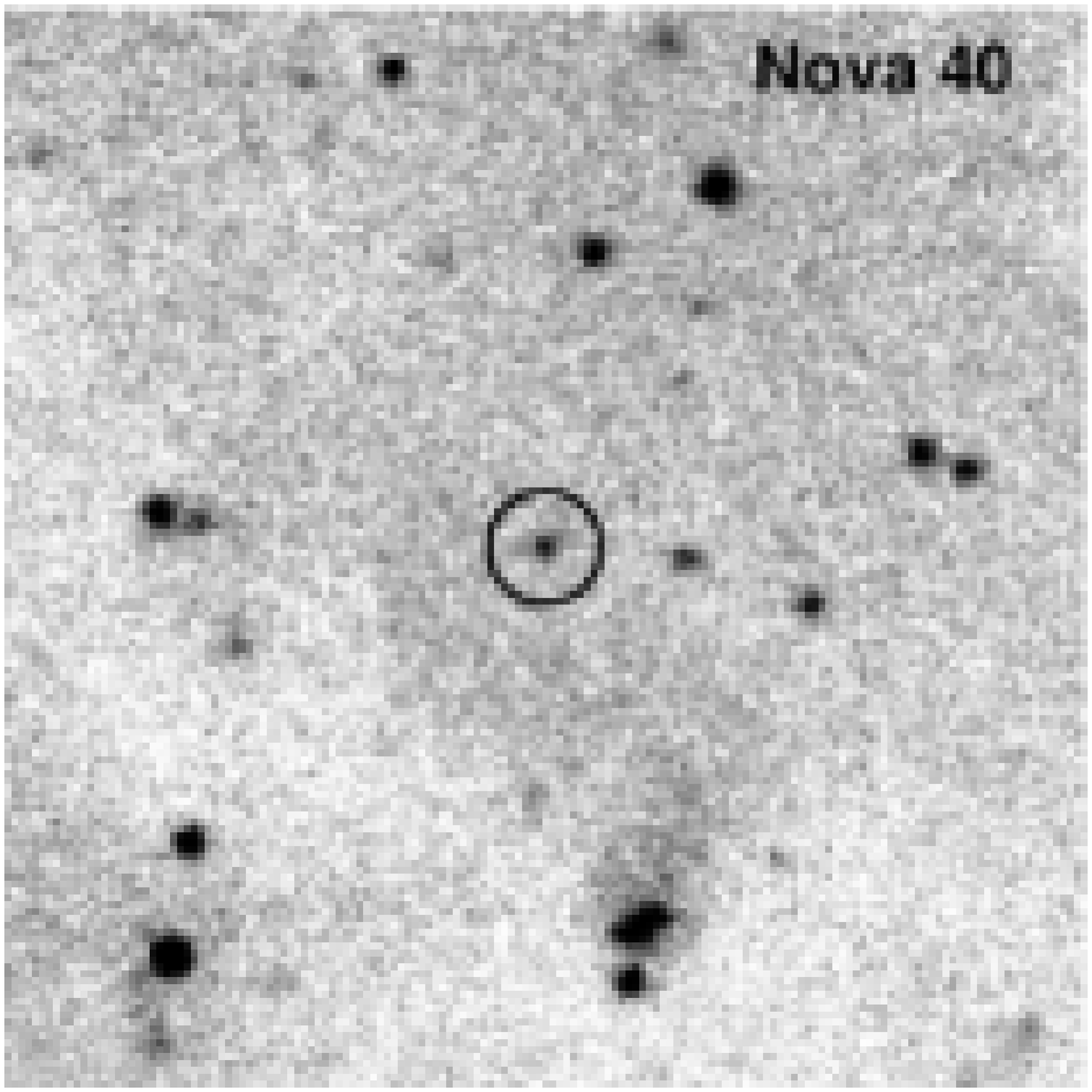}}\qquad
\subfigure[Nova 41 (B)]{\includegraphics[scale=.26, angle=0]{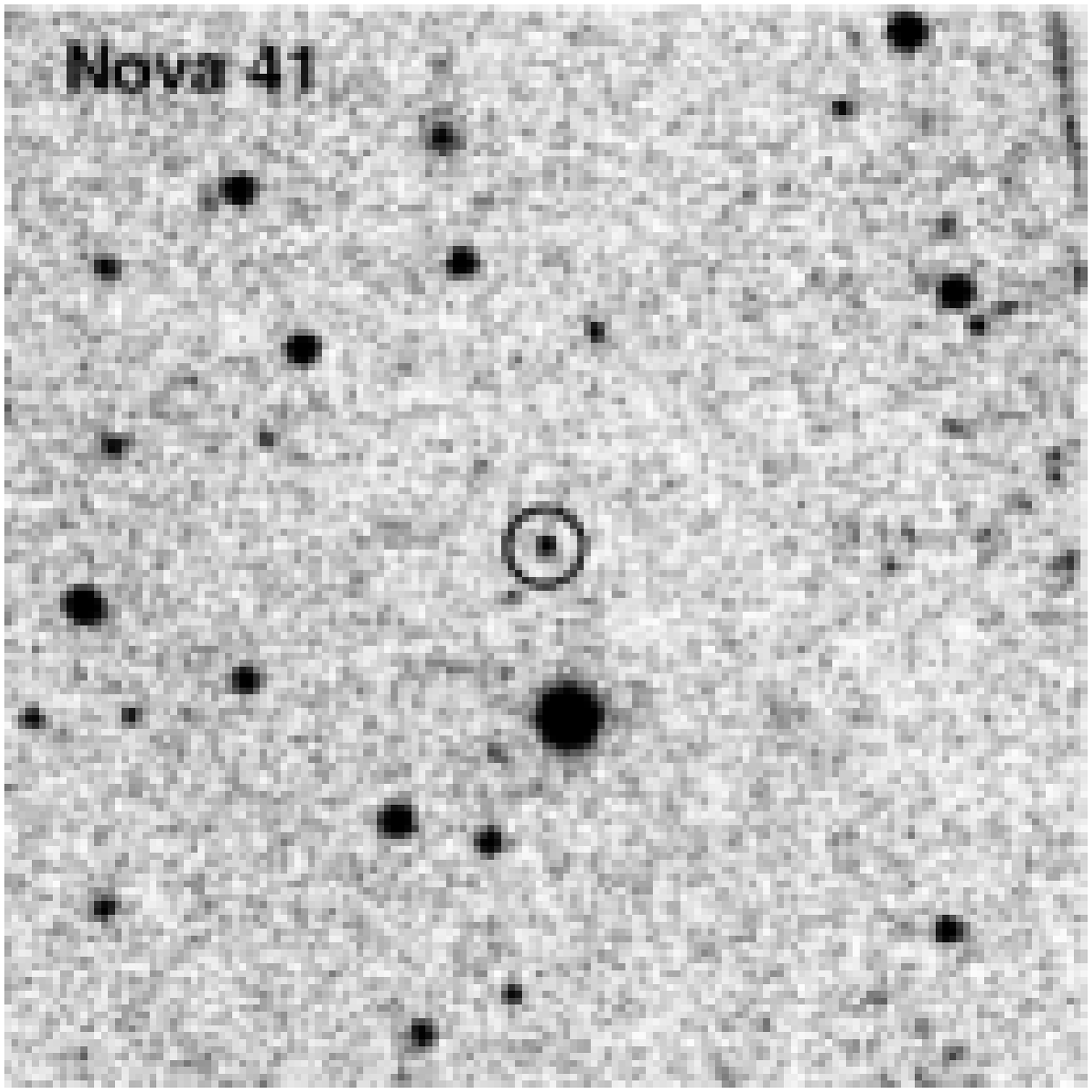}}\qquad
\subfigure[Nova 42 (B)]{\includegraphics[scale=.26, angle=0]{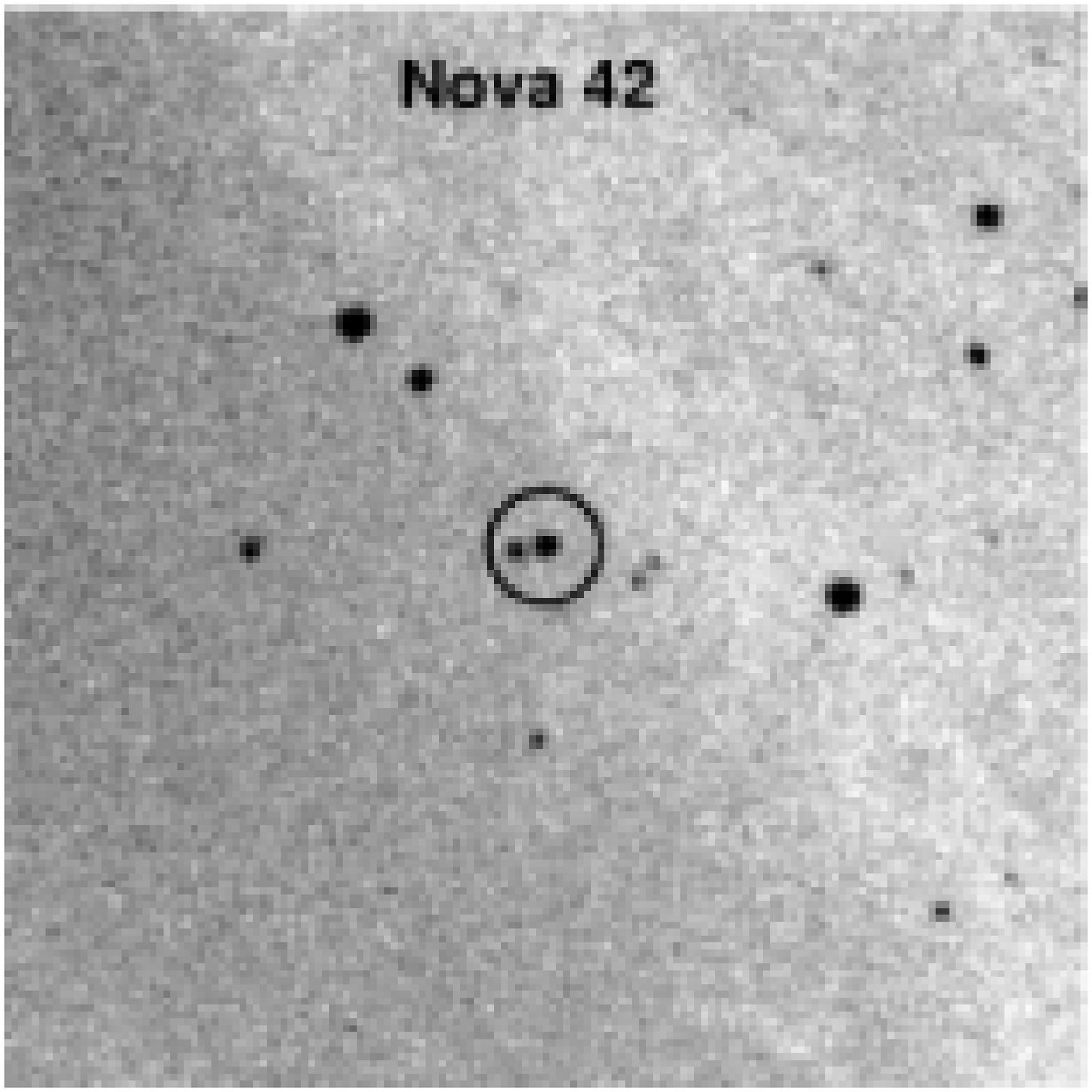}}\\
\subfigure[Nova 43 (B)]{\includegraphics[scale=.26, angle=0]{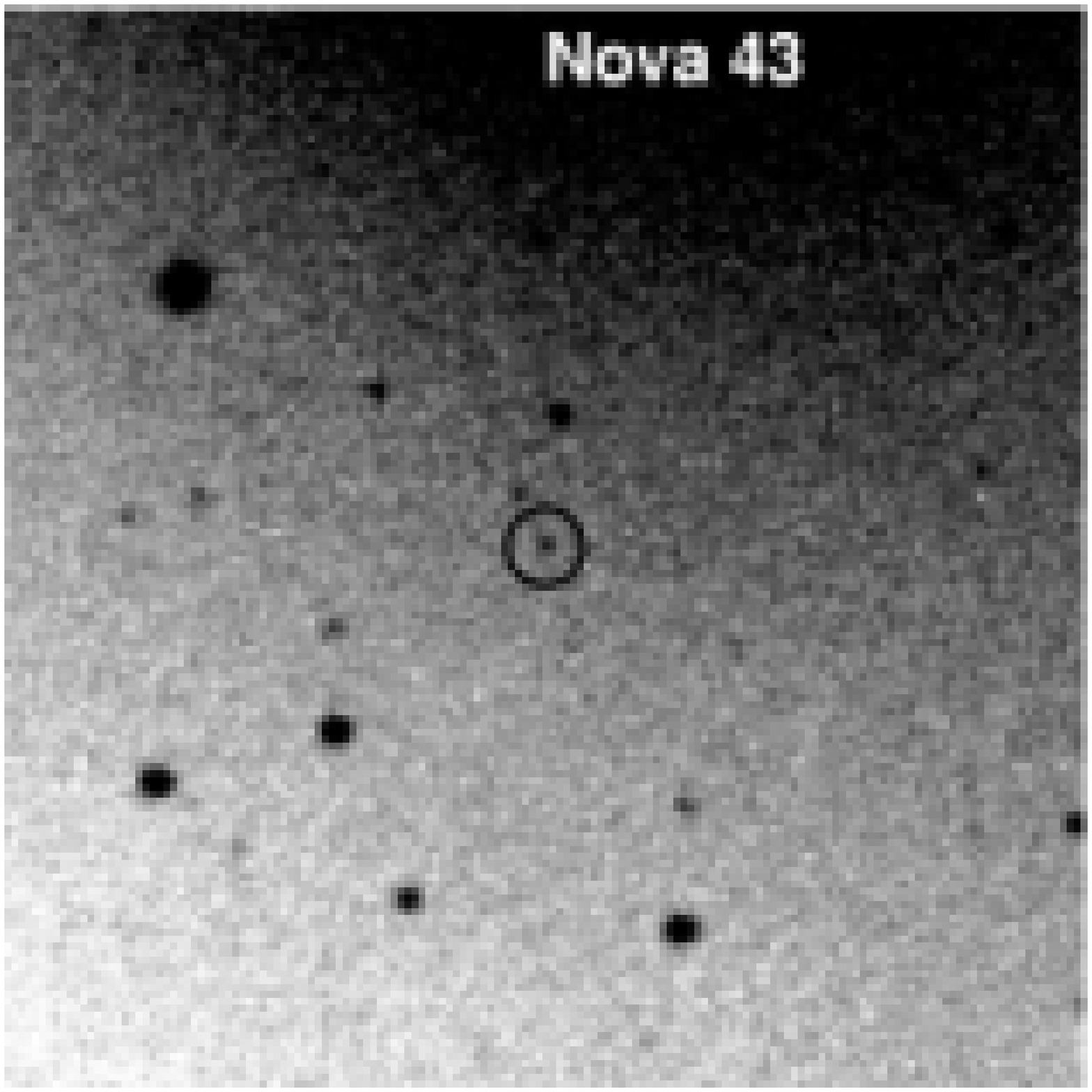}}\qquad
\subfigure[Nova 44 (V)]{\includegraphics[scale=.26, angle=0]{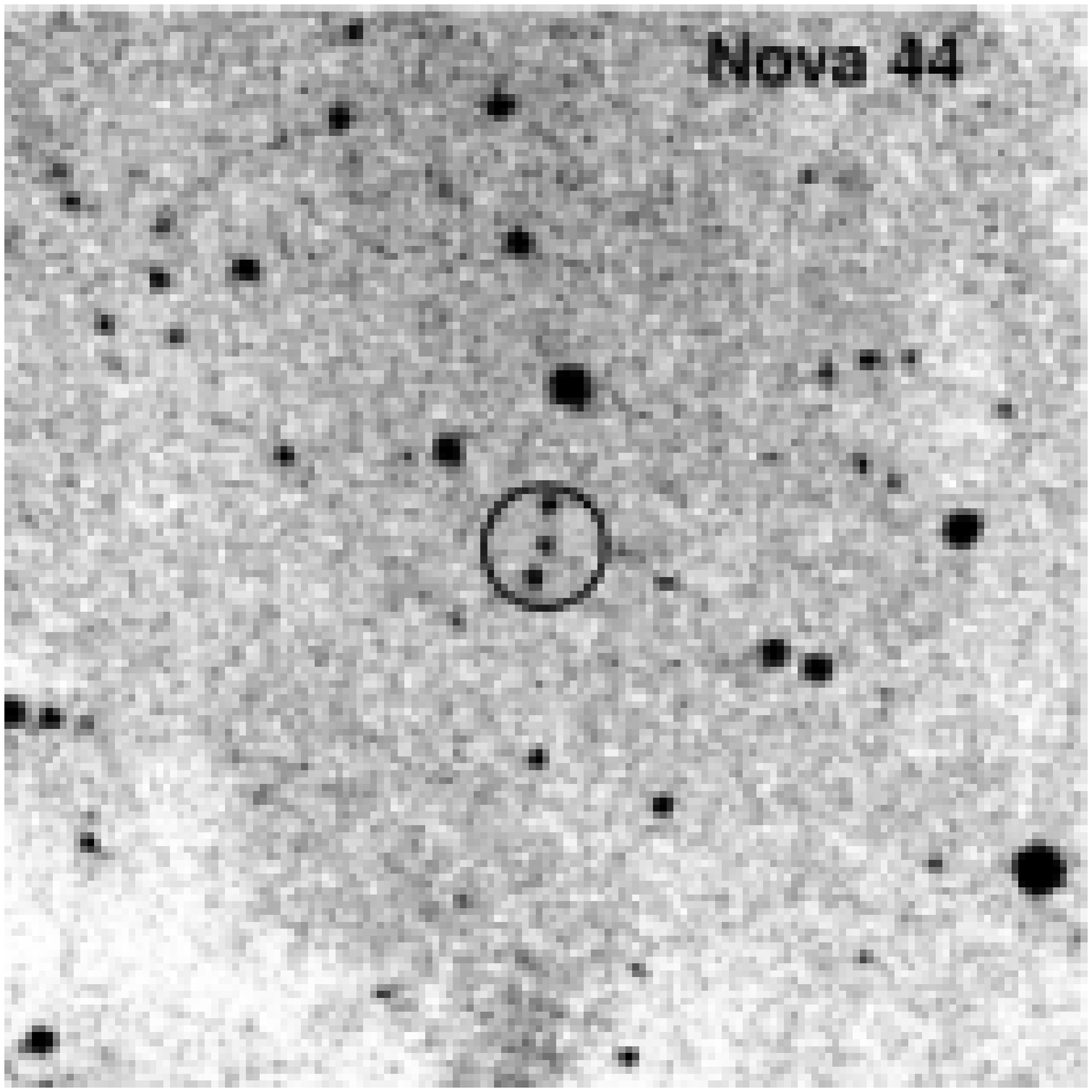}}\qquad
\subfigure[Nova 45 (B)]{\includegraphics[scale=.26, angle=0]{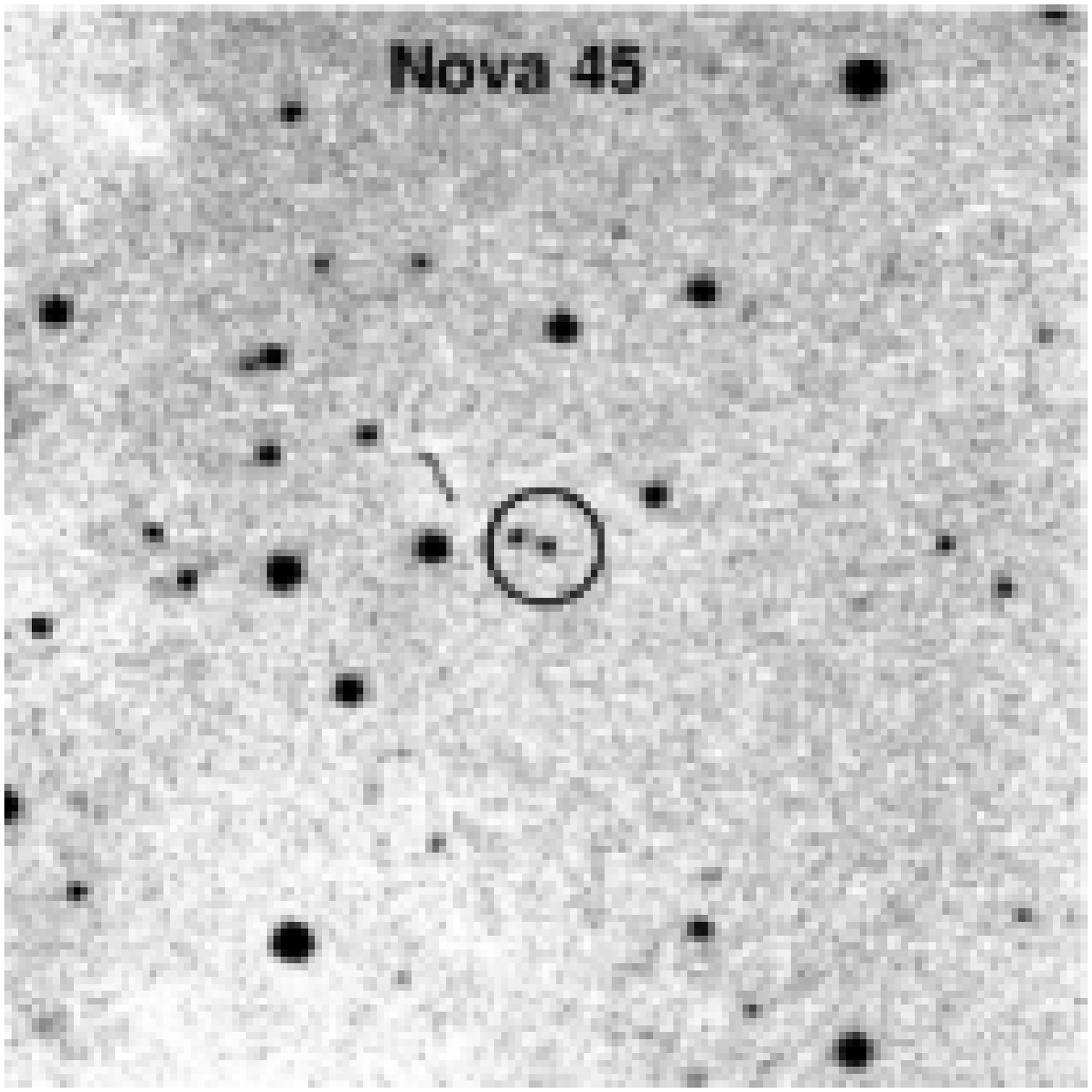}}\\
\subfigure[Nova 46 (B)]{\includegraphics[scale=.26, angle=0]{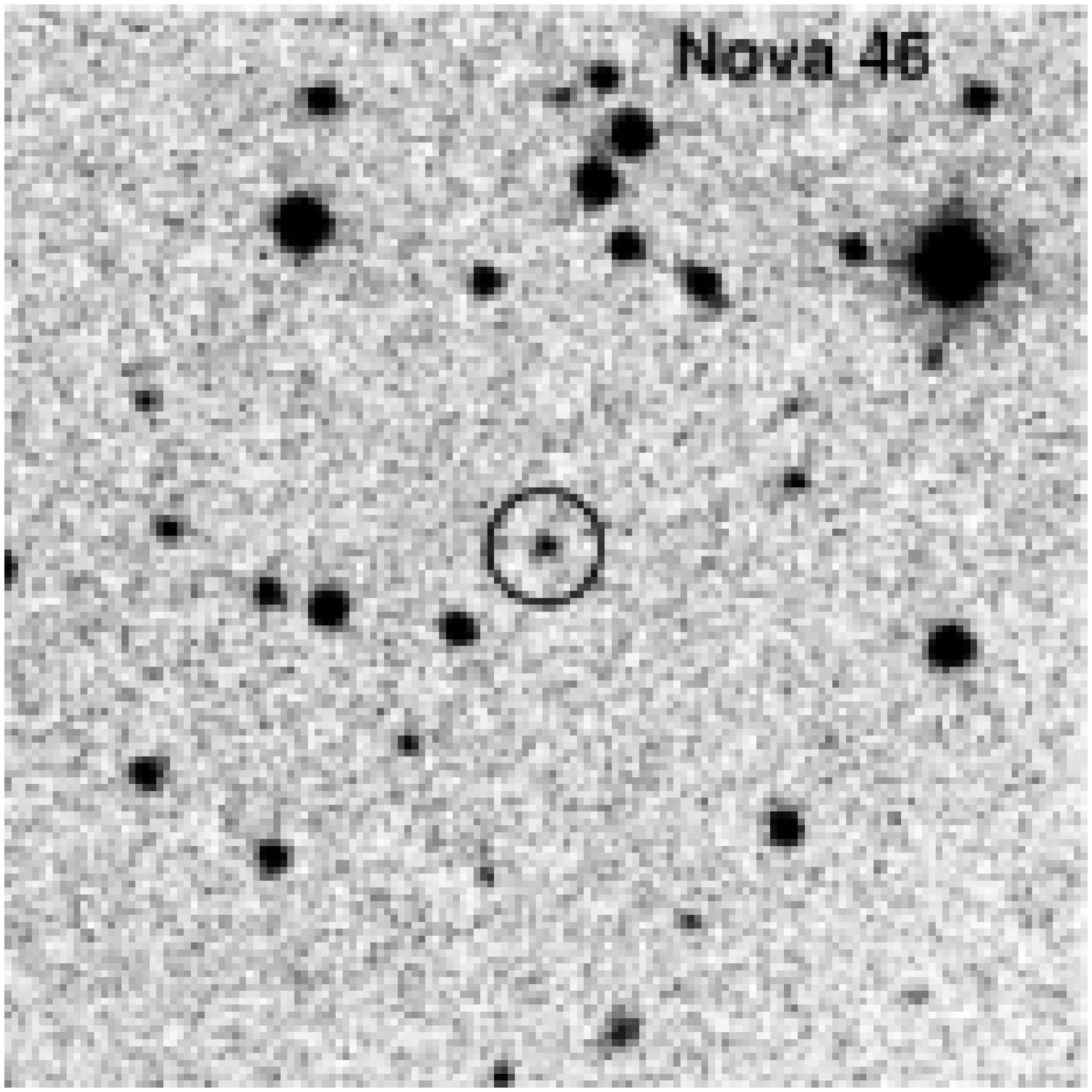}}\qquad
\subfigure[Nova 47 (B)]{\includegraphics[scale=.26, angle=0]{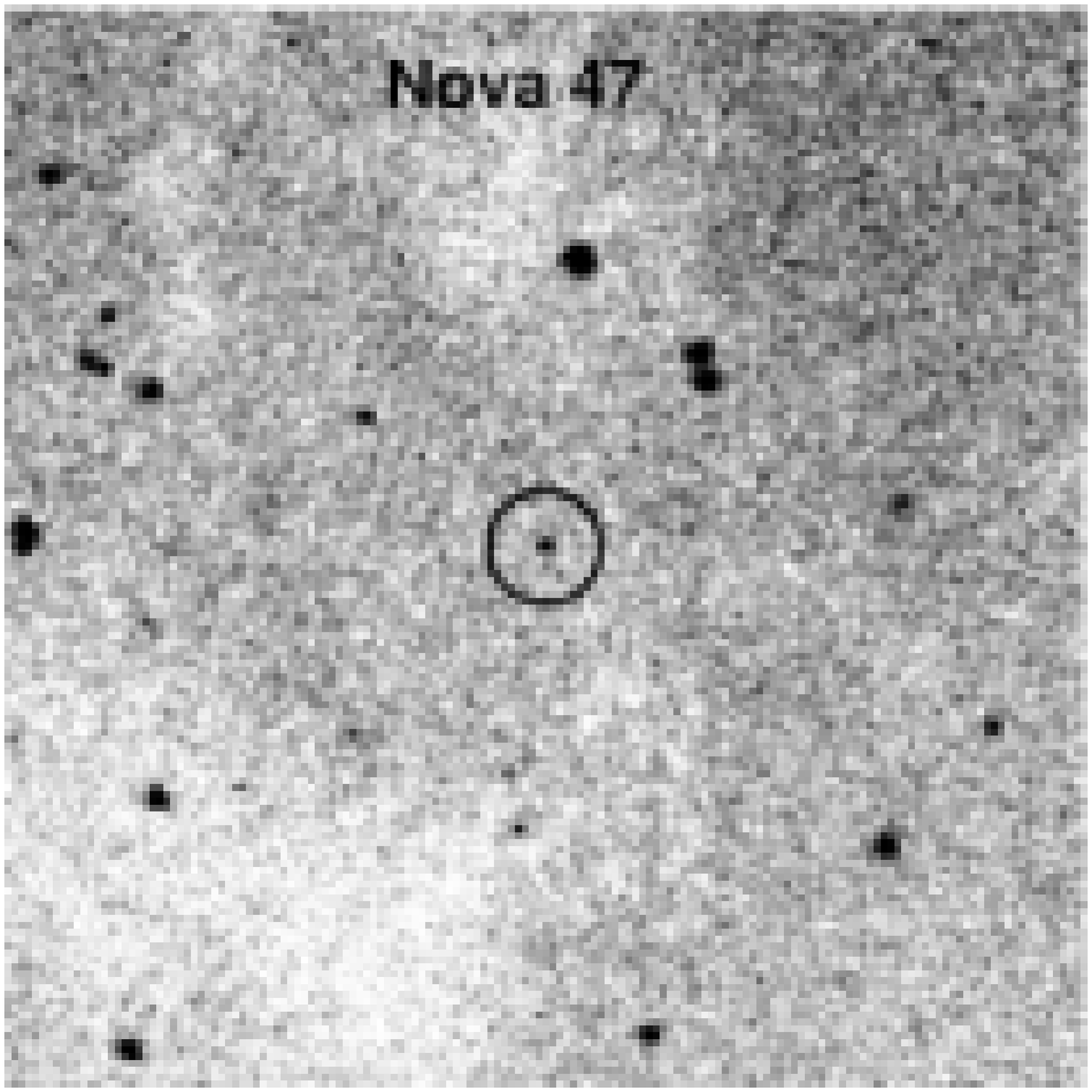}}\qquad
\subfigure[Nova 48 (B)]{\includegraphics[scale=.26, angle=0]{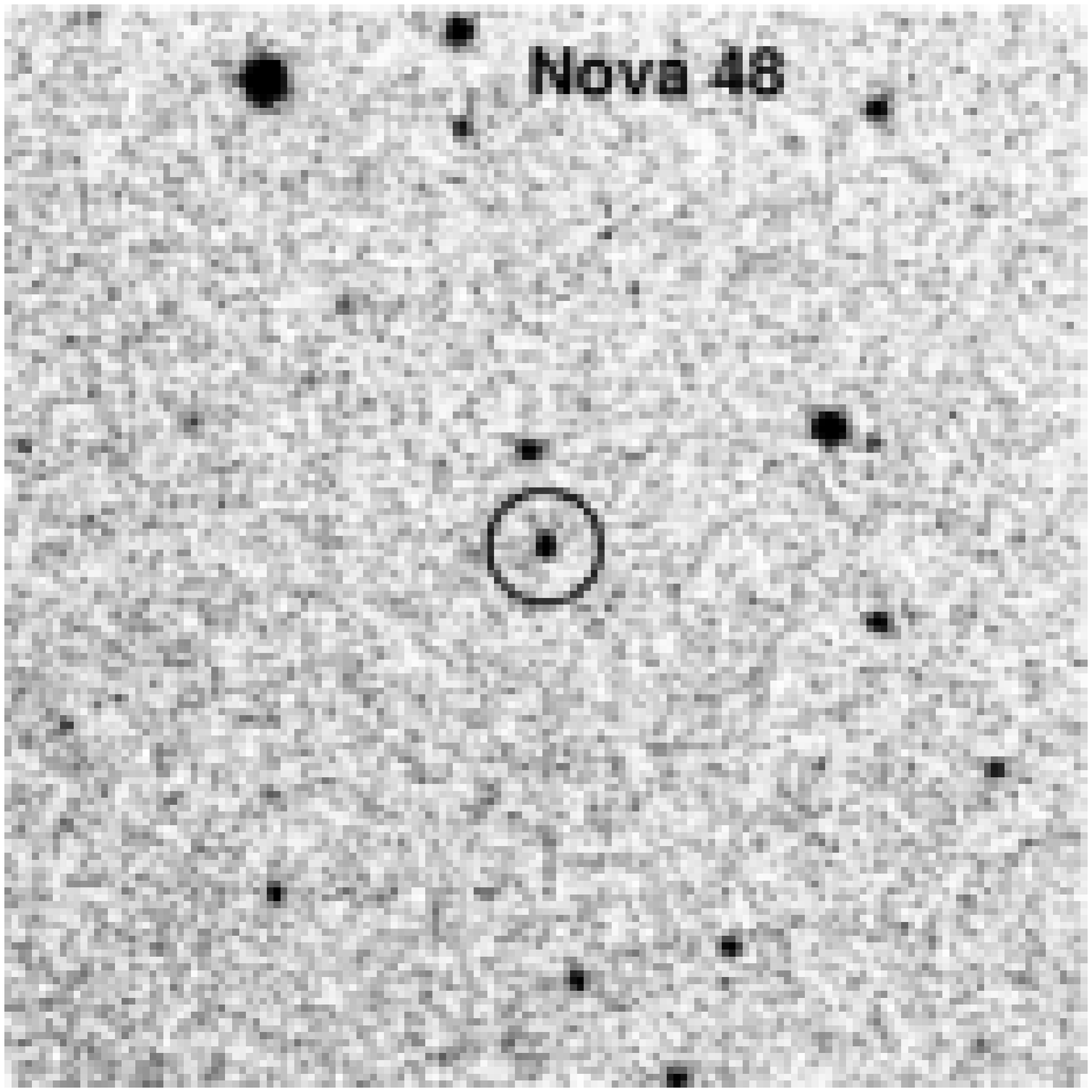}}
\caption{Finding charts for novae 37 - 48.}
\end{figure*}
}

\onlfig{12}{
\begin{figure*}[t]
\centering
\subfigure[Nova 49 (U)]{\includegraphics[scale=.26, angle=0]{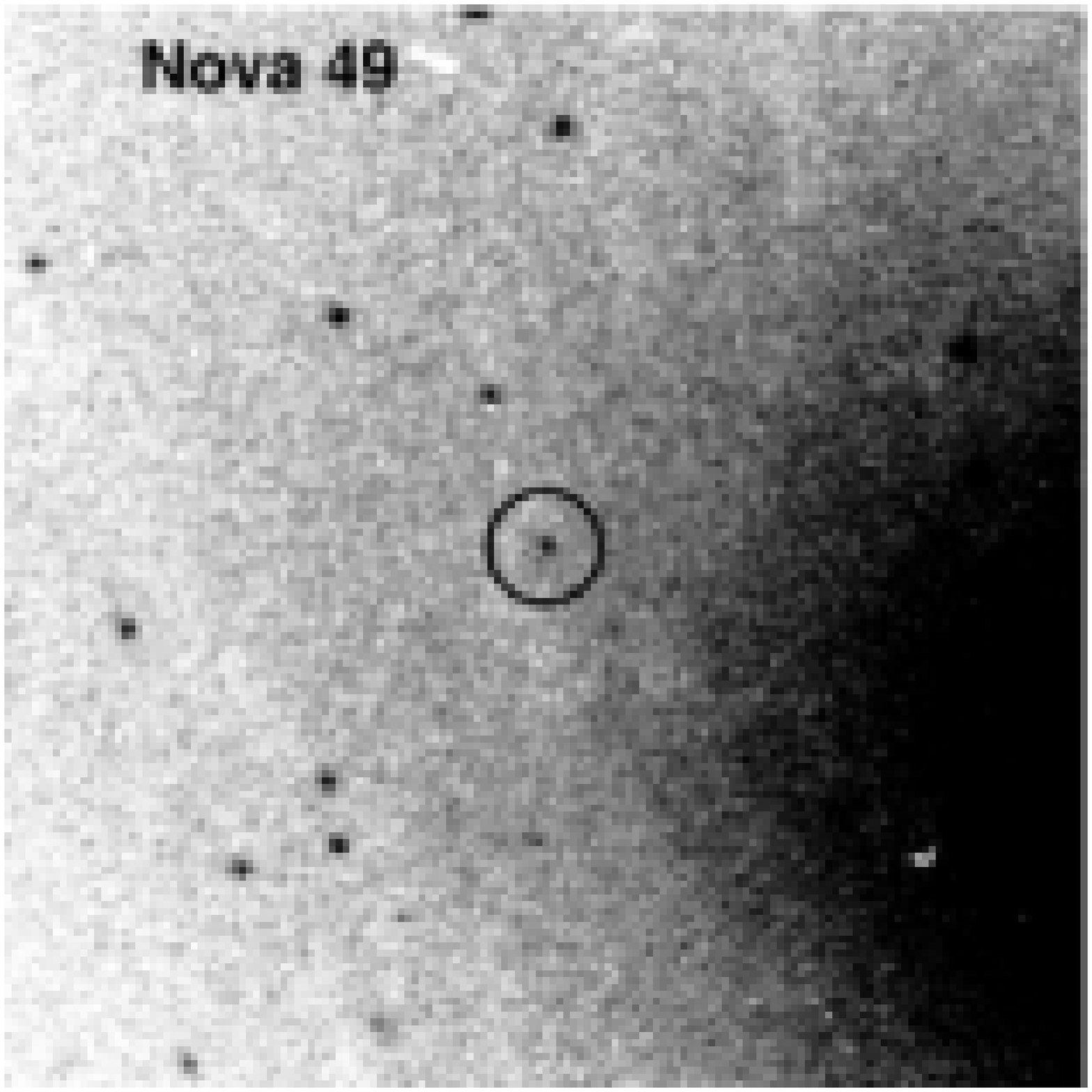}}\qquad
\subfigure[Nova 50 (B)]{\includegraphics[scale=.26, angle=0]{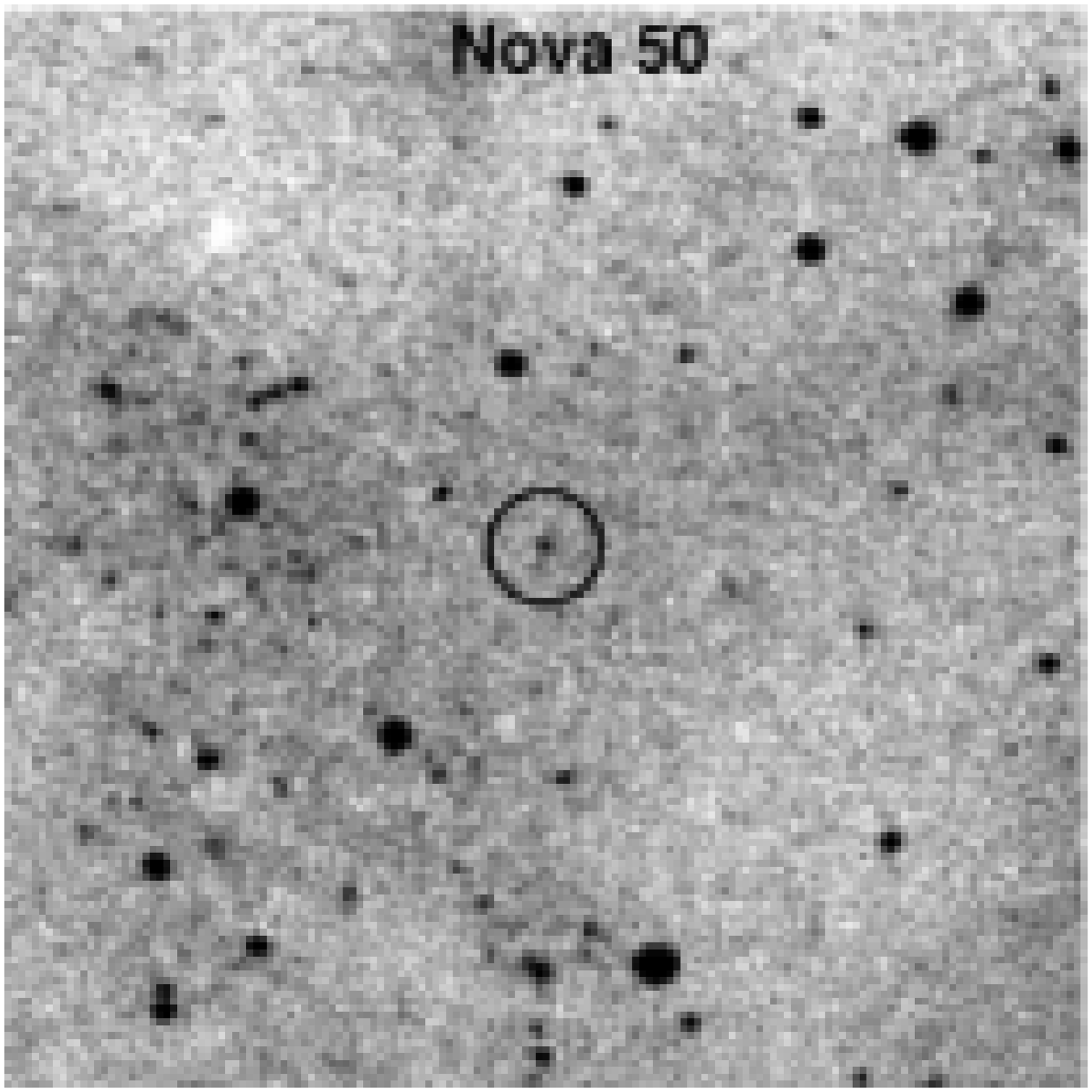}}\qquad
\subfigure[Nova 51 (B)]{\includegraphics[scale=.26, angle=0]{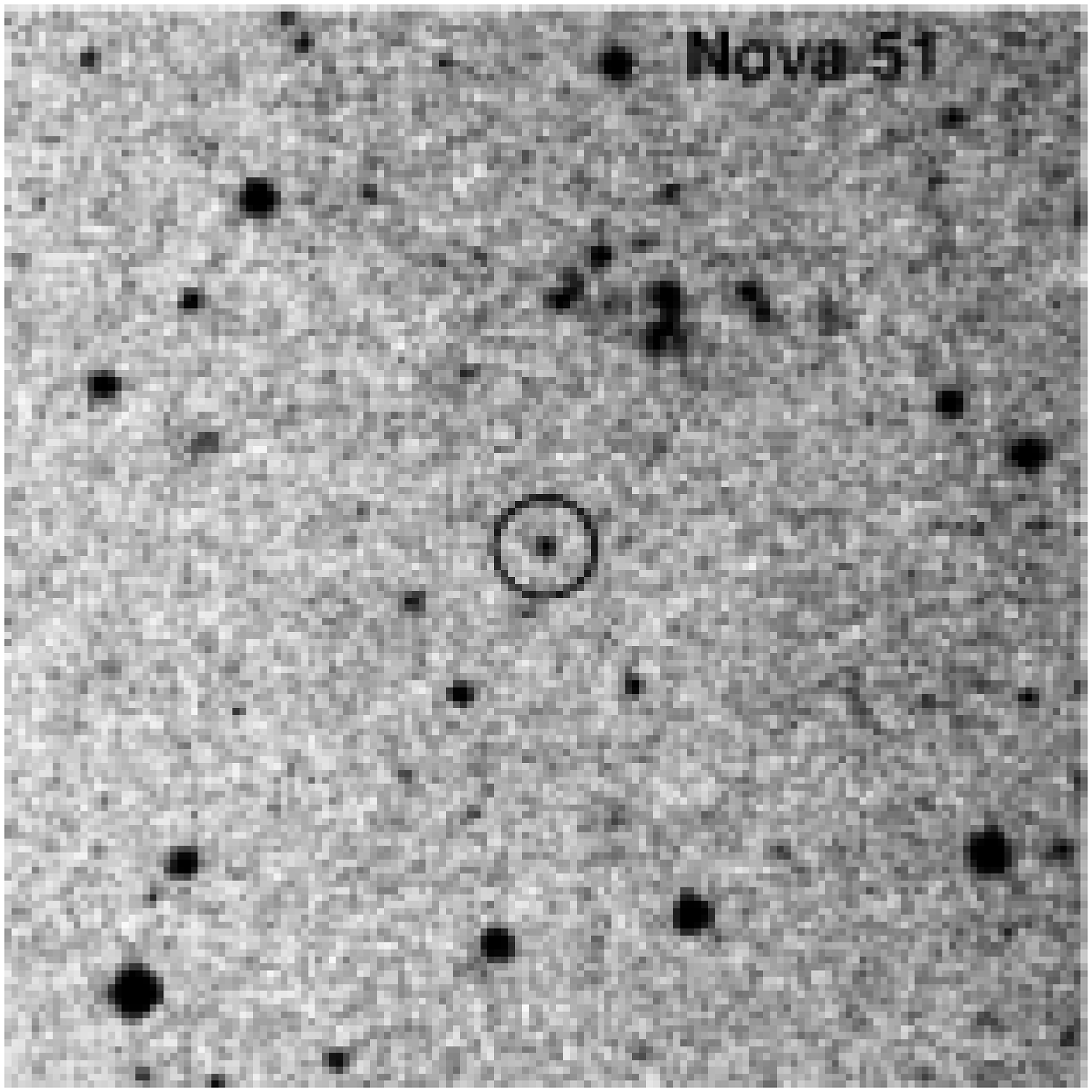}}\\
\subfigure[Nova 52 (B)]{\includegraphics[scale=.26, angle=0]{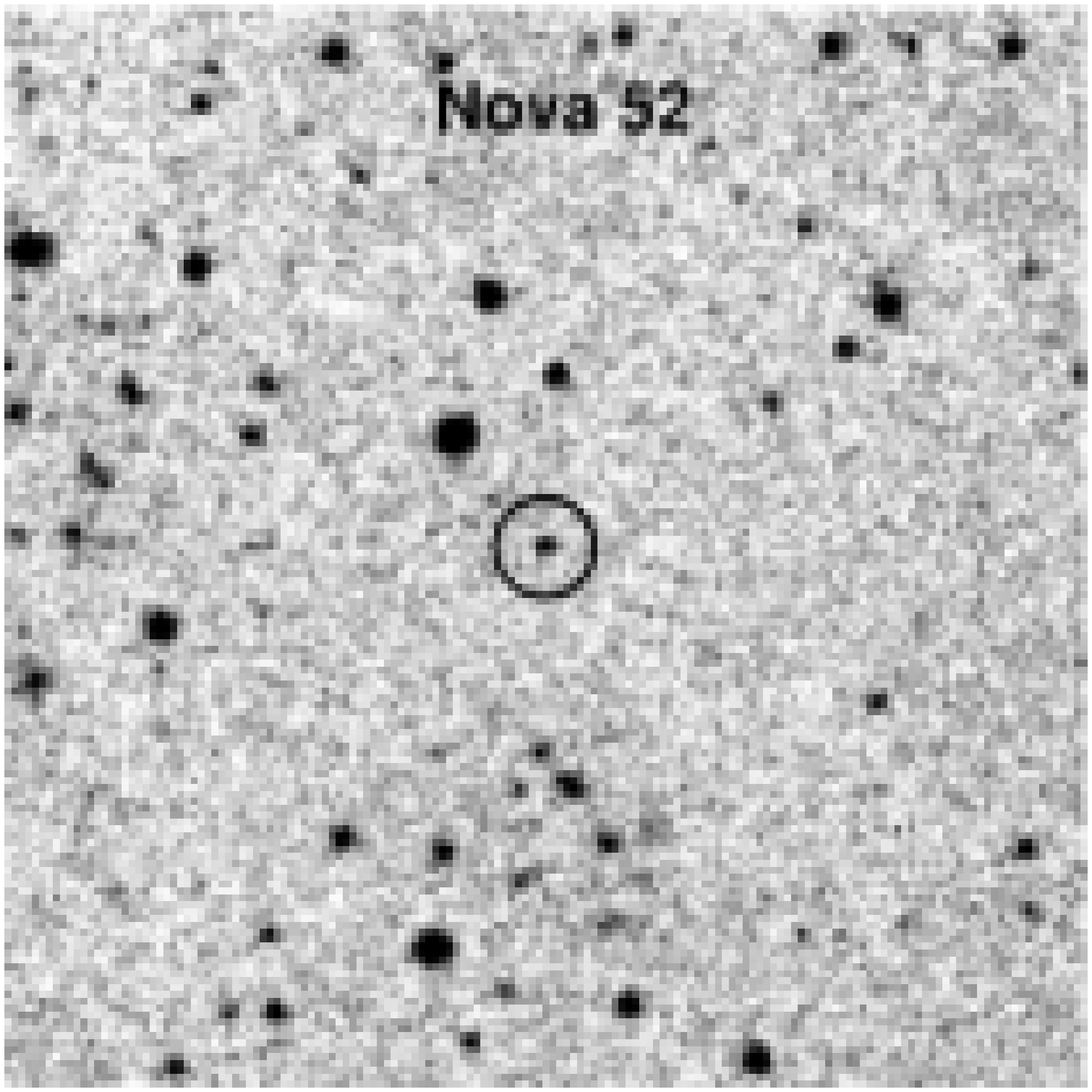}}\qquad
\subfigure[Nova 53 (U)]{\includegraphics[scale=.26, angle=0]{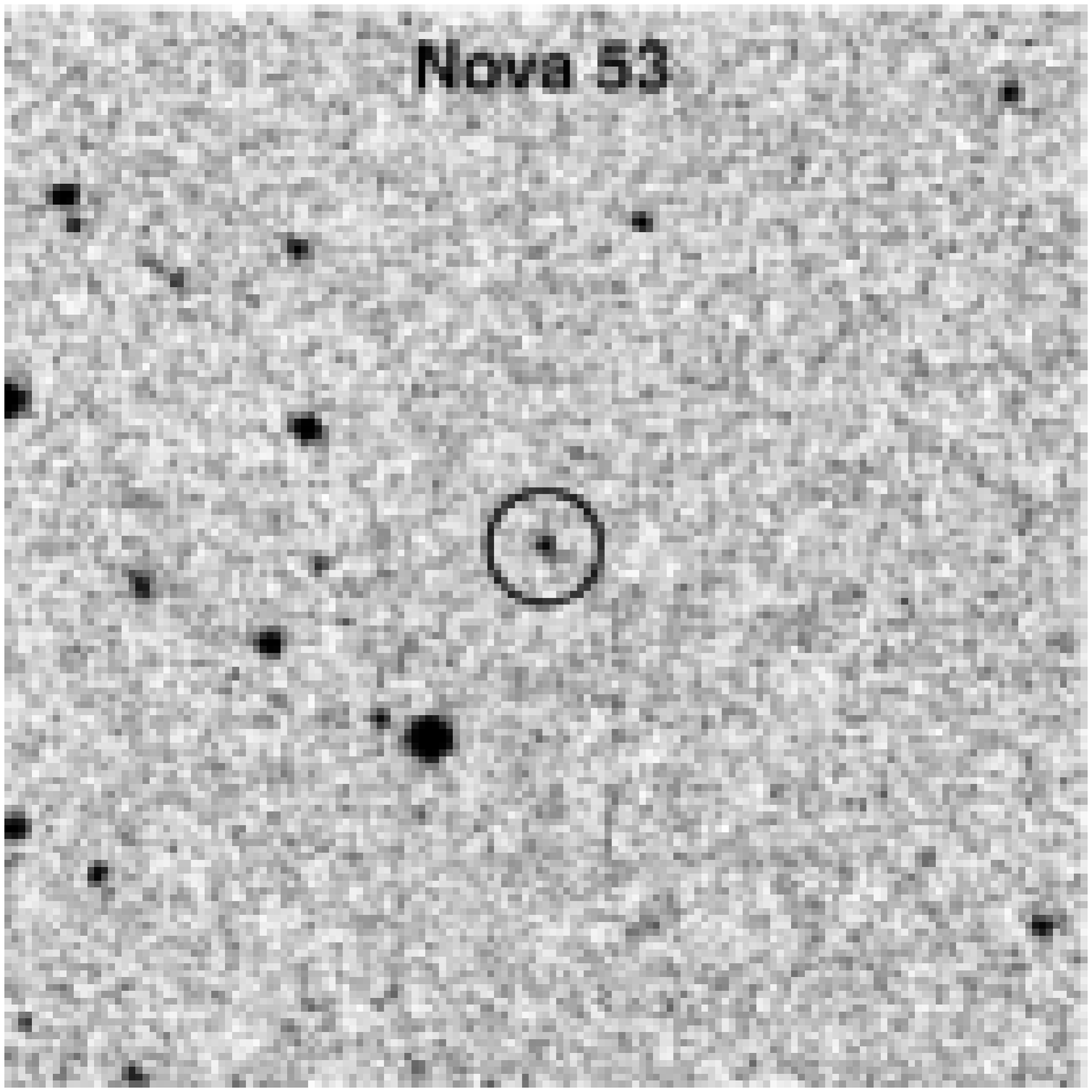}}\qquad
\subfigure[Nova 54 (B)]{\includegraphics[scale=.26, angle=0]{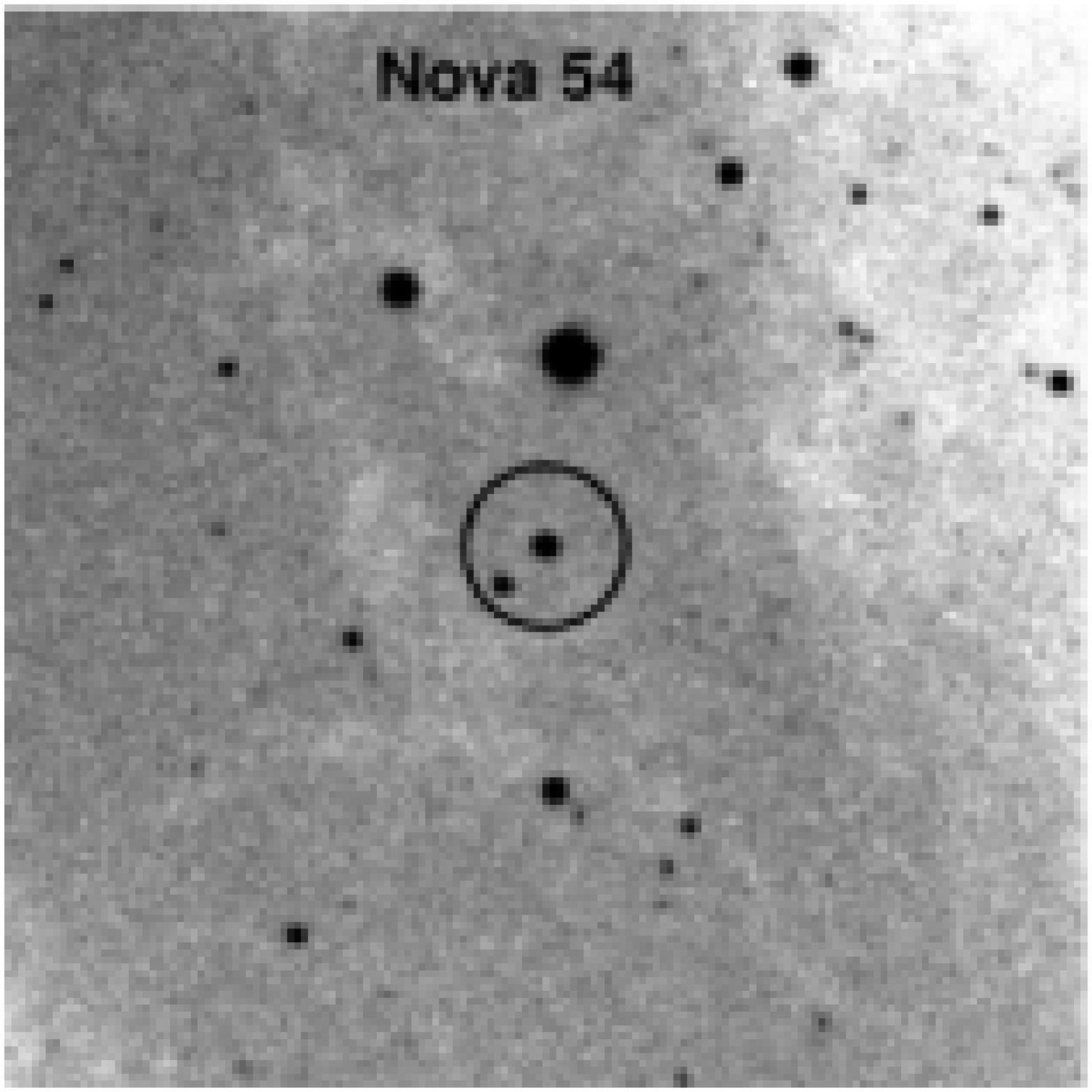}}\\
\subfigure[Nova 55 (B)]{\includegraphics[scale=.26, angle=0]{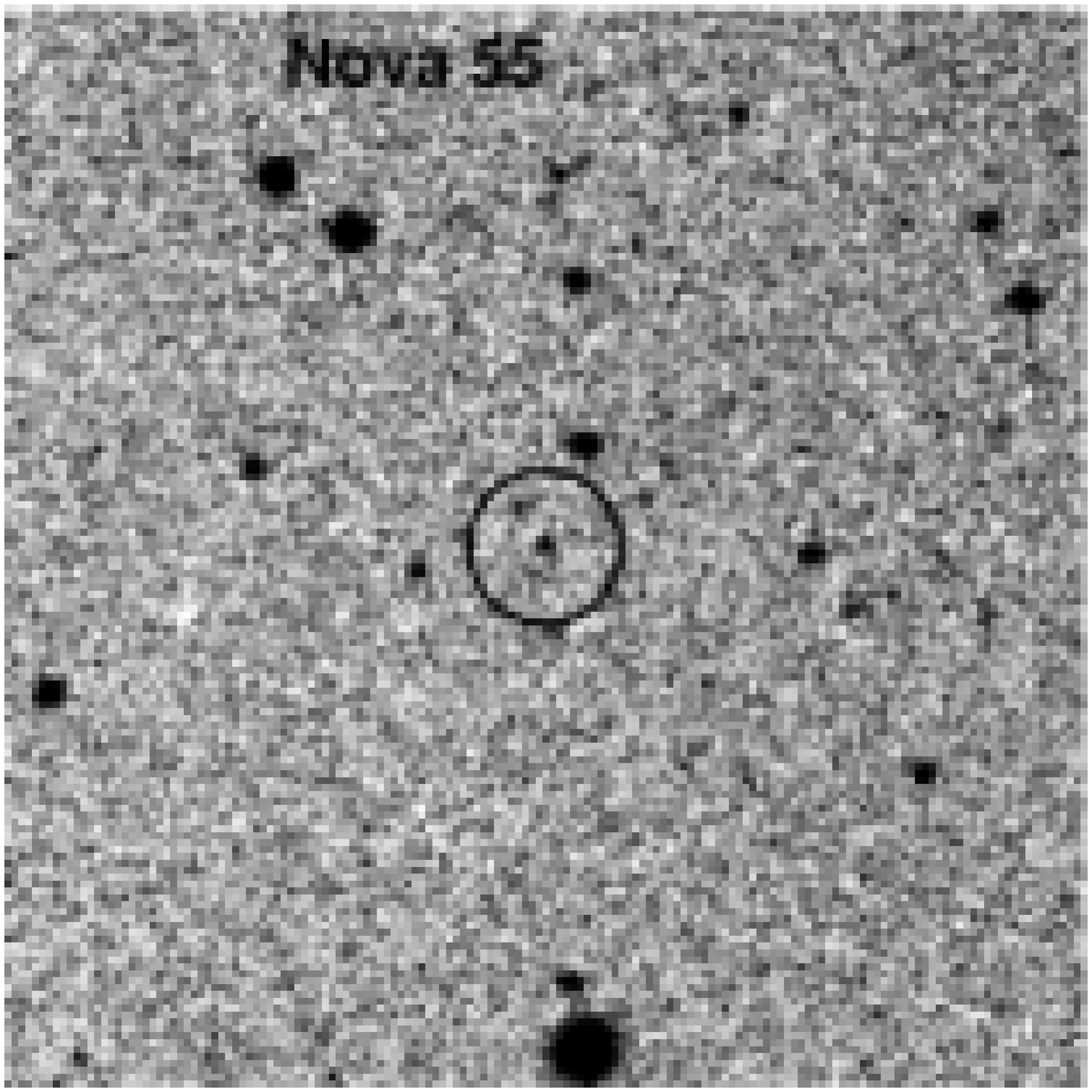}}\qquad
\subfigure[Nova 56 (B)]{\includegraphics[scale=.26, angle=0]{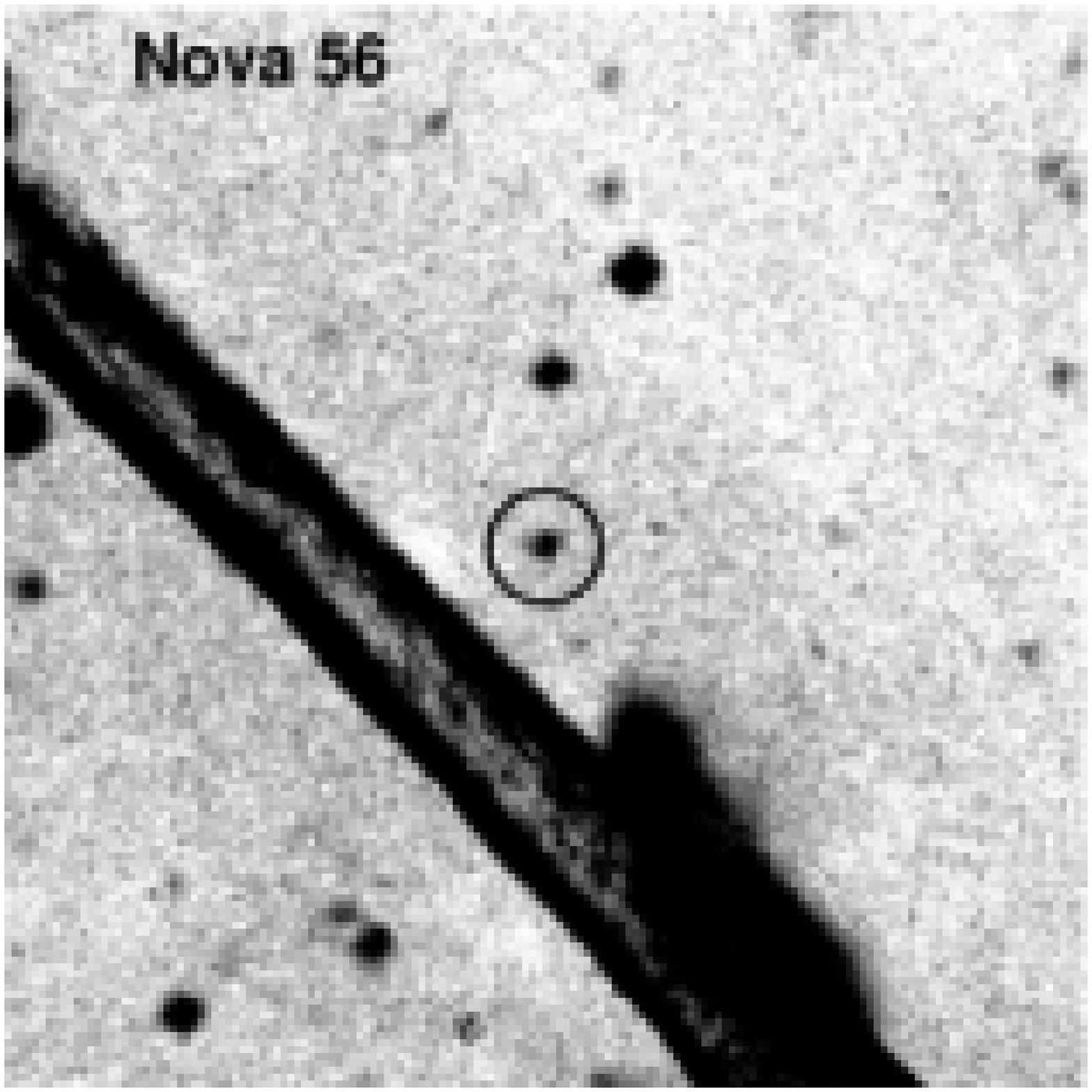}}\qquad
\subfigure[Nova 57 (U)]{\includegraphics[scale=.26, angle=0]{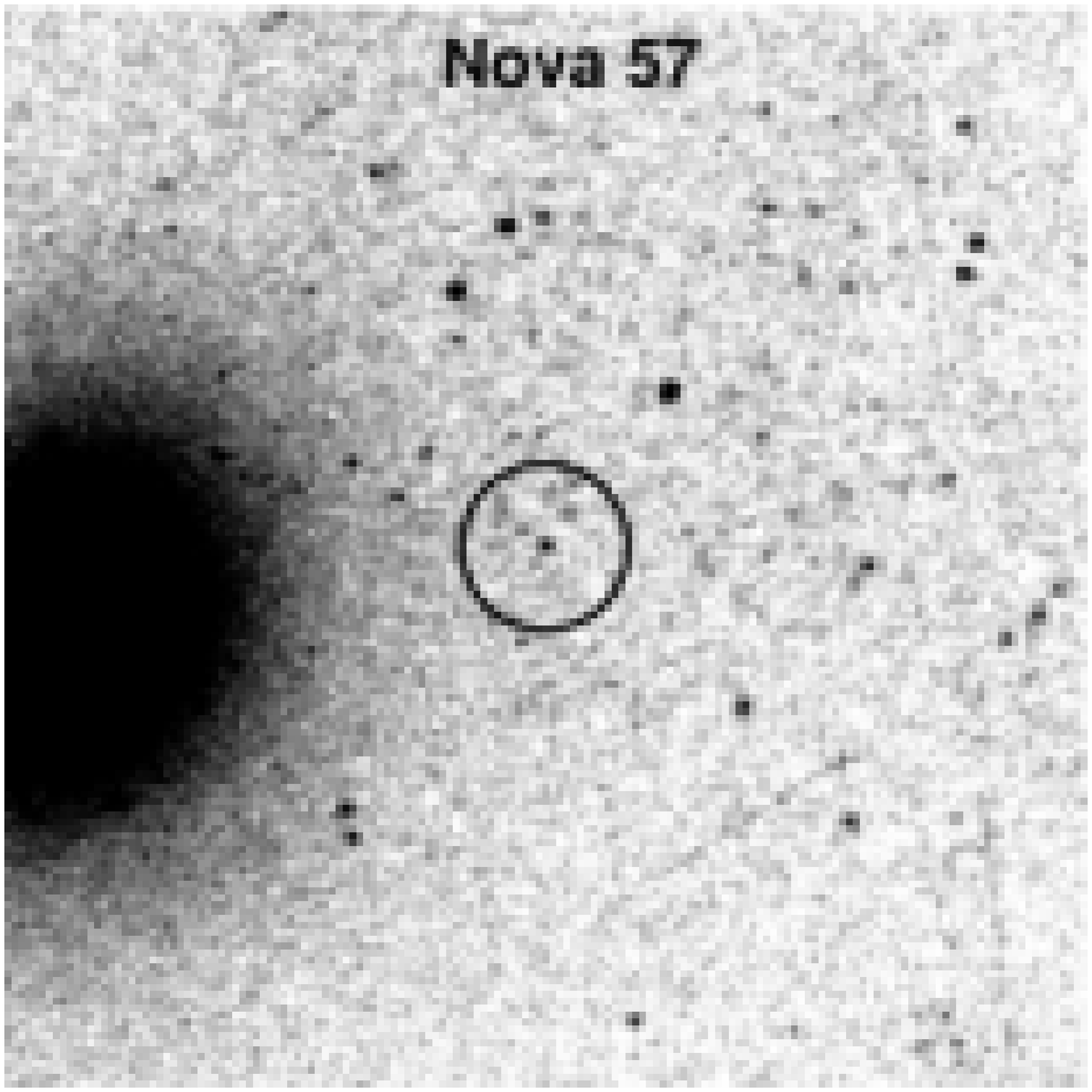}}\\
\subfigure[Nova 58 (U)]{\includegraphics[scale=.26, angle=0]{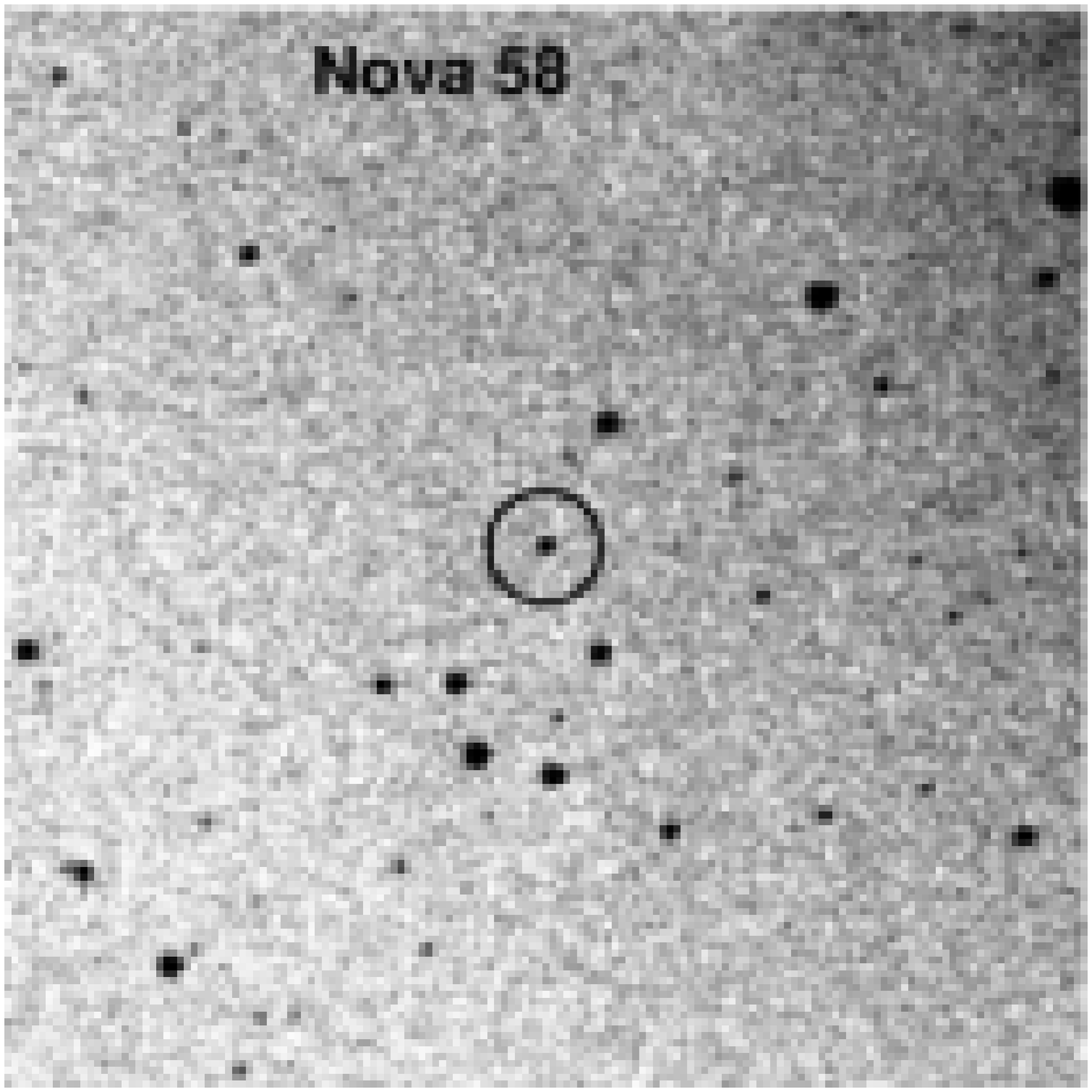}}\qquad
\subfigure[Nova 59 (B)]{\includegraphics[scale=.26, angle=0]{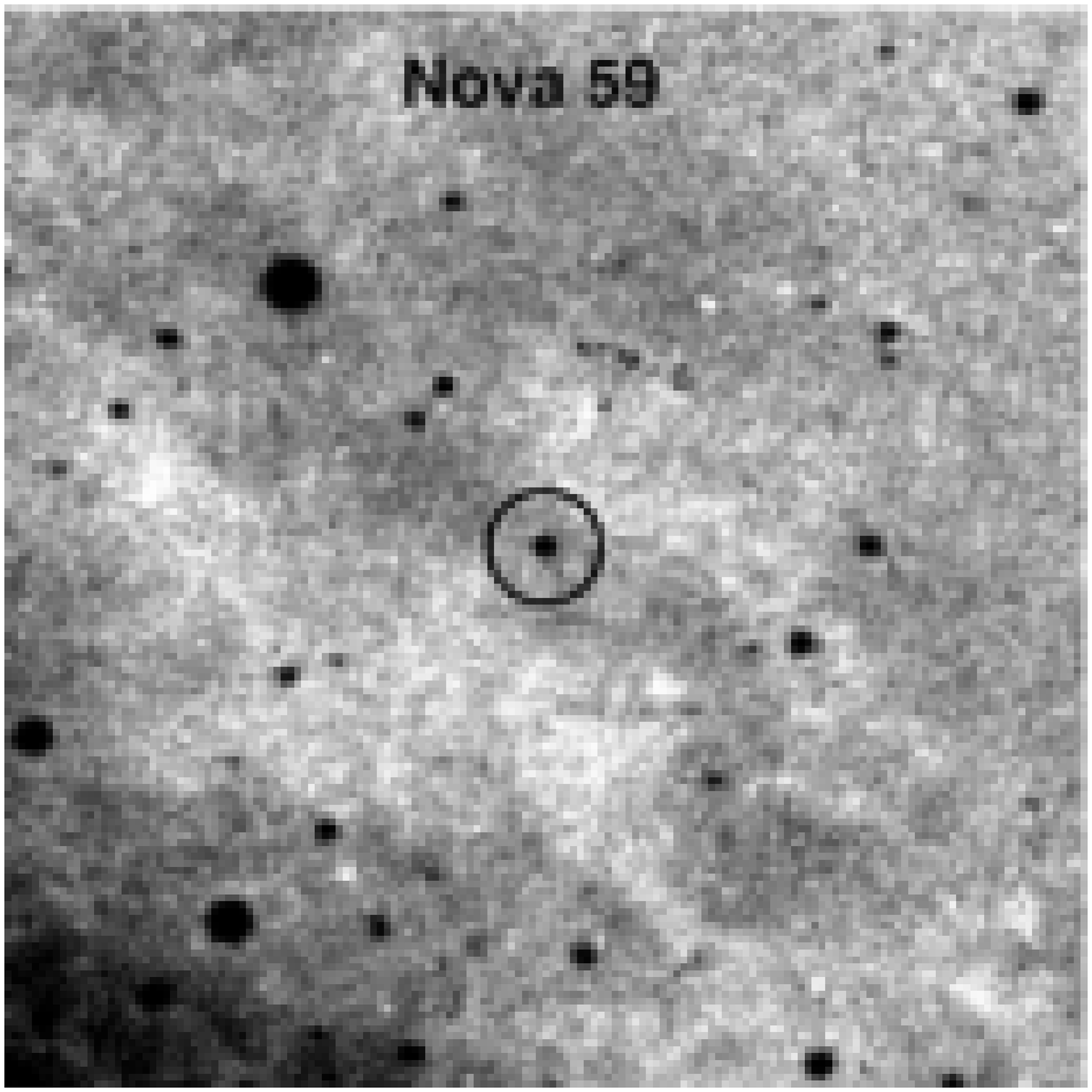}}\qquad
\subfigure[Nova 60 (U)]{\includegraphics[scale=.26, angle=0]{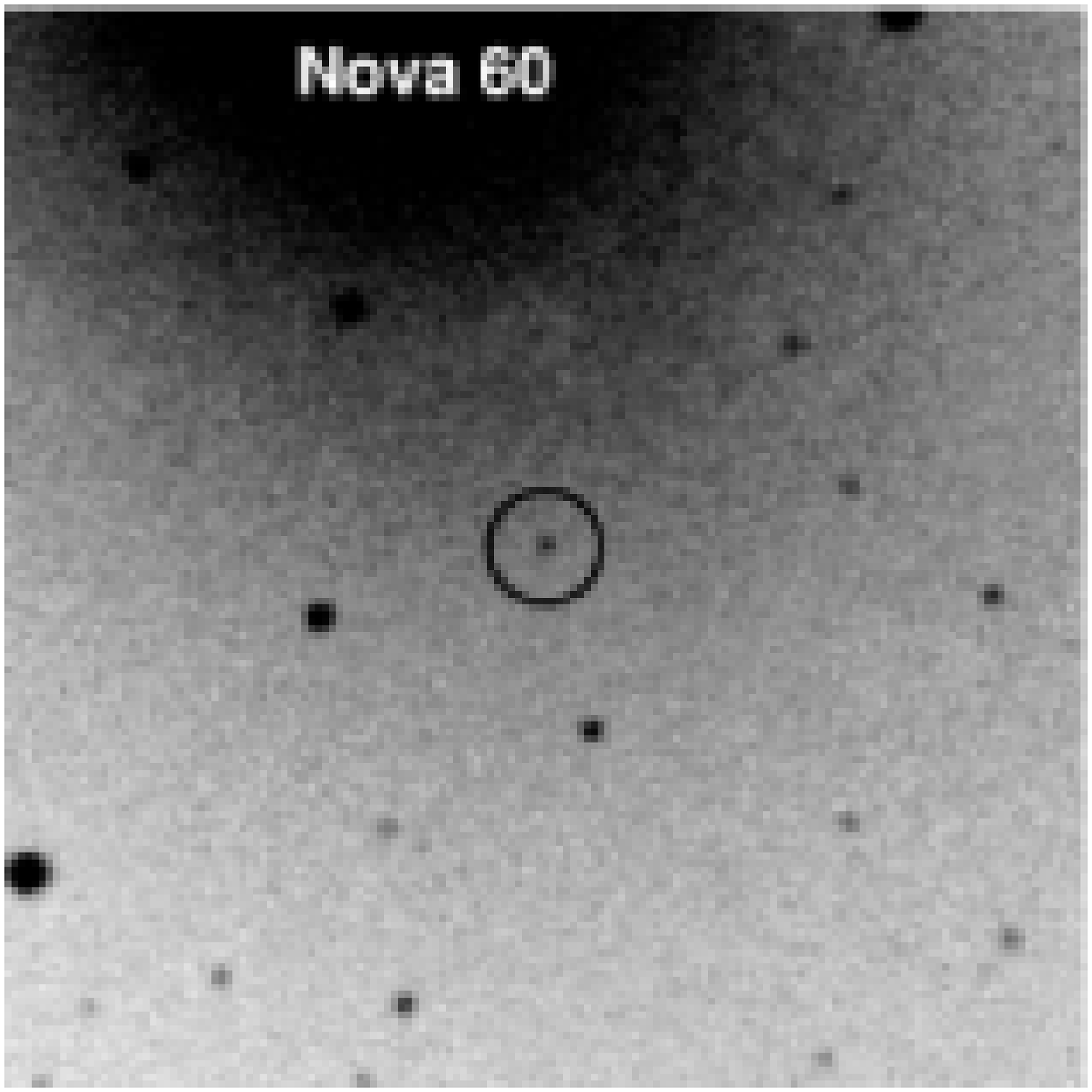}}
\caption{Finding charts for novae 49 - 60.}
\end{figure*}
}

\onlfig{13}{
\begin{figure*}[t]
\centering
\subfigure[Nova 61 (B)]{\includegraphics[scale=.26, angle=0]{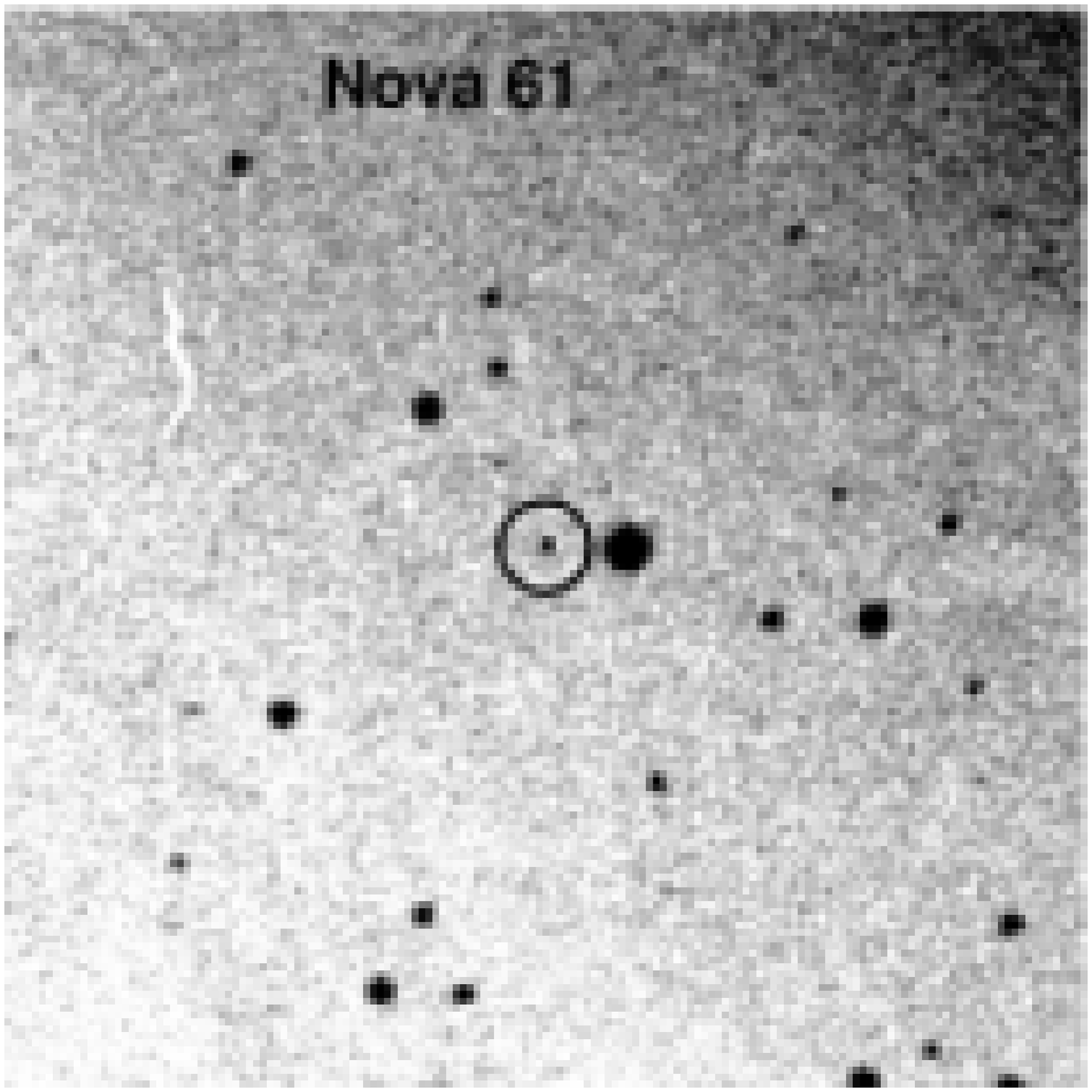}}\qquad
\subfigure[Nova 62 (B)]{\includegraphics[scale=.26, angle=0]{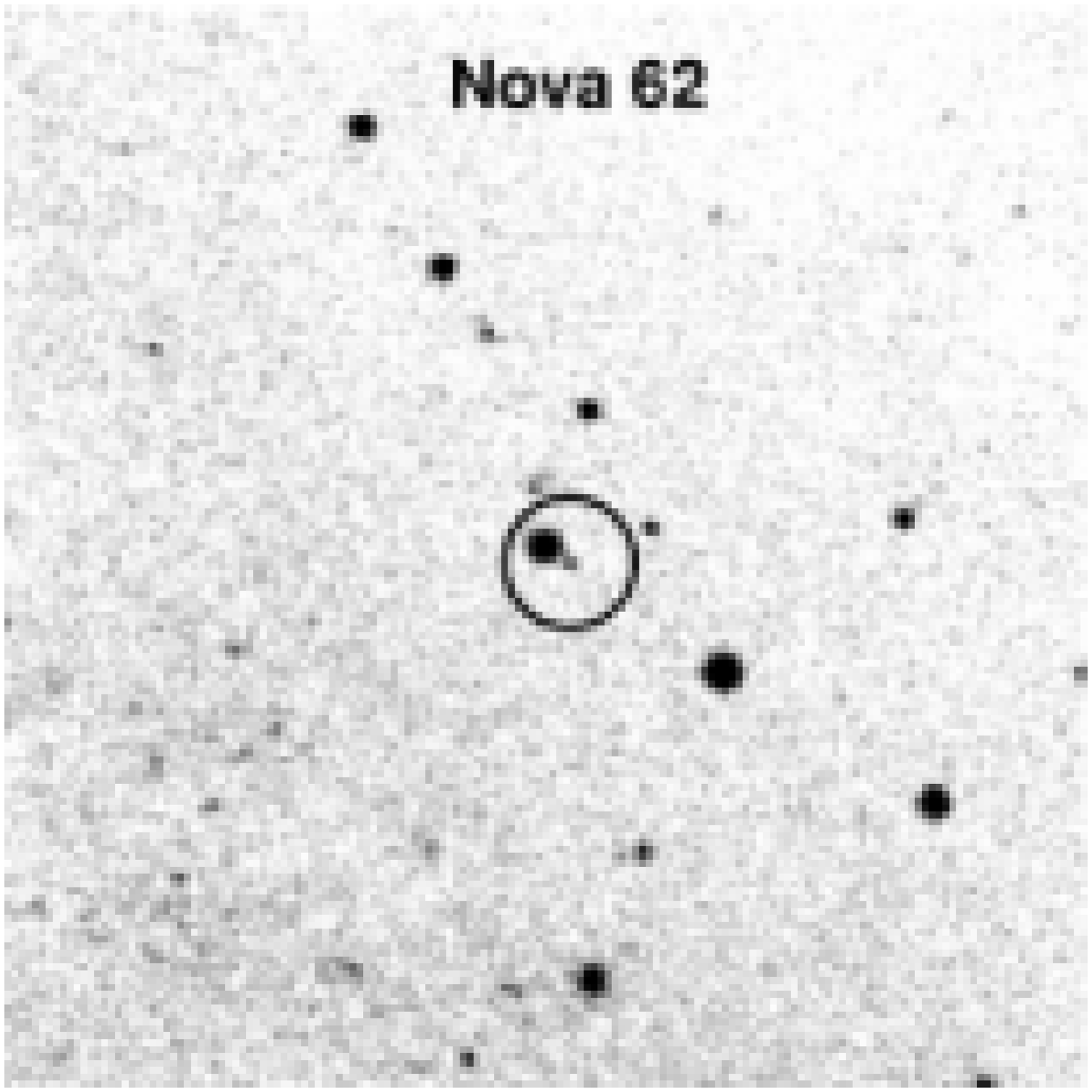}}\qquad
\subfigure[Nova 63 (B)]{\includegraphics[scale=.26, angle=0]{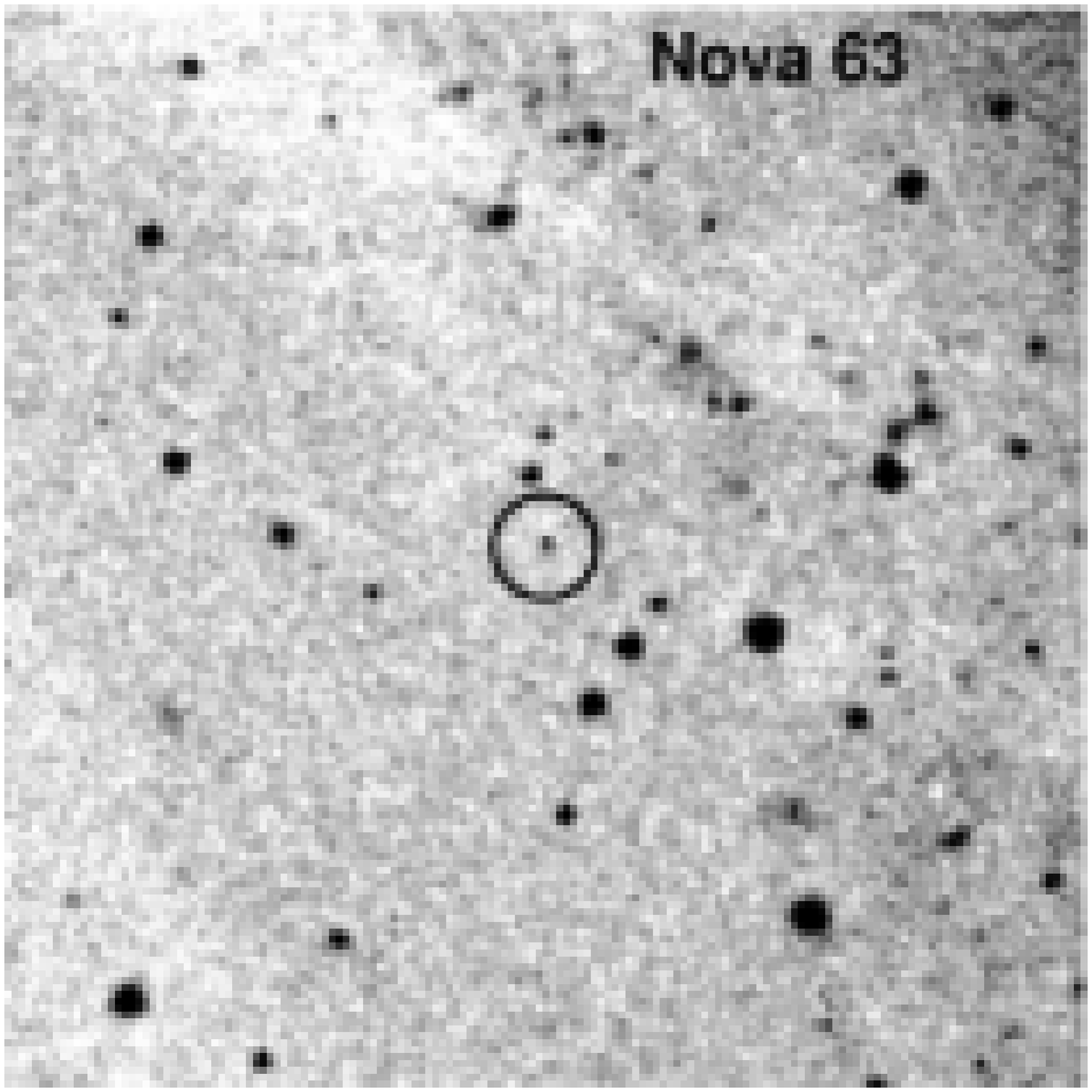}}\\
\subfigure[Nova 64 (U)]{\includegraphics[scale=.26, angle=0]{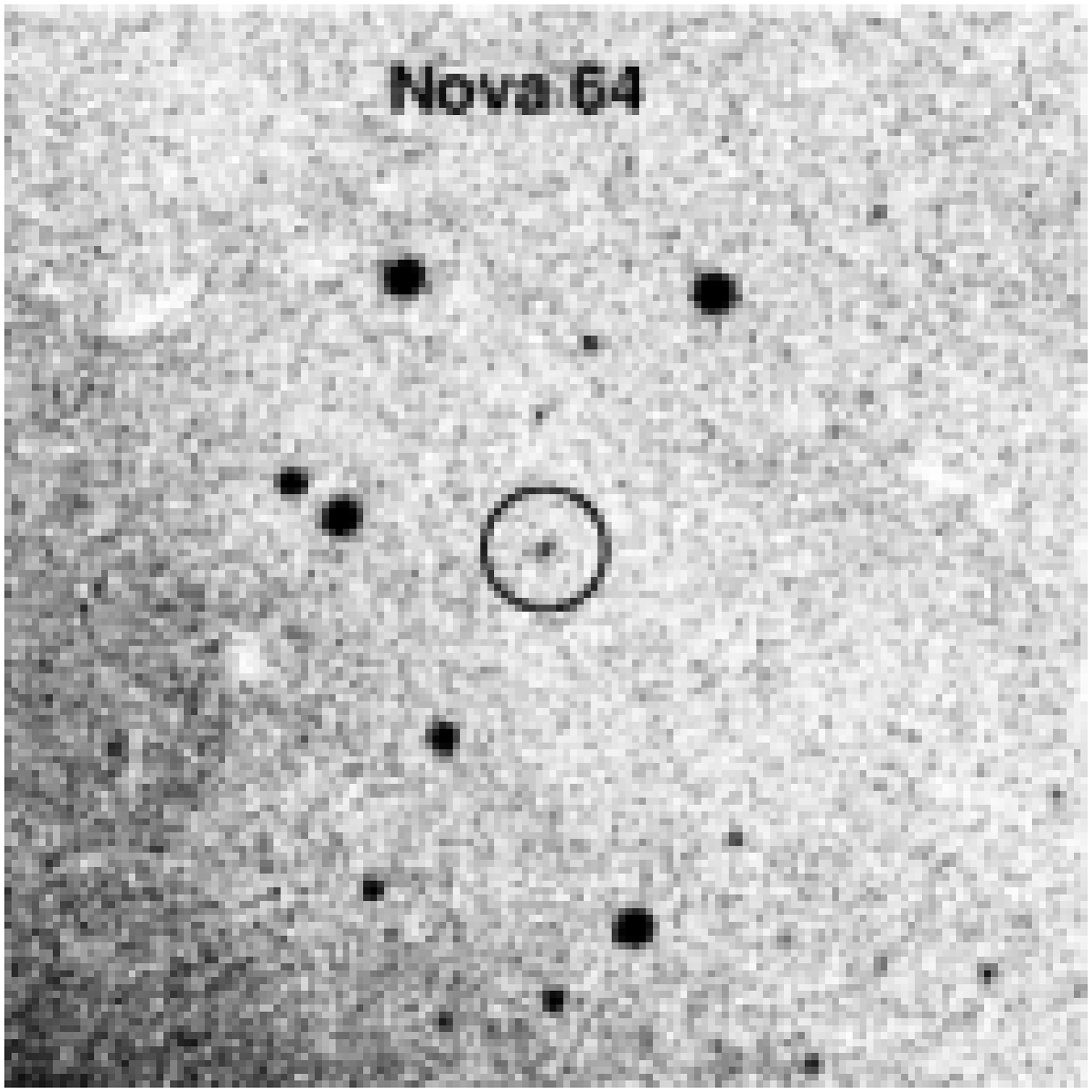}}\qquad
\subfigure[Nova 65 (B)]{\includegraphics[scale=.26, angle=0]{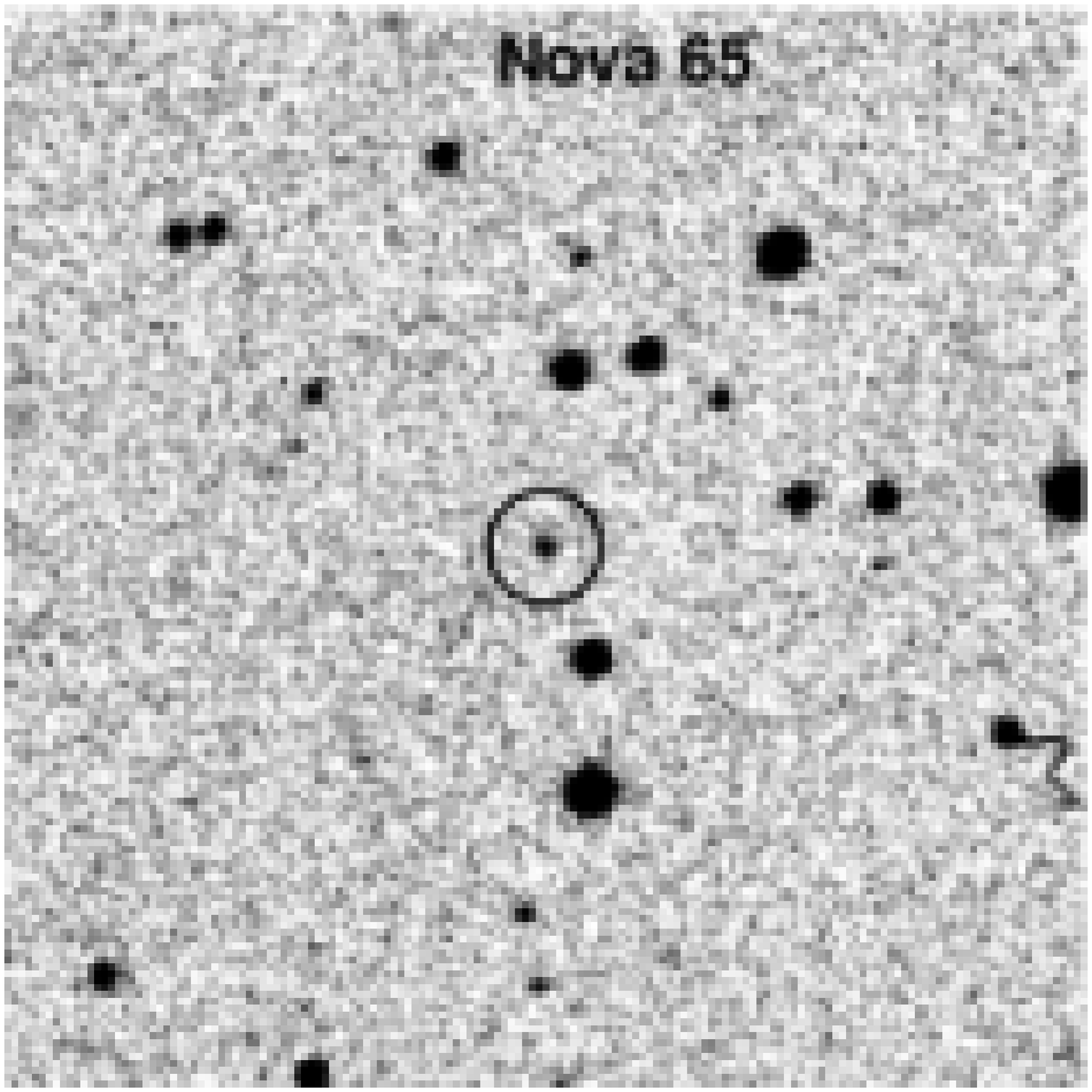}}\qquad
\subfigure[Nova 66 (U)]{\includegraphics[scale=.26, angle=0]{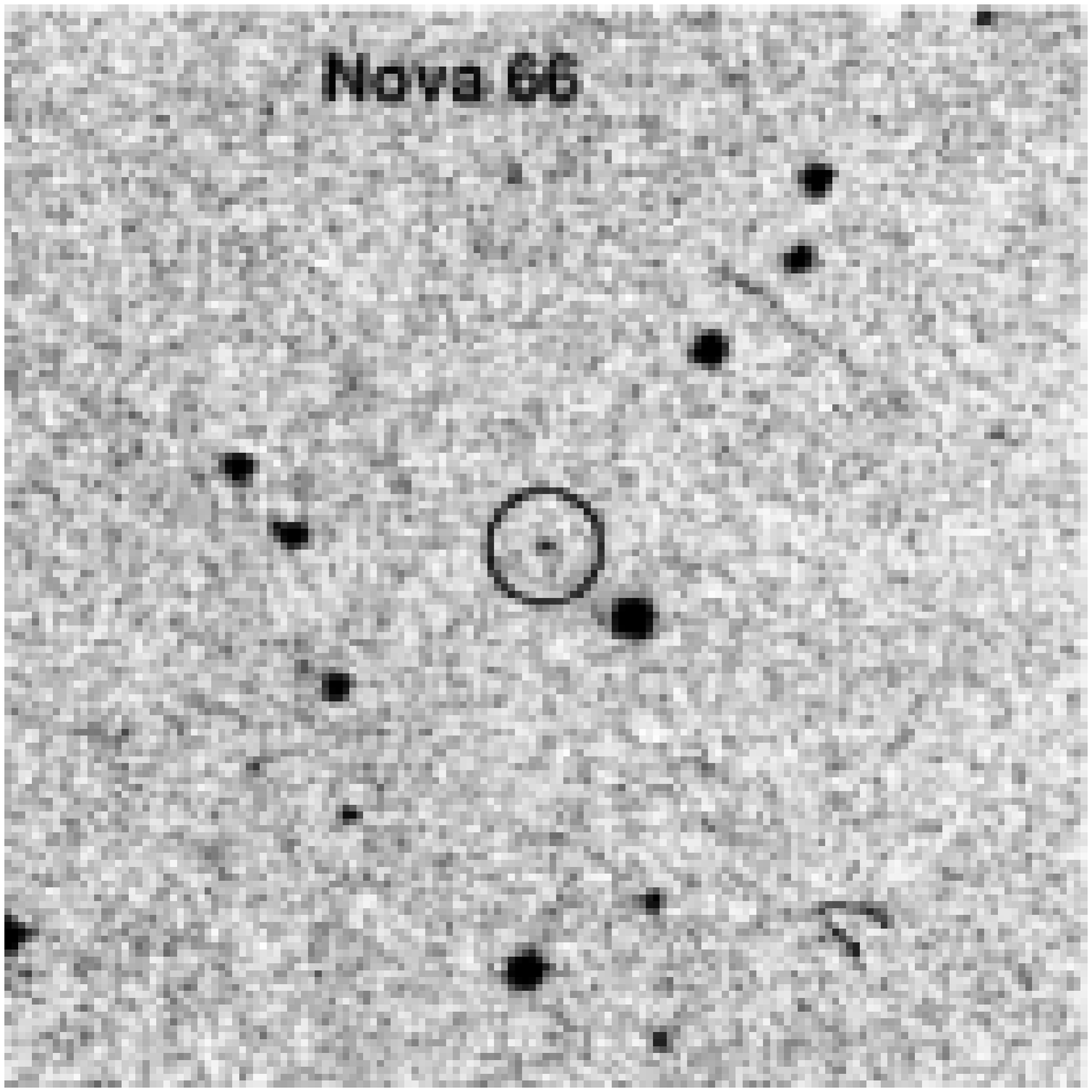}}\\
\subfigure[Nova 67 (V)]{\includegraphics[scale=.26, angle=0]{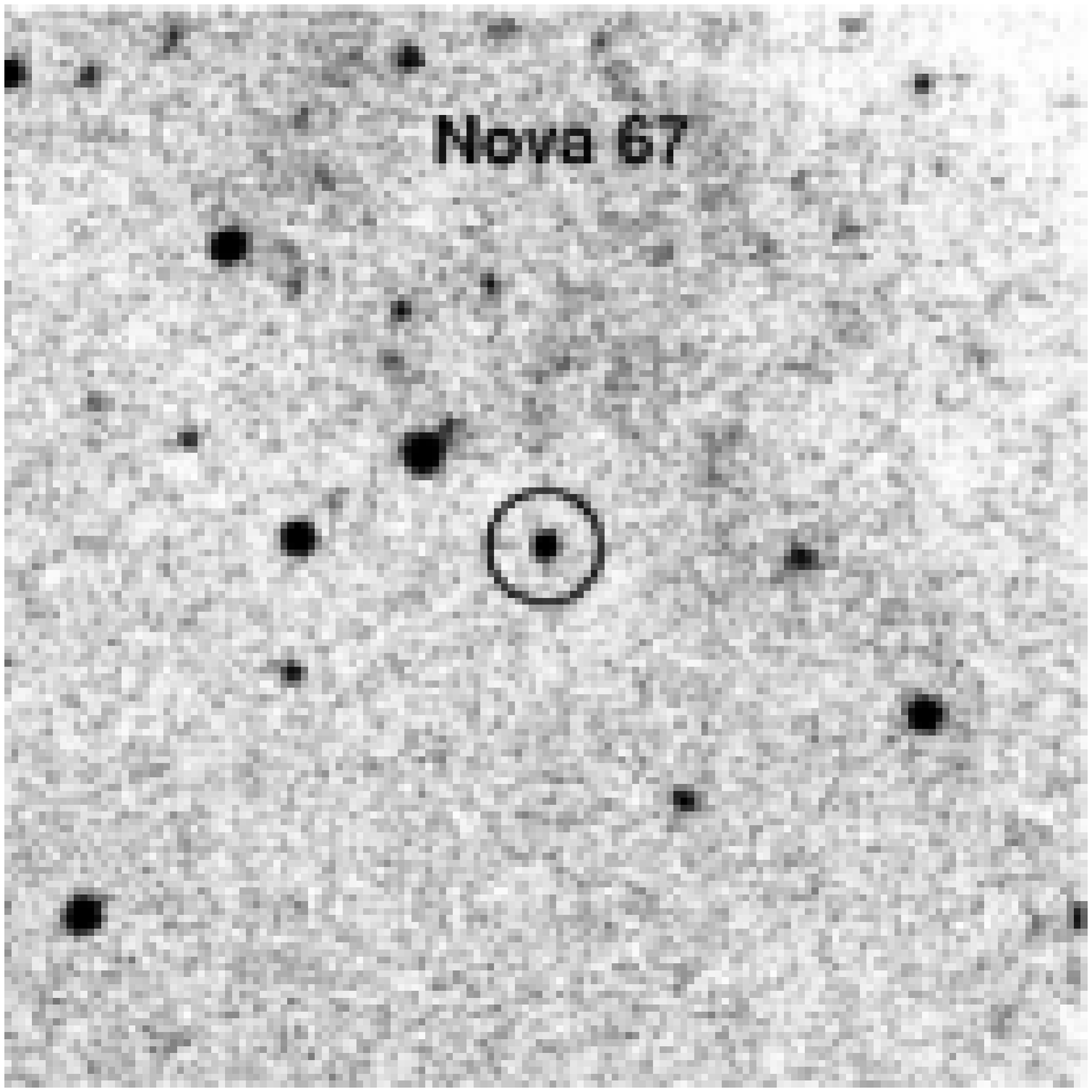}}\qquad
\subfigure[Nova 68 (U)]{\includegraphics[scale=.26, angle=0]{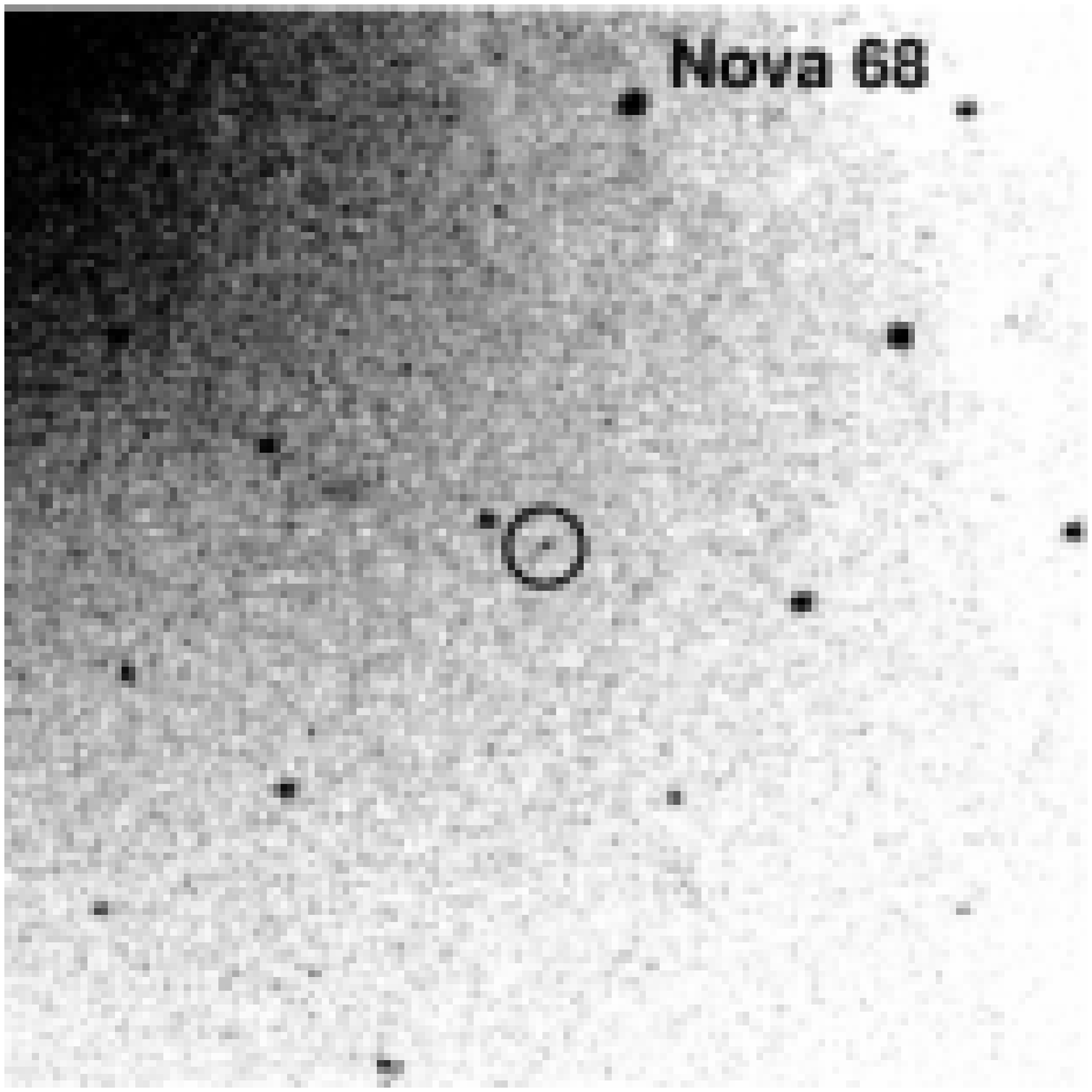}}\qquad
\subfigure[Nova 69 (U)]{\includegraphics[scale=.26, angle=0]{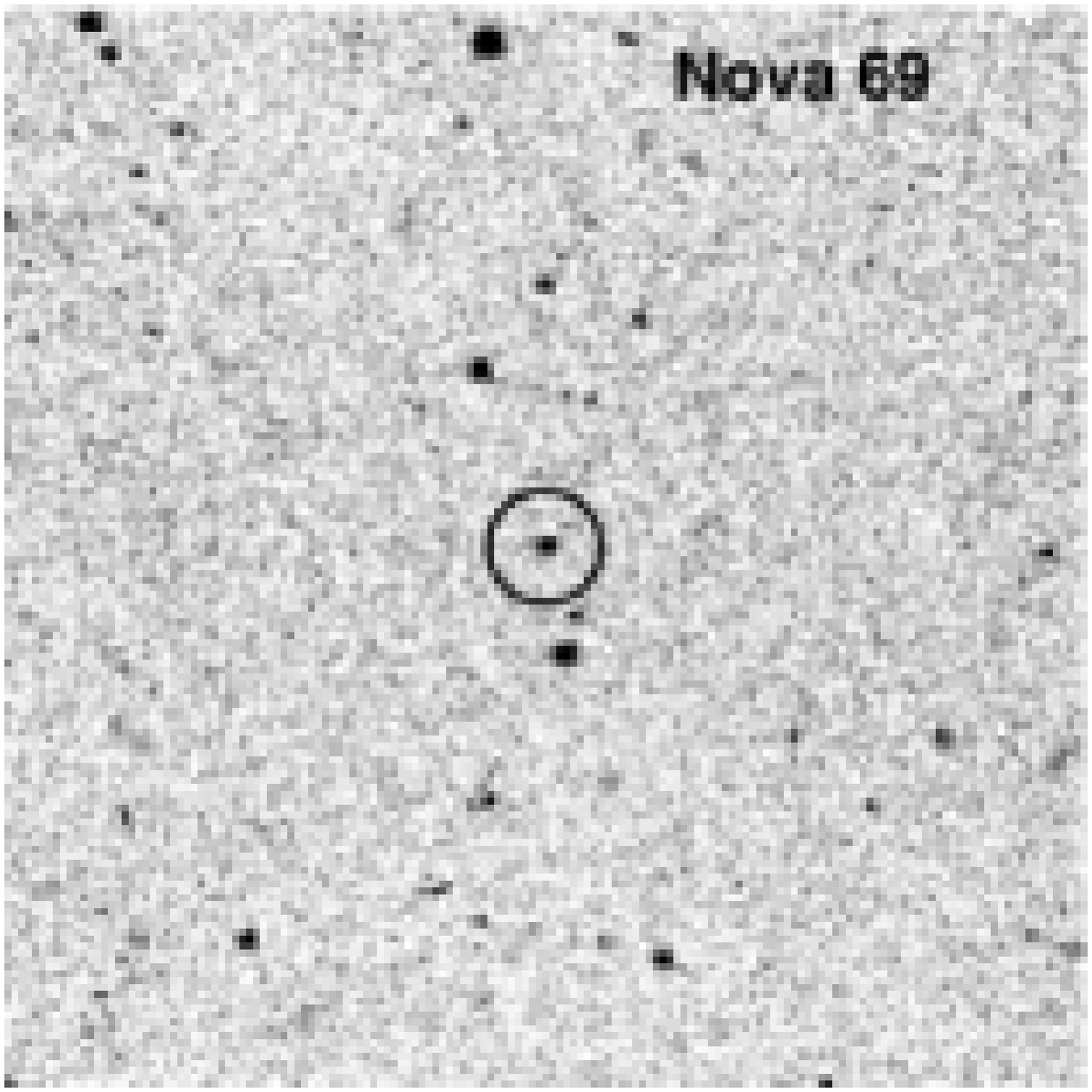}}\\
\subfigure[Nova 70 (U)]{\includegraphics[scale=.26, angle=0]{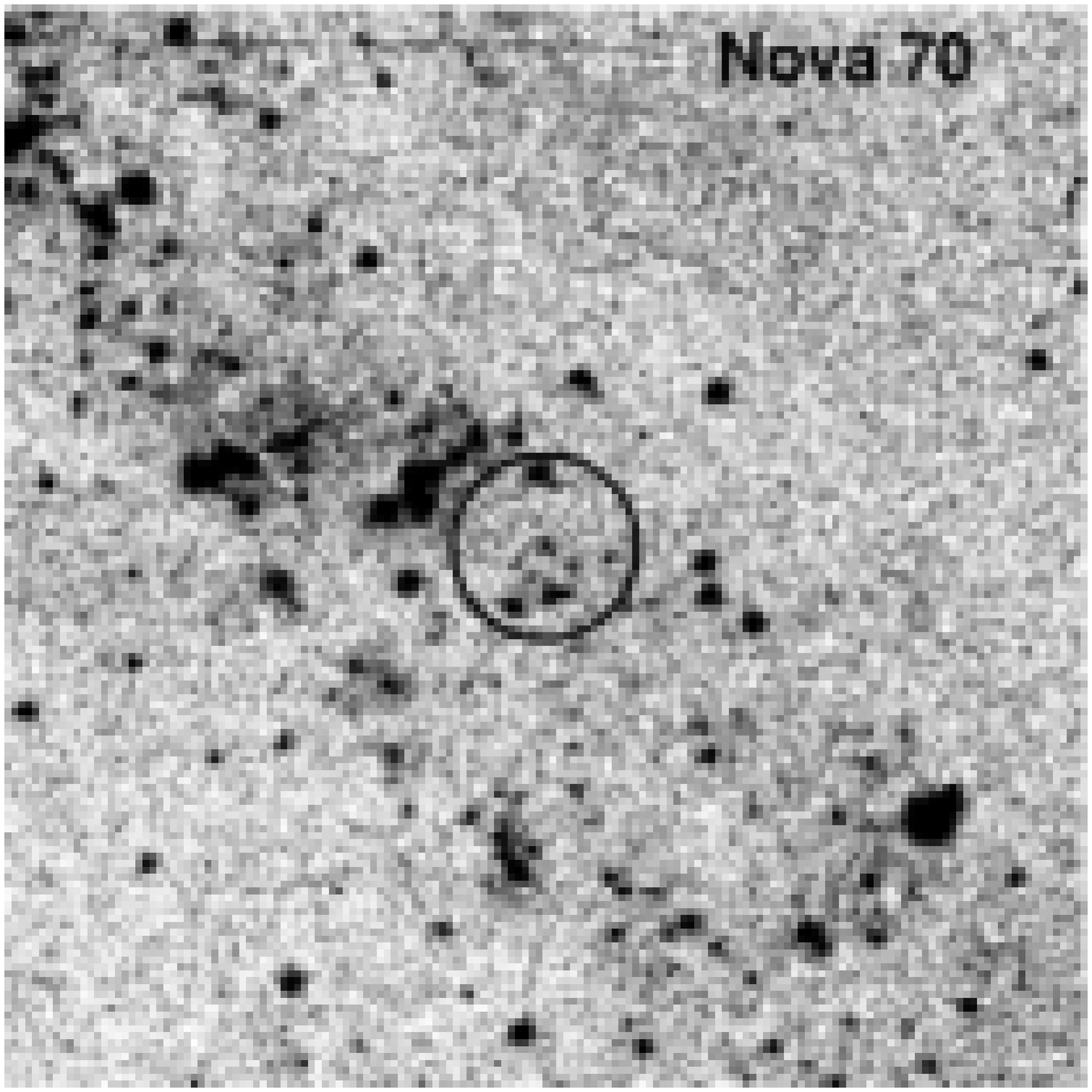}}\qquad
\subfigure[Nova 71 (U)]{\includegraphics[scale=.26, angle=0]{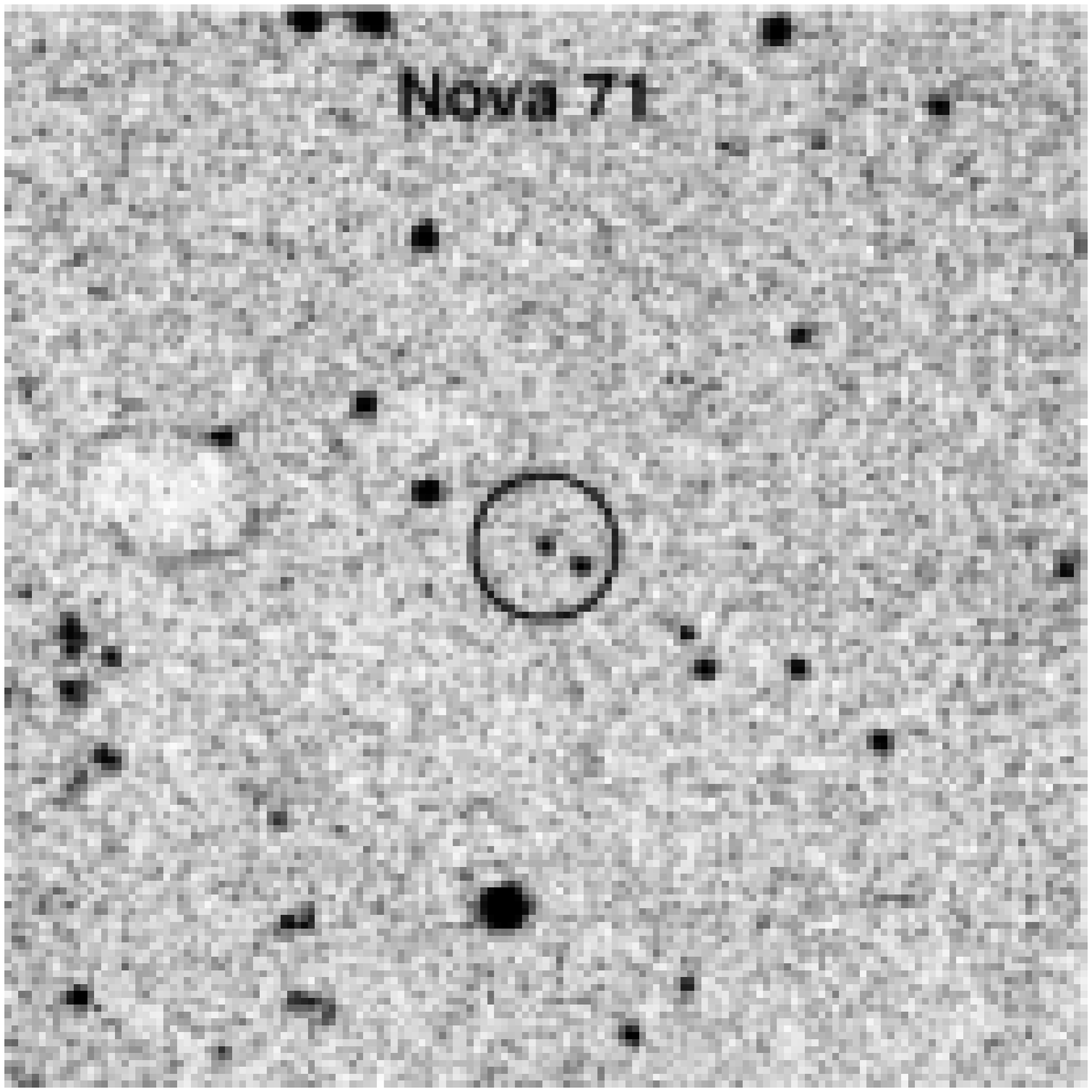}}\qquad
\subfigure[Nova 72 (B)]{\includegraphics[scale=.26, angle=0]{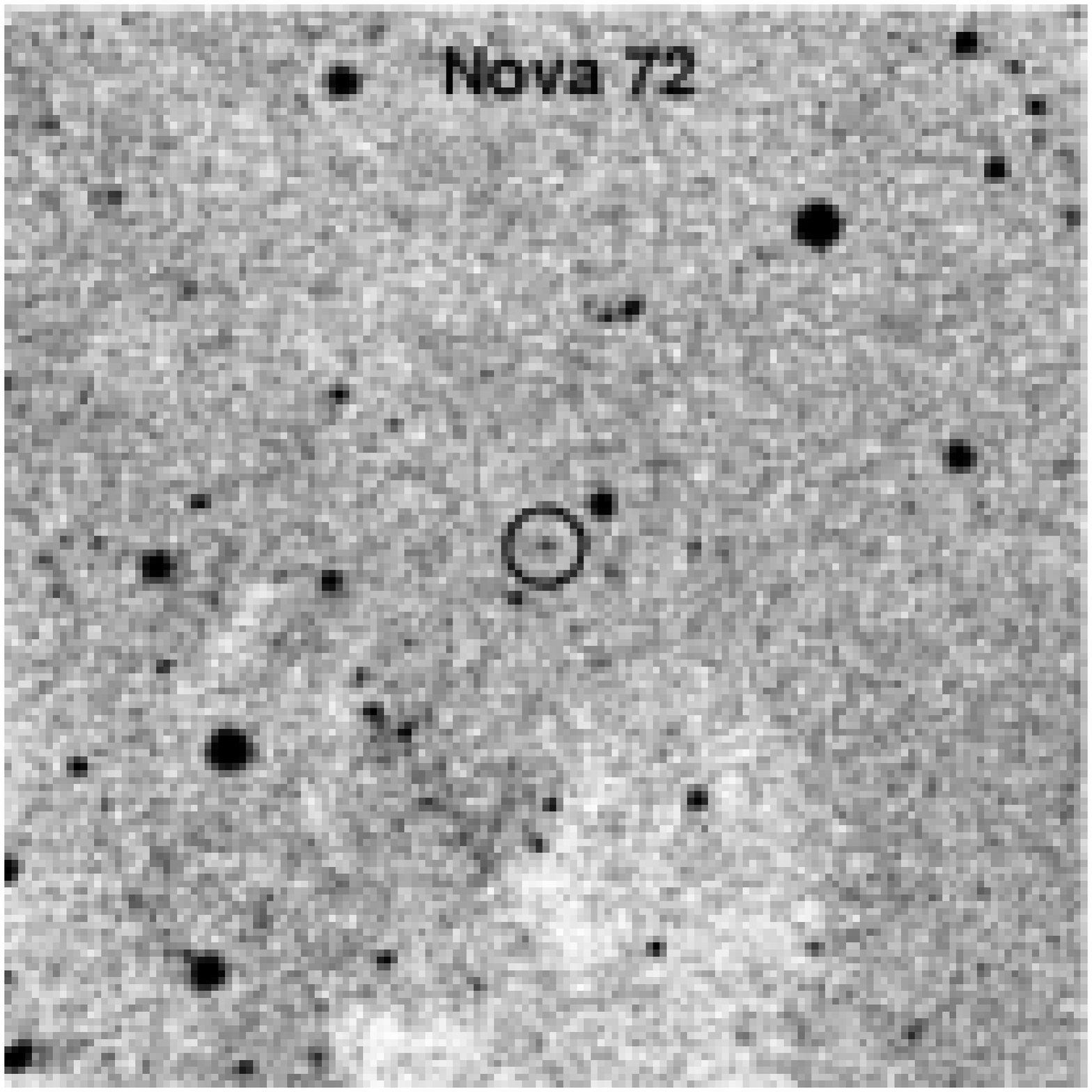}}
\caption{Finding charts for novae 61 - 72.}
\end{figure*}
}

\onlfig{14}{
\begin{figure*}[t]
\centering
\subfigure[Nova 73 (U)]{\includegraphics[scale=.26, angle=0]{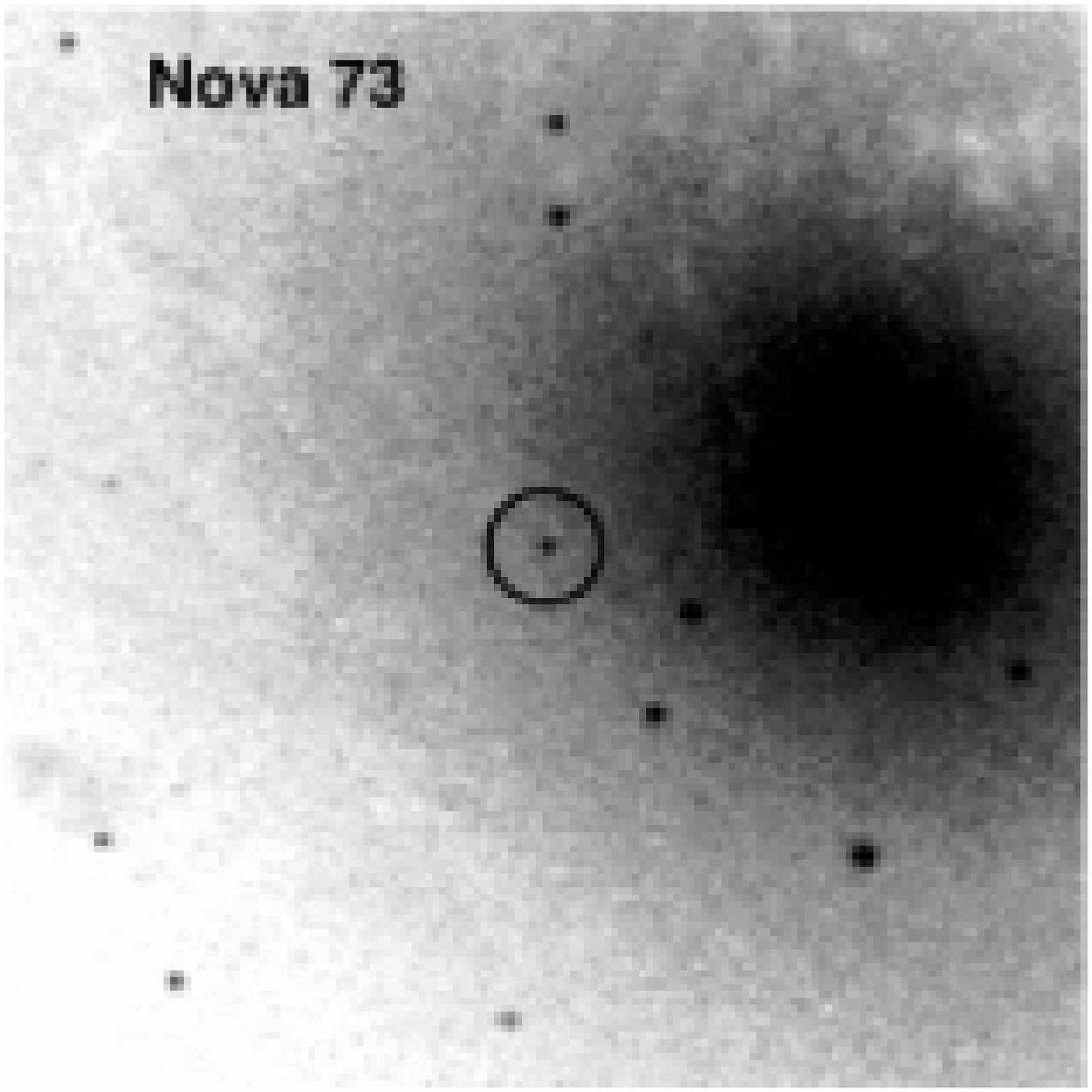}}\qquad
\subfigure[Nova 74 (B)]{\includegraphics[scale=.26, angle=0]{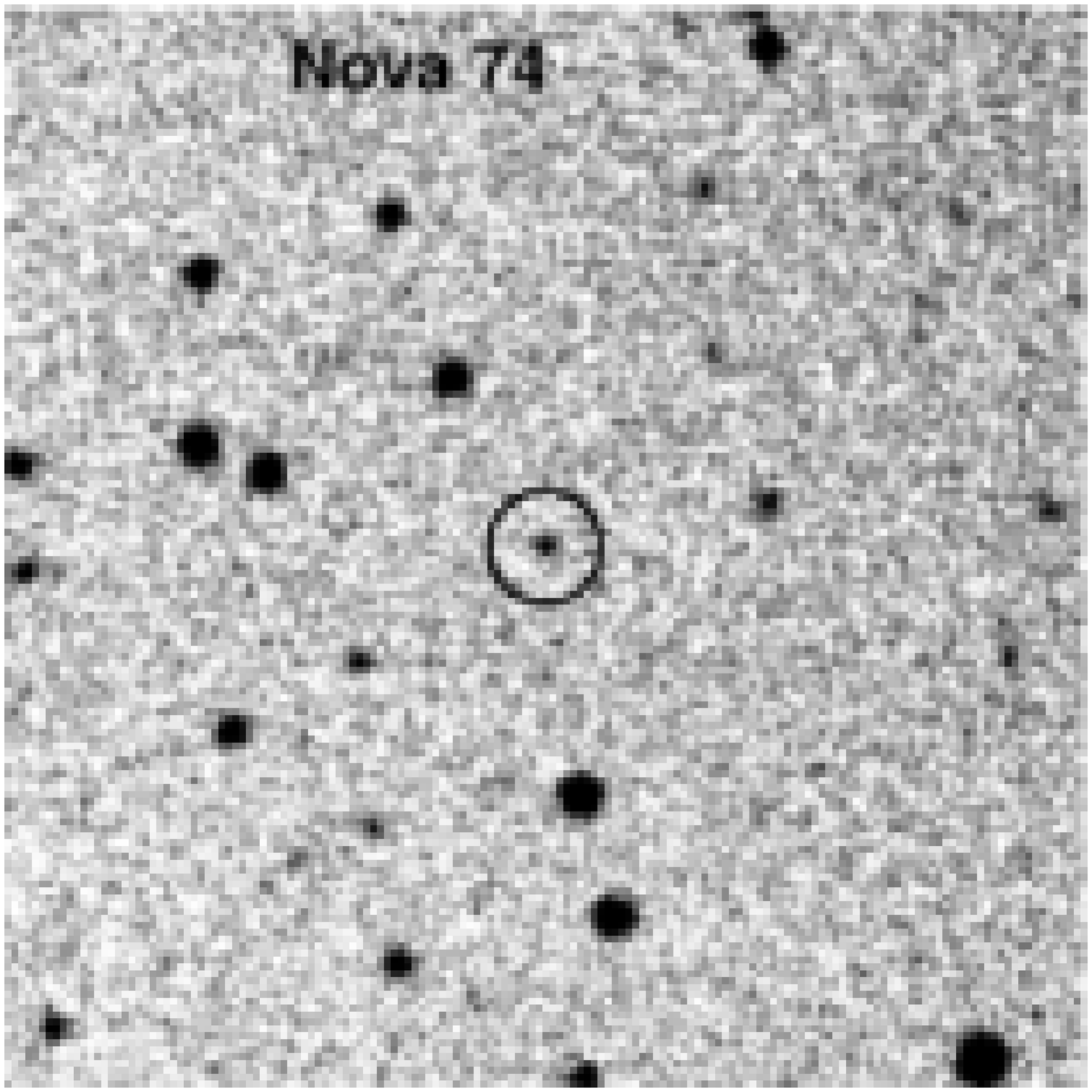}}\qquad
\subfigure[Nova 75 (B)]{\includegraphics[scale=.26, angle=0]{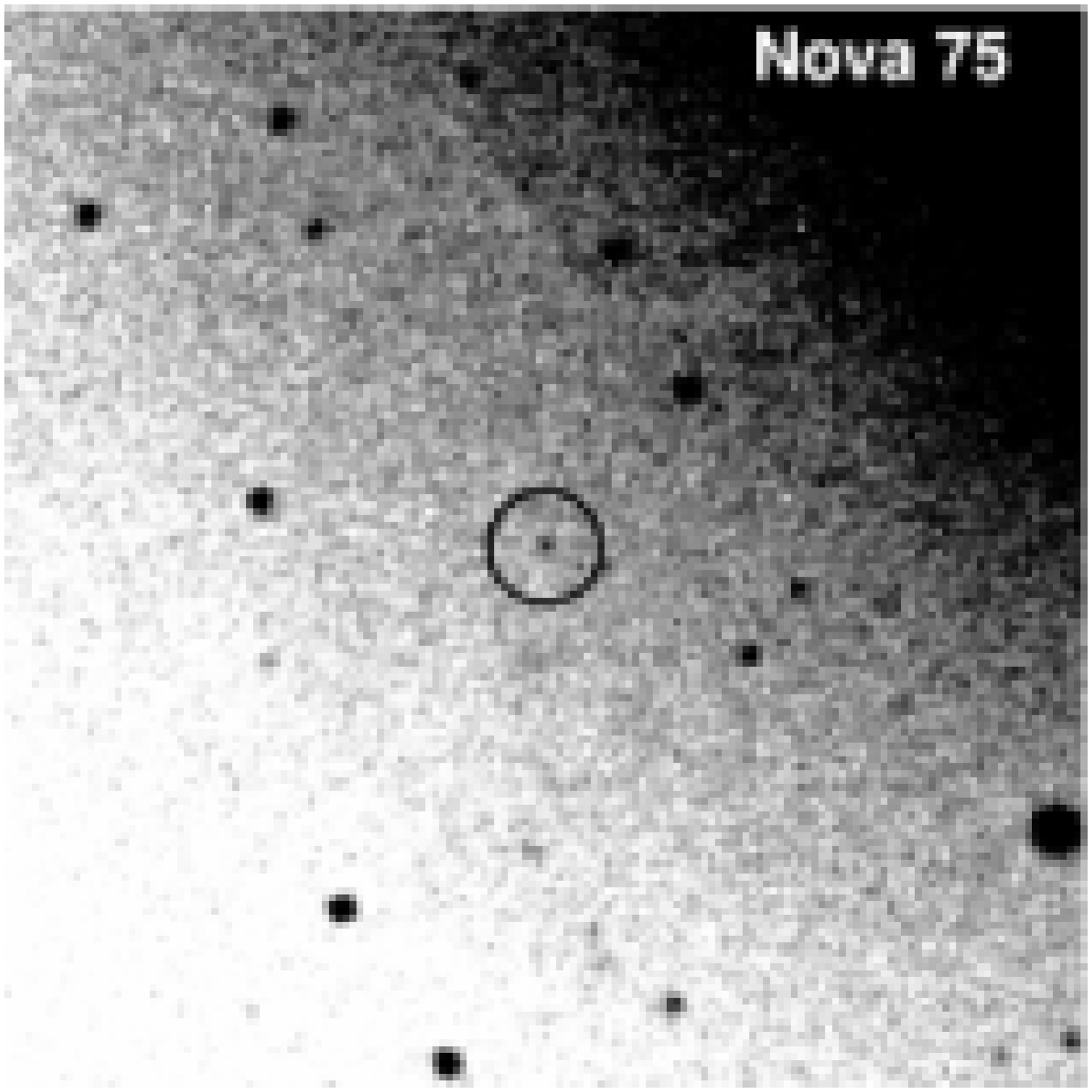}}\\
\subfigure[Nova 76 (B)]{\includegraphics[scale=.26, angle=0]{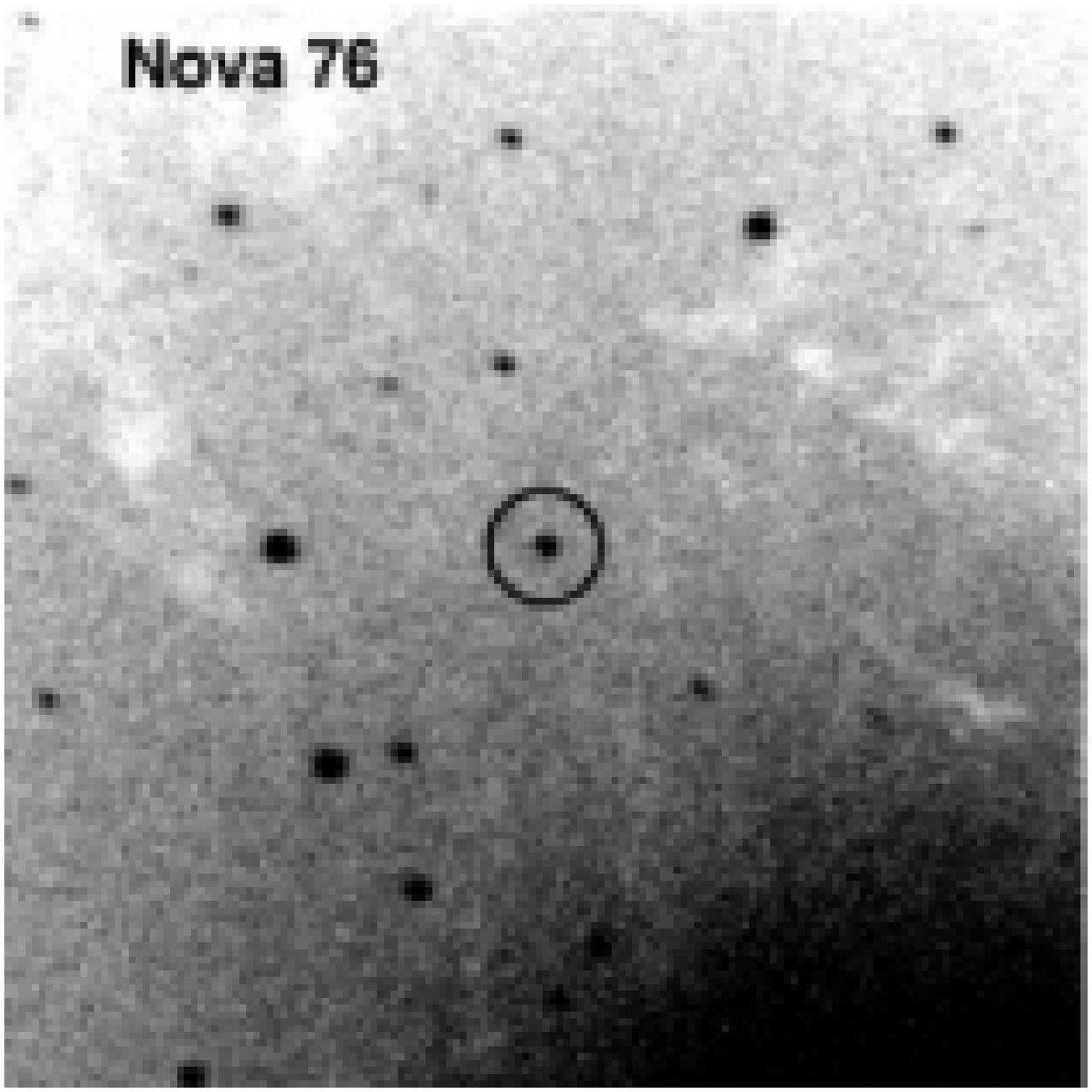}}\qquad
\subfigure[Nova 77 (U)]{\includegraphics[scale=.26, angle=0]{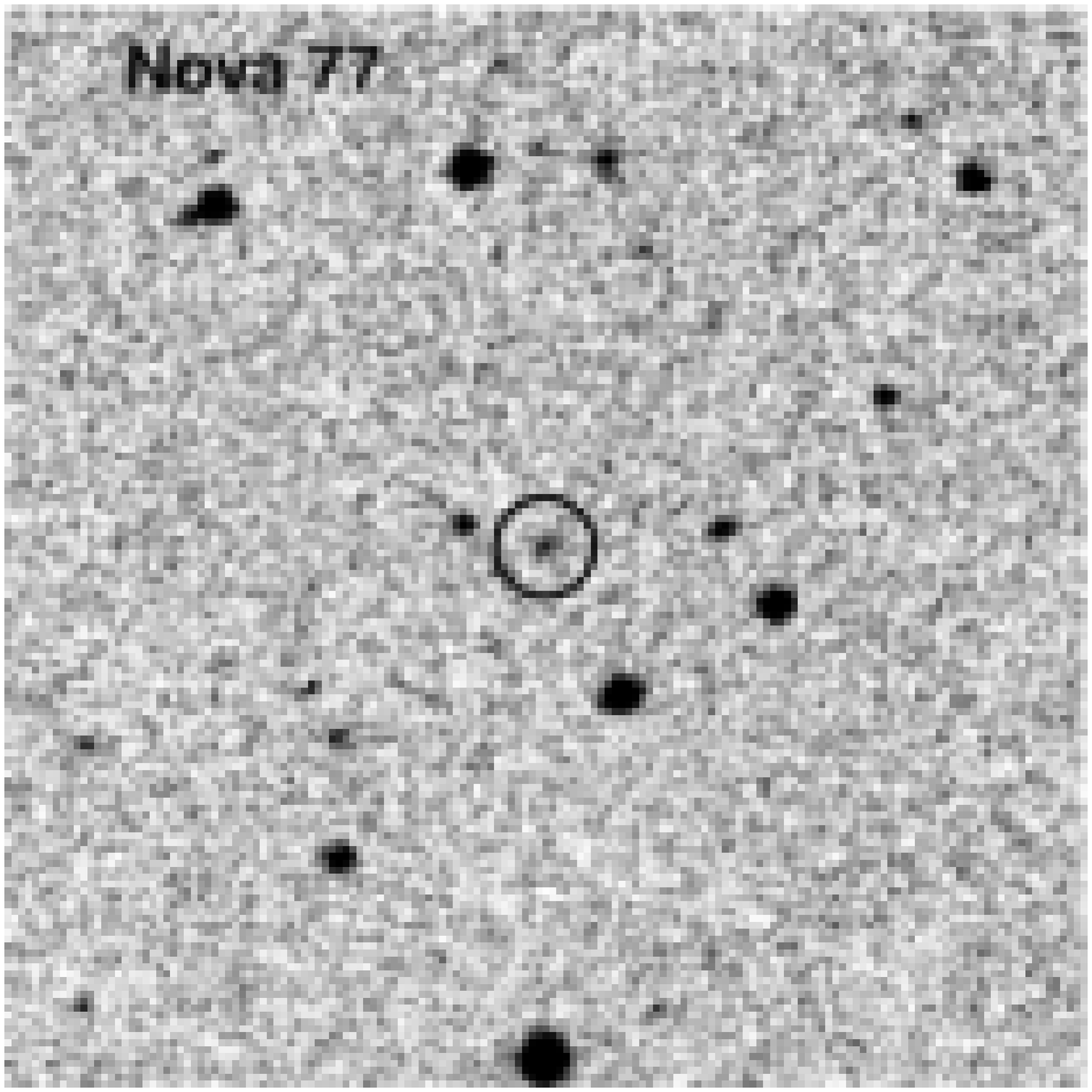}}\qquad
\subfigure[Nova 78 (U)]{\includegraphics[scale=.26, angle=0]{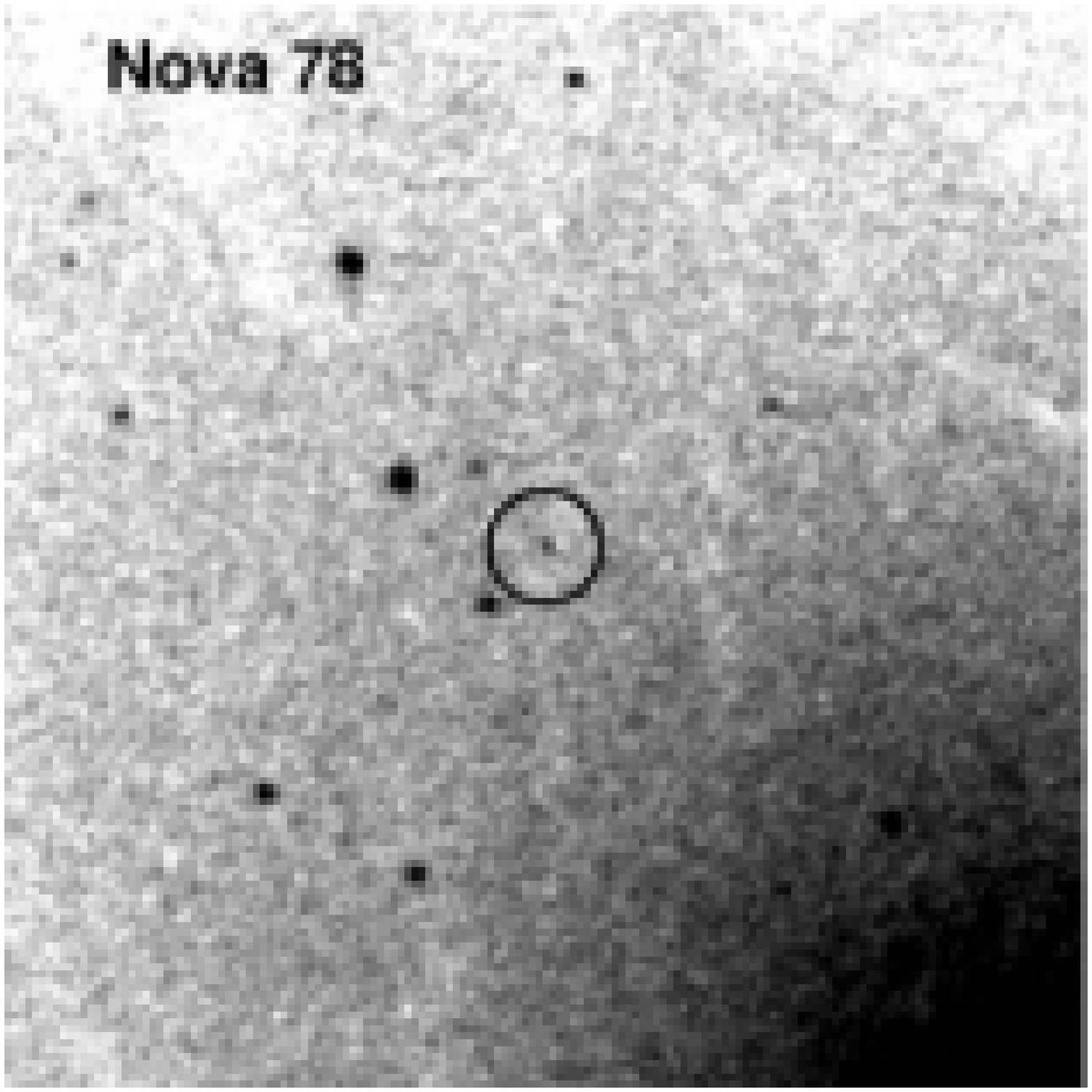}}\\
\subfigure[Nova 79 (U)]{\includegraphics[scale=.26, angle=0]{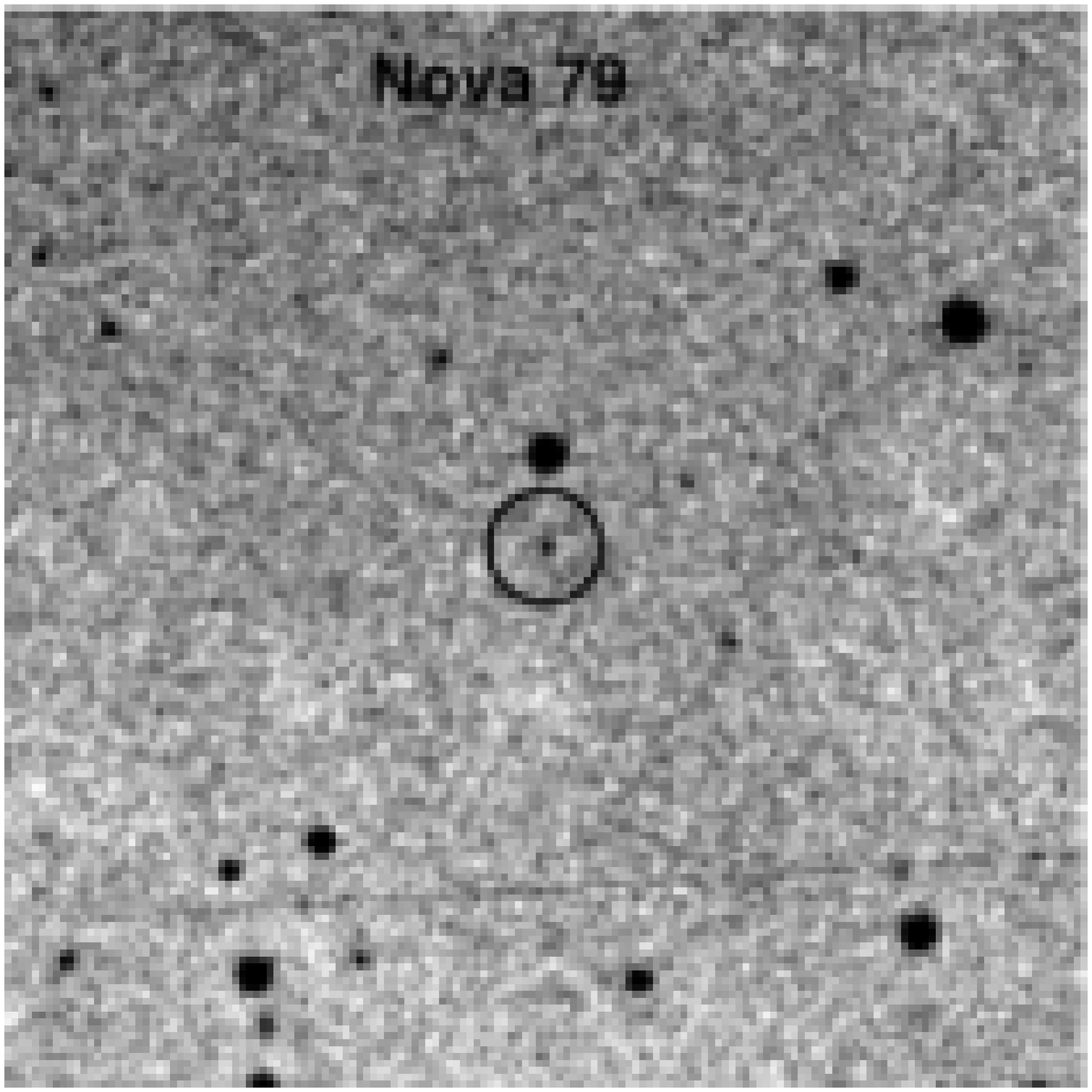}}\qquad
\subfigure[Nova 80 (U)]{\includegraphics[scale=.26, angle=0]{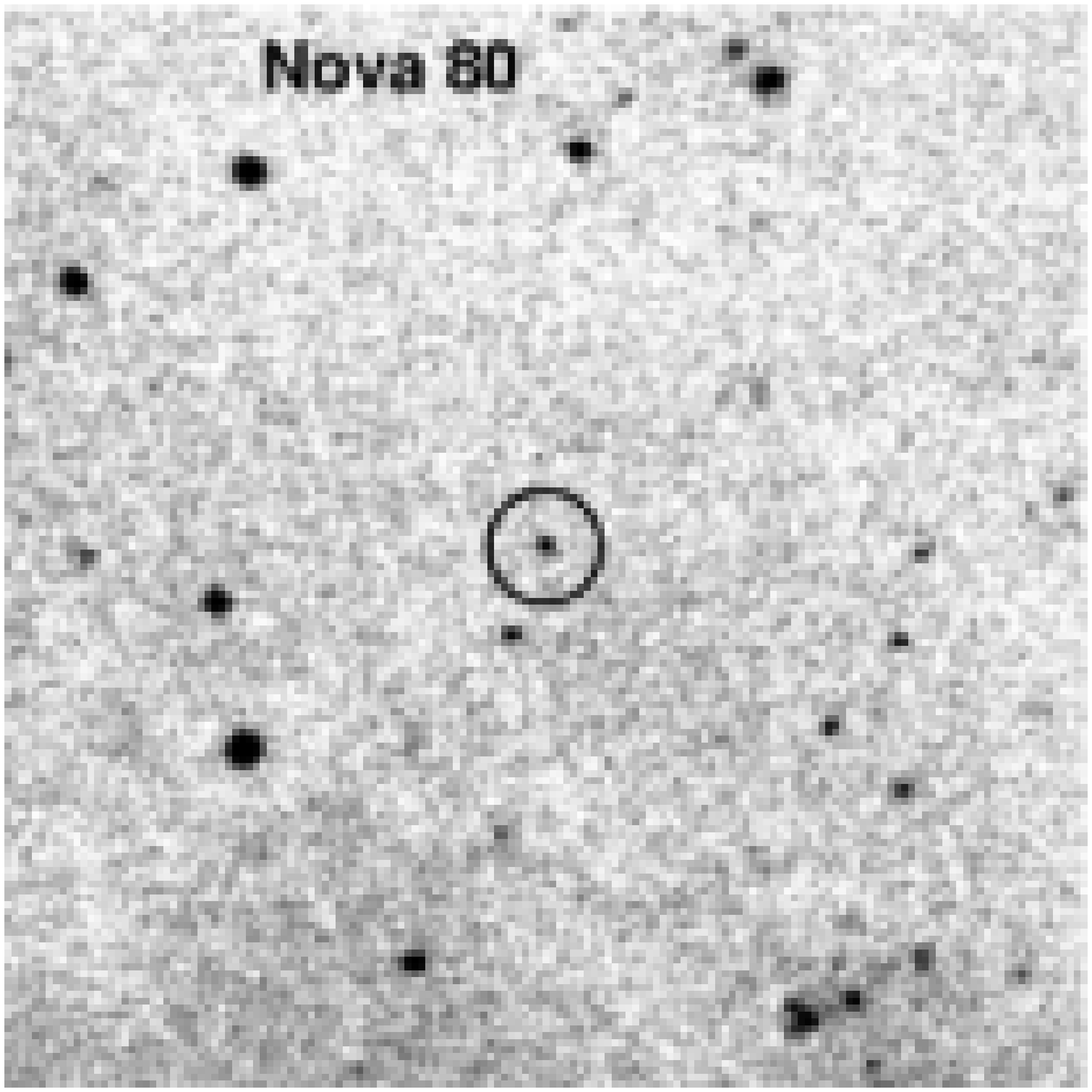}}\qquad
\subfigure[Nova 81 (V)]{\includegraphics[scale=.26, angle=0]{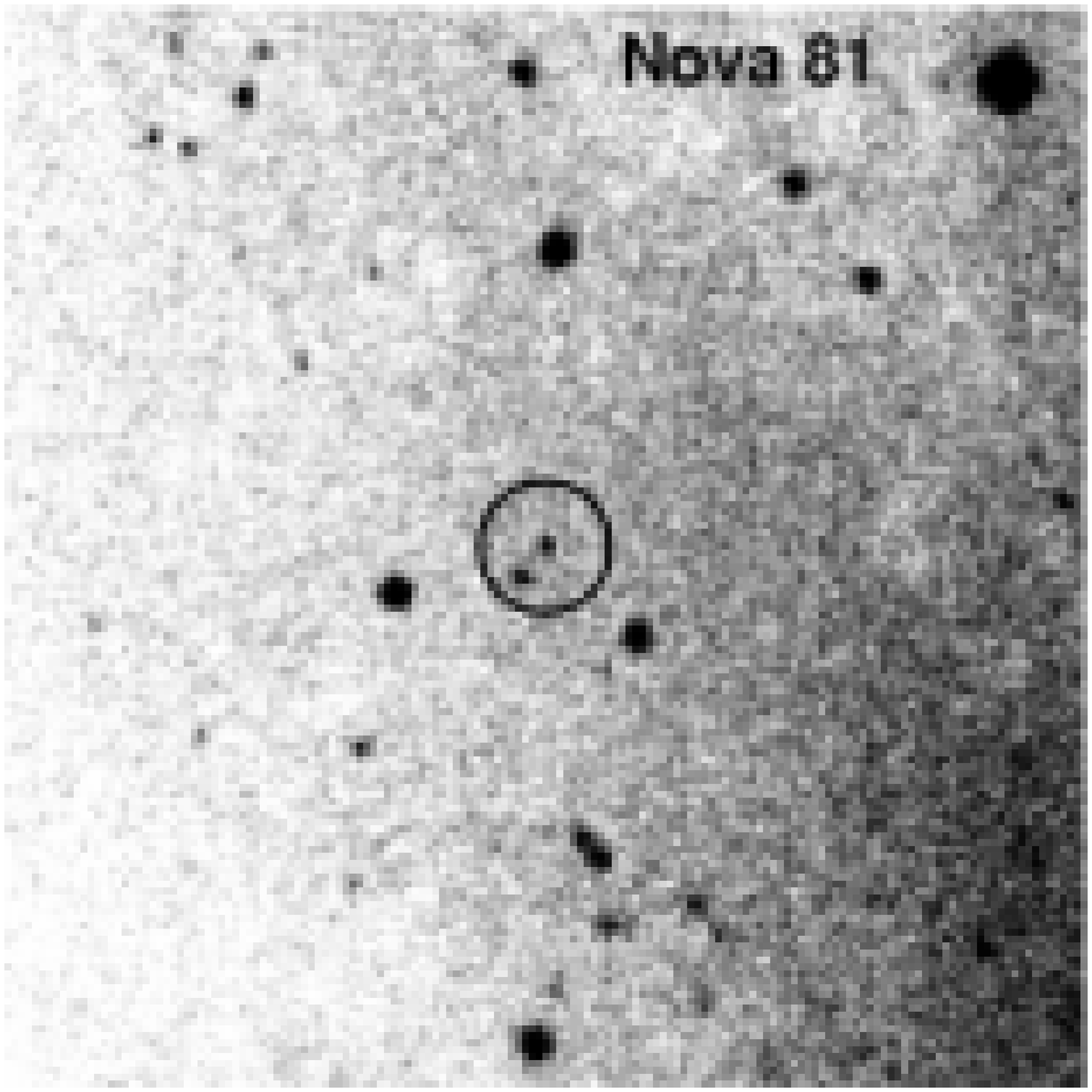}}\\
\subfigure[Nova 82 (U)]{\includegraphics[scale=.26, angle=0]{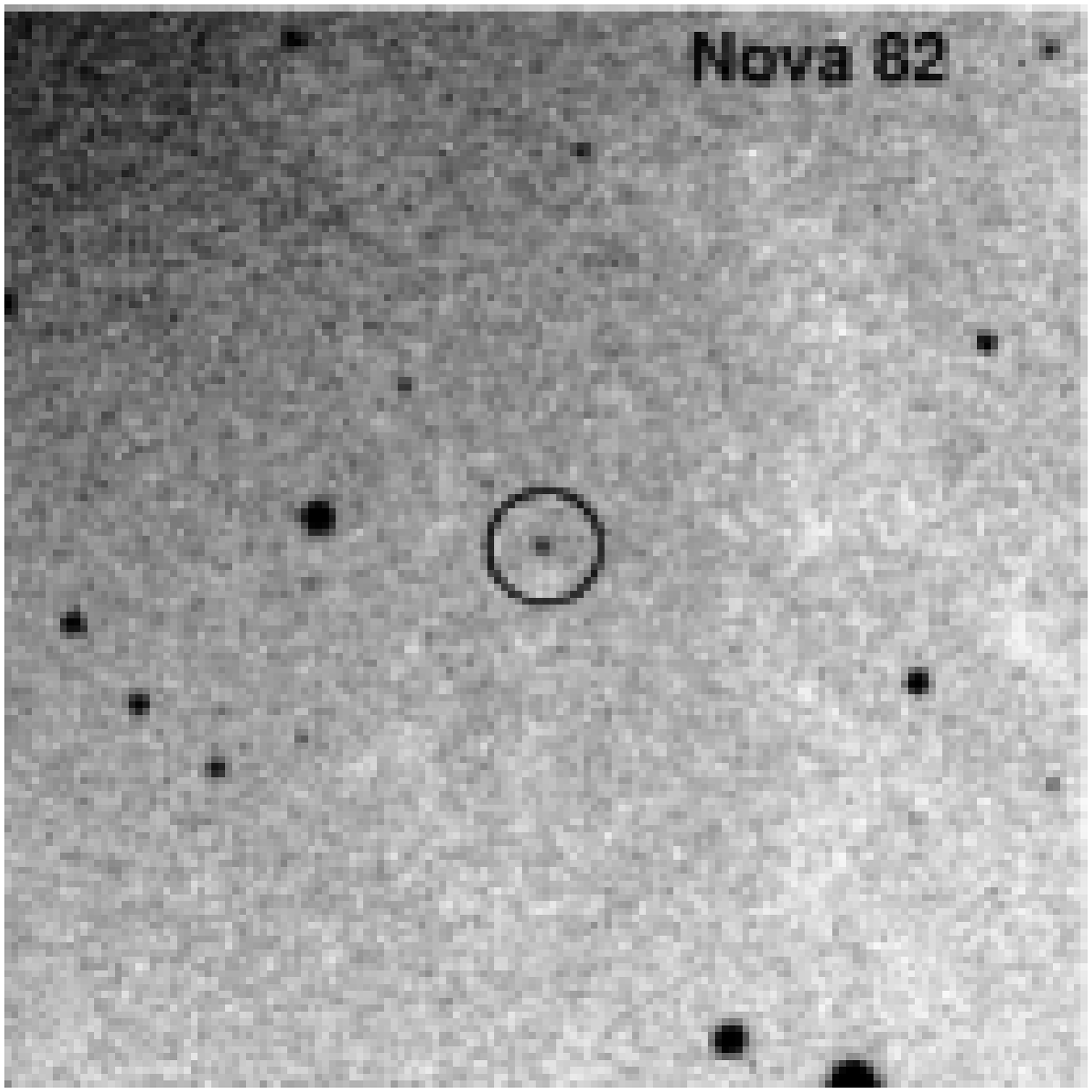}}\qquad
\subfigure[Nova 83 (B)]{\includegraphics[scale=.26, angle=0]{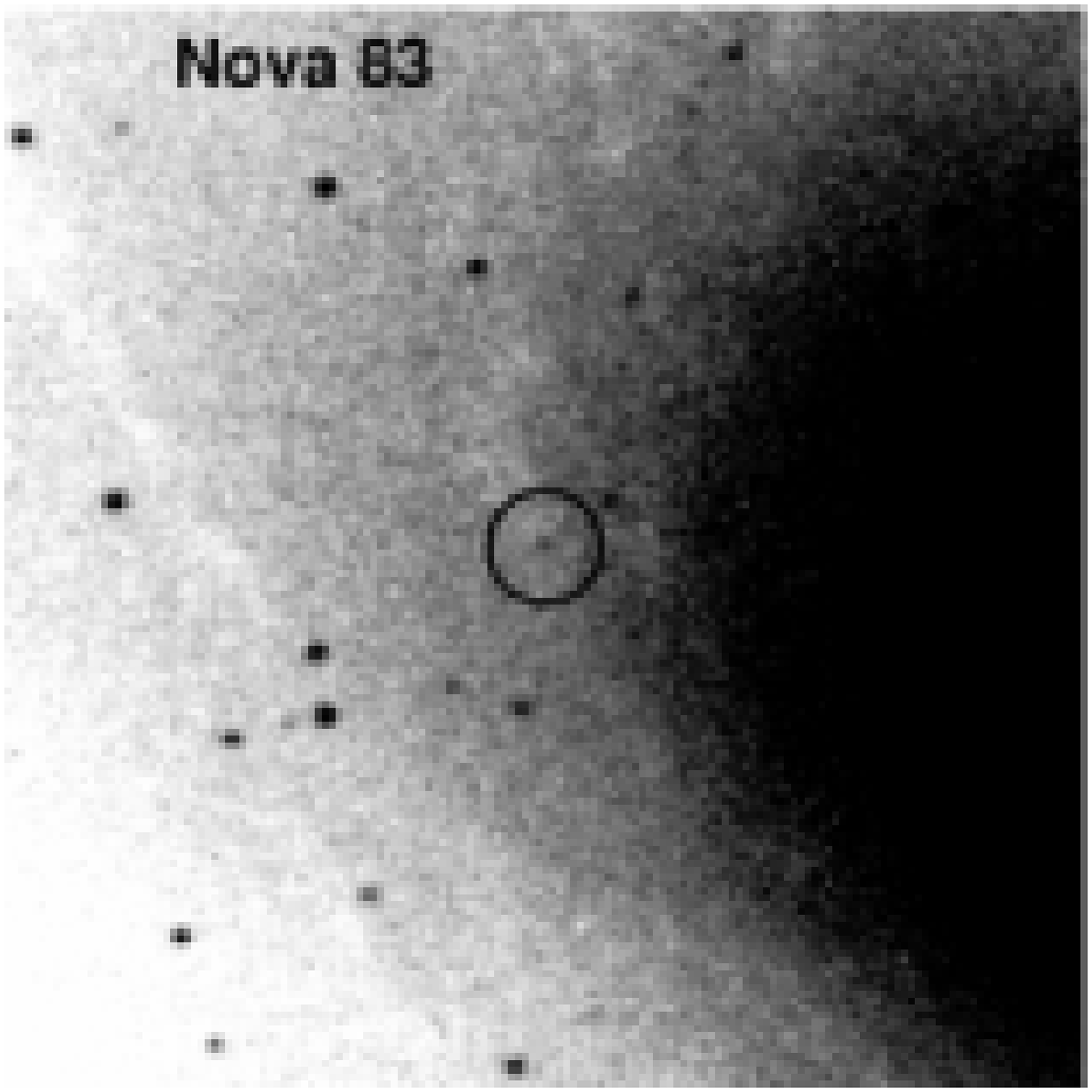}}\qquad
\subfigure[Nova 84 (B)\label{fig:chart84}]{\includegraphics[scale=.26, angle=0]{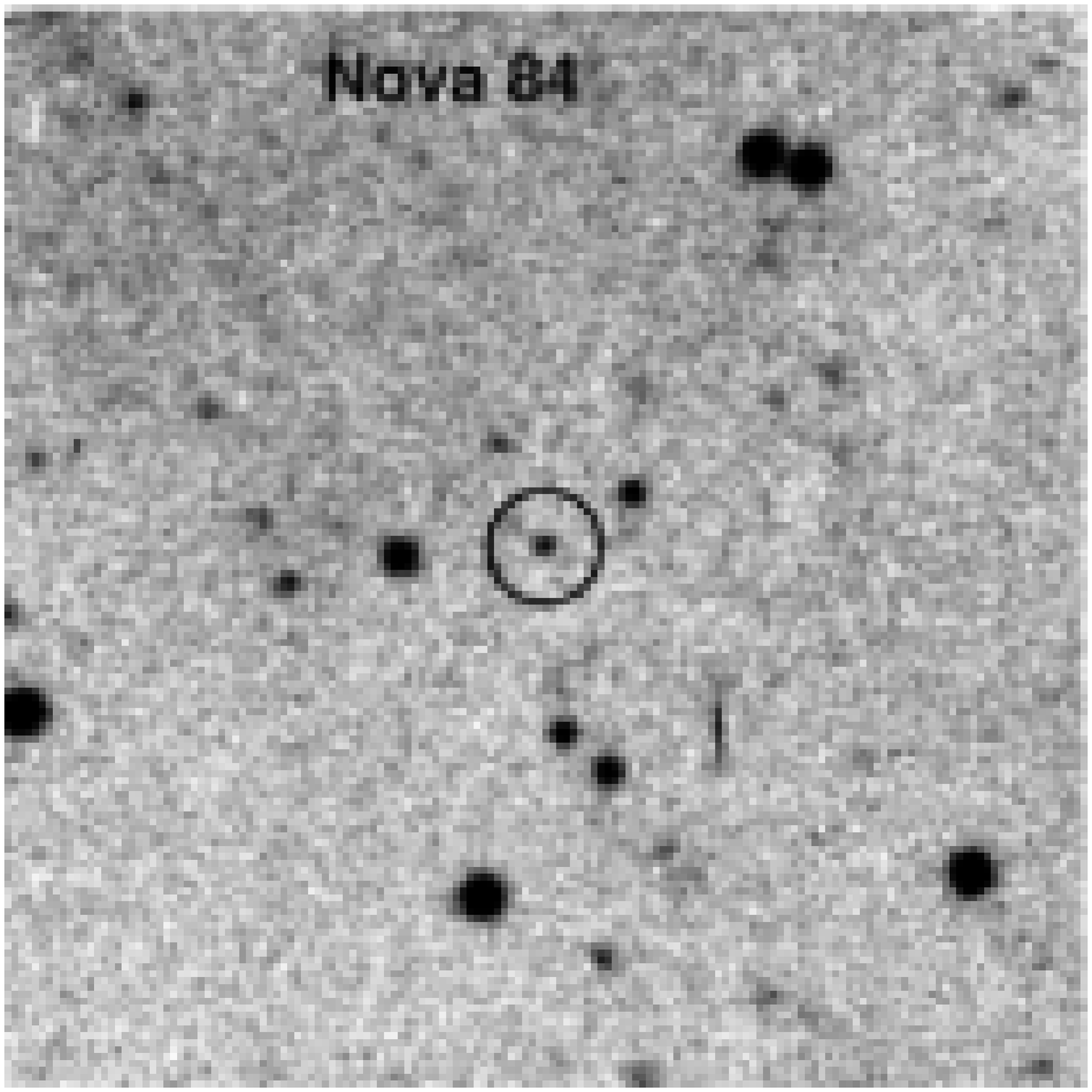}}
\caption{Finding charts for novae 73 - 84.}
\end{figure*}
}


\begin{thebibliography}{999}

\bibitem[2006]{Alksnis06} Alksnis, A. 2006, private communication
\bibitem[1994]{Andruk94} Andruk, V., Kharchenko, N., Schilbach, E. 1994, 
        Astron. Nachr. 315, 431
\bibitem[2004]{Ansari04} Ansari, R., Auriere, M., Baillon, P., et al. 2004
        A\&A, 421, 509
\bibitem[1956]{Arp56} Arp, H. 1956, AJ, 61, 15	
\bibitem[1964]{Baade64} Baade, W., Arp, H. 1964, ApJ, 139, 1027
\bibitem[2007]{Beaton07} Beaton, R. L., Majewski, S. R., Guhathakurta, P., et al. 2007, ApJ, 658, 91
\bibitem[1996]{Bertin96} Bertin, E., Arnouts S. 1996, A\&ASS, 117, 393
\bibitem[1968]{Boerngen68} B\"orngen, F. 1968, Mitteilungen des 
        Karl-Schwarzschild-Observatoriums Tautenburg Nr. 40
\bibitem[2001]{Brunzendorf01} Brunzendorf, J. 2001, PhD Thesis, 
        Friedrich-Schiller-Universit\"at Jena
\bibitem[1999]{Brunz99} Brunzendorf, J., Meusinger, H. 1999, A\&AS, 
        139, 141
\bibitem[2001]{Brunz01} Brunzendorf, J., Meusinger, H. 2001, A\&A, 373, 38
\bibitem[1987]{Ciardullo87} Ciardullo, R., Ford, H.C., Neill, J.D., et al.
        1987, ApJ, 318, 520 
\bibitem[2004]{Darnley04} Darnley, M.J., Bode, M.F., Kerins, A.M., et al.
        2004, MNRAS, 353, 51
\bibitem[2006]{Fliri06} Fliri, J., Riffeser, A., Seitz, S., et al. 2006
        A\&A, 445, 423	 
\bibitem[2006]{Galleti06} Galleti, S., Federici, L., Bellazzini, M., Buzzoni, A. \& Fusi Pecci, F. 2006, A\&A, 456, 985
\bibitem[1990]{Greiner90} Greiner, J., Wenzel, W., Degel, W. 1990, A\&A, 
        234, 251
\bibitem[1997]{Hatano97} Hatano, K., Branch, D., Fisher, A. 1997, 
        ApJ Lett., 487, L45
\bibitem[2006]{Henze06} Henze, M., Meusinger, H., Pietsch, W. 2006, IBVS, 5739, 1
\bibitem[2005]{Hernanz05} Hernanz, M. 2005, in The Astrophysics of Cataclysmic Variables and Related Objects, ed. J.-M. Hameury \& J.-P. Lasota, ASP Conf. Ser., 330, 265
\bibitem[1929]{Hubble29} Hubble, E. 1929, ApJ, 69, 103
\bibitem[2006]{Massey06} Massey, P., Olsen, K. A. G., Hodge, P. W., et al. 
        2006, AJ, 131, 2478
\bibitem[1975]{Meinunger75} Meinunger, L. 1975, MitVS, 5, 177
\bibitem[1967]{Moffat67} Moffat, A. F. J. 1967, AJ, 72, 10
\bibitem[2003]{Monet03} Monet D. G., Levine S.E., Casian B., et al. 2003, 
        AJ, 125, 984
\bibitem[2000]{Ochsenbein00} Ochsenbein F., Bauer P., Marcout J. 2000, A\&AS 
        143, 221
\bibitem[1957]{Payne-Gaposchkin57} Payne-Gaposchkin, C. 1957, The Galactic Novae, North-Holland, Amsterdam
\bibitem[2005]{Pietsch05} Pietsch, W., Fliri, J., Freyberg, M. J., et al. 2005, A\&A, 442, 879
\bibitem[2007]{Pietsch07} Pietsch, W., Haberl, F., Sala, G., et al. 2007, 
        A\&A, 465, 375-392 [PHS07]
\bibitem[2007]{Quimby07}Quimby, R., Mondol P., Wheeler, J. C. , et al. 2007, The Astronomer's Telegram, 1118, 1
\bibitem[1999]{Rector99}Rector, T. A., Jacoby G. H., Corbett D. L., Denham M., RBSE Nova Search Team 1999, Bull. Am. Astron. Soc., 31, 1420
\bibitem[2007]{Rector07}Rector, T. A. 2007, private communication
\bibitem[1964]{Rosino64} Rosino, L. 1964, Ann. d' Astrophys., 27, 498
\bibitem[1973]{Rosino73} Rosino, L. 1973, A\&AS, 9, 347
\bibitem[1989]{Rosino89} Rosino, L., Cappacioli, M., D`Onofrio, M., 
        Della Valle, M. 1989, AJ, 97, 83
\bibitem[1997]{Scholz97} Scholz, R., Meusinger, H., Irwin, M. 1997, 
        A\&A, 312, 833 
\bibitem[2001]{Shafter01} Shafter, A. W., Irby, B. K. 2001, ApJ, 563, 749 
\bibitem[1991]{Sharov91} Sharov, A. S., Alksnis, A. 1991, Ap\&SS, 180, 273
\bibitem[1992]{Sharov92} Sharov, A. S., Alksnis, A. 1992, Ap\&SS, 190, 119
\bibitem[1997]{Sharov97} Sharov, A. S., Alksnis, A. 1997, AstL, 23, 540
\bibitem[1964]{vandenBergh64} van den Bergh, S. 1964, AJ, 69, 610
\bibitem[1987]{Walterbos87} Walterbos, R. A. M., Kennicutt, R.C. 1987, A\&AS, 69, 311
\bibitem[1995]{Warner95} Warner, B. 1995, Cataclysmic Variable Stars, 
        Cambridge University Press
\bibitem[1997]{Yungelson97} Yungelson, L., Livio, M., Tutukov, A. 1997,
        ApJ, 481, 17	

\end{thebibliography}
\end{document}